\documentclass[english, oneside]{sapthesis}
\usepackage[utf8]{inputenx}
\usepackage{indentfirst}
\usepackage{microtype}
\usepackage{lettrine}
\usepackage{slashed}
\usepackage[nottoc, notlof, notlot]{tocbibind}
\usepackage{hyperref}
\hypersetup{
			hyperfootnotes=true,			
			bookmarks=true,			
			colorlinks=true,
			linkcolor=red,
                        linktoc=page,
			anchorcolor=black,
			citecolor=red,
			urlcolor=blue,
			pdftitle={A sample Bachelor's thesis for Sapienza Università di Roma},
			pdfauthor={FirstName LastName},
			pdfkeywords={thesis, sapienza, roma, university}
 }

\title{Fragmentation measurements with the FOOT\\ experiment}
\author{Luana Testa}
\IDnumber{1913445}
\course{Fisica }
\courseorganizer{Facolt\`a di Scienze Matematiche Fisiche e Naturali}
\submitdate{2023/2024}
\copyyear{2025}
\advisor{Prof. Giacomo Traini}
\coadvisor{Prof. Marco Toppi}
\authoremail{testa.1913445@studenti.uniroma1.it}
\examdate{27/01/2025}
\examiner{Prof. Andrea Pelissetto} 
\examiner{Prof. Maria Grazia Betti} \examiner{Prof. Andrea Crisanti} \examiner{Prof. Giulio D’Agostini}  \examiner{Prof. Cecilia Voena}  \examiner{Prof. Marco Merafina} \examiner{Prof. Francesco Macheda}  

\begin{document}

\frontmatter
\maketitle
\dedication{A mamma, papà,\\ Luca e Leonardo}
\begin{abstract}
Particle Therapy (PT) has emerged as a powerful tool in cancer treatment, leveraging the unique dose distribution of charged particles to deliver high radiation levels to the tumor while minimizing damage to surrounding healthy tissue. Despite its advantages, further improvements in Treatment Planning Systems (TPS) are needed to address uncertainties related to fragmentation process, which can affect both dose deposition and effectiveness. These fragmentation effects also play a critical role in Radiation Protection in Space, where astronauts are exposed to high level of radiation, necessitating precise models for shielding optimization.

The FOOT (FragmentatiOn Of Target) experiment addresses these challenges by measuring fragmentation cross-section with high precision, providing essential data for improving TPS for PT and space radiation protection strategies.

This thesis  contributes to the FOOT experiment in two key areas. First, it focuses on the performances of the vertex detector, which is responsible  for reconstructing particle tracks and fragmentation vertexes with high spatial resolution. The study evaluates the detector's reconstruction algorithm and its efficiency to detect particles.  Second the thesis present a preliminary calculation of fragmentation cross section, incorporating the vertex detector for the first time in these measurements.  

\end{abstract}

\tableofcontents

\mainmatter
\chapter*{Introduction}
\addcontentsline{toc}{chapter}{Introduction}
Radiotherapy (RT) plays a central role in cancer treatment by using external beam radiation to damage the DNA of cancer cells, ultimately leading to their destruction. Since Robert Wilson's 1946 proposal to utilize charged particles, such as protons or ions, for tumor treatment,  Particle Therapy (PT) has gained traction and is increasingly being recognized for its effectiveness, particularly in the treatment of deep-seated solid tumors. The primary advantage of PT lies in the dose distribution profile of charged particles, characterized by minimal dose deposition in the entrance channel and maximum dose deposition at a specific depth, known as the Bragg Peak. This contrasts with the exponential attenuation observed in photon therapy. PT enables the precise targeting of tumors by centering the Bragg Peak within the tumor, thereby delivering a high dose directly to the malignant cells while minimizing exposure to surrounding healthy tissue and critical organs. Additionally, ions are advantageous for treating radioresistant tumors due to their enhanced biological effectiveness compared to other forms of RT.

Although PT has become a well-established approach in clinical settings, there remains a need for further improvement in the accuracy of Treatment Planning Systems, particularly by incorporating advanced Monte Carlo (MC) models into dose calculations. Despite the above mentioned advantages, the use of PT remains limited compared to photon RT, primarily due to ongoing concerns regarding the contribution of nuclear fragmentation processes in this therapy.

Nuclear interactions between the PT beam and patient tissue can result in the fragmentation of both projectiles and target nuclei. In proton treatments, the creation of short-range target fragments can lead to a non-negligible dose deposition, especially in the entrance channel. In ion therapy, there is also projectile fragmentation that can result in the production of long-range fragments that may release dose even beyond the Bragg Peak.

The study of nuclear fragmentation is also of significant interest in the field of Radiation Protection in Space (RPS), where high-energy charged particles represent the primary source of radiation absorbed by astronauts. In order to enable long-duration space missions, such as a journey to Mars, it is crucial to optimize the shielding of spacecraft. This is particularly important because such missions expose both the spacecraft and its occupants to prolonged periods of elevated radiation, requiring advanced protection strategies to ensure crew safety.

To optimize both PT and RPS, it is crucial to accurately understand nuclear fragmentation processes to improve treatment plans and spacecraft shielding. However, the current lack of complete fragmentation cross-sections for many nuclear processes, combined with the limited precision of existing data due to a scarcity of experimental results, poses a significant challenge in achieving the desired level of accuracy.

The \textbf{FOOT} (FragmentatiOn Of Target) experiment aims to address this issue by measuring the double differential cross-sections of nuclear fragmentation reactions with respect to their emission angle and kinetic energy, with a maximum uncertainty of $5\%$. FOOT is a fixed-target experiment designed to track and measure the kinematic properties of all charged particles, including both primary particles and fragments. The experimental program plans to use ion beams such as $^4He$,  $^{16}O$, $^{12} C$ in the energy range of $100-800 MeV/u$, interacting with targets made of carbon or polyethylene ($C_2H_4$), to extract cross-sections for the primary elements of the human body, namely hydrogen, carbon, and oxygen.

This thesis focuses on 2 main aspects:
\begin{itemize}
    \item The study of the vertex detector, which is positioned downstream of the target and is capable of reconstructing the tracks of both primary particles and fragments exiting from the target as they pass through its four layers. In particular, the thesis examines the reconstruction alghorithm and efficiency of this detector.
    \item The calculation of a preliminary cross-section, where the vertex detector is employed for the first time to contribute to this measurements.
\end{itemize}
Chapter \ref{chap:1} will introduce the fundamental principles of radiation interaction with matter for charged particles, that will provide a better understanding of the concepts underlying PT and RPS.  Chapter \ref{chap:2} will provide a detailed overview of the experimental setup used in the FOOT experiment. The various detectors and components will be described, with their individual roles and specific configurations, with attention on the resolution they need to achieve to reach the FOOT goals. Also a description of the trigger and data acquisition system. 
Chapter \ref{chapter 3} will be focused on the study of the vertex detector, analyzing its reconstruction algorithm. The chapter will examine the algorithm's ability to accurately reconstruct the fragmentation vertexes.  Chapter \ref{chapter 4} will delve into the performance evaluation of the detector in detecting particles. This chapter will include an optimization of the threshold for each detector layer, aiming to maximize detection efficiency while minimizing noise. And at the end, chapter \ref{chapter 5} will focus on the calculation of the cross section.

\chapter{Charged particle interaction with matter }
\label{chap:1} 
According to IARC statistics from 2022 \cite{ref:IARC}, approximately 20 million new cancer cases and 9.7 million cancer-related deaths occurred globally. Among the various cancer treatment modalities, alongside chemical and surgical approaches, radiotherapy (RT) plays a crucial role. This technique utilizes ionizing radiation beams to damage cancer cells, thereby halting or inhibiting their uncontrolled proliferation, either through direct or indirect interactions with their DNA. One of the most promising forms of RT is Particle Therapy (PT), which employs protons or ions as therapeutic beams. PT is particularly effective for treating deep-seated tumors located near vital organs. Its advantage lies in the unique dose distribution, characterized by a finite range and a pronounced dose peak at a specific depth, along with a lower dose in the entrance channel.

The study of charged particle interactions with matter is also of significant interest in the field of Radiation Protection in Space (RPS), as astronauts and spacecraft are continuously exposed to cosmic radiation. The primary sources of this space radiation are protons and helium ions generated during Solar Particle Events (SPEs) and from Galactic Cosmic Rays (GCR).

This chapter provides an overview of the physics underlying PT and RPS.

\section{Interaction with matter}
\label{sec:Interacton}
When a charged particle enters and is absorbed by an object, it interacts with the material in different ways. In the energy range of $100-800 MeV/u$ , which is relevant for PT and RPS, the primary interactions of charged particles include:
\begin{itemize}
    \item \textbf{Inelastic collisions with atomic electrons} which are the main source of energy loss and define the energy deposition profile and range of ions.
    \item \textbf{ Multiple Coulomb Scattering (MCS)} which refers to the elastic Coulomb scattering off nuclei in the material, determining the lateral spread of primary particles.
    \item \textbf{ Nuclear interactions} with the nuclei of the medium, with both elastic and inelastic collisions.
\end{itemize}

The first two processes are governed by electromagnetic forces, while nuclear interactions are mediated by the strong nuclear force.

\subsection{Inelastic collision with $e^-$}
\label{subsec:inelastic coll.}
Among the processes governed by electromagnetic forces, inelastic collisions are primarily responsible for the energy loss of charged particles as they traverse matter. When interacting with the electromagnetic fields of the electrons within the material, charged particles transfer a portion of their kinetic energy to these electrons, which can lead to ionization or excitation. Ionization, the principale outcome, occurs when an electron acquires sufficient energy to escape from its atomic orbital, resulting in the formation of an ion pair comprising the ejected electron and the remaining positively charged atom. Conversely, if the transferred energy is less than the electron’s binding energy, the electron may be promoted to a higher energy level. Eventually, the excitation energy is released, either as electromagnetic radiation or as Auger electrons.

The amount of energy transferred in each collision is generally a small fraction of the particle’s total kinetic energy. However, due to the high frequency of collisions per unit path length, the cumulative energy loss becomes significant. Since both the energy transferred and the number of inelastic collisions have a statistical nature, it is customary to consider the average energy loss per unit path length, denoted as $\left < \frac{dE}{dx} \right >$, also known as the stopping power.

Initially estimated by Bohr using classical methods, the correct quantum-mechanical description of this energy loss was later provided by Bethe and Bloch. The mean energy lost by a particle with charge $z$ through electromagnetic collisions within a homogeneous material of density $\rho$ is given by the following Bethe-Bloch formula.\cite{ref:passage}

\begin{equation}\label{eq:bethe-bloch}
    \left < -\frac{dE}{dx} \right>=4\pi r_e^2 N_A m_e c^2 z^2 \frac{Z \rho}{A} \frac{1}{\beta^2}\left[\frac{1}{2} ln \frac{2m_ec^2 \beta^2 \gamma^2 W_{max}}{I^2}-\beta^2-\frac{\delta(\beta \gamma)}{2}-\frac{C}{Z}\right]
\end{equation}

where:
\begin{itemize}
\item $m_e c^2$ is the electron mass $\times c^2$  ($m_e c^2 =0.511 Mev$) \cite{ref:PDG}
\item $r_e$ is the electron radius ($r_e=2.817 fm$)\cite{ref:PDG}
\item $N_A$ is the Avogadro's number ($N_A=6.022$x$ 10^{23}$)
\item $Z$ is the target atomic number
\item  $A$ is the target mass number
\item $\rho$ is the target density
\item $z$ is the incident particle charge
\item $\beta$ and $\gamma$ are the Lorentz factors for the incident particle 
($\beta=\frac{v}{c}$ , $\gamma=\frac{1}{\sqrt{1-\beta^2}}$)
\item $W_{max}$ is the maximum energy that the incident particle can transfer to an electron of the material 
\begin{equation}
    W_{max}=\frac{2m_e c^2 \beta^2 \gamma^2}{1+2 \gamma m_e/M+ (m_e/M)^2}
\end{equation}
where M is the mass of the incident particle. In the “low-energy” approximation ($2\gamma m_e \ll M$) $W_{max} = 2m_ec^2\beta^2 \gamma^2$ 
\item $I$  the mean excitation potential of the absorber (target)
\item $\delta(\beta \gamma)$ is density effect correction to ionization energy loss, which becomes significant for high energies of the incident particle. This effect is a consequence of the fact that the electric field of the particle tends to polarize the atoms along its path. So electrons far from the
path of the particle will be shielded from the full electric field intensity
\item $C$ is the shell correction, which becomes important when the velocity of the incident particle is low enough to be comparable with the one of orbital electrons, in this case the assumption that the electron is stationary with respect to the incident particle is no longer valid
\end{itemize}

The equation \ref{eq:bethe-bloch}, expressed in units of $MeV \textit{ }g^{-1} cm^2$, is accurate to within a few percent over the entire range $ 0.1 \le \beta \gamma \le 1000$, therefore this equation reliably describes the energy loss processes relevant for PT and RPS. In practice, the linear stopping power, $\rho \times \left< \frac{dE}{dx}\right >$, is also frequently employed, where $\rho$ represents the target density in $g/cm^3$.

At low energies (below 10 MeV/u), when the velocity of the charged particle is comparable to the orbital velocity of the target electrons, it is essential to account for corrections arising from electron capture, which becomes increasingly significant. To accurately describe the ion's behavior during its passage through matter the charge $z$ is replaced by an effective charge $z_{eff}$, which accounts for the reduction in the projectile's charge due to ionization and recombination processes along its path. There exist several higher-order corrections in $z$, like Barkas, Bloch, and Mott corrections. The Barkas equation for example is the following \cite{ref:nuclear phy}:

\begin{equation}
    z_{eff}=z\left [ 1- e^{-125 \beta z ^{-2/3}}\right ]
\end{equation}

\begin{figure}[h!]
\begin{center}
  \includegraphics[width=9cm]{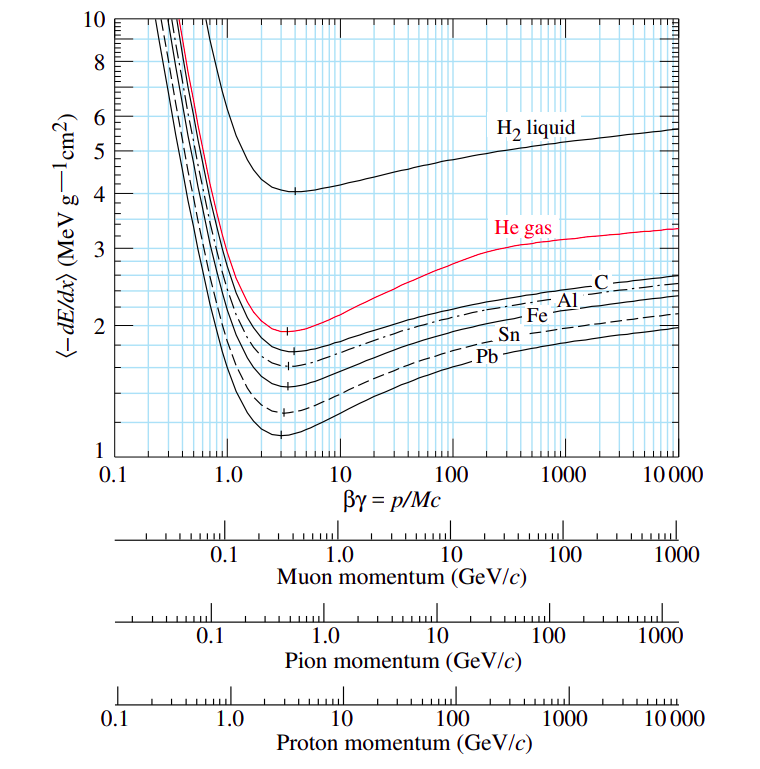}\\
  \caption{Stopping power $\left< \frac{dE}{dx}\right>$ described by Bethe-Bloch curves for different particles and different material.}
  \label{fig:bethe-bloch}
\end{center}
\end{figure}

When analyzing the dependence of stopping power on $\beta \gamma$ (as shown in Figure \ref{fig:bethe-bloch}), one observes an initial decreasing region, characterized by a behavior proportional to $1/\beta^\alpha$, where $\alpha$ ranges from 1.4 to 1.7, depending on the incident particle and on the charge of the target. The stopping power reaches a minimum at approximately  $\beta \gamma \sim 3$ after which it begins to increase due to the logarithmic term, leading to the so-called relativistic rise, up to a density effect plateau. \\

\subsection{Multiple Coulumb Scattering}
\label{subsec:MCS}
Another process governed by electromagnetic forces is Multiple Coulomb Scattering (MCS). In this case, the charged particle, due to Coulumb interactions, undergoes elastic scattering with the nuclei of the material through which it travels. The result of each collision is a slight deflection of the particle at a small angle. However over the path lengths in the material the particle experiences numerous such events. The primary consequence of MCS is the lateral spread of the particle beam.

\begin{figure}[h!]
\begin{center}
\includegraphics[width=9cm]{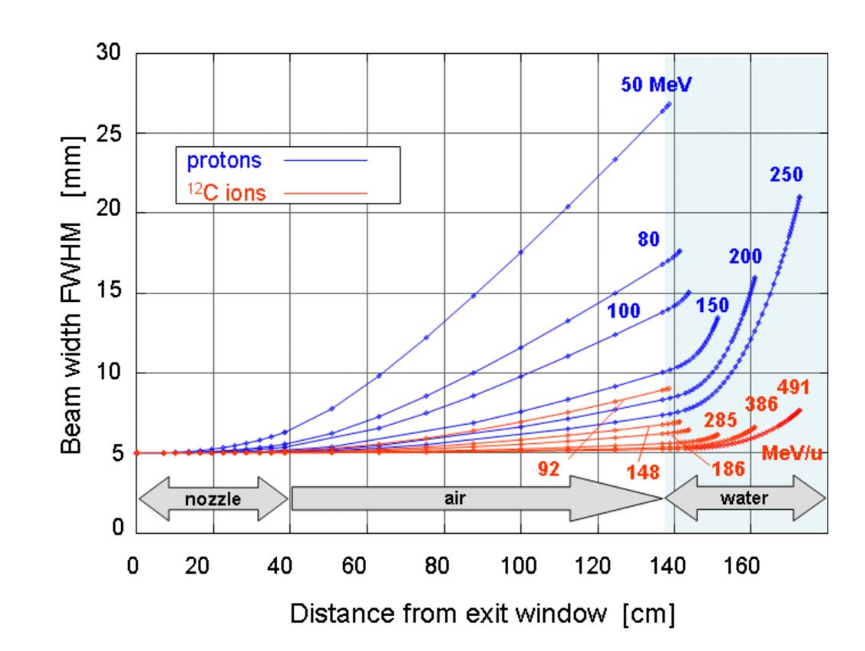}
\caption{Calculated beam spread for $^{12}C$ ions and protons in a typical treatment beam line. \cite{ref:heavy-ion}}
\label{fig:MCS}
\end{center}
\end{figure}

The MCS is well described by the theory of Molière. The angular distribution of the charged particle as a function of the penetration depth($d$) can be described by a Gaussian distribution\cite{ref:nuclear phy}:
\begin{equation}
    F(\theta, d)=\frac{1}{\sqrt{2 \pi \sigma_{\theta}}} e^{\frac{-\theta^2}{2 \sigma_{\theta}}}
\end{equation}
The standard deviation of the distribution, first obtained by Highland, is:
\begin{equation} \label{eq:MCS sigma}
    \sigma_{\theta}[rad]=\frac{14.1 MeV}{\beta c p} z \sqrt{\frac{d}{X_0}} \left [ 1 + \frac{1}{9} ln \left (\frac{d}{X_0} \right )\right]
\end{equation}
Where $p$, $\beta c$, and $z$ are the momentum, velocity, and charge number of the incident particle, $d$ is the thickness of the material and $X_0$ the radiation length\footnote{The radiation length of a material is the mean length (in cm) to reduce the energy of an electron by the factor $1/e$.} of the material.
In PT the lateral spreading of the beam caused by MCS can take place before the beam enters the patient and within the patient. To minimize the former effect, which is predominant at lower energies, all materials in front of the patient must be kept as thin as possible. At higher energies, the scattering within the patient becomes more significant. This phenomenon is illustrated in figure \ref{fig:MCS}, where the beam spreading for carbon ions and protons is shown. In the example, a particle beam with an initial full width at half maximum (FWHM) of 5 mm was used. At low energies, scattering is more pronounced in the pre-water zone, while at higher energies, the scattering predominantly occurs within the water.

\subsection{Nuclear interactions}\label{subsec:nuclear int}
Beyond electromagnetic interactions, when a charged particle travels through matter, it can undergo strong nuclear interactions. Although these interactions are less probable, they contribute significantly and must be considered in PT and RPS. Specifically, interactions with the nuclei of the material have minimal effects on energy loss but significantly impact the penetration of the particle.

Nuclear interactions can be classified into elastic and inelastic interactions:

\begin{itemize}
\item \textbf{Elastic Collisions}: These do not lead to a loss of kinetic energy but, similar to MCS, they contribute to the deviation of the particle. These interactions increase the lateral spread of the beams.
\item \textbf{Inelastic Collisions}: These lead to nuclear fragmentation (fragmentation of the target for protons beam and in the case of heavy ions beam also of the projectile). This results in the emission of lighter particles. Nuclear excitation can also occur, with the consequent emission of prompt $\gamma$ rays (approximately $0-10 MeV$) after the relaxation of the nuclei. 
\end{itemize}

\begin{figure}[h!]
\begin{center}
\includegraphics[width=10cm]{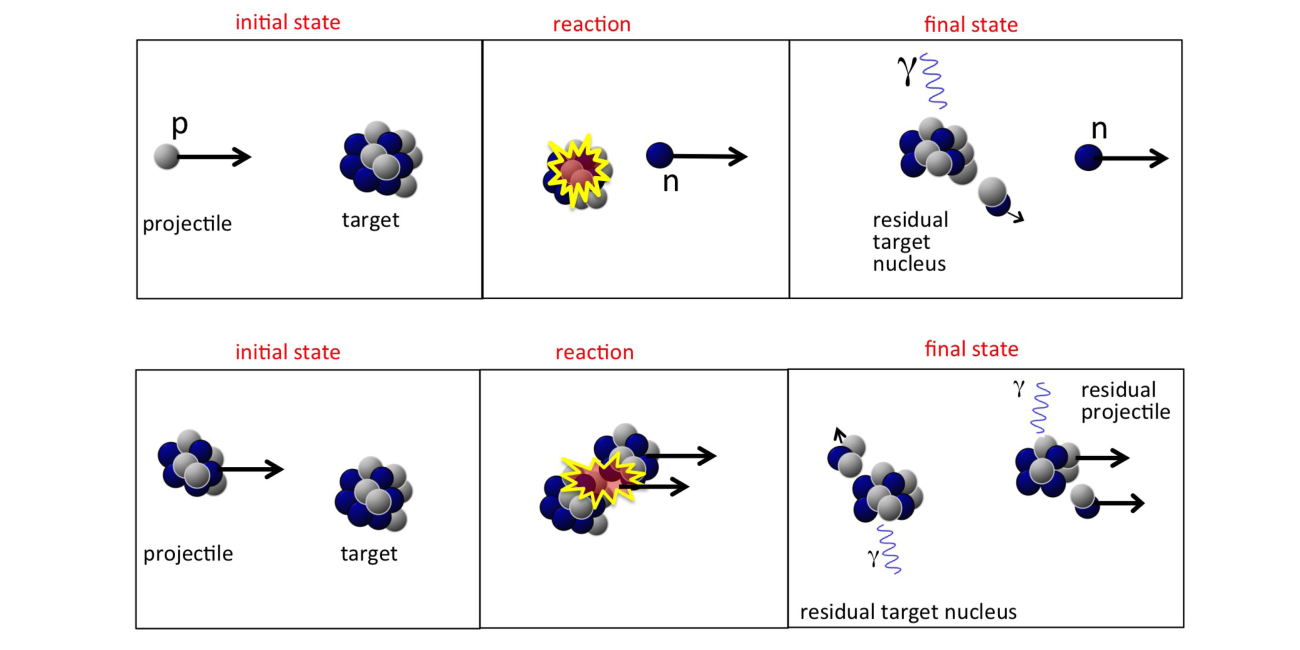}
\caption{The first sequence depicts a possible nucleon-nucleus reaction in proton therapy, resulting in neutron creation. The second sequence illustrates a nucleus-nucleus reaction in heavy ion therapy, leading to the creation of light fragments. \cite{ref: range v}}
\label{fig:nuclear_int}
\end{center}
\end{figure}

The study of nuclear inelastic collisions has been the main subject of many theoretical and experimental works. These interactions can be modeled as a multiple stage process. \textit{Intra-Nuclear Cascade} that occur over a timescale of approximately $10^{-23} - 10^{-22} s$, this stage includes all interactions between the projectile and the target nucleons. During this phase, high-energy emissions of protons, neutrons, and light nuclear fragments are possible. The fragments generated in this step have high energy and are predominantly emitted in the direction of the incident particle. \textit{Pre-Equilibrium Stage}, after the cascade, the energy of the interacting particles decreases to a lower limit, typically a few tens of $MeV$, but the nucleus has not yet reached thermal equilibrium. In this stage, nucleons interact with each other and redistribute the excitation energy, with the possibility of additional particle emission. And at the end the \textit{slow stage} with a characteristic timescale longer than $10^{-18}s$, this phase involves the de-excitation of the residual nuclear products. Depending on their mass and energy, the nuclei can emit low-energy fragments via nuclear evaporation or fission. These fragments are generally emitted in an isotropic manner and have significantly lower energy than those from earlier stages. The residual nuclei may also return to their ground state through gamma emission, typically within $10^{-16} s$. These processes are illustrated in figure \ref{fig:nuclear_int}, depicting interactions for both proton and nucleus projectiles. The primary difference between the two scenarios lies in the fact that, in the case of protons, only the target nucleus fragments, while for nucleus projectiles, fragmentation of both the projectile and target can occur. From the point of view of the model the substantial difference lies in the fact that in nucleus-nucleus reactions the incoming nucleons are not free.

Various models have been developed to describe all the stages, depending on the system's center-of-mass energy. Accurately modeling the entire process is highly complex, as it involves multi-body interactions within both electromagnetic and nuclear potentials. Typically, comprehensive modeling is achieved by integrating a model that describes the initial state, considering the probability of a nuclear event, with a model that captures the subsequent reaction phase.

\section{Radiobiology}
According to the definition provided by the World Health Organization (WHO), \textit{cancer is a large group of diseases that can originate in almost any organ or tissue of the body, characterized by the uncontrolled growth of abnormal cells }\cite{ref: WHO}.

Among the various treatment options for cancer, one approach involves the use of ionizing radiation. The goal of radiotherapy is to cause irreparable damage to the DNA of cancer cells, thereby halting their ability to reproduce. The study of the effects of ionizing radiation on tissues falls under the domain of radiobiology. The subsequent sections will go into the specifics of radiobiology for PT.

\label{sec:radiobiology}
\subsection{Dose and Bragg Peak}
In medical applications, it is crucial to accurately measure the radiation dose to which a patient or a specific organ is subjected. The typical quantity used for this purpose is the absorbed dose ($D$), which is defined as the energy deposited by the ionizing radiation in the mass element:

\begin{equation}
D=\frac{dE}{dm} 
\end{equation}
The absorbed dose is measured in Gray (Gy), where $1  Gy = 1J/kg $.

\begin{figure}[h!]
\begin{center}
\includegraphics[width=8cm]{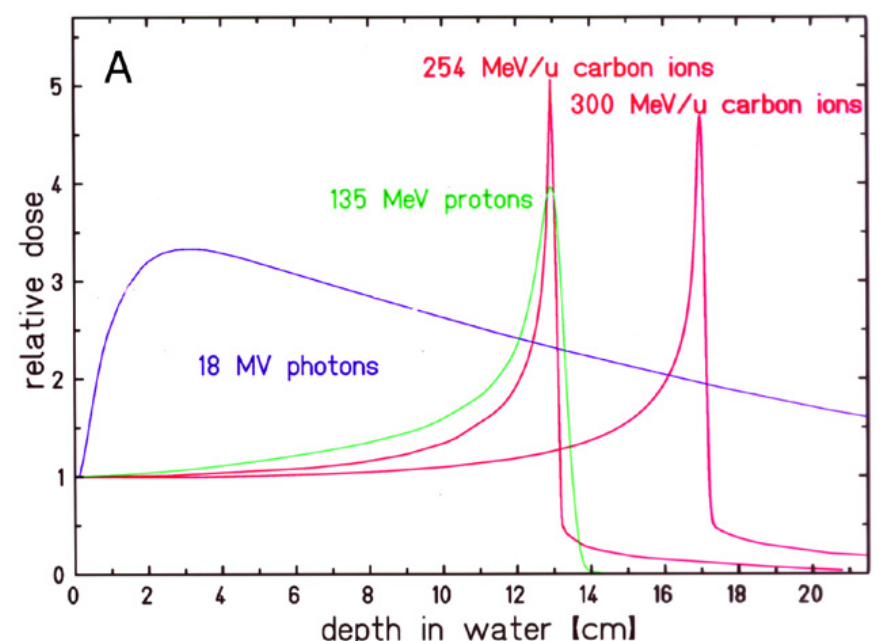}
\caption{Comparison of depth-dose profile for carbon ions, protons and photons.\cite{ref:nuclear phy}}
\label{fig:bragg-paak}
\end{center}
\end{figure}

The primary advantage of using charged particles in radiation therapy lies in their distinctive dose-depth profile. As illustrated in Figure \ref{fig:bragg-paak}, the depth-dose profile for photons is characterized by a high dose in the entrance channel, followed by an exponential decrease. In contrast, the profile for PT is significantly more favorable (as shown for carbon ions and protons in the figure). In this case, only a small dose is released in the entrance channel (typically from $10\%$ to $20\%$ of the maximum), and there is a pronounced peak at a specific depth, commonly known as the Bragg Peak. Additionally, the dose deposited by charged particles drops sharply after the peak, reducing the risk of damage to tissues located beyond the target area.

This advantageous dose distribution enables the alignment of the peak with the tumor, maximizing the dose delivered to the tumor while minimizing exposure to the surrounding healthy tissues. This characteristic typically allows for more conformal dose coverage of the tumor using beams from a reduced number of angles, making it particularly well-suited for treating deep-seated solid tumors.

This behaviour for the beams used in PT is a consequence of the factor $\frac{1}{\beta^2}$ that appear in the Bethe-Bloch equation (\ref{eq:bethe-bloch}), this meaning that the particle loses more energy as it slows down, until it stops. 
The different beams used in PT have also same differences in the depth-dose profile. As illustrated in Figure \ref{fig:bragg-paak}, $^{12}C$ ions exhibit a higher and narrower Bragg peak compared to protons, infact, in principle, heavier ions can deliver a more concentrated dose to the tumor with greater precision. However, the situation is not straightforward. First, the equipment required for heavier ions typically demands more space and  higher cost. Secondly, for heavier ions, there is often a noticeable dose tail beyond the Bragg peak. This aspect will be explored in more detail later.

\begin{figure}[h!]
\begin{center}
\includegraphics[width=8cm]{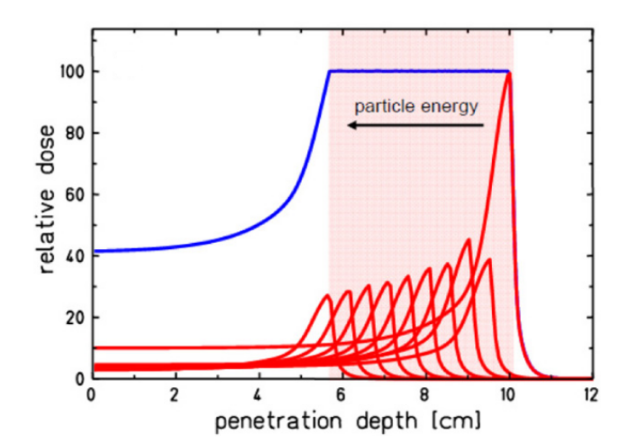}
\caption{Example of Spread-Out Bragg Peak. \cite{ref:nuclear phy}}
\label{fig:sobp}
\end{center}
\end{figure}

Given the narrow nature of the particle beam, typically is not possible to cover the entire tumor with a single Bragg Peak. To achieve a uniform dose across the entire tumor volume, the energy and intensity of the beams are modulated. This process, known as the Spread-Out Bragg Peak (SOBP), creates a flattened dose distribution that ensures the tumor receives the desired uniform dose. In figure \ref{fig:sobp} there is an example of the SOBP. The flattened dose is centered on the tumor; however, it is important to consider that during treatment, the patient is not perfectly immobilized. Natural physiological movements, such as breathing and heartbeats, can cause slight shifts in the tumor's position. If the radiation beam does not account for these movements in real-time, there is a risk that the high dose may inadvertently affect healthy tissue.

The Bragg Peak is located almost at the end of the path length, defined as the actual distance travelled by the radiation in matter, that for a particle with energy $E$, is defined:
\begin{equation}\label{eq:total path}
    R(E)=\int_0^E \left ( \frac{dE'}{dx} \right ) ^{-1} dE'
\end{equation}

Since heavy ions travel nearly straight through matter due to minimal scattering, the range, defined as the thickness over which the radiation travelled in matter, closely approximates this path length. For heavy ions with the same energy, the mean range in water is proportional to $\frac{A}{Z^2}$.

When analyzing the depth-dose profile, it is important to note that it can be derived from the Bethe-Bloch equation (Eq. \ref{eq:bethe-bloch}), which represents the average energy loss per unit length. Due to the statistical nature of this quantity, fluctuations in energy loss occur over a large number of collisions, leading to a broadening of the Bragg Peak. These fluctuations are described by the Vavilov distribution, which, in the limit of many collisions, approximates a Gaussian distribution:
\begin{equation}
    f(\Delta E)=\frac{1}{\sqrt{2 \pi \sigma_E}} e^{\frac{(\Delta E - <\Delta E>)^2}{2 \sigma_E^2}}
\end{equation}
whit:
\begin{equation}
\sigma_E=4 \pi z_{eff} Z e^4 N \Delta x \left[ \frac{1 - \frac{\beta^2}{2}}{\-\beta^2}\right]
\end{equation}

\begin{figure}[h!]
\begin{center}
\includegraphics[width=8cm]{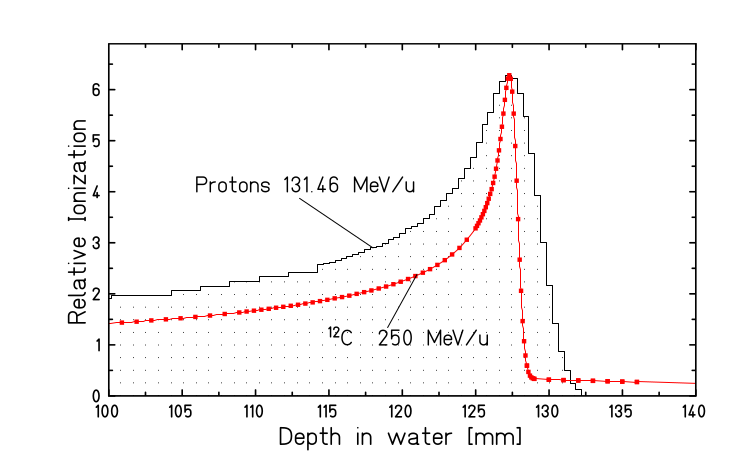}
\caption{Measured Bragg peaks of protons and $^{12}C$ ions having the same mean range in water. \cite{ref:heavy-ion}}
\label{fig:broad bragg-paak}
\end{center}
\end{figure}

As a result, the variance of the total path length ($\sigma_R$) also depends on $\sigma_E$:
\begin{equation}
    \sigma_R=\int_0^E \left ( \frac{d \sigma_E}{d x} \right) \left ( \frac{dE'}{dx}\right ) ^{-3} dE'
\end{equation}
Taking the ratio :
\begin{equation}
    \frac{\sigma_R}{R}= \frac{1}{\sqrt{M}} f \left ( \frac{E}{Mc^2} \right )
\end{equation}

where $f$ is a slowly varying function depending on the absorber, and $E$ and $M$ are the particle's energy and mass. The $1/\sqrt{M}$ factor is more significant for lighter ions compared to heavier ones (e.g., approximately a factor of 3.5 for protons compared to $^{12}C$, see Fig.: \ref{fig:broad bragg-paak}). However practically the broadening of the Bragg peak profile is primarily due to density inhomogeneities within the penetrated tissue.

\label{subsec:dos and bp}
\subsection{Biological effectiveness}
\label{subsec:biological eff}
The advantages of PT compared to conventional radiotherapy with photons are not solely related to the improved depth-dose profile but also to the higher biological effectiveness.

The ability to damage DNA depends on the absorbed dose but also on various physical and biological factors, such as the type of radiation, energy, and tissue type. DNA damage can occur through direct or indirect mechanisms. Direct DNA damage results when particles ionize the DNA molecule, breaking one or both of its helices, while indirect DNA damage is caused by radiation-induced radicals.

\begin{figure}[h!]
\begin{center}
\includegraphics[width=10cm]{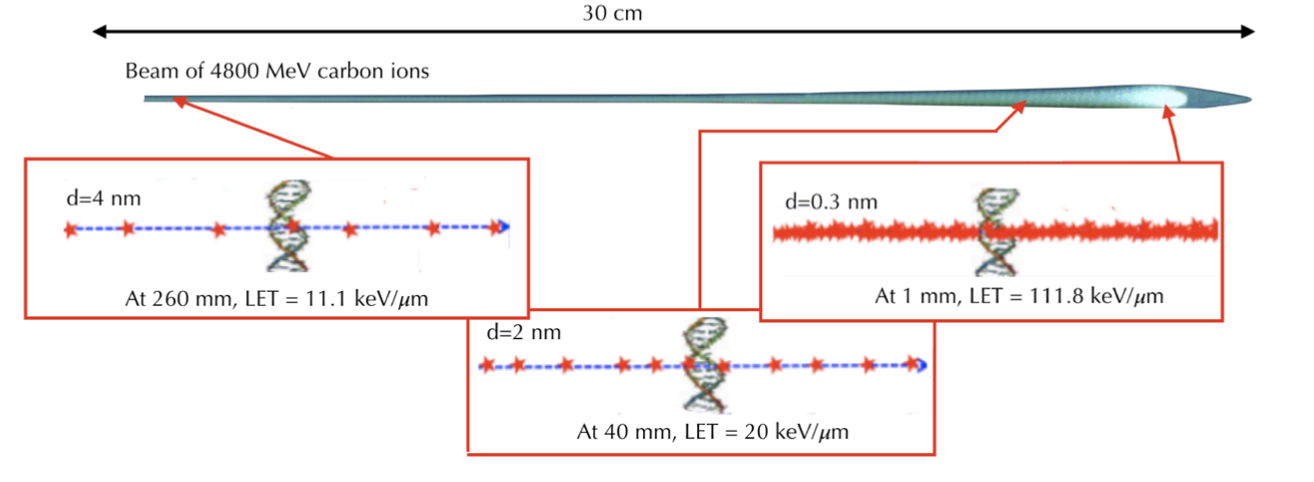}
\caption{Ionization density for different distance from the Bragg Peak placed at $30cm$ of penetration depth. The data are obtained considering a Carbon ion beam at $4800MeV$ that crosses a water target. $d$ indicates the distance between two consecutive ionization events.}
\label{fig:let}
\end{center}
\end{figure}

For this reason, it is essential to study the ionization density caused by the radiation. The first quantity to introduce is the Linear Energy Transfer (LET), a measure of the average energy deposited locally in the absorber per unit path length by the incident radiation. Although similar to the stopping power defined in eq.:\ref{eq:bethe-bloch}, LET specifically refers to the energy directly transferred to the medium via ionization and excitation, excluding the energy lost through radiative processes and secondary electrons known as $\delta$-rays. In the process of ionization, most electrons created are stopped near their point of emission. However, some electrons, known as $\delta$-rays, acquire sufficient energy to travel further and deposit energy at a distance from the ionizing track. Thus, the energy deposited locally in the material is not identical to the stopping power, as it excludes the kinetic energy carried away by $\delta$-rays.

\begin{equation}
    LET=\left( \frac{dE}{dx} \right )- \sum E(\delta_e)
\end{equation}
If X-rays are considered low-LET (sparsely ionizing) radiation, hadron beams are classified as high-LET (densely ionizing) radiation. As a consequence of the LET’s dependence on penetration depth, the dose released along the charged particle path is not uniform, with ionization density increasing nearby the Bragg peak (Fig.: \ref{fig:let}). This characteristic makes PT more likely to damage or destroy DNA compared to conventional radiotherapy, particularly at the Bragg peak. For low-LET radiation, indirect damage contributes more ($\sim 70 \%$) compared to direct hits ($\sim 30 \%$). However, with high-LET radiation, such as carbon ions, the contribution of direct hits increases, making PT particularly effective against radioresistant tumors.

One way to analyze the different effects of radiation on tumors is by using survival curves. These curves are produced by analyzing cell survival 1-2 weeks after irradiation. Cells are considered to have survived if they form a colony with more than 50 daughter cells. The surviving fraction is normalized to the number of seeded cells. Typically, cell survival is parameterized using a linear-quadratic model:
\begin{equation}
    S(D)=e^{-\alpha D- \beta D^2}
\end{equation}
where $D$  is the absorbed dose, and $\alpha$ and $\beta$  are parameters determined experimentally. While survival curves are an important tool for analyzing radiation effects, it is crucial to remember that the biological response to radiation is highly complex, involving many factors. 

Another useful quantity is the Relative Biological Effectiveness (RBE). The RBE is defined as the ratio of the dose of a reference source (typically $\gamma$ rays) to the dose of the radiation under study required to produce the same level of damage as the reference dose (isoeffect).

\begin{equation}
    RBE=\frac{D_{ref}}{D_{ion}}
\end{equation}

\begin{figure}[h!]
\begin{center}
\includegraphics[width=7cm]{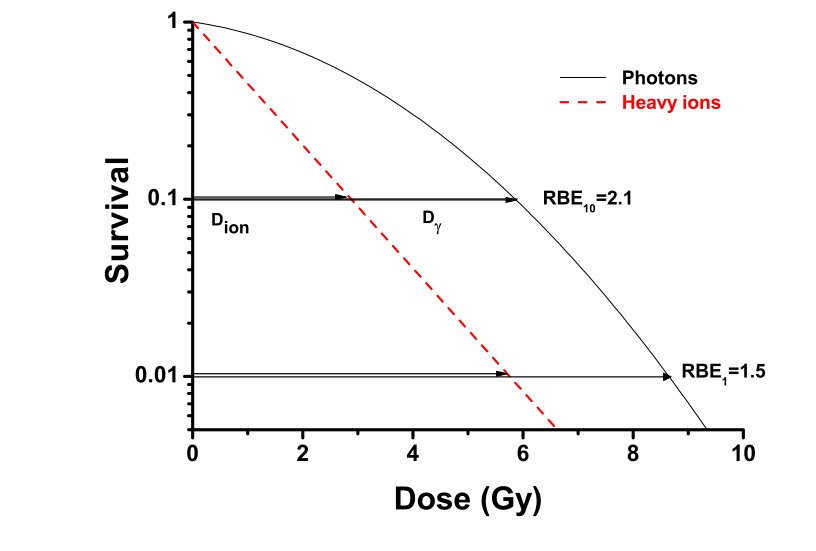}
\caption{Determination of RBE for $10\%$ and $1\%$ survival level for a typical heavy irradiation respect to x-rays irradiation\cite{ref:heavy-ion}}
\label{fig:survival}
\end{center}
\end{figure}

The advantage of using RBE is that it accounts for both the effectiveness of the radiation and the tissue-specific response. An example of RBE calculation using a cell survival curve is shown in Fig. \ref{fig:survival}. It is notable that, to achieve the same damage, a lower dose is needed in PT, and the RBE also depends on the dose level, or equivalently, on the survival fraction one aims to achieve.

\begin{figure}[h!]
\begin{center}
\includegraphics[width=8cm]{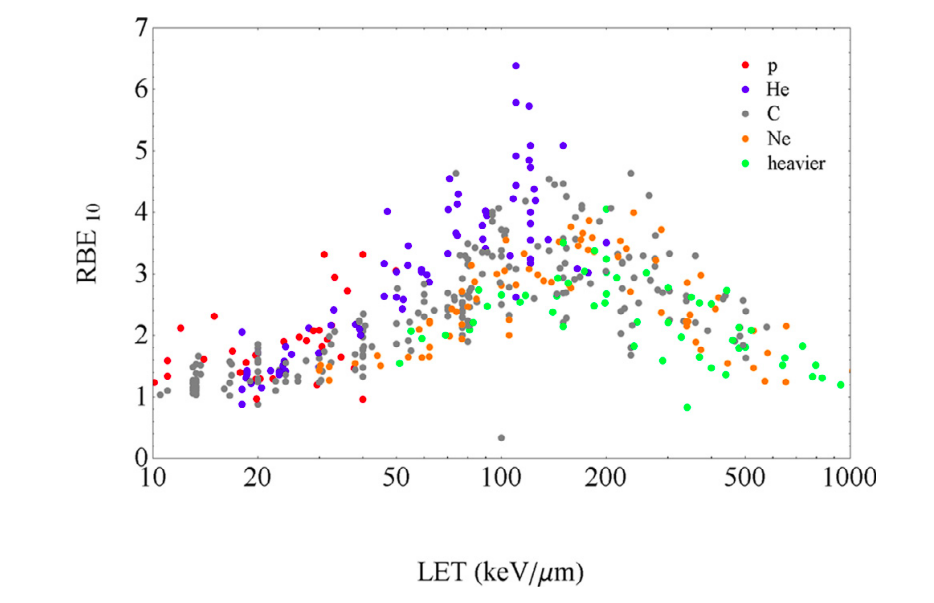}
\caption{RBE compared with LET from published experiments on in vitro cell lines. RBE is calculated at $10\%$ survival. Data points are extracted from the particle radiation data ensemble (PIDE) database (Friedrich et al 2013\cite{ref:pide}). \cite{ref: Durante}}
\label{fig:RBE vs LET}
\end{center}
\end{figure}

Naturally, RBE depends on various factors such as tissue type, dose, fractionation of treatment, LET, and others. The Figure \ref{fig:RBE vs LET} illustrates the behavior of RBE as a function of LET for different particles. It demonstrates that RBE steadily increases with rising LET, (ionization density increase and consequently improve the biological effects) reaching a peak at approximately $100-200 keV/\mu m$. Beyond this peak, the RBE begins to decrease due to overkill effects, where cells are damaged more than necessary to cause their death, resulting in tissues receiving an unnecessary dose. The exact position of this peak varies depending on the type of primary particle, with heavier ions causing the peak to shift towards higher LET values.

\subsubsection{Oxygen effect}
\begin{figure}[h!]
\begin{center}
\includegraphics[width=6cm]{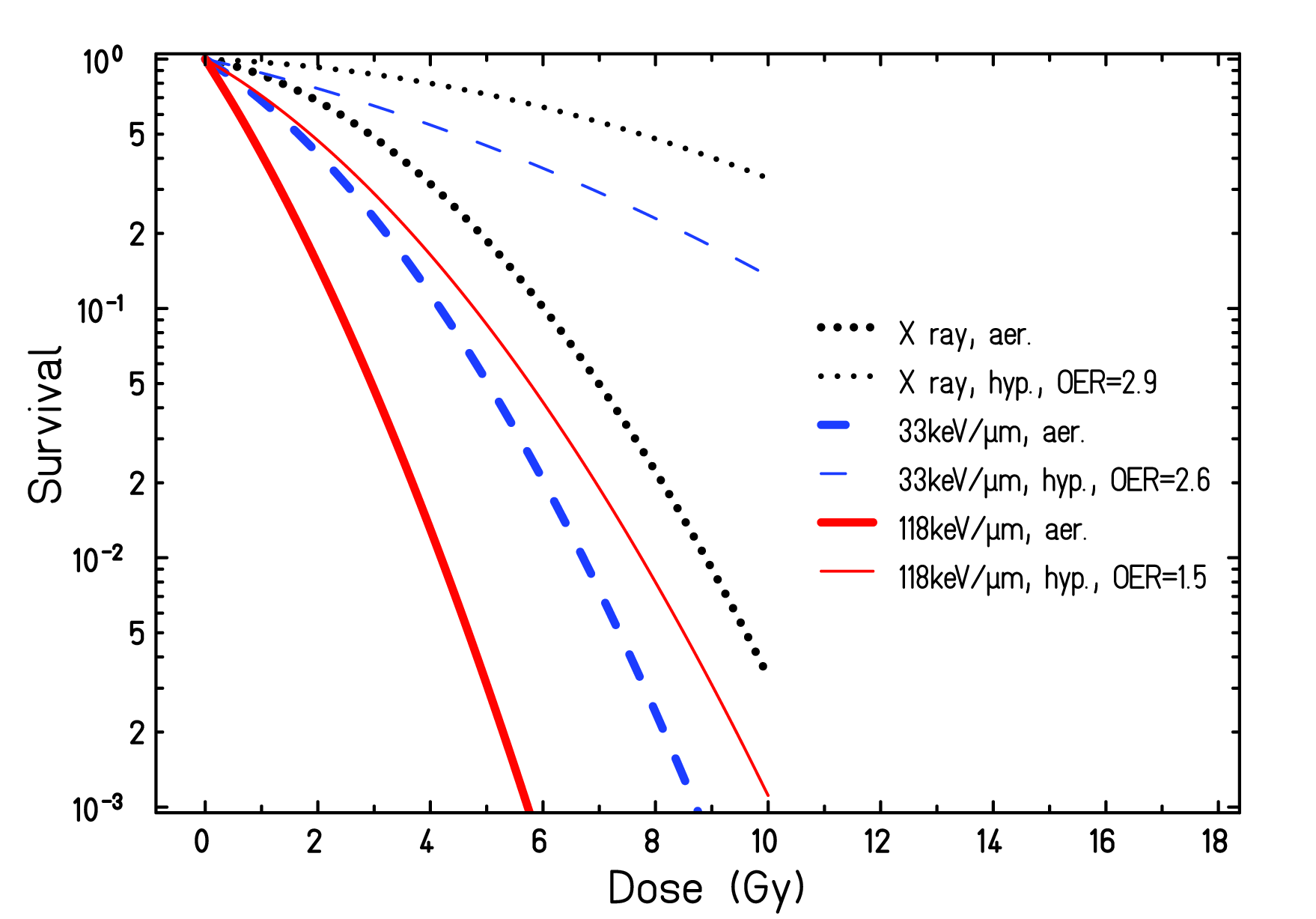} \quad \includegraphics[width=6cm]{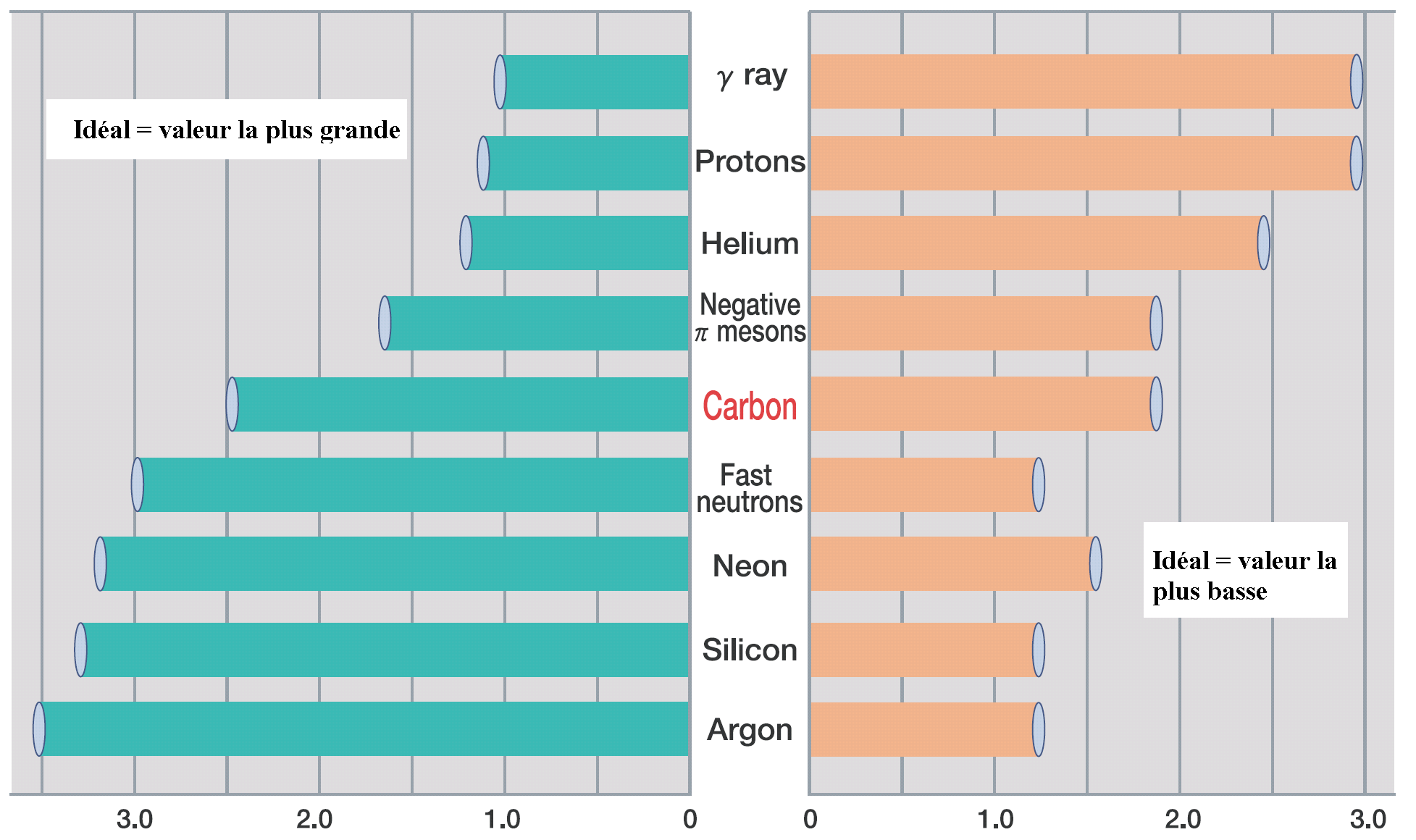}
\caption{\textbf{left:}Influence of the oxygen level on cell survival of human kidney T-1 cells for carbon ions with different LET. \textbf{right:} RBE (left) and OER (right) values of different particles.}
\label{fig:OER}
\end{center}
\end{figure}

Another effect to consider is the Oxygen Effect. Cancerous tumor cells are often hypoxic, meaning they have a low oxygenation rate, which makes them more resistant to radiation. To quantify this resistance, the Oxygen Enhancement Ratio (OER) is used, defined as:
\begin{equation}
    OER=\frac{D_{hypoxic}}{D_{aerobic}}
\end{equation}
where $D_{hypoxic}$ is the dose required for heavy ions to kill a fixed number of hypoxic cells, and $D_{aerobic}$ is the dose needed to kill the same number of aerobic cells (cells with a normal oxygenation rate). The left figure of \ref{fig:OER} illustrates the comparison between the survival curves of aerobic and hypoxic cells for different LET values. It is evident that the survival rates vary with LET, indicating that OER is also LET-dependent. The figure clearly shows that aerobic cells are easier to destroy compared to hypoxic cells. In the right panel, a summary of RBE and OER values for different particles is presented. The data reveals that RBE increases with the mass of the ion, while OER decreases. This suggests that using heavier ions could improve the effectiveness of tumor treatment, as they require lower doses to kill cancerous cells and are also effective against hypoxic, radioresistant cells. These benefits must be weighed against the logistical and financial challenges they present, infact heavier ions are more challenging to produce, requiring larger and more complex machinery and the economic cost associated with their use is substantially higher.

\subsection{Impact of nuclear fragmentation}
\label{subsec:impact of nuclear frag}

\begin{figure}[h!]
\begin{center}
\includegraphics[width=6cm]{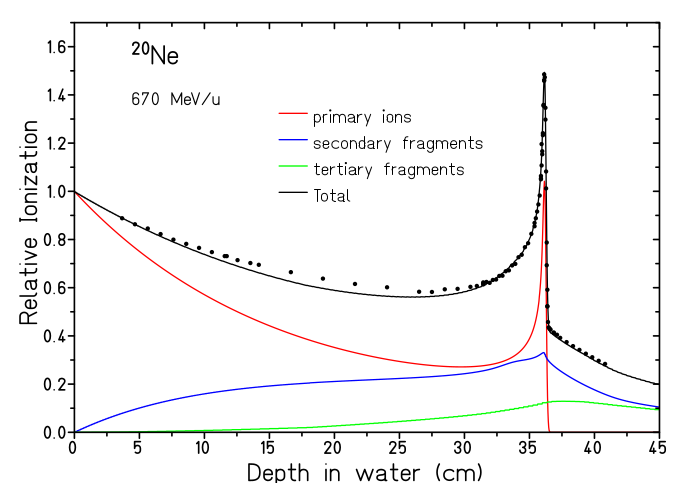} \quad \includegraphics[width=6cm]{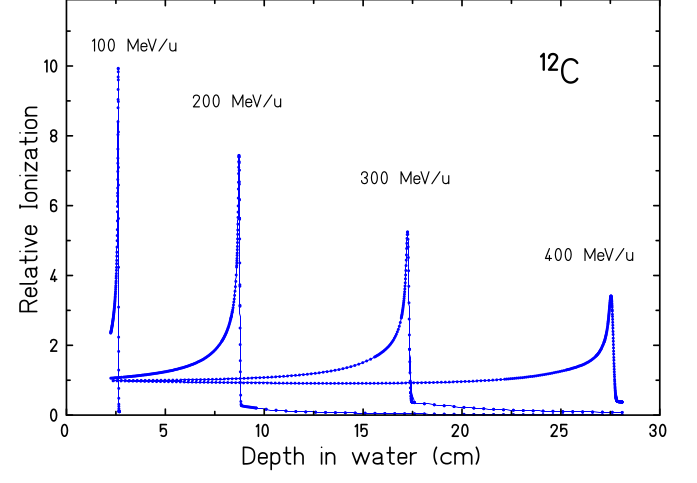}
\caption{ \textbf{left:} Bragg peak for $^{20}Ne$ with contributions of primary ions, secondary and tertiary fragments. \textbf{right:}Bragg peak for $^{12}C$ stopping in water for differnt energy.}
\label{fig:fragments}
\end{center}
\end{figure}

How seen in section \ref{subsec:nuclear int}, although less probable, nuclear interactions can occur when a charged particle travels through a material. In particular, these interactions can result in the fragmentation of both the target and projectile. When such processes occur, they lead to a significant loss of primary fluency, potentially reducing the Bragg peak, while simultaneously producing fragments that can deposit dose far from the peak. The projectile fragments maintain nearly the same velocity as the primary particle but having a lower $Z$, and becouse of the $\frac{A}{Z^2}$ scaling of the range, the frgments possess greater ranges, thereby creating a dose tail beyond the peak. Instead the target fragments and the decay and evaporation products (essentially protons, neutrons and pions) are isotropically distributed in space and have a low kinetic energy and therefore small range, follow as a consequence the almost absence of tail for protons beam. The angular distribution of the fragments depends on the kinematics of the interaction, but they undoubtedly contribute to the lateral spread of the beam. In Fig. \ref{fig:fragments}, the left panel illustrates how these fragments release dose beyond the Bragg peak, showing the contributions of the primary particles and those of fragments. Another important aspect, as seen in the right panel, is that the ratio of peak to entrance dose becomes less favorable as the penetration depth increase.

\begin{figure}[h!]
\begin{center}
\includegraphics[width=6.6cm]{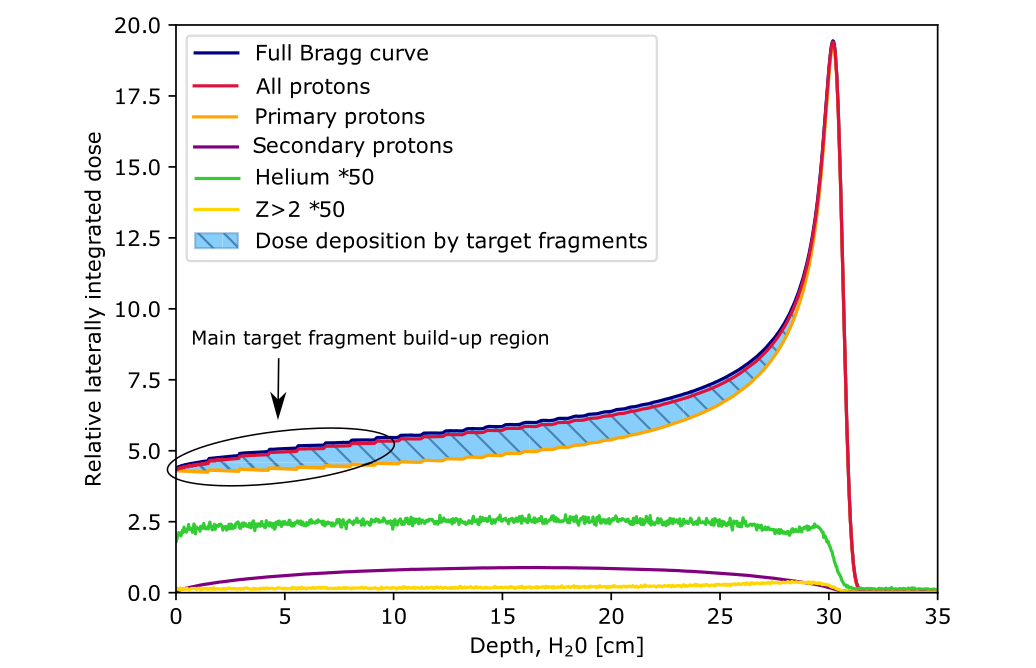} \quad \includegraphics[width=6.5cm]{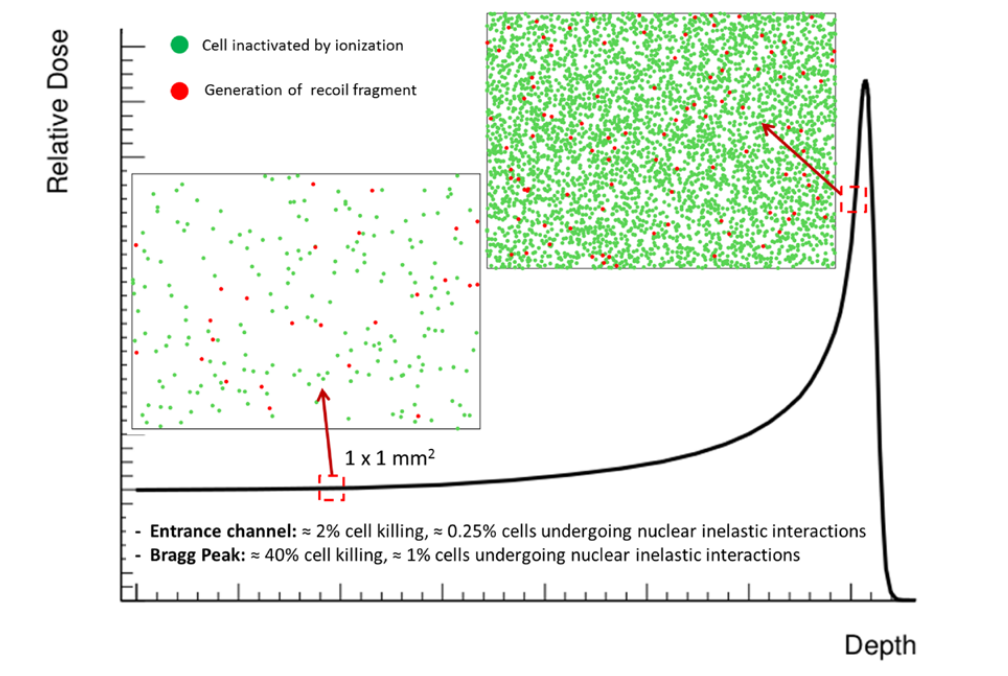}
\caption{\textbf{left:} FLUKA simulation of the contribution of different particle species to the shape of a 220 MeV proton Bragg curve \cite{ref: dose}. \textbf{right:} Cell killing by electromagnetic interaction between target electrons and the primary proton with the cell killing due to target fragmentation, assuming that each cell where a fragmentation occurs is killed by the recoil fragments, but no neighbor cells are interested. \cite{ref:nuclear phy in pt}}
\label{fig:fragment. brag peak}
\end{center}
\end{figure}

The damage to DNA caused by fragments is less significant in the Bragg peak region compared to other parts of the dose-deposition curve. The  right panel of Fig. \ref{fig:fragment. brag peak}, show the cell killing by primary particles and the cell killing by fragments at different points along the curve. Near the Bragg peak, the impact of the primary particles is so dominant that it renders the effects of fragmentation negligible (the ratio of cell killed by fragmentation reactions respect to ionization is $\sim 1/40$). However, in the entrance channel, where the cell killing effect of primary particles is less pronounced, the contribution of fragmentation becomes more relevant (the ratio is $\sim 1/10$) this meaning a bigger contribute to the dose to healthy tissues by the fragments. The left panel instead show the contribution to the Bragg curve of primaries and secondary particles for a proton beam. To notice that in the entrance channel the contribution of target fragments to the dose is quite relevant.

\begin{figure}[h!]
\begin{center}
\includegraphics[width=12cm]{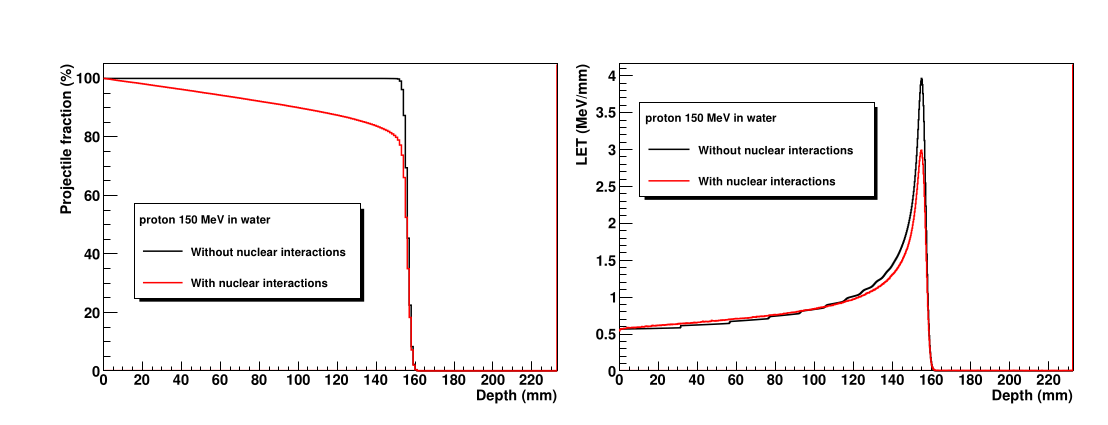}
\caption{\textbf{left:} Evolution of the primary protons ratio, \textbf{right:} the mean LET for protons with the penetration depth for $150 MeV$ protons according to GEANT4 simulations. \cite{ref:nuclear phy}}
\label{fig: proton fragm}
\end{center}
\end{figure}

\begin{figure}[h!]
\begin{center}
\includegraphics[width=12cm]{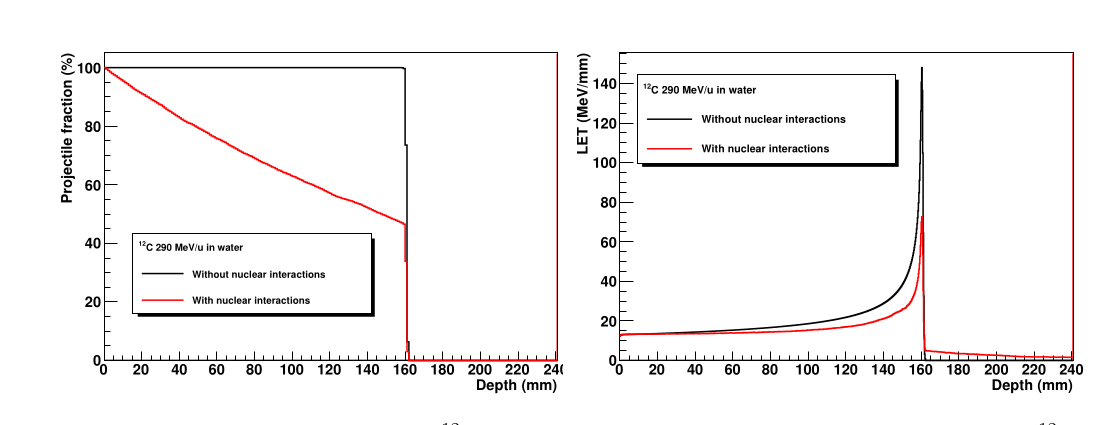}
\caption{\textbf{left:} Evolution of the primary protons ratio, \textbf{right:} the mean LET for protons with the penetration depth for $290 MeV/u$ $^{12}C$ according to GEANT4 simulations. \cite{ref:nuclear phy}}
\label{fig: C fragm}
\end{center}
\end{figure}

At this point, it is important to understand how the effects of fragmentation manifest. The figure \ref{fig: proton fragm} compares simulations conducted with GEANT4\footnote{GEANT4 is a toolkit for the simulation of the passage of particles through matter.}, that either include or exclude nuclear interactions, for the evolution of the primary protons ratio and the mean LET as a function of penetration depth in liquid water. Examining the left side of the figure, it becomes clear that when only electromagnetic interactions are considered, all primary particles reach the Bragg Peak. However, when nuclear fragmentation is included, only $80\%$ of the particles reach the Bragg Peak. This results in a $20\%$ reduction in the Bragg Peak's intensity, as shown in the right panel. It is important to note that the position of the Bragg Peak remains unchanged, as it is determined solely by the Bethe-Bloch equation. Additionally, the tail is not observed in this case, which is attributed to the fact that protons are used in the simulation. Since only target fragmentation occurs for protons and the resulting fragments have very low velocities, their range does not exceed a few $\mu m$, causing them to deposit their energy near the collision site. Another observation that can be done is about the integral of the LET (right panel), because energy is required for fragment production, when nuclear interactions are included it is $97\%$ of the integral when only electromagnetic interactions are considered. The situation is slightly different when particles like $^{12}C$ ions are used, as shown in the figure \ref{fig: C fragm} where $290 MeV/u$ $^{12}C$ ions are simulated. In this case, only $50\%$ of the primary particles reach the Bragg Peak when nuclear interactions are accounted for. This leads to a $50\%$ reduction in the Bragg Peak's intensity, but because projectile fragmentation also occurs, a tail appears after the peak. In this latter scenario, the integral of the LET, when nuclear interactions are included, is $93\%$ of the integral without them.

\section{Radioprotection in space}
Will it ever be possible to reach Mars? 

One of the significant challenges associated with a mission to Mars, as with all long-duration space missions, is the exposure to space radiation. Unlike on Earth, where the atmosphere provides a protective shield against many of these harmful particles, the absence of such a barrier in space results in much higher levels of radiation exposure. This increased exposure dramatically elevates the risks associated with space travel, including potential damage to astronauts' health and the integrity of the spacecraft's systems. This is where Radio Protection in Space (RPS) becomes crucial. The primary goal of radiation safety studies is to enable humans to undertake such missions while maintaining an acceptable level of risk, ensuring that the health of the crew and the functionality of the spacecraft are preserved throughout the journey.

\begin{figure}[h!]
\begin{center}
\includegraphics[width=8cm]{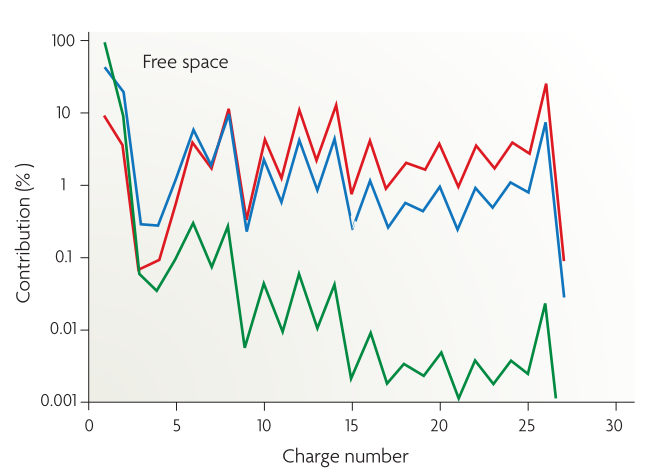}
\caption{The contribution in fluence (green), dose (blue), and dose equivalent (red) of different nuclei in galactic cosmic radiation. \cite{ref: space exploration}}
\label{fig:GCR fluence and dose}
\end{center}
\end{figure}

During space travel, a spacecraft encounters three primary sources of radiation:
\begin{itemize}
    \item \textbf{Solar Particle Events (SPEs)} These events are primarily composed of protons emitted by the Sun, with energies that can reach up to the order of GeV. SPEs are unpredictable and can occur suddenly, leading to intense bursts of radiation.
    \item \textbf{Galactic Cosmic Rays (GCR)} Are high-energy particles primarily composed of protons ($\sim 86\%$), helium nuclei ($\sim 12\%$), and a small fraction of heavier nuclei ($\sim 1\%$). These particles originate from supernovae within the Milky Way Galaxy. Their energy spectrum ranges from $MeV$ to $TeV$, with a peak around $100 -800 MeV/u$.
    \item \textbf{Geomagnetically trapped particles} These particles are confined by the Earth's magnetic field and consist of protons (with energies up to a few hundred MeV) and electrons (with energies up to 100 keV). While these particles are less of a concern during interplanetary travel, they are significant during missions within or near Earth's magnetic environment.
\end{itemize}
In Figure \ref{fig:GCR fluence and dose}, the contributions to fluence, dose, and equivalent dose \footnote{the dose do not take into account the nature of the radiation or the type of interaction between radiation and matter, it is therefore preferable to define the \textbf{equivalent dose.} Equivalent dose is calculated for individual organs. It is based on the absorbed dose to an organ, adjusted to account for the effectiveness of the type of radiation. It is expressed in Sieverts ($1 Sv = J/kg$) to an organ. Equivalent dose: $H= D \cdot Q \cdot N$, where $D $ is the absorbed dose, $Q$ the quality factor (1 for $X$ and $\gamma$, $\sim 10$ for neutrons and
protons, $\sim 20$ for a and heavy ions); N the factor that takes into account the mode of
esposure, e.g., fractional, intensity, etc..} for GCR ions are presented as a function of their charge. It is evident that, despite the relatively low fluence of high charge and high energy (HZE) nuclei, their contribution to dose and equivalent dose is significant, this is a consequence of the dose’s $Z^2$ dependence. HZE radiation is characterized by high LET, which implies not only greater penetration power but also a substantial impact on biological effectiveness.

Understanding the spectrum of this radiation is essential for studying its interactions during space missions and for mitigating the exposure of astronauts and spacecraft systems. The dose astronauts receive can increase the risk of various health issues, including both acute and long-term effects such as cancer, central nervous system damage, cataracts and so on. These risks could potentially compromise the success of the mission. Additionally, radiation poses a significant threat to the spacecraft’s electronic systems and instrumentation, which are crucial for mission reliability and safety. 

To reduce radiation exposure, three strategies are typically considered: increasing the distance from the radiation source, minimizing exposure time, and providing adequate shielding for both astronauts and equipment. The first two methods are impractical for space missions because cosmic radiation is isotropic, making it impossible to increase distance from the source, and long-duration missions are a key goal of future space exploration. Consequently, shielding becomes the primary solution. The challenge is to identify optimal materials that can protect against both ionizing energy loss and nuclear fragmentation. While low-energy ions can be stopped with modest amounts of shielding material, high-energy particles are more likely to penetrate the shield. These high-LET particles are particularly harmful. However, HZE particles that pass through shielding can undergo nuclear interactions, resulting in fragmentation. As previously discussed, the lighter fragments produced in this process have a longer range compared to the primary particles and can reach astronauts. At the same time, these secondary particles are generally less biologically damaging, which suggests that fragmentation reactions could be exploited to reduce the risks associated with radiation exposure in space \cite{ref:shielding}.

The radiation dose expected for a Mars mission is substantial, making it critical to optimize the shielding. Some studies indicate that lighter materials perform better in terms of radiation shielding \cite{ref:shielding}. Typically, different material combinations are studied, and their effectiveness is assessed through MC simulations. However, the accuracy of these simulations depends heavily on the availability of precise cross-section data, including those for nuclear fragmentation. The scarcity of comprehensive cross-section data limits the ability to fully optimize spacecraft shielding and ensure the highest level of safety.

\section{Fragmentation cross sections} \label{sec: cross section}
A crucial aspect of PT is the precise definition of the beam parameters to deliver the correct dose to the tumor. These parameters are determined using Treatment Planning Systems (TPS). After analyzing the patient's anatomical data, obtained through imaging techniques such as Computed Tomography (CT) or Magnetic Resonance Imaging (MRI), and consulting with the medical team, the appropriate dose to be delivered to the tumor is established. At this stage, it is necessary to provide the beam control system with key information such as position, intensity, and direction to ensure the correct dose is delivered to the tumor while minimizing exposure to the surrounding healthy tissue. The accuracy of the TPS is of paramount importance in this process. MC simulations, particularly Fast MC simulations, are typically employed to achieve the required accuracy by providing a realistic assessment of the patient’s anatomy (derived from CT, MRI, etc.) in a short timeframe, thus preventing tumor growth during the planning phase. To achieve this level of precision, the MC simulations must utilize accurate cross-sections. However, the lack of comprehensive fragmentation cross sections for many nuclear processes, combined with the limited precision of existing data due to a scarcity of experimental results, presents a significant challenge in achieving the desired level of accuracy.

This issue is also prevalent in RPS. To simulate the interaction of radiation with a spacecraft and optimize the shielding required to protect astronauts and equipment, accurate cross-sectional data are essential.

\begin{figure}[h!]
\begin{center}
\includegraphics[width=8cm]{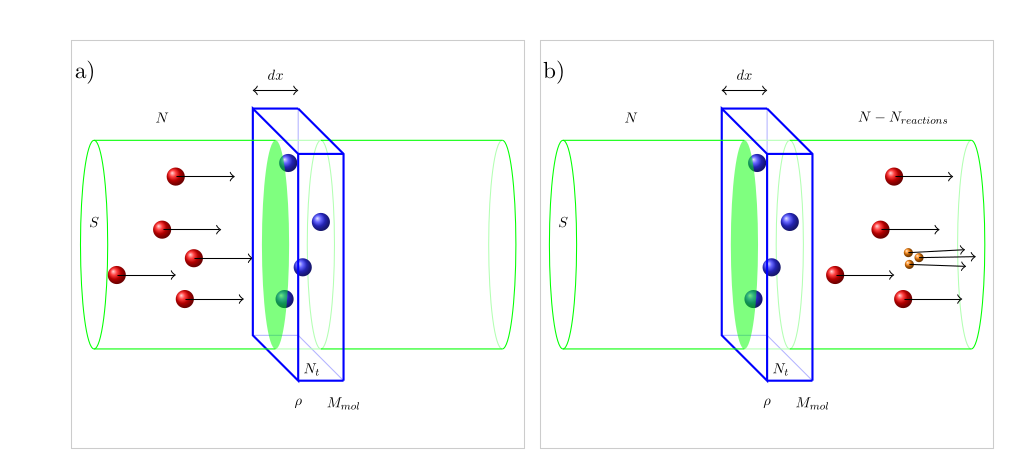}
\caption{Schematic view of the interaction of a beam of ions crossing a piece of absorber, in detail is show an event of projectile fragmentation.}
\label{fig:cross section}
\end{center}
\end{figure}

The cross sections provide the probability of interactions between the particles in the beam and the target material. If $\frac{dN_p}{dt}$ is the number of projectiles per unit time, $\frac{dN_r}{ dt}$ is the number of reactions per unit time, $n_b$ is the target density, and $d$ is its thickness, the number of reactions depends on the number of projectiles and the number of target particles through the cross section ($\sigma$):
\begin{equation}
    \frac{dN_r}{dt}=\sigma n_B d \frac{dN_p}{dt}
\end{equation}
Figure \ref{fig:cross section} shows a schematic view of the interaction of an ion beam passing through a piece of absorber. By accurately measuring the cross section, it is possible to predict the final outcome of such interactions.

\begin{figure}[h!]
\begin{center}
\includegraphics[width=8cm]{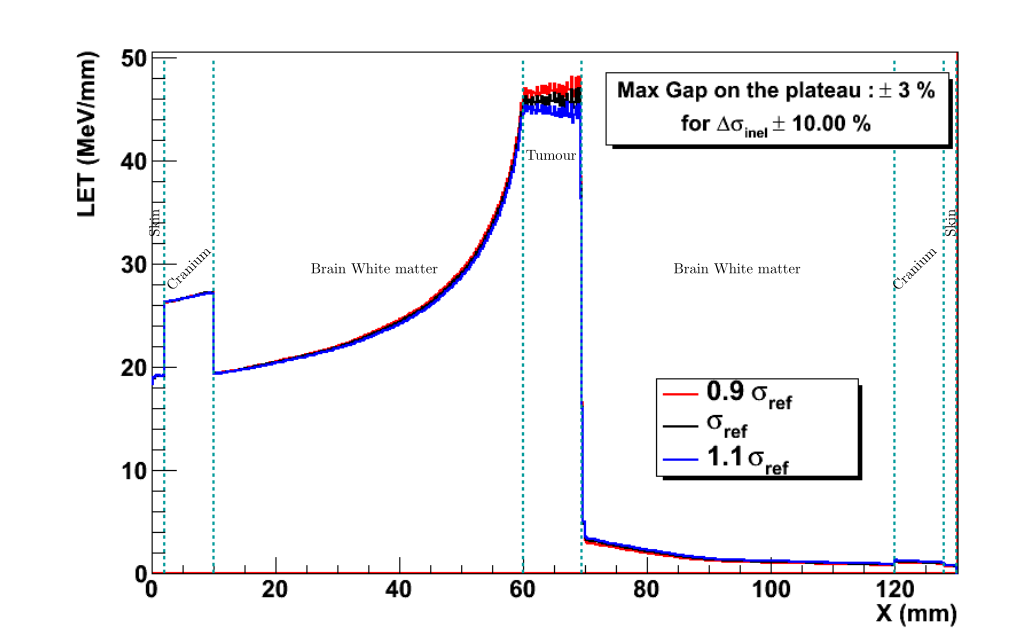}
\caption{LET as function of penetration depth for three GEANT4 simulations with the usual total reaction cross section \textcolor{black}{$\sigma_{ref}$}, \textcolor{red}{$0.9 \cdot\sigma_{ref}$} and \textcolor{blue}{$1.1 \cdot \sigma_{ref}$}.\cite{ref:nuclear phy} }
\label{fig_geant4-cross}
\end{center}
\end{figure}

Given the limitations in cross-section data, it is crucial to understand how variations in cross sections impact the planning of astronaut shielding or TPS. According to \cite{ref:nuclear phy}, there is a study that demonstrates the dependence of energy deposition on variations in the $\sigma$ value. The figure \ref{fig_geant4-cross} illustrates LET as a function of penetration depth for $^{12}C$ ions in a stack of materials (skin, cranium, brain white matter, cancerous tumor) obtained with GEANT4. The incident energy distribution has been adjusted to achieve a SOBP corresponding to the tumor size. The black curve represents the mean LET with the standard total cross section. Two additional simulations were performed: one with all cross-section values increased by $10\%$ (blue LET curve) and another with all values decreased by $10\%$ (red LET curve). It was observed that such variations in cross-section values result in a $\pm 3 \%$ variation in the LET value within the tumor. This implies that reaction cross sections must be known within $10\%$ to achieve the $3\%$ accuracy in dose computation. However, such precision in experimental cross-section measurements has not yet been achieved.

\chapter{The FOOT experiment}
\label{chap:2}

As discussed in the previous chapter, the lack of comprehensive cross-section data, including nuclear fragmentation in the energy range of $100-800 MeV/u$, presents a significant challenge.  This knowledge gap is a major issue in both PT and RPS. Accurate cross-sections are essential for MC simulations, which currently cannot achieve the $3\%$ accuracy required for radiotherapy applications. Similarly, the lack of this data hampers the optimization of spacecraft shielding, making it difficult to ensure astronaut safety during long-term space missions.

In response to this challenge, the \textbf{FOOT} (FragmentatiOn Of Target) experiment was initiated. Founded by INFN (Istituto Nazionale di Fisica Nucleare, Italy), through collaboration among researchers from France, Italy, Germany and Japan, the primary goal of FOOT is to measure the double differential cross-section with respect to the kinetic energy and production angle of emitted fragments within the energy range relevant to PT and RPS. The experiment utilizes $^{12}C$, $^{16}O$, $^4 He$ beams impinging on carbon and hydrogen-rich targets to study the interactions of ions with the principal components of the human body, such as oxygen, carbon, and hydrogen atoms.

\section{Experimental requirements}

The final goal of the FOOT experiment is to measure differential cross-sections with respect to the kinetic energy ($\frac{d \sigma}{dE_{kin}}$) for target fragmentation processes with an accuracy better than $10\%$, and double differential cross sections ($\frac{d^2\sigma}{d \Omega dE_{kin}}$) for projectile fragmentation processes with an accuracy better than $5\%$ \cite{ref:mesuring}. To achieve this level of precision, INFN has developed a fixed-target experiment designed to detect, track, and identify all charged particles (both primary and fragments) exiting the target, as well as to track all primary particles impinging on the target. Achieving these goals requires a charge identification accuracy of approximately $2-3\%$, and an isotope identification accuracy of around $5\%$.

Another critical requirement for the FOOT experiment is the need for a portable setup, as the experiment is conducted in various locations: at CNAO (Centro Nazionale di Adroterapia Oncologica) in Pavia, Italy; GSI in Darmstadt, Germany; and the Heidelberg Ion Therapy Center (HIT) in Germany.

The experimental program of FOOT involves a series of measurements using different beams and targets. A summary of the physics program for the experiment is presented in Table \ref{tab: program}.
\begin{center}
\begin{tabular}{|c c c c c |} 
 \hline
 Fragmentation  & Beam & Target & Upper Energy $[MeV/u]$  & Interaction process \\ [0.5ex] 
 \hline\hline
Target   & $^{12}C$ & $C$, $C_2H_4$ & 200 & $H+C$\\
Target   & $^{16}O$ & $C$, $C_2H_4$ & 200 &  $H+O$\\
 \hline
Beam   & $^4He$ & $C$, $C_2H_4$ & 250  & $\alpha +C$, $ \alpha + H$\\
Beam   & $^{12}C$ & $C$, $C_2H_4$ & 400  & $C +C$, $ C + H$\\
Beam   & $^{16}O$ & $C$, $C_2H_4$ & 500  & $O +C$, $ O + H$\\
\hline
Beam   & $^4He$ & $C$, $C_2H_4$ & 800  & $\alpha +C$, $ \alpha + C_2H_4$, $ \alpha + H$\\
Beam   & $^{12}C$ & $C$, $C_2H_4$ & 800  & $C +C$, $ C + C_2H_4$, $ C + H$\\
Beam   & $^{16}O$ & $C$, $C_2H_4$ & 800  & $O +C$, $ O + C_2H_4$, $ O + H$\\
 \hline
\end{tabular} \label{tab: program}
\captionof{table}{Summary of the experimental program of the FOOT collaboration}
\end{center}

\section{The experimental methods}
\begin{figure}[h!]
\begin{center}
\includegraphics[width=11cm]{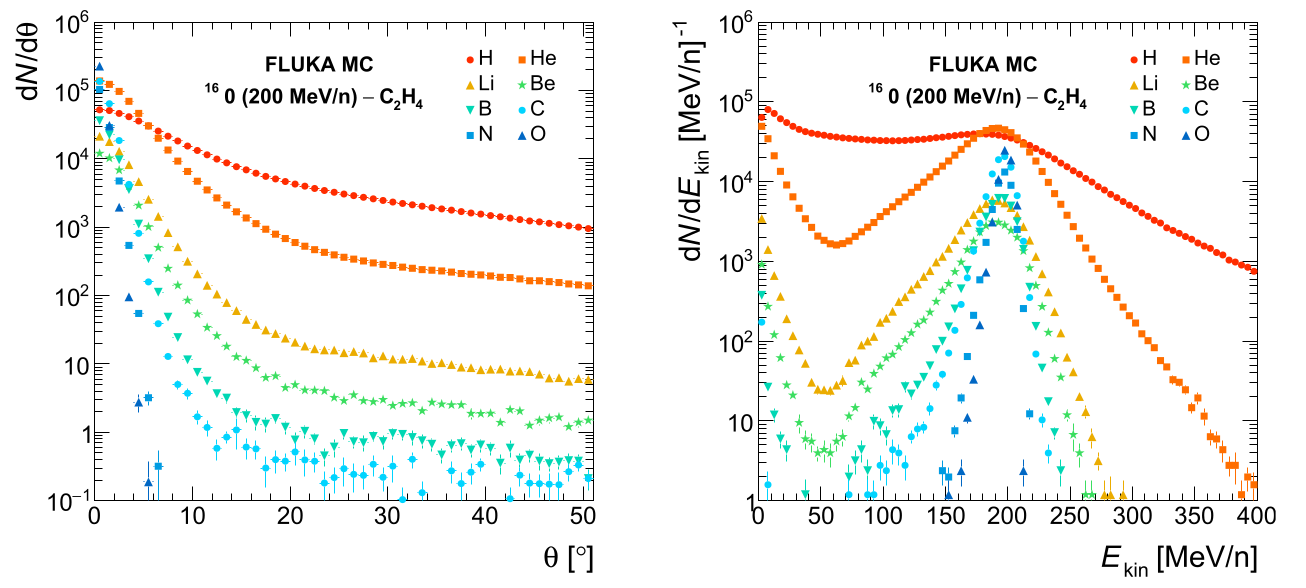}
\caption{MC calculation of the angular (left) and kinetic energy (right) distributions of different fragments produced by a $200 MeV/u$ $^{16}O$ beam impinging on a $C_2H_4$ target. \cite{ref:mesuring}}
\label{fig: fluka.foot}
\end{center}
\end{figure}

To enhance the design of the detector, a Monte Carlo simulation (with a FLUKA\footnote{FLUKA is a fully integrated particle physics MC simulation package. It has many applications in high energy experimental physics and engineering, shielding, detector and telescope design, cosmic ray studies, dosimetry, medical physics and radio-biology.\url{https://fluka.cern/}} code \cite{ref:fluka} \cite{ref:the Fluka}) of $^{16}O$ impinging on $C_2H_4$ was conducted. The simulation results, shown in the figure \ref{fig: fluka.foot}, depict the produced fragments as a function of angle and kinetic energy. It is evident that heavier fragments are predominantly produced in the forward direction (until $\sim 10^{\circ}$), with kinetic energies peaking around the primary energy. In contrast, the distribution of lighter fragments is broader, both in terms of angle and energy.

Due to these differences, and the need to develop an efficient tracking system with limited dimensions while maintaining a good angular acceptance for all the fragments involved, the FOOT experiment is organized into two distinct setups:
\begin{itemize}
    \item \textbf{Magnetic Spectrometer}: This setup is focused on characterizing heavier nuclear fragments (with $3 \le Z \le 8$) and has an angular acceptance of up to $10^{\circ}$ from the beam axis.
    \item \textbf{Emulsion Spectrometer}: This setup is based on nuclear emulsion films and is optimized for studying lighter fragments ($Z \le 3$), with an angular acceptance of nearly $70^{\circ}$.
\end{itemize}

In this thesis, the focus is on the Magnetic Spectrometer setup. 

Between the challenges of the FOOT experiment there is the study of target fragmentation. Unlike projectile fragments, which are typically emitted at the same velocity as the primary beam, target fragments have much lower energy and, therefore, a much shorter range. The path length expected for target fragments produced by a typical proton therapy beam inside a patient is on the order of $10-100 \mu m$, leading to a low probability of these fragments escaping the target, even if it is very thin. In fact, even a target just a few millimeters thick would stop the fragments, rendering their detection impossible. Furthermore, using an even thinner target introduces additional challenges, such as mechanical instability and reduced reaction rates, leading to longer data acquisition times.  To overcome these limitations, an inverse kinematic approach is employed. For a more detailed explanation, see the next section (\ref{subsec:inverse kinematic}).

\subsection{Inverse kinematic method}\label{subsec:inverse kinematic}
The FOOT experiment employs an inverse kinematic approach to address the challenge of studying target fragmentation.

\begin{figure}[h!]
\begin{center}
\includegraphics[width=11cm]{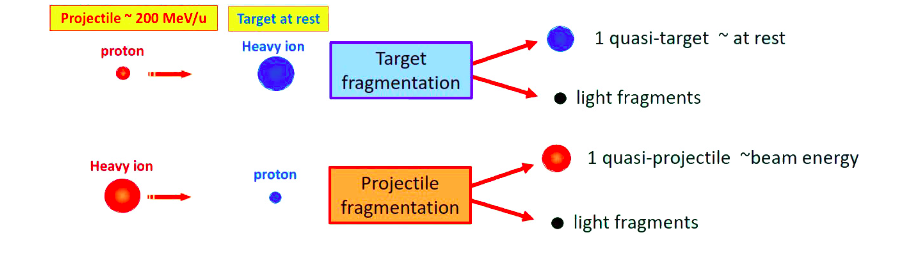}
\caption{ Scheme of fragmentation for direct and indirect kinematic approach}
\label{fig: inversion}
\end{center}
\end{figure}

To provide an idea of the ranges of target fragments, the table \ref{table: data proton} presents data obtained using a $180 MeV$ proton beam in water. As shown, the ranges are of the order of micrometers. To study processes like $p + X \rightarrow p + X'$, the roles of the projectile and target are reversed. Instead of using a proton beam on a target with a composition similar to human tissue, the experiment utilizes tissue-like ion beams, ($C$ and $O$ the main constituents of the human body) that impinge on hydrogen-rich targets. By maintaining the same velocity for the ion beam as would be used for protons (same kinetic energy, and so higher total energy due to the greater mass of the ions), the two systems are related through a Lorentz transformation. In this configuration, the fragments produced are more energetic and can escape from the target, allowing the experimental setup to detect them ( a schematic view of this inversion is shown in Fig. \ref{fig: inversion}). This inverse kinematic approach has been successfully used in other experiments since the 1990s (\cite{ref: 1990}).

\begin{center}
\begin{tabular}{c c c c} 
 \hline
Fragments & $E[MeV]$ & $LET[keV/\mu m]$ & Range [$\mu m$]\\
 \hline\hline
$^{15}O$ & 1.0 & 983 & 2.3 \\
$^{15}N$ & 1.0 & 925 & 2.5\\
$^{14}N$ & 2.0 & 1137 & 3.6\\
$^{12}C$ & 3.8 & 912 & 6.2\\
$^{11}C$ & 4.6 & 878 & 7.0\\ 
$^{10}B$ & 5.4 & 643 & 9.9\\
$^{8}Be$ & 6.4 & 400 & 15.7\\
$^{6}Li$ & 6.8 & 215 & 26.7\\
$^{4}He$ & 6.0 & 77 & 48.5\\
$^{3}He$ & 4.7 & 89 & 38.8\\
$^{2}H$ & 2.5 & 14 & 68.9\\ 
 \hline
\end{tabular}
\captionof{table}{Average data for target fragments from a $180 MeV$ proton beam in water. \cite{ref:conceptual design}}\label{table: data proton}
\end{center}

Considering an ion beam moving along the positive $z-axis$ at constant velocity $\beta$ towards a proton, two reference frames can be defined: $S$ (the laboratory frame), where the ion is moving and the proton target is at rest, and $S'$ (the patient frame), where the
situation is reversed, i.e. the ion is at rest and the proton is moving with speed $\beta$ in the negative $z$ direction. Let $P=(\frac{E}{c}, p)$ represent the four-momentum of the ion in the S frame, where $E$ is the energy and $p$ is the three-momentum of the ion. Similarly, $P'=(\frac{E'}{c}, p')$  represents the four-momentum of the proton in the $S'$ frame. The relationship between the four-momenta $P$ and $P'$ in the two frames is given by the Lorentz transformation matrix ($\Lambda$), which can be expressed as:

\begin{equation}
\Lambda(\beta)=
\begin{pmatrix}
\gamma & 0 & 0 &- \beta \gamma \\
0 & 1 & 0 & 0 \\
0 & 0 &1 &0 \\
- \beta \gamma & 0 & 0 & \gamma\\
\end{pmatrix}
\end{equation}
 and so $  P'=\Lambda(\beta) P $, that in a more explicit way is:
\begin{equation}
\begin{pmatrix}
    E'/c \\ p'_x \\ p'_y \\ p'_z
\end{pmatrix}
=
\begin{pmatrix}
\gamma & 0 & 0 &- \beta \gamma \\
0 & 1 & 0 & 0 \\
0 & 0 &1 &0 \\
- \beta \gamma & 0 & 0 & \gamma\\
\end{pmatrix}
\cdot
\begin{pmatrix}
    E/c \\ p_x \\ p_y \\ p_z
\end{pmatrix}
= \begin{pmatrix}
    \gamma E/c - \beta \gamma p_z \\ p_x \\ p_y \\ -\beta \gamma E/c + \gamma p_z
\end{pmatrix}
\end{equation}

Is true also the inverse $P=\Lambda(\beta)^{-1} P'$, with $\Lambda(\beta)^{-1}$ the following inverse matrix:
\begin{equation}
\Lambda(\beta)^{-1}=
\begin{pmatrix}
\gamma & 0 & 0 & \beta \gamma \\
0 & 1 & 0 & 0 \\
0 & 0 &1 &0 \\
\beta \gamma & 0 & 0 & \gamma\\
\end{pmatrix}
\end{equation}

\subsection{Target}
As a consequence of using the inverse kinematic approach, achieving a cross section with a maximum uncertainty of $5\%$ requires a few percent level of accuracy in the measurements of the energy and momentum of the produced fragments, as well as a resolution in the emission angle of the order of a few $mrad$. To obtain such precision in the angle, it is crucial to have high accuracy in tracking both the primary particles and the fragments, while also minimizing MCS as much as possible. To achieve this latter objective, a thin target is necessary. However, it is important to note that using a thin target also reduces the probability of fragmentation events.

The target materials are selected to simulate human tissue, with carbon, oxygen, and hydrogen being of primary interest. Since the experiment has been conceived to acquire data at relatively low beam rates ($\sim 5-10 kHz$), a gaseous target would imply very low reaction rates and, consequently, lead to excessively long acquisition time. Moreover, the FOOT experiment usually takes place in clinical facilities and it would not be trivial to employ an hazardous material target in such structures.  The same considerations apply to the use of liquid hydrogen and oxygen targets, which would also require a cryogenic system. This represents an important issue for measurements on $H$ and $O$ targets in both direct and inverse kinematics.

The solution proposed by the FOOT collaboration is to employ both mono-atomic (e.g. graphite, $C$) and composite targets, like PolyEthylene ($C_2 H_4$ ) or PolyMethylMethAcrylate (PMMA, $C_5 O_2 H_8$ ), and then extract single cross sections through the subtraction method. This involves taking data with two different targets: for example one made of carbon and the other made of polyethylene. The hydrogen cross section is then determined by the following equation:
\begin{equation}
    \sigma(H)=\frac{1}{4} ( \sigma(C_2H_4) -2 \sigma(C))
\end{equation}
 This is also true for the differential cross sections:
\begin{equation}
   \frac{ d\sigma(H)}{dE}=\frac{1}{4} \left ( \frac{d\sigma(C_2H_4)}{dE}-2 \frac{d\sigma(C)}{dE} \right )
\end{equation}
\begin{equation}
   \frac{ d\sigma(H)}{d\Omega}=\frac{1}{4} \left ( \frac{d\sigma(C_2H_4)}{d\Omega}-2 \frac{d\sigma(C)}{d \Omega} \right )
\end{equation}

\begin{figure}[h!]
\begin{center}
\includegraphics[width=11cm]{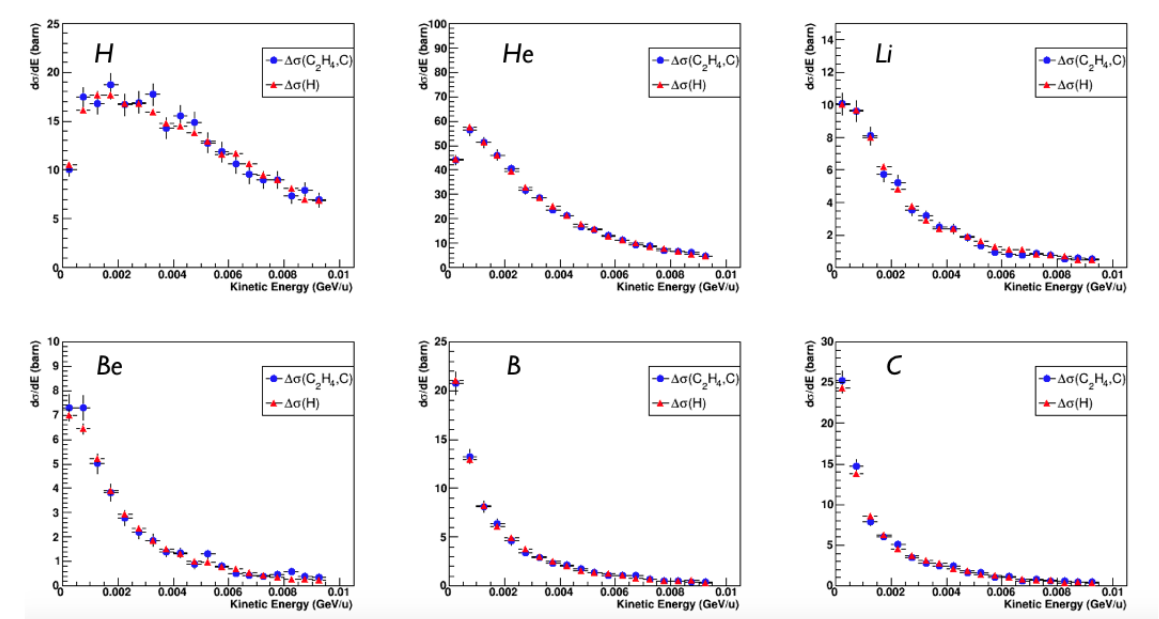}
\caption{ Energy differential cross-section of a $200 MeV/u$ $^{12}C$ beam on hydrogen target in inverse kinematics. The estimations performed with the $\Delta \sigma (C_2H_4, C)$ and $\sigma(H)$methods are reported as blue dots and red triangles respectively. \cite{ref:conceptual design}}
\label{fig: subtraction}
\end{center}
\end{figure}

This subtraction method has been previously employed in other experiments (example: \cite{ref: subtraction cross section} ). To verify the validity of this approach, a simulation was conducted comparing the cross section on a hydrogen target with the cross section obtained using the subtraction method. The results, shown in Figure \ref{fig: subtraction}, demonstrate good agreement between the two, thereby confirming the reliability of the method. The primary concern with this technique is that uncertainties might become more significant. However, the detectors used in this experiment are designed for high precision, so the impact of these uncertainties is expected to be minimal.

\section{Magnetic Spectrometer}
The FOOT experiment with the magnetic spectrometer is optimized for heavier fragments (heavier than $^4 He$) and has an angular acceptance up to a polar angle of approximately $10^{\circ}$ with respect to the beam axis. The setup can be divided into three distinct regions:
\begin{itemize}
    \item \textbf{Upstream Region}: The pre-target region, used to monitor and track the primary beam.
    \item \textbf{Interaction and Tracking System}: The region encompasses the target and subsequent detectors placed both upstream, between and downstream of two permanent magnets. Its primary purpose is to track the fragments produced in the target.
    \item \textbf{Particle Identification (PID) Region}: The final region, responsible for measuring the kinetic energy of the fragments.
\end{itemize}

\begin{figure}[h!]
\begin{center}
\includegraphics[width=11cm]{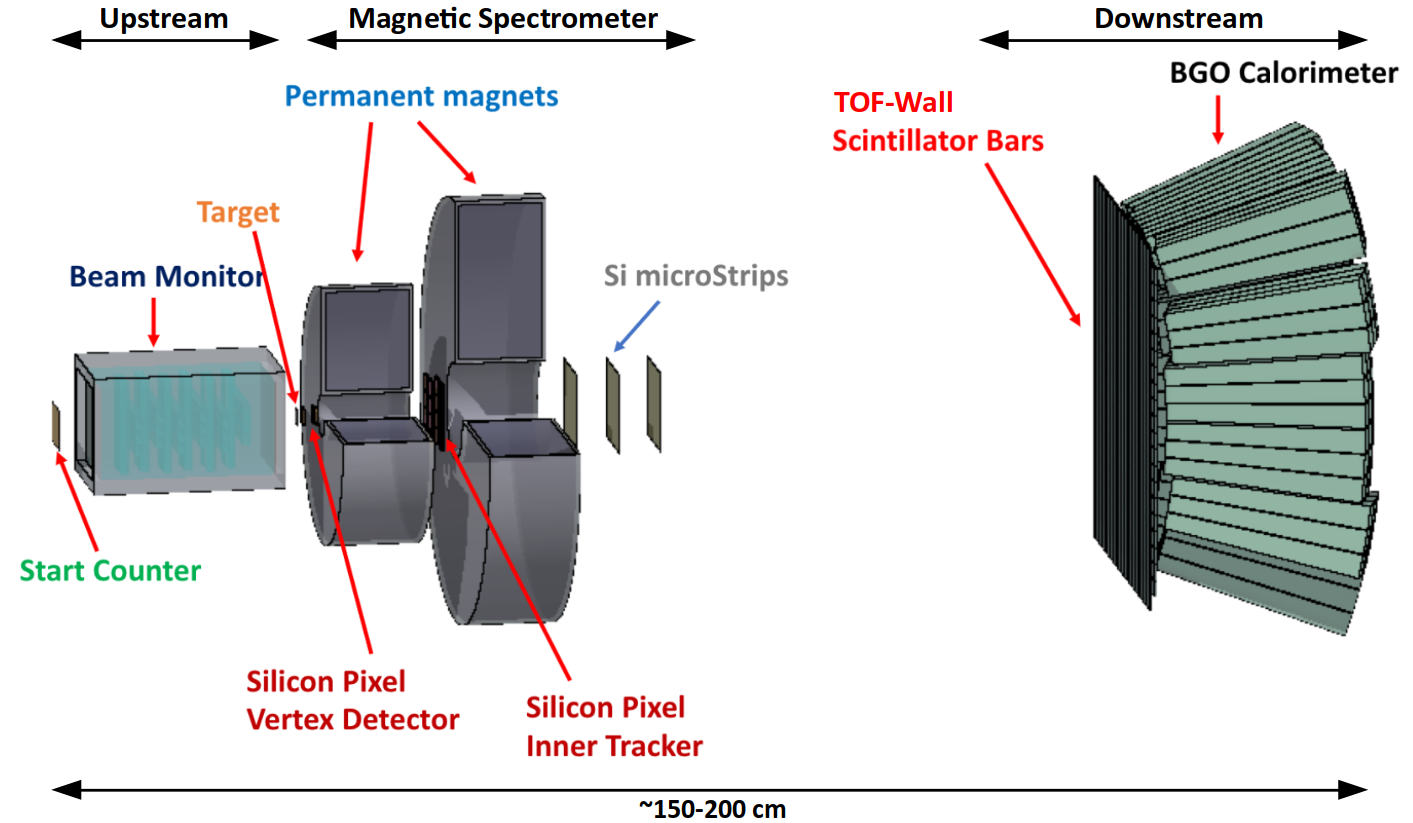}
\caption{Schematic view of foot (Magnetic Spectrometer setup). \cite{ref:mesuring}}
\label{fig: foot setup}
\end{center}
\end{figure}

The setup has been designed to be sufficiently compact, allowing it to be transported to different facilities. The entire experiment is contained within a range of 2-3 meters. The final configuration ( reported in Fig.:\ref{fig: foot setup}) is the result of studies using FLUKA simulations to optimize the dimensions of the detectors. These simulations were crucial in ensuring that the setup meets the required angular acceptance.

A crucial aspect of the FOOT experiment is the accurate identification of the charge and isotope of the produced fragments. In particular the possibility to use inverse kinematic approach to perform cross section measurements relies on a very good accuracy in the reconstruction of the trajectories of primaries before the interaction and of the produced fragments. As stated in \cite{ref:conceptual design}, the correct application of the Lorentz transformation is only possible if the emission angles of fragments can be measured with a maximum uncertainty at the level of $1 mrad$. To achieve the required level of accuracy, the experiment employs multiple methods for particle identification, to determine charge and mass of the fragments in different ways to keep the systematic errors in the calculations as low as possible. The setup includes Time of Flight (TOF) measurements, a calorimeter for fragment energy measurements $E_k$. These measurements are complemented by data from the magnetic spectrometer,indeed with the magnetic field is possible to measure the fragments' rigidity ($\frac{p}{Z}$).

\begin{itemize}
    \item The \textbf{mass number A} of the particles is determined with three different approaches based on the concurrent measurements of momentum $p$, $TOF$ and kinetic energy $E_k$. These quantities can be combined two-by-two to obtain three measurements of the particle mass number
    \begin{equation}
        A_1=\frac{E_k}{U c^2 (\gamma -1)} \quad , \quad A_2= \frac{p^2c^2-E_k^2}{2Uc^2E_k} \quad , \quad A_3=\frac{p}{Uc\beta \gamma}
    \end{equation}
with $U=931.5 meV$ is the Unified Atomic Mass and  $\beta$ and $\gamma$ the Lorentz factors are evaluated from TOF measurements. The final $A$ will be calculated through a fitting procedure, like $\chi ^2$ minimization or an Augmented Lagrangian Method \cite{ref:optimas}. The expected resolution for mass measurement with this approach ranges from about $3\%$ to $6\%$ \cite{ref:conceptual design}.

\item The \textbf{charge Z} identification is performed through the measurements of the energy loss ($\Delta E$) of the fragments that reach the TOF-wall and of the TOF.
Knowing the path length ($L$) it is possible to determine the velocity ($\beta c= L/TOF$). The atomic number Z of the particle is then calculated from the Bethe-Bloch formula \ref{eq:bethe-bloch}. The redundancy of charge identification is achieved by measuring the energy loss in other detectors, the silicon trackers. The final resolution expected on $Z$ should range from $6\%$ for $^1H$ to about $2\%$ for $^{16} O$ nuclei.
\end{itemize}

With the mass ($m$) and charge ($Z$) identification, the fragment is uniquely identified.

To achieve the precision required for accurate cross-section measurements, it is essential to meet the following experimental resolution \cite{ref:mesuring} :
\begin{itemize}
    \item $\sigma(p)/p$ at level of $4-5 \%$
    \item $\sigma(TOF)$ at level 100 ps
    \item  $\sigma(E_k/E_k)$ at level $1-2\%$
    \item  $\sigma(\Delta E) /\Delta E$ at level of $5\%$
\end{itemize}

\subsection{Upstream region}
The upstream region, located before the target, plays a crucial role in counting the number of incoming ions, determining their direction, and identifying their point of incidence on the target. To achieve these objectives, two detectors are employed: the \textbf{Start Counter} and the \textbf{Beam Monitor}. Special care is taken in designing this region to minimize the amount of material that the beam crosses, thereby reducing MCS and minimizing any pre-target fragmentation.

\subsubsection{Start Counter}
\begin{figure}[h!]
\begin{center}
\includegraphics[width=6cm]{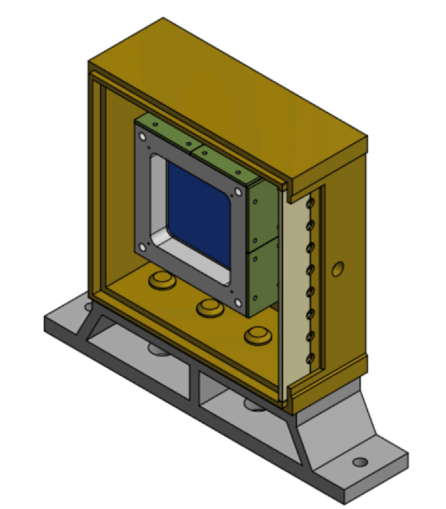} \quad \includegraphics[width=6cm]{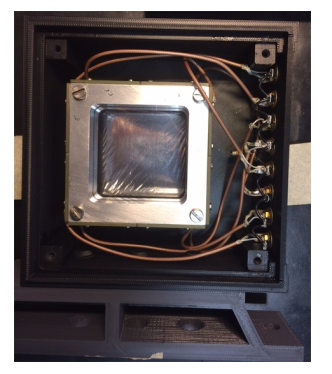}
\caption{ \textbf{left:} Schematich view of the Start Counter detector  \textbf{right:} Picture of the Start Counter detector in its mechanical frame and outer casing.}
\label{fig: start counter}
\end{center}
\end{figure}

The Start Counter (SC) (Fig.:\ref{fig: start counter}) consist of a square foil of EJ-228 plastic scintillator, $250 \mu m$ thick, manufactured by Eljen Technology (Sweetwater, Texas). The scintillator is held in place by an aluminum frame enclosed within a black 3D-printed box. This box has two square windows aligned with the scintillator and is important also to shield the detector and the SiPMs from environmental background light. The active surface area of the SC is $5\times 5 cm^2$, designed to cover the typical beam size used in the experiment. It is placed at about $20-30 cm$ upstream of the target.

When a particle passes through the scintillator, the light it generates is collected by 48 Multi-Pixel Photon Counter (MPPC) SiPMs, arranged 12 per side and each having a surface area of $3 \times 3 mm^2$. These SiPMs are grouped into six and are read by electronic channels. The readout and power supply for the SiPMs are handled by the WaveDAQ system \cite{ref:WaveDAQ}. Tests conducted at GSI with a carbon beam at $700 MeV/u$ have demonstrated a time resolution on the order of $ \sigma _t \simeq 60 ps$. To ensure consistent performance in terms of time efficiency and resolution with different beam particles, the SC can be equipped with different thicknesses of scintillator, ranging from $250 \mu m$ to $1 cm$.

This detector is responsible for measuring the incoming flux (with an efficiency close to 1 up to $1kHz$) and providing the reference time for other detectors. In conjunction with the TOF-detector (TOF Wall, see \ref{subsubsect:TOF Wall}), it measures the time of flight (TOF) and serves as the minimum bias trigger for the experiment.

The SC was developed through a collaboration between the University La Sapienza, Centro Fermi, and the INFN section of Rome.

\subsubsection{Beam monitor}
\begin{figure}[h!]
\begin{center}
\includegraphics[width=6cm]{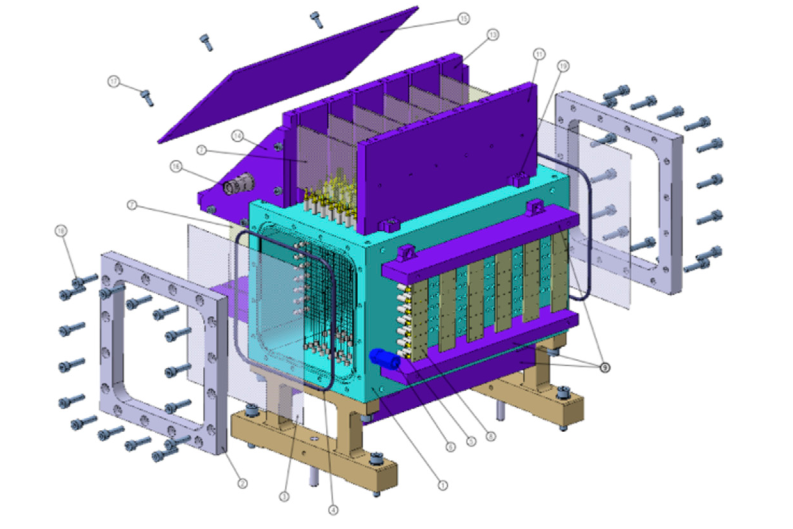} \quad \includegraphics[width=6cm]{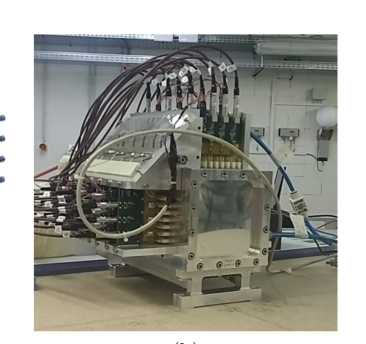}
\caption{ \textbf{left:}Technical drawing of the Beam Monitor drift chamber. The two orthogonal views x–y of the wires are clearly visible. Two enclosing mylar windows held by aluminum
frames are shown as well. \textbf{right:} picture of the
system during a data acquisition campaign.}
\label{fig: beam monitor}
\end{center}
\end{figure}

The second detector in the upstream region is the Beam Monitor (BM)( Fig.:\ref{fig: beam monitor}), a drift chamber consisting of 12 layers  of alternating horizontal and vertical wires, with each layer containing three drift cells. The drift cells are rectangular, measuring $16 mm \times 11 mm^2$, and to resolve left-right ambiguity, two consecutive layers are staggered by half a cell.  The chamber has a transverse active area of $5.6 \times 5.6 cm^2$, and a total length along the beam line of $21 cm$ with an active region of $13 cm$. The beam entrance and exit windows are made of $100 \mu m$ thick mylar foils. The BM is filled with an $80/20\%$ mixture of $Ar/CO_2$ at approximately $0.9 bar$.  The design of this detector was motivated by the need to minimize the amount of material before the target.

The BM was accurately characterized during a dedicated data taking in 2020 at the Trento Proton Therapy Center. In this occasion, proton beams with energy ranging from 80 to 220 $MeV$ were used to evaluate the detector's spatial and angular resolution. The data acquired show that the BM has a hit detection efficiency of approximately $93\%$. The spatial resolution of the drift chamber in the central part of the cell ranges from 150 to 300 $\mu m$, which corresponds to an angular resolution of 1.6 to 2.1 $mrad$ for the highest and lowest beam energies, respectively. A detailed description of the analysis performed to extract the BM performance is provided in \cite{ref:BM}. Since these results have been obtained using proton beams, the BM is expected to work at least at the same level of accuracy with FOOT primaries, which are typically $^{12}C$ and $^{16} O$.

Positioned after the SC and before the target, the BM is crucial for determining the direction and impact point of the primary particle on the target to properly apply the Lorentz boost needed for inverse kinematics measurements. Additionally, the BM plays a key role in discriminating events such as pre-target fragmentation (i.e., fragmentation occurring in the SC or the first layers of the BM). It is also essential for identifying pile-up events in the vertex by matching the BM track with the reconstructed vertex. In fact the BM read-out time, of the order of $1 \mu s$ or less, is fast enough to ensure that tracks belonging to different events are not mixed, unlike the VTX detector which has a readout time of approximately $ 187\mu s$.

The BM has been inherited from the FIRST experiment at GSI \cite{ref:FIRST}, and like for the SC they have been already employed in several data takings and are now in their final configuration.

\subsection{Interaction and Tracking System}
The magnetic spectrometer of FOOT is responsible for tracking and momentum determination of the produced nuclear fragments. Placed after the target, it consists of three distinct parts, categorized based on their location relative to the two permanent magnets (upstream, between and downstream of the two magnets). Beginning from the \textbf{target} (TG), the system is composed of the following components: \textbf{Vertex Detector} (VTX), first magnet, \textbf{ Inner Tracker} (ITR), second magnet and \textbf{ Micro Strip Detector} (MSD). The main objective of this section is to extract the momentum of fragments passing through the setup by analyzing how their trajectories are deflected by the magnetic field.

\begin{figure}[h!]
\begin{center}
\includegraphics[width=11cm]{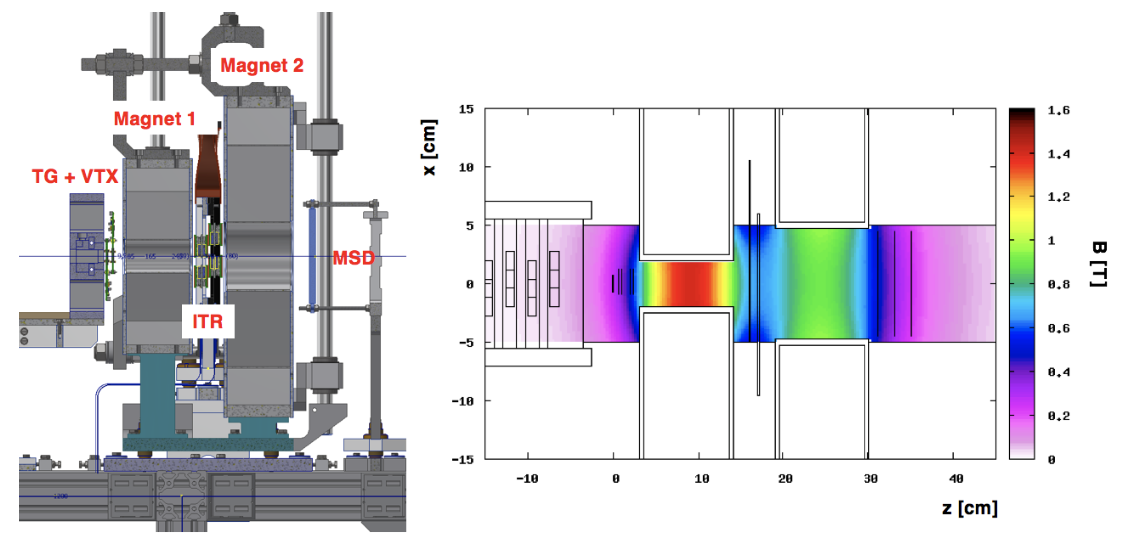}
\caption{ \textbf{left}: Technical design of the interaction and tracking regions: the vertical axis is the y axis, while the horizontal axis is the z axis. The beam coming from the left, along the z-axis, cross sequentially TG, VTX, moves into the magnets region and crosses the ITR and, immediately after the second magnet, passes through the MSD. \textbf{right:} Computed magnetic ﬁeld map produced by the FOOT magnets in Halbach conﬁguration. The magnetic ﬁeld intensity $B$, shown in the palette, is referred to its y-axis component.}
\label{fig: interaction region}
\end{center}
\end{figure}

The layout of these regions is illustrated in the figure \ref{fig: interaction region}.

\subsubsection{The magnetic system}
\begin{figure}
\includegraphics[width=0.40\textwidth]{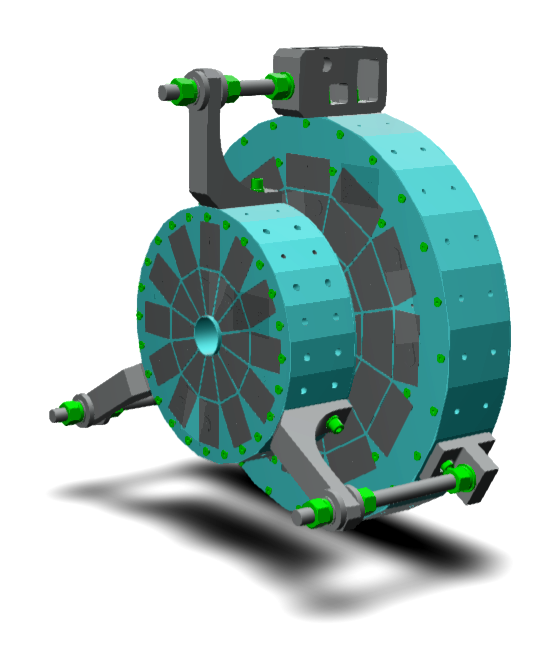} 
\quad \quad
\includegraphics[width=0.35\textwidth]{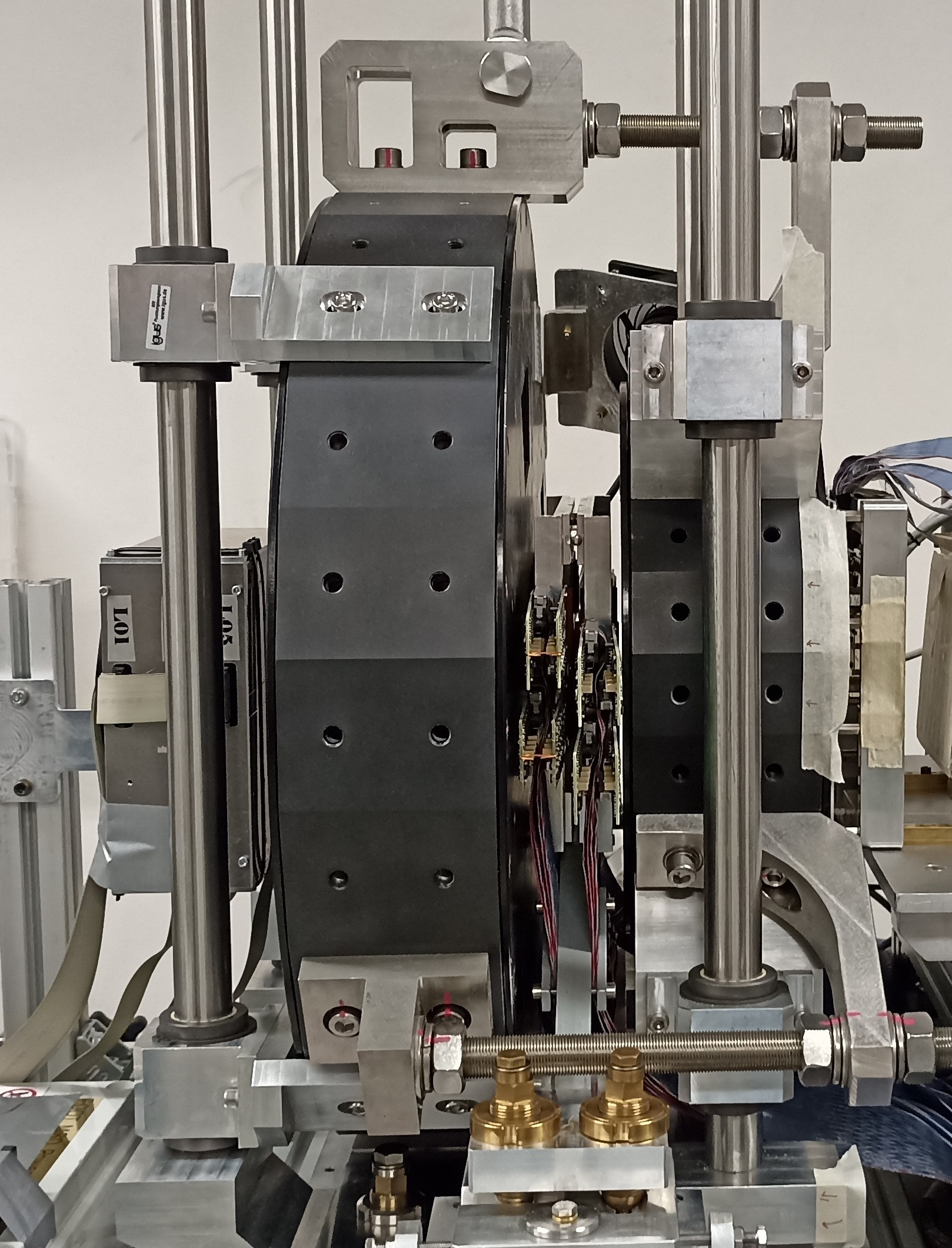}

\caption{ \textbf{left}: Technical drawing of the magnetic system, with the 12 pices organized in the Halbach configuration. \textbf{right}: Picture of the system. }
\label{fig: magnets}
\end{figure}

A key element of the FOOT spectrometer is its magnetic system (Fig.: \ref{fig: magnets}), designed to bend the fragments produced in the target and enable the extraction of their momentum by analyzing the deflection of their trajectories within the magnetic field. The spectrometer's design balances the need for precise momentum resolution with the requirement of maintaining a compact and portable apparatus.

For a particle of charge $q$ traveling through a magnetic field $B$ over a region of length $L$ , the variation of transverse momentum ($\Delta p _T$) is given by:
\begin{equation} \label{eq:delta p}
    \Delta p _T= q \int_0 ^L B dl
\end{equation}
The resolution of momentum measurements improves as the variation in transverse momentum $\Delta p _T$ increases.

Preliminary feasibility studies using MC simulations resulted in the configuration shown in Fig.: \ref{fig: magnets}. The final design includes two magnets made of twelve single pieces arranged in a Halbach configuration. This geometry has the advantage of minimizing the magnetic field outside the spectrometer while ensuring a nearly uniform field along the transverse  x-y planes. As a matter of fact, the field has been measured to be uniform at the percent level up to a distance of $3 cm$ from the centers of the magnets. The magnets are constructed from $SmCo$ (Samarium-Cobalt), a material that retains its magnetic properties even in high-radiation environments. To meet the required momentum resolution and maintain an angular acceptance of $10^\circ$ for the emitted fragments, two different magnet sizes were selected. The first magnet has a gap diameter of $\sim 3 cm$ and can generate a maximum magnetic field of $1.4 T$ within its cylindrical hole. The second magnet, with a gap diameter of $\sim 6 cm$, provides a field intensity of up to $0.9 T$. The entire system cover a longitudinal distance of approximately $ 30 cm$, with a distance between the two magnet of $5 cm$, enough to host the Inner Tracker detector in between, which will be subjected to a magnetic field of $\sim 0.6 T$. Te magnetic field is oriented along the positive Y-axis, as shown in the computed magnetic map (Fig.:\ref{fig: interaction region}, right). 

In the final configuration the magnets weigh between 200 and $300 kg$. A mechanical support has been developed to withstand the magnetic forces and ensure precise alignment with the tracking stations. This support system is designed with the ability to move the magnets out of the beam line for alignment studies. The magnets and their mechanics were completed in late 2023 and have been tested and employed in a data taking campaign at the CNAO facility.

\subsubsection{Vertex detector}\label{subsubsection: vertex}
Both the Vertex Detector (VTX) and the Target (TG) are housed within a mechanical structure designed to accommodate up to five different targets.

\begin{figure}[h!]
\begin{center}
\includegraphics[width=8cm]{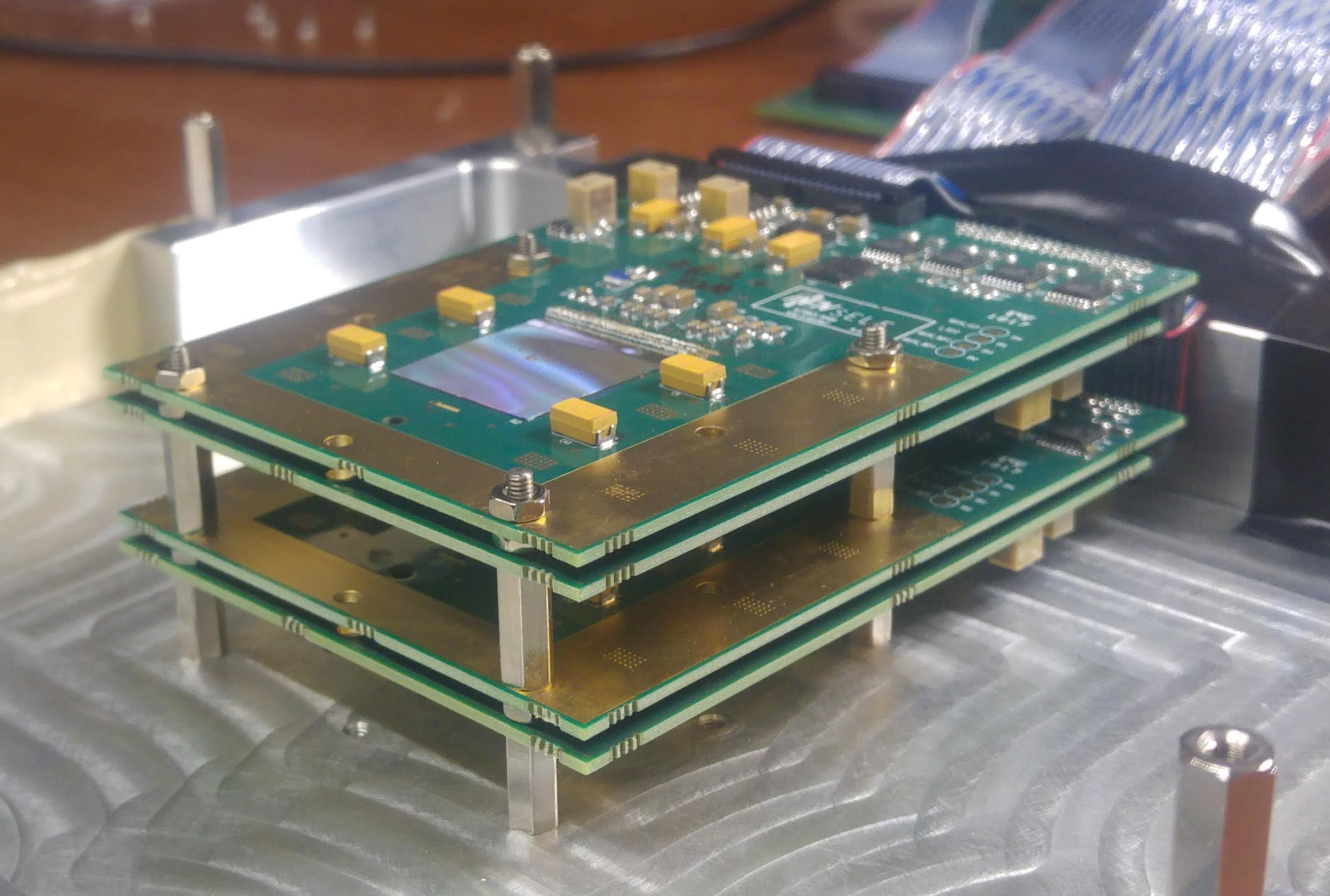}
\caption{ Picture of the detecting planes of the Vertex showing the M28 chips.}
\label{fig: vtx}
\end{center}
\end{figure}

The VTX consist of four layers (Fig.:\ref{fig: vtx}), each with dimensions of $2\times 2cm^2$. To fulfill the stringent requirements for a low material budget, high precision, and efficiency, the technology of MIMOSA-28 (M28) Monolithic Active Pixel Sensors (MAPS) (chips developed by the Strasbourg CNRS PICSEL group \cite{ref: PICSEL}) has been selected. Each M28 sensor features a matrix consisting of 928 rows by 960 columns, with a pixel pitch of $20.7 \mu m$, covering a total active area of $20.22 mm \times 22.71 mm$. The epitaxial layer has a thickness of $15 \mu m$, and each sensor is thinned to $50 \mu m$, resulting in a total material budget of $200 \mu m$ for the entire VTX. 

The four layers are organized into two groups, each consisting of two layers. The distance between the two groups is $\sim 15 mm$, while the separation between the two layers within each group is $\sim 3 mm$. This configuration allows for an angular acceptance of about $40 ^\circ$ from the beam axis for the emitted fragments. The sensor operate using a rolling shutter readout technique with a $185.6 \mu s$ frame readout time.

The goal of the VTX is to accurately track the particles exiting from the target. As particles pass through the VTX, they are detected by the four layers of sensors. By analyzing the hits recorded on these layers, particle tracks can be reconstructed with high precision. The VTX achieves a remarkable spatial resolution of $5 \mu m$ \cite{ref: cmos alignement}, which, when combined with the information from the BM, ensures that the angular measurements are sufficiently precise to meet the experimental requirements. This high level of precision also allow to minimizing the effects of multiple scattering, thanks to the reduced material budget of both the BM and the VTX.

\subsubsection{Inner Tracker}

Between the two magnets lies the Inner Tracking (ITR). It is designed with two planes of pixel sensors to accurately track fragments within the magnetic region, providing information on the direction and transverse position of particle. The detector consist of 32 M28 chips, the same type used in the VTX. This sensors  are expected to maintain their tracking performance even in the residual magnetic field present between the permanent magnets. However, unlike the VTX, the larger detector area of the ITR necessitates a mechanical support structure, which slightly increases the overall material budget.

\begin{figure}[h!]
\begin{center}
\includegraphics[width=7cm]{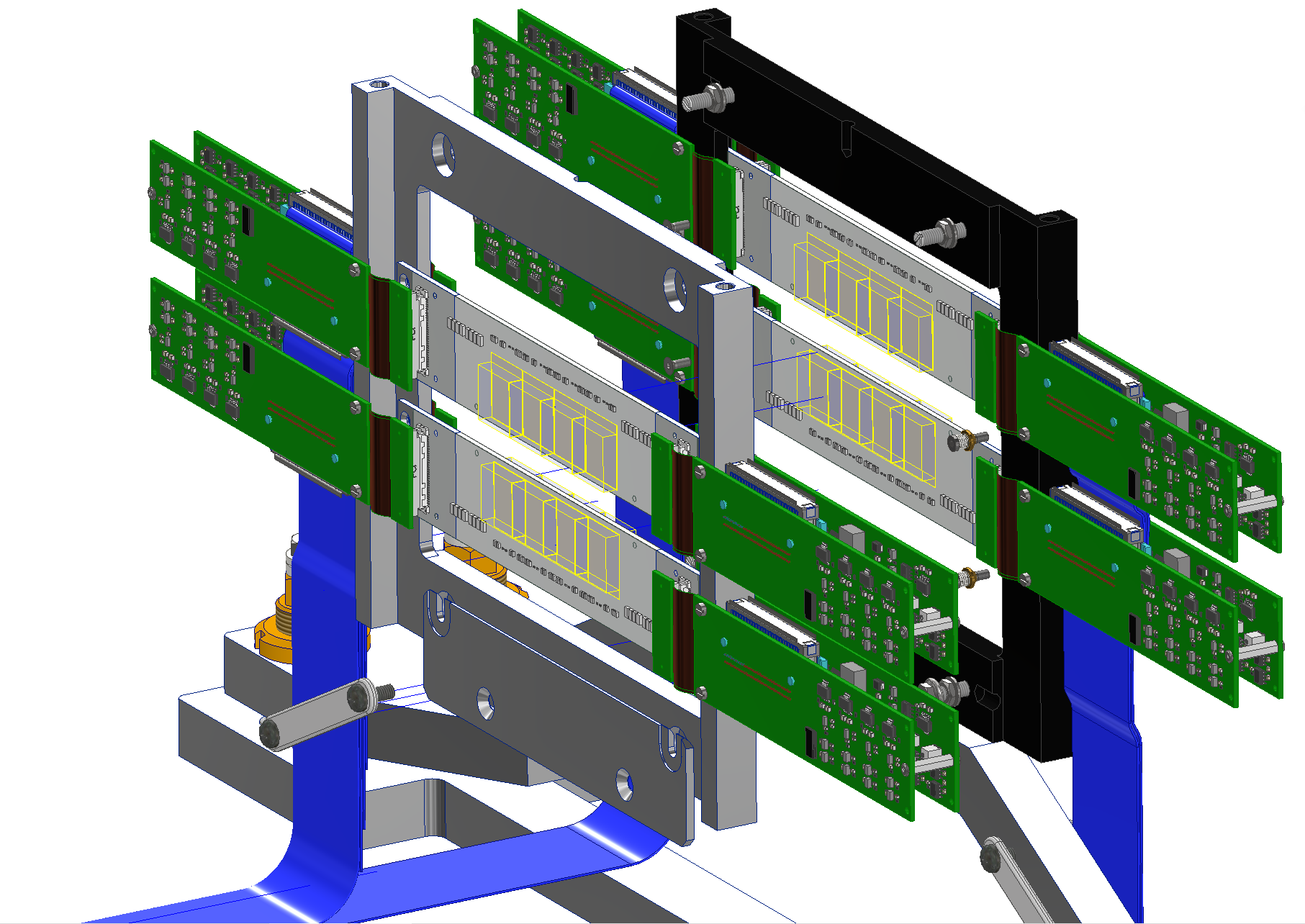}
\caption{Technical drawing of the Inner Tracker in its mechanical structure.}
\label{fig: itr}
\end{center}
\end{figure}

The ITR will be constructed using ladders similar to those developed in the PLUME project \cite{ref:PLUME}. Each ladder features a double-sided layout, consisting of two modules of M28 sensor layers attached to opposite sides of a support structure. This support structure is a $2 mm$ thick low-density silicon carbide (SiC) foam, chosen for its minimal impact on the material budget. Each module consist of four M28 sensors, which are glued and bonded to a flexible Kapton-metal cable. The ITR will consist of four ladders, two for each plane, mounted on an aluminum  frame that supports the entire tracking system (Fig.: \ref{fig: itr}).  To prevent the overlap of dead zones, the two planes of each ladder are laterally staggered. As mentioned earlier, the M28 chips have an active area of $2 \times 2 cm^2$ ,  giving the ITR a total transverse area of $8\times 8 cm^2$. This configuration balances the required angular acceptance, granularity, and tracking performance while adhering to the low material budget constraints. 

The detector has been completed and tested for the first time at the Beam Test Facility in Frascati, using $e^-$ beams, and at the CNAO facility in 2023, with $p$ and $^{12}C$ ions. The full characterization of the device is currently ongoing.

At present the VTX and IT are employed in FOOT only for particle tracking. According to the study reported in \cite{ref: M28}, the M28 chips demonstrate a precise correlation between the energy deposited in the active layer and the number of pixels that are fired. This characteristic suggests that these sensors could potentially be used for particle charge identification,which is currently performed by other detectors. In the future, this feature of the M28 chips could be exploited as an additional tool for the experiment.

\subsubsection{Micro Strips detector}
The Microstrip Silicon Detector (MSD) is the final detector in this section, positioned after the second magnet. Its primary function is to track the particle fragments after they pass through the magnetic region, which is crucial for measuring their momentum. Additionally, the MSD is used for the matching of reconstructed tracks with hits in the Tof-Wall (TW) and the calorimeter. MSD also provide a redundant measure, the energy loss per unit distance ($dE/dx$), which is used for identify the charge ($Z$) of the fragments, complementary to the charge identification performed by the TW.

\begin{figure}[h!]
\begin{center}
\includegraphics[width=6cm]{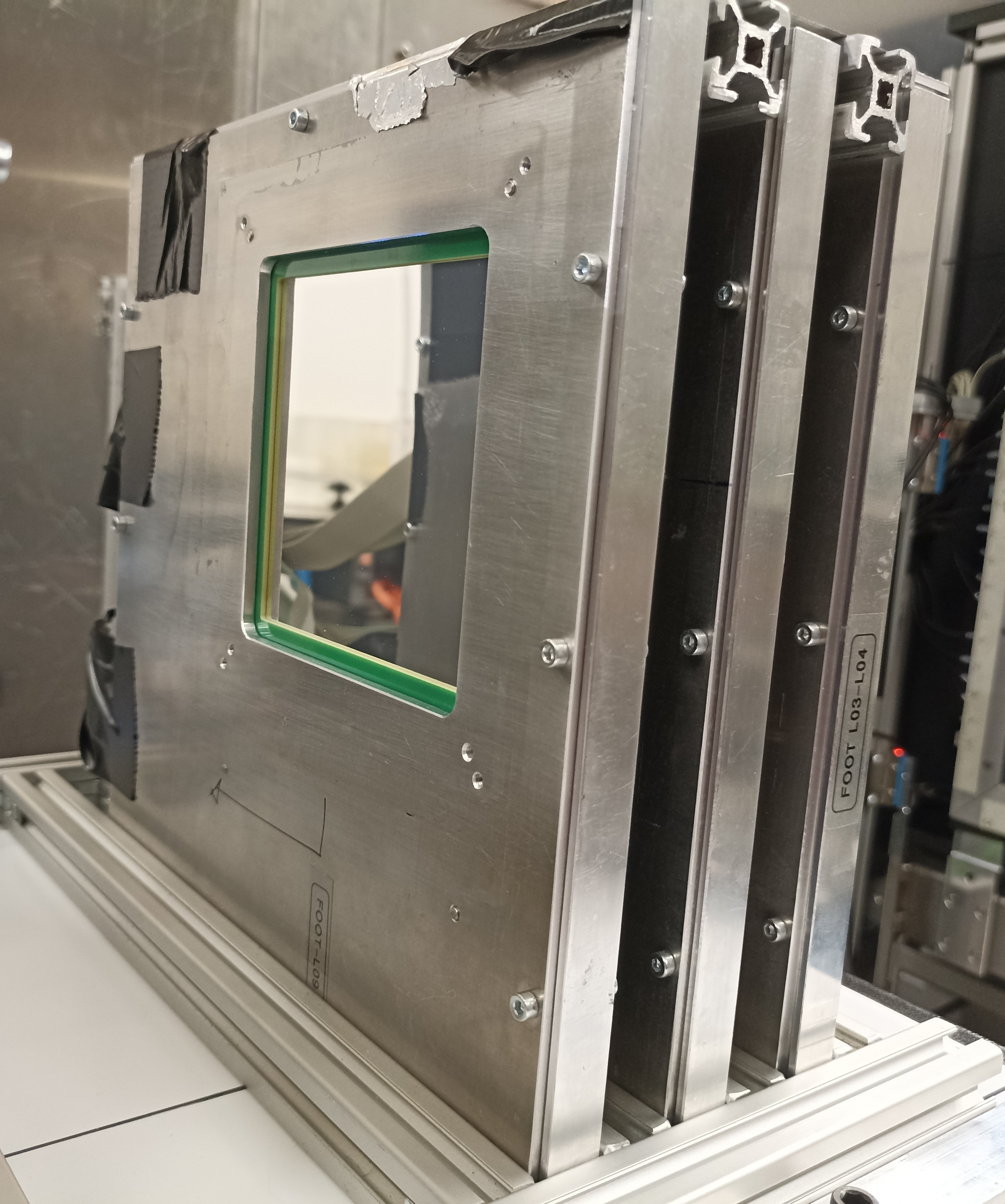}
\caption{Picture of the MSD detector}
\label{fig: msd}
\end{center}
\end{figure}

The detector (Fig.: \ref{fig: msd}) is composed of six Single-Sided Silicon Detectors (SSSD) grouped in three stations of alternatively orthogonal layer, each with an active area of $9.6 \times 9.6 cm^2$. The layers are separated by a $ \sim 2 cm$ gap along the beam direction, ensuring sufficient angular acceptance to detect ions with $Z>2$, as predicted by FLUKA simulations. Each SSSD have a thickness of $150\mu m $ thick, for a total of $900 \mu m$ of silicon on the beam line. These sensors, manufactured by Hamamatsu Photonics, are mounted on a hybrid Printed Circuit Board (PCB) that provides both mechanical support and an interface for readout.

Each layer is divided in 1920 strips with a width of $50 \mu m$. To reduce the number of readout channels while maintaining good spatial resolution ($\sim 43 \mu m$), the MSD employs a floating strip readout method. In this approach, only one out of every three strips is connected to the readout electronics, resulting in a final readout pitch of $150 \mu m$.

The MSD detector has been completed and employed in several data takings. Beam tests using proton, $^{12}C$and $^{16}O$ beams have demonstrated that the current detector configuration achieves a spatial resolution ranging from $10$ to $35 \mu m$, surpassing the expected performance \cite{ref: msd}.

\subsection{Particle Identification Region}
The final part of the FOOT setup is the particle identification (PID) region, located at least 1 meter downstream from the target. This region plays a crucial role in identifying and characterizing the particles produced during the experiment. It consists of two main detectors: the \textbf{Time-of-Flight Wall} (Tof-Wall) detector and a \textbf{calorimeter}.

\subsubsection{Tof-Wall Detector}\label{subsubsect:TOF Wall}
The Time-of-Flight Wall (TW) detector is designed to achieve three main objectives: measuring the energy deposited ($\Delta E$) by particles, calculating their time of flight (TOF) using the initial timing reference provided by the SC, and determining the precise hit position. By simultaneously measuring $\Delta E$ and TOF, the TW enables accurate identification of the charge $Z$ of the ions that impact the detector. 

\begin{figure}[h!]
\begin{center}
\includegraphics[width=4cm]{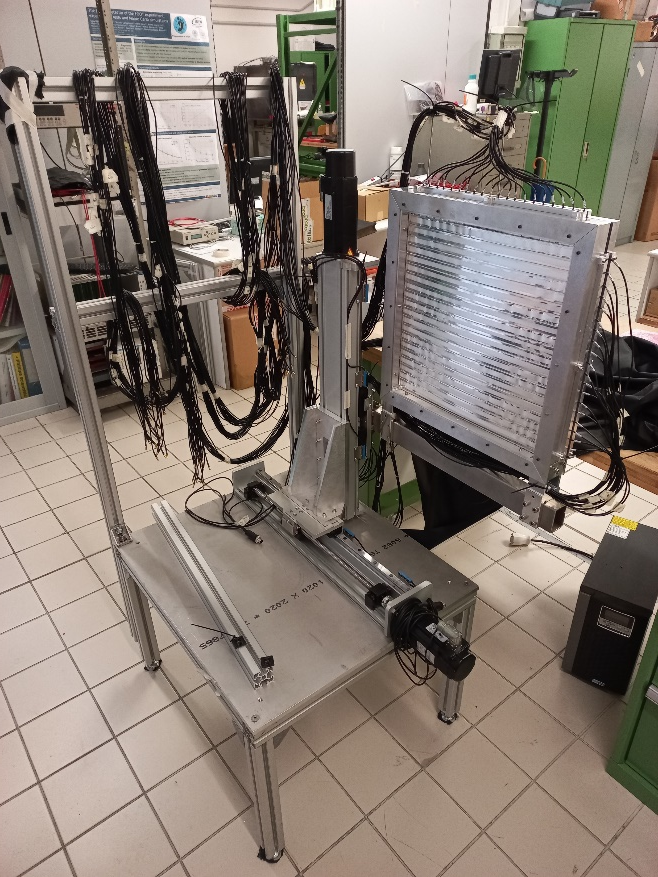} \quad \includegraphics[width=5cm]{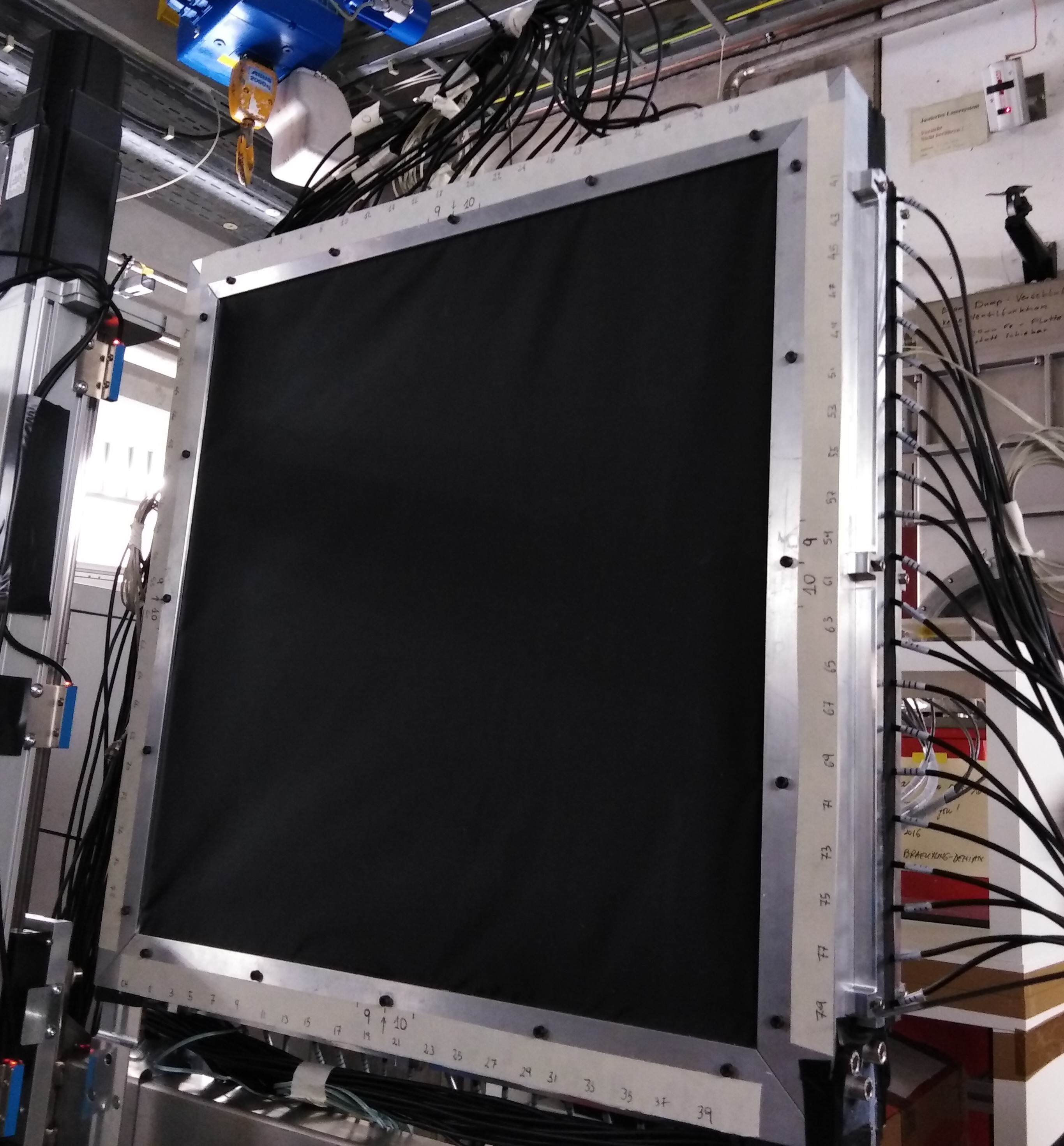}
\caption{ \textbf{left:} picture of the TW with its mechanical support and \textbf{right:} detector on the beam line with its shielding from background light.}
\label{fig: tw}
\end{center}
\end{figure}

The TW is composed of two orthogonal layers, each consisting of 20 plastic scintillator bars made from EJ-200 material, provided by Eljen Technology. These bars are  wrapped in reflective aluminum and darkening black tape to optimize the collection of scintillation light and to shield it from background light(Fig.: \ref{fig: tw}). Each bar measures $0.3 cm$ in thickness, $2 cm$ in width, and $44 cm$ in length. The two layers, arranged perpendicularly to form an x-y grid, create a $40 \times 40 cm^2$ active detection area, covering approximately $\sim 10 ^{\circ}$ angular acceptance at $1 m$ from the target. The resulting $2 \times 2 cm^2$ granularity matches that of the downstream BGO crystals in the calorimeter and keeps the occurrence of multiple fragments hitting the same bar below $1\%$. The thickness of the scintillator bars was selected as a compromise between two key factors: generating a strong scintillation signal for improving both timing and energy resolution, and minimizing the probability of secondary fragmentation within the bars, which could otherwise interfere with accurate particle identification and tracking. Each end of the bars is coupled with four Silicon Photomultipliers (SiPMs) (model MPPC S13360-3025PE2), which feature a $3 \times 3 mm^2$ active area and a $25 \mu m$ microcell pitch. For each signal the entire waveform is recorded, allowing for precise offline extraction of both time and charge information. 

The TW has been designed to meet performance criteria, including a time-of-flight resolution of $\sim 70 ps$ and an energy loss resolution of approximately $4\%$ for the heaviest fragments ($C$ or $O$ at $200 MeV/u$). The maximum uncertainty on the hit position is simply dictated by the detector's granularity. However, it has already been shown that the time information given by the signals collected at each side of the bars can be exploited to improve the accuracy on the particle hit position, it can reach a positional accuracy better than $8 mm$ \cite{ref: de-tof}.

TW data are also critical for the global reconstruction of particle tracks as the bending of a fragment's trajectory is dependent on its charge (see eq.: \ref{eq:delta p}). The information from TW is used as a seed for track extrapolation and fitting in the global reconstruction algorithm. Furthermore an accurate charge measurement is essential for proper momentum evaluation.

Previous studies, such as \cite{ref: charge} and \cite{ref: calib tw}, have demonstrated that the TW meets the experimental requirements, even reaching TOF resolution of the order of $50 ps$ for the heavier ions.

The TW is in its final configuration and has been already employed in all acquisition campaigns.

\subsubsection{Calorimeter}
The final detector in the FOOT experiment is the calorimeter (CALO), whose primary objective is to measure the kinetic energy of the fragments. This measurement is crucial for determining the mass number $A$ of the particles.

Different processes dominate the interaction depending on the energy range. At higher energy levels ($700-800 MeV/u$, relevant for RPS), the pion production threshold is exceeded, leading to hadronic showering. Since fully containing these showers within a reasonably sized calorimeter is not feasible, the energy resolution at such high energies is worst. For fragmentation studies involving $^{12}C$ and $^{16}O$ at energies up to $200 MeV/u$, the dominant mechanism is electromagnetic interactions with target electrons and nuclei. In this case, proper containment of the fragments can be achieved, and consequently improved energy resolution. It should be noted, however, that in all cases, neutron production occurs in a fraction of events, leading to energy leakage and a systematic error that degrade the overall energy resolution. The impact of this effect can be mitigated by using the redundant information provided by other detectors in the setup.

\begin{figure}[h!]
\begin{center}
\includegraphics[width=6cm]{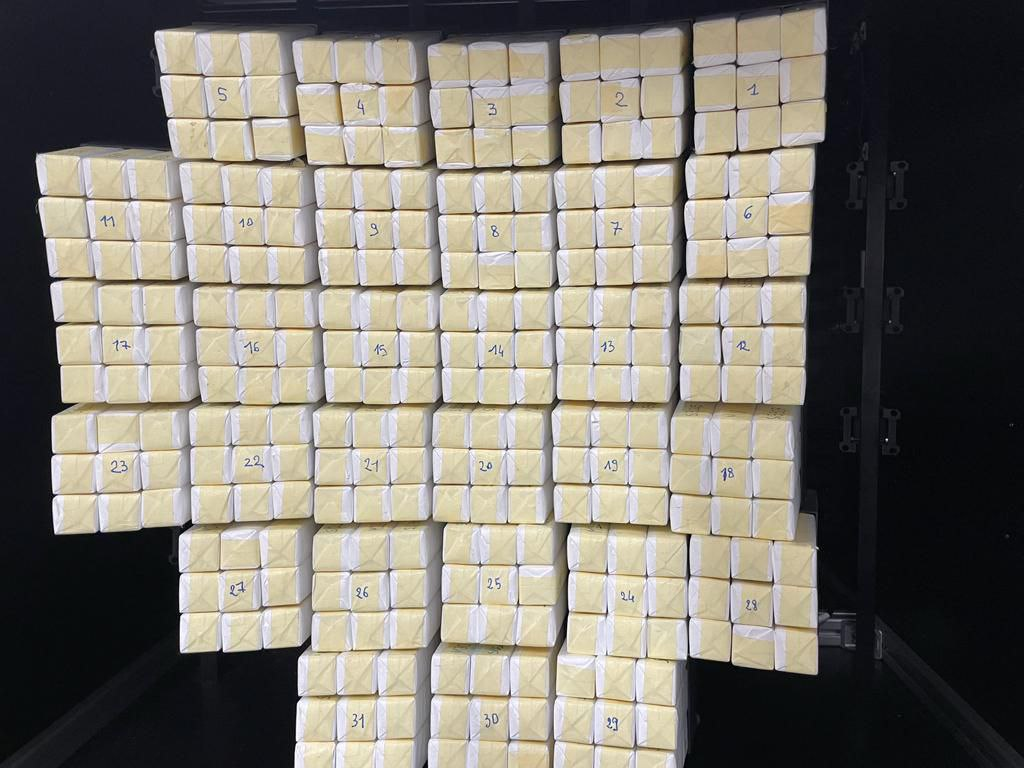} \quad \includegraphics[width=6cm]{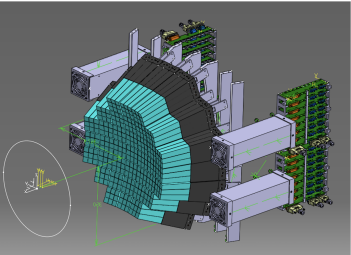}
\caption{ \textbf{left:} Picture of the FOOT calorimeter assembly during a data taking at CNAO in 2023, showing the completed 3x3 modules, \textbf{right:} sketch of the final geometry .}
\label{fig: calorimeter}
\end{center}
\end{figure}

For the FOOT experiment, BGO (bismuth germanate: $Bi_4Ge_3O_{12}$) crystals were identified as the optimal material for the calorimeter. The high density of BGO ($\rho \sim 7.13 g/cm^3$) provides excellent stopping power, and its light yield of $\sim 10 photons/keV$ ensures the required energy resolution. The calorimeter will consist of 320 BGO crystals arranged in a disk-like configuration with a radius of around $20 cm$, and the crystals grouped into modules of 3 x 3 for easier handling. Each crystal has a truncated pyramid shape with a front face of $2 \times 2 cm^2$, a back face of $3 \times 3 cm^2$, and a length of $24 cm$ (Fig.: \ref{fig: calorimeter}). The transverse size of the BGO crystals is comparable to the granularity of the TW. With this granularity, the pile-up probability due to multi-fragmentation events is kept below $1-2 \%$, depending on the beam energy and the experimental room configuration. The crystal depth was chosen to minimize energy leakage, particularly from neutrons escaping the calorimeter.

The scintillation light produced in the BGO is collected at the downstream side of each crystal by a $5\times 5$ matrix of silicon photomultipliers (SiPMs) with an active surface of $2 \time 2 cm^2$ and a microcell pitch of $15 \mu m$. This configuration ensures a linear response for energies up to approximately $10 GeV$. The SiPM matrix is connected to a readout board. The front-end board interfaces with the WaveDAQ system, and also reads the SiPM temperature sensor to compensate for temperature-induced variations and equalize the calorimeter response during offline analysis.

Several beam tests were conducted over a wide energy range (from $70 MeV$ protons to $400 MeV/u  ^{12}C$) to identify the optimal combination of SiPM array, readout configuration, and BGO wrappings. These tests confirmed excellent linearity across the entire energy range, with an energy resolution $\sigma(E_{kin})/E_{kin}$ below $2\%$, meeting the experimental requirements for the heavier fragments \cite{ref: Scav} \cite{ref: Scav2}. The CALO is currently nearing completion, with the final design expected to be ready and fully functional for the 2024 data acquisition campaigns.

\subsection{Trigger and data acquisition}
The FOOT experiment is equipped with a TDAQ (Trigger and Data AcQuisition) system designed to ensure precise data collection within a controlled and continuously monitored environment. The TDAQ is based on the one used in ATLAS experiment at LHC (Large Hadron Collider, CERN) \cite{ref: ATLAS} and it is maintained and developed by the University and INFN section of Bologna.

The TDAQ architecture is a flexible, hierarchical and distributed system based on Linux PCs, detector readout systems, VME crates and boards connected via standard communication links like USB, Ethernet and optical fibers. 

The Storage PC, which is kept in the experimental room during data takings, manages the whole acquisition. It is equipped with two network interfaces, one for internal communication with the other components of the TDAQ and one going to the outside network. The different parts of the internal network are linked to each other via an Ethernet switch with $1 GbE$ and $10 GbE$ ports. The two fastest ports are reserved for the Storage PC and the NAS (Network Attached Storage). The switch collects all the data coming from detector readouts: 20 $DE10nano$ or $DE10 Terasic$ boards for the tracking system and the WaveDAQ \cite{ref:WaveDAQ} for the SC, TW and CALO detectors. Moreover, an optical fiber link connects the Storage PC to a VME crate hosting the V2495 trigger board and all the readout electronics of the BM. The main Ethernet switch is also connected to two additional PCs located in the control room during data takings. The Control PC is used to connect to the Storage PC and manage the acquisition while the Monitor PC is reserved to data monitoring with online and quasi-online software processing tools. A detailed description of the TDAQ system can be found in \cite{ref: tesi_Ridolfi}.

\begin{figure}[h!]
\begin{center}
\includegraphics[width=9cm]{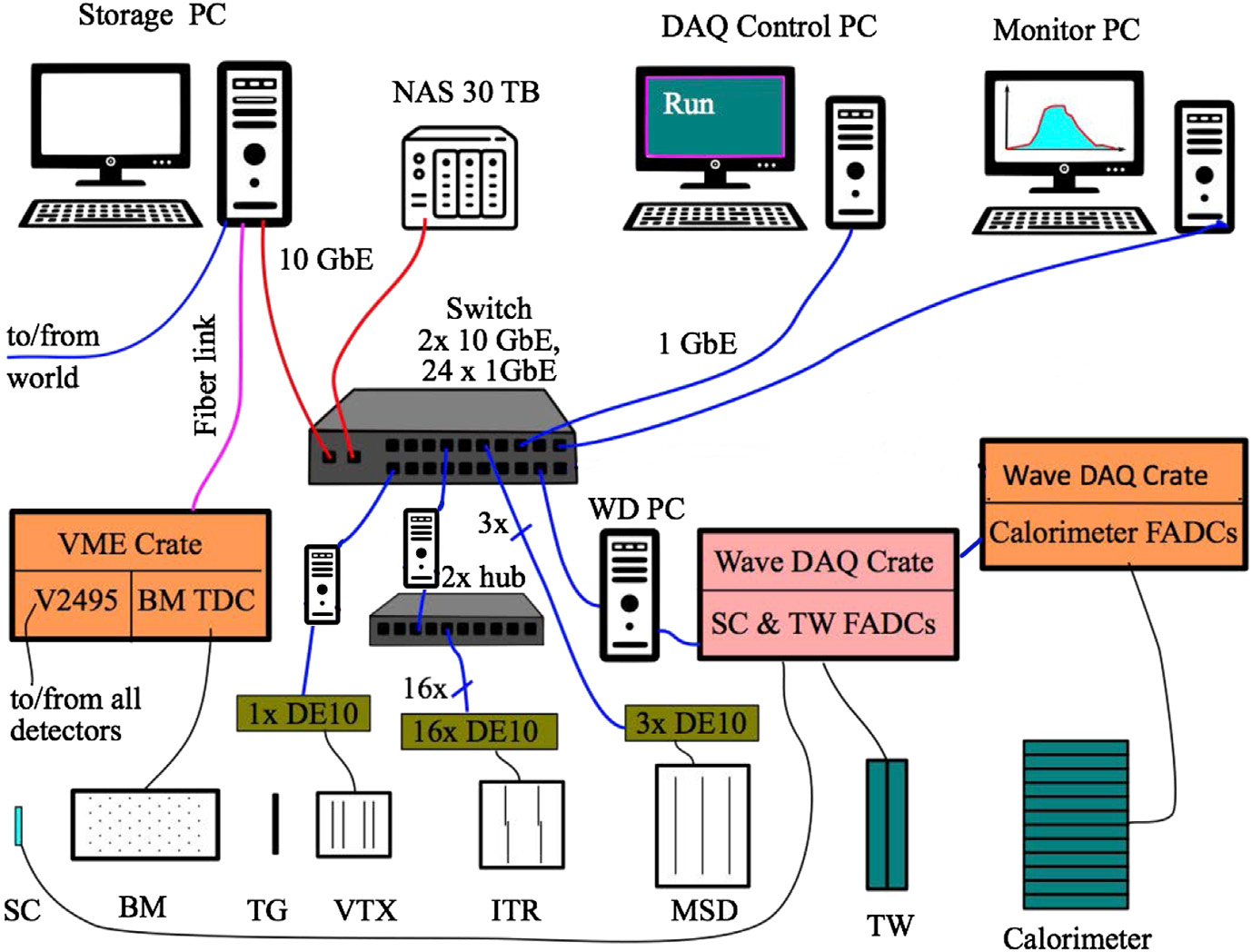}
\caption{ FOOT TDAQ infrastructure}
\label{fig: TDAQ}
\end{center}
\end{figure}

\begin{figure}[h!]
\begin{center}
\includegraphics[width=9cm]{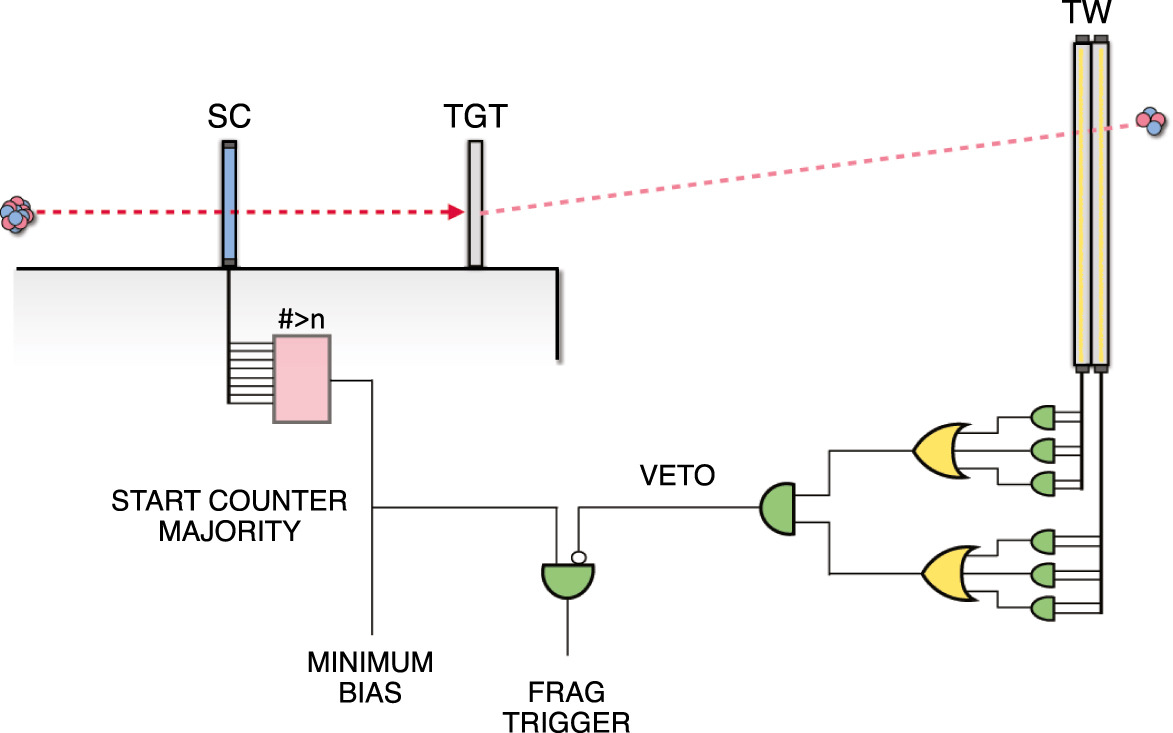}
\caption{Logical scheme of the FOOT Minimum Bias and fragmentation triggers, highlighting also the logical scheme for the generation of the VETO signal in the TW. \cite{ref: frag trigger}}
\label{fig: trigger}
\end{center}
\end{figure}

Given that only a small fraction of events in FOOT involve fragmentation reactions, the DAQ system has been equipped with a set of specialized triggers to optimize data acquisition. The various triggers employed are as follows:
\begin{itemize}
    \item  \textbf{Minimum Bias (MB) Trigger}: This trigger is based on signals provided by the SC when a particle crosses it. It verifies that a minimum number of SC channels register a signal above a predefined amplitude threshold (a majority trigger), effectively minimizing false triggers caused by electronic noise. An additional feature of the MB trigger is the inclusion of a clipping logic, which accepts a trigger as valid only if no other trigger signal has been generated within a predefined preceding time window. For FOOT, this requires two consecutive MB triggers to be spaced by at least $10 \mu s$. 

    \item \textbf{Fragmentation Trigger}: This trigger is activated when both the MB condition is met and there is no VETO signal from the TW. The VETO signal is generated whenever two central bars of the TW (one per layer) register a pulse height consistent with the passage of a primary particle through the setup. This mechanism helps eliminate events that do not involve fragmentation reactions. Considering that only a few percent of primaries are expected to undergo a nuclear interaction in the target, the usage of this trigger logic can significantly enhance the number of fragmentation events observed. The area where the fragmentation trigger logic looks for a possible VETO signal in the TW considers the three central bars of each layer, i.e. 9 possible crossings in a $6 \times 6 cm^2$ surface.
    
    The main advantage of this type of trigger is that it allows to also measure fragments impinging at the center of the TW. This enables the possibility to accurately extract particle yields also at low emission angles, which is a crucial characteristic since projectile fragments are expected to be emitted mainly in the forward direction. At the same time, the chosen VETO logic is solely based on the energy loss in the central portion of the TW. This has two important implications: on the one side, the thresholds for the rejection of primaries have to be set each time the beam is changed. However, this does not pose a problem since they can be calculated during the acquisition with the MB trigger. On the other hand, as shown \ref{fig: fluka.foot}, the detectable fragments are expected to have a velocity very close to that of the primary particle, especially in the case of heavier fragments. In this situation the fragmentation trigger can not distinguish between primaries (e.g. $^{16} O$) and fragments (e.g. $^{15}O$) with the same charge, leading to a loss of selection efficiency for these specific particles.

    Both fragmentation and Minimum bias trigger logical scheme are shown in Fig. \ref{fig: trigger}.

    \item \textbf{TWalone and CALOalone Triggers:} These triggers are used mainly during calibration runs for the TW and the calorimeter (CALO). The TWalone trigger is a logical OR of all the 40 TW bars, generating a trigger whenever a bar is hit by a particle. This condition is satisfied when both corresponding channels register a pulse above the Zero-Suppression threshold. The CALOalone trigger operates similarly, generating a trigger each time one of the calorimeter crystals registers a pulse above the Zero-Suppression threshold.
\end{itemize}

The technology used to implement the trigger logic is provided by the CAEN V2495 board, featuring a fully programmable FPGA and internal logic.  The TDAQ has been designed to withstand the maximum acquisition rate achievable by the FOOT setup with the MB trigger, which is only dictated by the incoming beam rate and the dead time of the slowest detector. The slowest detectors in the experiment are the MIMOSA-28 chips used in the pixel trackers (VTX and ITR), they have a frame readout time of $185.6 \mu s$ meaning a maximum theoretical
acquisition rate of $1.8 kHz$. In order to minimize pile-up effects in the MIMOSA chips, the actual trigger rate will be limited to around $1 kHz$.

\section{Emulsion Spectrometer}
The Emulsion Spectrometer in the FOOT experiment has been designed to specialize in the  characterization of low-Z ($Z\le 3$) fragments produced during fragmentation reactions. This setup offers a significantly wider angular acceptance compared to the Electronic Setup, reaching up to $70^{\circ}$ from the primary beam direction, compatible with the wider emission angles expected
for lighter fragments. The decision to utilize nuclear emulsion films stems from their exceptional spatial resolution in tracking ionizing particles, capable of resolving trajectories at sub-micrometric scales. These films consist of a gelatin-based material embedded with tiny $AgBr$ crystals. When ionizing particles pass through, the crystals are sensitized, creating a latent image along the particle's trajectory. This image can be enhanced through a chemical development process and then analyzed using optical microscopes.

Nuclear emulsion systems enable the construction of compact setups, where detection layers are interleaved with target material. Additionally, recent advancements in automated scanning techniques facilitate fast and accurate analysis of large data sets, making these systems highly efficient. \cite{ref: scan emulsion}

The nuclear emulsion films employed in the FOOT experiment are produced by Nagoya University (Japan) and the Slavich Company (Russia). Each film consists of two $70 \mu m$-thick emulsion layers coated on either side of a $210 \mu m$ plastic base. The sensitivity of these films corresponds to $30 AgBr$ grains per $100 \mu m$ of track length for a minimum ionizing particle (MIP). \cite{ref:emulsion}

The FOOT Emulsion Spectrometer setup is made of two main components. The upstream region, which consists of the same Start Counter and Beam Monitor detectors of the Electronic Setup, used only to monitor the beam flux to avoid spatial pile-up of primarie. The second component is the Emulsion Cloud Chamber (ECC), that performe the measurements of the cross section.

\begin{figure}[h!]
\begin{center}
\includegraphics[width=9cm]{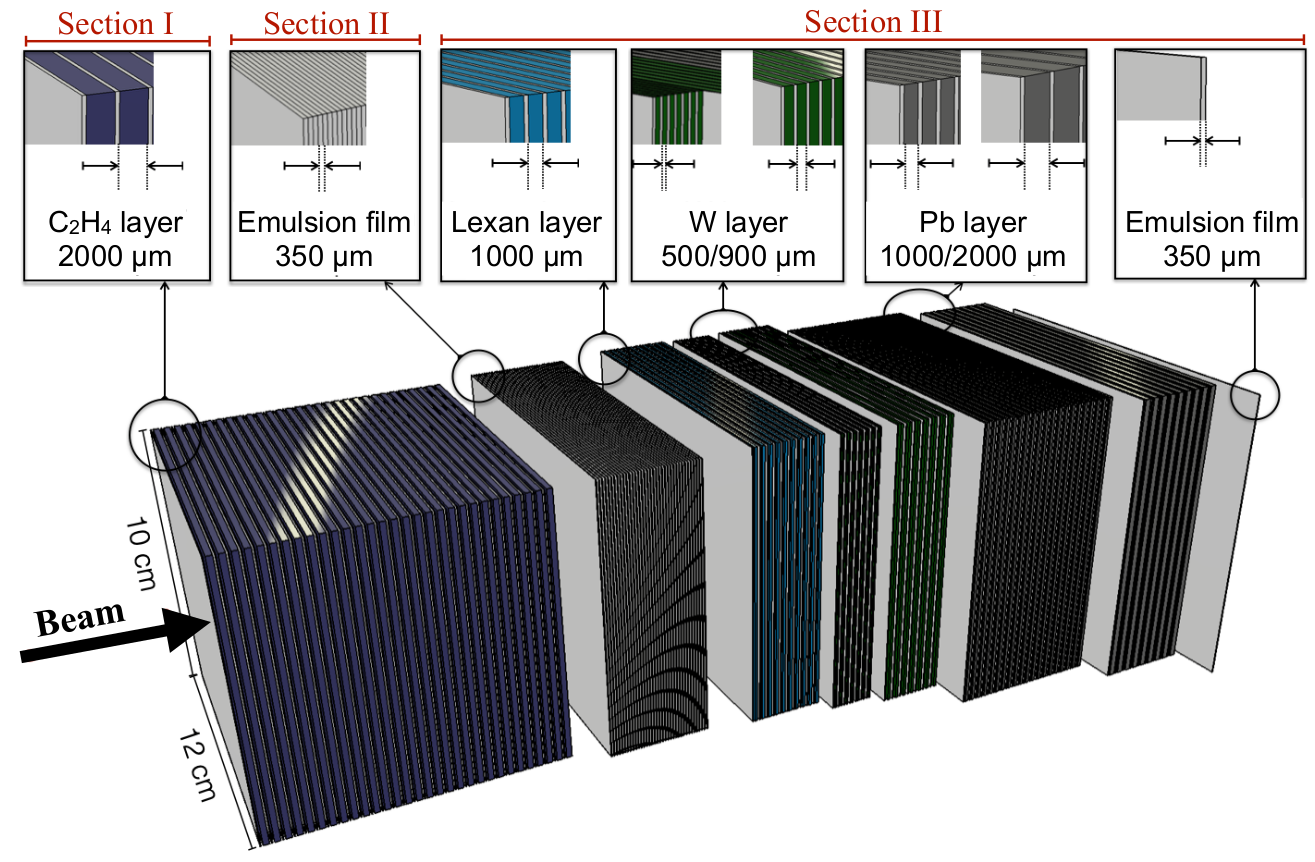}
\caption{Schematic view of the FOOT Emulsion Cloud Chamber setup.}
\label{fig: emulsion}
\end{center}
\end{figure}

The ECC  is divided in three main sections. The first section is made of 30 nuclear emulsion films alternated with thin ($\sim 1-2mm$) layers of target material, i.e. polyethylene or carbon. This section is devoted to the identification of interaction points between primaries and target. The total length of this section has been chosen so that the Bragg peak of primary particles is completely included in it, meaning that only nuclear fragments can reach the downstream sections of the ECC. Moreover, the depth at which the fragmentation occurs makes it also possible to accurately reconstruct
the energy of the primary at the moment of the interaction. This feature allows, in principle, to perform cross section measurements for different energies of the impinging beam with a single exposure.
The second section of the ECC is entirely made of nuclear emulsion layers and is dedicated to charge identification of fragments. The 36 emulsion films in this section are divided into 9 cells of 4 layers. The four films of each cell are thermally treated at different temperatures to enhance their sensitivity to ionization, extending their dynamic range and enabling the possibility to discriminate particles with different charge. The third and last section of the ECC is made of 55 emulsion films interleaved with layers of lexan and high density materials, such as tungsten and lead. This section is
dedicated to the measurement of fragment momentum, which can be extracted from particle range and multiple Coulomb scattering. The entire setup is shown in figure \ref{fig: emulsion}.

The Emulsion Spectrometer has been employed in several experimental campaigns since 2019. The data acquired in this occasion were used to perform the first assessment of charge identification capabilities with the ECC. The work reported in \cite{ref:emulsion} shows that, using a mix of simple cuts and some analysis techniques, the ECC is able to correctly identify the charge of lighter fragments emitted in fragmentation reactions inside the detector.

\chapter{Vertex Detector: Data and MC Analysis} \label{chapter 3}

In the previous section \ref{subsubsection: vertex} the vertex detector (VTX) was described, including its configuration, technology, and role in tracking particles exiting from the target with high precision. This detector is essential for the reconstruction of the fragmentation vertexes.

In this chapter, studies conducted on the vertex detector are presented. The focus is on an examination of the reconstruction algorithms used for track reconstruction and vertex identification, and an evaluation of their performance. These studies aim to assess the accuracy and efficiency of the algorithm in reconstructing the fragmentation vertexes, with particular attention to its ability to handle pile-up events. 
Indeed the long dead time (625 $\mu$s) of the VTX readout limits the FOOT data acquisition speed. A beam rate of about 1~kHz is employed during data acquisition in order to minimize the pile-up in VTX detector. In the next session the working principles of the VTX detector are reported and a summary of the clustering, tracking and vertexing algorithms used in FOOT for the VTX detector is reported. A detailed description of these algorithms is given in~\cite{ref: vtx performance}. Before that, some useful definitions of reference frames in FOOT are discussed.

\section{References frames}
In the FOOT experiment multiple reference frames are employed, to describe the positions and orientations of the detectors. The global reference frame is defined with its origin at the center of the target. This frame serves as the primary coordinate system, and the positions of all detectors are specified relative to this origin. In addition, each detector has it own local reference frame. For instance, the VTX detector uses a local reference frame with origin in the center of the VTX itself. Similar definition apply to the local frame of other detectors. Knowing the position and orientation of each component is possible to pass from one reference frame to another.

\begin{figure}[h!]
\begin{center}
\includegraphics[width=12 cm]{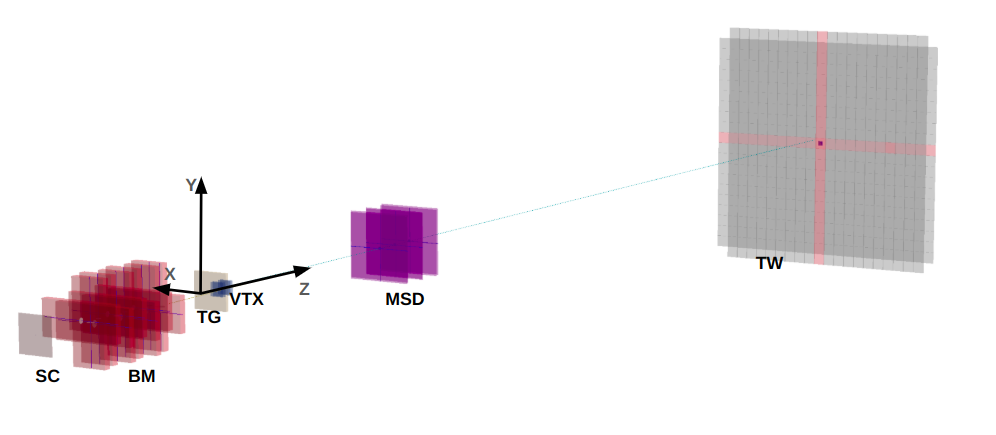}  
\caption{Schematic view of the apparatus, with a particle track passing through th Start Counter, Beam Monitor, Target, Vertex, MSD and TofWall. The axes of the reference system are also shown: $z$ along the beam direction, $y$ as the vertical axis and $x$ as the transverse axis, all referenced to the global coordinate system centered at the TG.}
\label{fig: axis}
\end{center}
\end{figure}

In all reference frames, the z-axis is considered to be the direction of the beam, while the y-axis is the vertical axis and the x-axis is the transverse axis, (Fig.: \ref{fig: axis}).

\section{VTX detector reconstruction algorithms}
\subsection{M28 working principles and clustering algorithm}
\begin{figure}[h!]
\begin{center}
\includegraphics[width=6.7 cm]{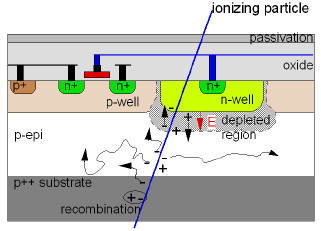}  
\caption{Schematic representation of the detection mechanism in Mimosa-28 sensors, highlighting the P- and N-doped regions and the PN-junctions involved in signal collection.}
\label{fig: cmos}
\end{center}
\end{figure}

The detection mechanism of the Mimosa-28 is based on a moderately P-doped epitaxial layer, sandwiched between a highly P-doped substrate and P-well implant layers. Each pixel contains a collection diode formed by a PN-junction between the epitaxial layer and an N-well region (Fig.: \ref{fig: cmos}). When a charged particle traverses the sensor, it generates electron-hole pairs proportional to the particle's energy and ion species. The electron motion arises from the internal electric fields generated by the built-in voltage between layers of differing doping concentrations. These electric fields accelerate electrons in the depleted regions, effectively guiding them to the collection diode. Since the epitaxial layer is not fully depleted, electrons that are created by a charged particle may also undergo some thermal diffusion before being collected by the diode. As a consequence of this diffusion the electrons generated by a single ion can be collected by multiple adjacent pixels (referred to as cluster). The spatial distribution of the detected signals depends on several factors, such as the current induced by the built-in voltage and the depleted region, but also on the particle species and its energy ~\cite{ref: M28}.

To recognize and reconstruct clusters associated with the same particle, the algorithm identifies adjacent active pixels using the first neighbors method, where two pixels are considered neighbors if they are contiguous in either row or column. In detail, it begins by selecting a pixel from the list of active pixels. It checks whether any of its first neighbors are also active, if active neighbors are found, they are added to the current cluster. Each newly added pixel then serves as a reference point for further searches, extending the cluster until no additional active neighbors are identified. The process is repeated for any remaining unclustered pixels in the list, starting a new cluster each time. 

After clustering, the algorithm proceeds to reconstruct the tracks and subsequently analyzes whether these tracks converge, forming a fragmentation vertex.

\subsection{Tracking Algorithm}

Using cluster information from the four planes of the VTX detector, the algorithm reconstructs particle tracks.

For the local reconstruction of tilted tracks, two different algorithms, "Full" and Standard ("Std"), can be applied.

In the Full algorithm, the process begins by selecting for each event a cluster from the last layer which is the furthest layer of the VTX from the TG. All the possible combinations of this cluster with the clusters in the two intermediate layers are considered for tracking. For each combination, a preliminary track is reconstructed by fitting a three-dimensional line to the cluster coordinates. This track is then projected onto the remaining layers, where the algorithm searches for additional clusters within a specified "Search Hit Distance" (SHD). 
Each time a new cluster is added, the track is re-fitted to update its slope and offset parameters. When the first layer is reached, if a track with at least three clusters is identified and validated through a $\chi^2$ selection, the track is saved, and its clusters are removed from further tracks searching. This process iterates until all clusters in the last layer are processed. Given the three-clusters requirement for a track, the algorithm also checks clusters not yet tracked in the penultimate layer, attempting to form tracks starting from these.

The Standard (Std) algorithm also begins with clusters from the last layer but uses the BM track, projecting it onto the center of the target, as an initial guide. The Std algorithm continues layer-by-layer, checking for clusters within the SHD along the line that connects the BM-projected point and the initial cluster in the last layer. As with the Full algorithm, each additional cluster triggers a re-fit of the track parameters. This iterative process continues until all clusters in the last layer, as well as any unassigned clusters in the second-to-last layer, are examined.

For straight tracks, the algorithm starts with clusters in the last layer and extrapolates tracks aligned parallel to the Z-axis of the global reference frame. Similar to the other algorithms, it checks each layer for clusters within the SHD, re-fitting the track parameters with each new cluster. If a track meets the necessary criteria (including $\chi^2$ and a minimum of 3 clusters), it is saved, and the clusters that compose it are excluded from further track searching.

\subsection{Vertex Reconstruction Algorithm}
After reconstructing the particle tracks for each event, the algorithm evaluates the presence of a fragmentation vertex. To achieve this, the algorithm scans the thickness of the target, searching for the point that maximizes the probability of multiple tracks converging.

The vertex reconstruction algorithm proceeds through a series of steps. First, it iterates over the tracks in each event, pairing them two by two. For each pair of tracks, the algorithm searches for a potential vertex within a region defined by two planes positioned at fixed Z-values $([a, b]$) which surrounds the target.

\begin{figure}[h!]
\begin{center}
\includegraphics[width=8 cm]{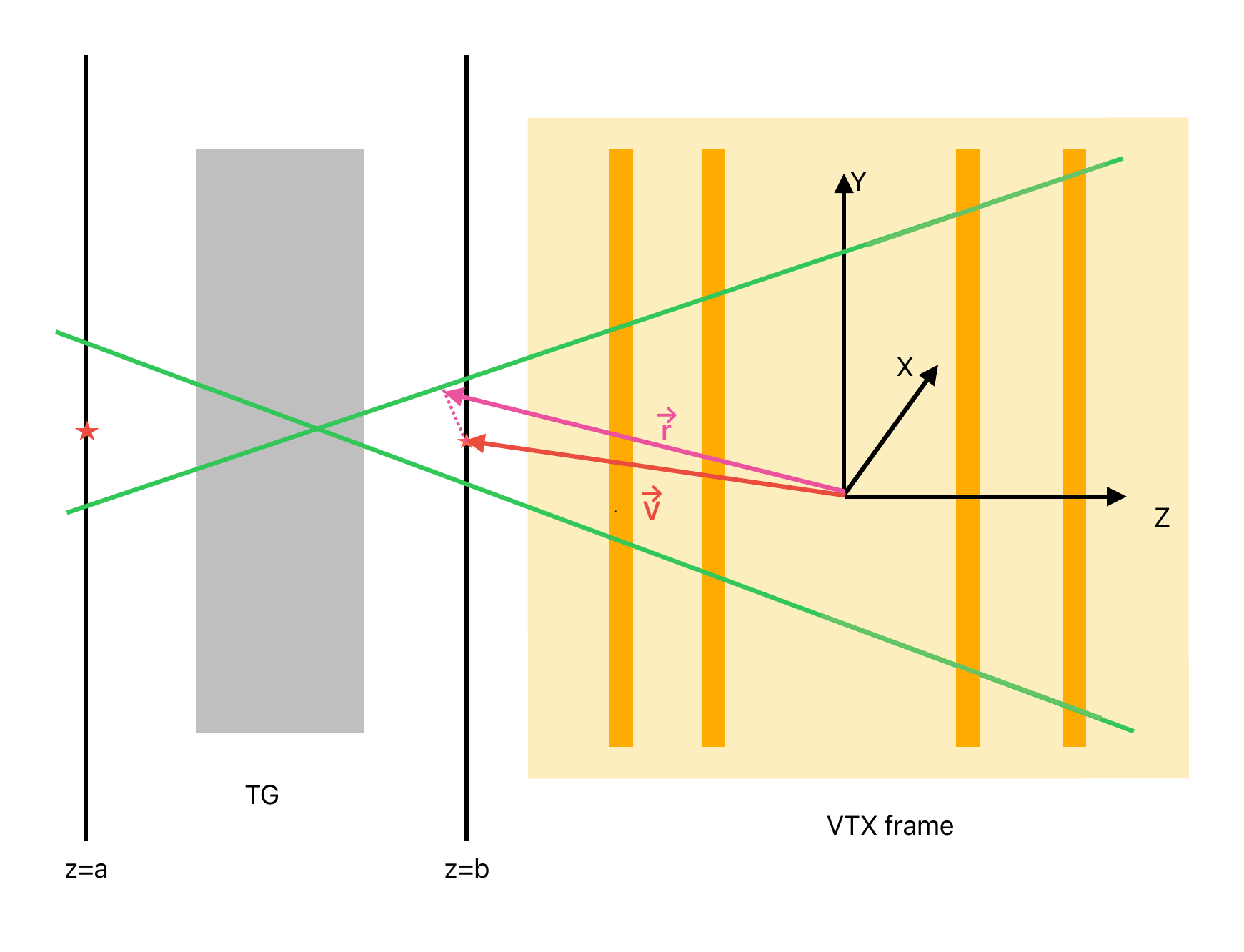}  
\caption{Schematic representation of the vertex reconstruction algorithm. The two tracks intersect on the target. The algorithm utilizes two z-planes to reconstruct the vertex position. On the 2 planes, the red stars mark the initial reconstructed vertexes. The 2 vectors $\vec{v}$ and $\vec{r}$ are shown, in the local frame of the VTX detector. The picture is not in scale.}
\label{fig: schematic alg.}
\end{center}
\end{figure}

The algorithm identifies two potential vertex positions by calculating the intersections of the two tracks with the two Z-planes and taking the midpoint of these intersections. To account for the uncertainty in track positions, a Gaussian probability density function is used to model the tracks' deviations. This results in the creation of a "Gaussian probability density tube" around each track, where the width of the tube is determined by the track's dispersion, in detail:

\begin{equation}
    f_i(\vec{v})=exp\left( -\frac{1}{2} (\vec{v}- \vec{r_i}) V_i^{-1} (\vec{v}-\vec{r_i})\right)
\end{equation}
In this equation, $\vec{v}$ represents the vector pointing to the vertex, $\vec{r_i}$  is the vector corresponding to the point of closest approach of track $i$ to the identified vertex (Fig.: \ref{fig: schematic alg.}), and $V_i$ is the covariance matrix of the track at $r_i$. In our case, the covariance matrix contains errors on track position, and is assumed to be diagonal. Details can be found in~\cite{ref: vtx performance}.

At this point, the algorithm calculates a probability for each of the two potential vertex positions, i.e. from Fig.~\ref{fig: schematic alg.} $P(b)=f_i(b)\times f_j(b)$ where i and j are the two tracks converging in vertex at position z=b.
These values are used to calculate the slope:
\begin{equation}
    slope=\frac{P(b)-P(a)}{b-a}
\end{equation}

If $slope>0$, the range $[a,b]$ is reduced by redefining $a$, narrowing it to half of the current range. If $slope<0$, $b$ is redefined instead. This process is repeated iteratively until the range becomes smaller than $1\mu m$, at which point the vertex is defined as the average of the coordinates of the last two remaining vertex candidates.

The vertex probability is defined as: $P=\frac{P(a)+P(b)}{2}$. This probability is used to determine whether the two tracks indeed converge to form a fragmentation vertex or if they belong to tracks that do not converge, using a probability threshold (value 0.1).

After all pairs of tracks have been evaluated, the algorithm checks if there are any tracks that do not converge with any other tracks. These tracks are labeled as not valid vertexes, meaning they are vertex composed of single tracks.  To these tracks, the algorithm assigns a validity flag \textbf{0}.


If one or more vertexes pass the probability threshold, the algorithm uses maximization of the function $Q(\vec{v})$ for the selection of the best vertex~\cite{ref: vtx performance}.

\begin{equation}
    Q(\vec{v})= \sum_{i=1}^{N} f_i(\vec{v}) + \frac{\sum_{i=1}^{N} f_i^2(\vec{v}) }{\sum_{i=1}^{N} f_i(\vec{v}) }
\end{equation}

Once the vertex is identified, a check is performed on all remaining valid tracks. If their distance from the identified vertex is less than the Impact Parameter ($250 \mu m$), these tracks are added to the valid vertex; otherwise, they are classified as invalid. Valid vertexes are exclusively those composed of two or more tracks and are assigned a flag of \textbf{1}.
\subsubsection{Debugging}
During the study of the vertex reconstruction algorithm, a small bugs was identified.

This issue pertains to the calculation of the $slope$. In some cases, the $slope$ was found to be zero, which led to the redefinition of the $b$ value. However, a $slope$ of zero is not meaningful in this context, and it can result in selecting the incorrect half of the range, causing the algorithm to fail in identifying that the tracks converge. To address this, a third option was introduced: if the slope is equal to zero, both $a$ and $b$ are restricted by a small amount ($0.01 cm$). This modification allows the loop to continue enabling the algorithm to proceed as expected. 


\section{Monte Carlo analysis for vertexes reconstruction}
Monte Carlo simulations play a crucial role in evaluating the expected results, configuration-specific differences, and overall efficiencies of vertex reconstruction algorithm in the FOOT experiment. For each FOOT data-taking corresponding MC campaigns have been created to replicate the precise geometry of the detectors, including passive materials, closely approximating real experimental conditions. These detailed simulations capture effects such as scattering, energy loss, and particle interactions within non-active or structural components. Additionally, the beam characteristics (center and spread) in each simulation are calibrated to match those observed in the corresponding experimental data. This alignment ensures a reliable basis for comparing simulation and experimental results.

The term "campaign" refers to a specific data-taking, characterized by the specific beam and experimental setups. For example, the GSI2021 campaign corresponds to data collected at GSI in 2021, while CNAO2022 and CNAO2023 correspond to data acquired at CNAO in 2022 and 2023, respectively. Each campaign features distinct configurations that influence the experimental conditions and subsequent analyses.

In the following an analysis of the VTX detector performance in the different MC campaigns is reported. The focus is on the VTX performance in reconstructing vertexes, while the VTX performances about tracking and clustering have been already studied in other works~\cite{{ref: vtx performance}}. This study is fundamental for the final FOOT purpose, reminding that the total fragmentation cross section is proportional to the ratio between the measured vertexes with respect to the number of incident primaries on the TG. Comparing MC simulations from different campaigns reveals how variations in geometry, detector positioning, and beam characteristics affect vertex reconstruction. This analysis helps identify campaign-specific conditions that influence algorithm performance.
The FOOT campaigns considered in this analysis are CNAO2023, CNAO2022, and GSI2021. For GSI2021, the setup involves an oxygen beam of $400 MeV/u$ on a $5mm$ carbon target without a magnetic field. CNAO2022 employs a carbon beam of $200 MeV/u$ on a $5mm$ carbon target, also without a magnetic field. In contrast, CNAO2023 uses a carbon beam of $200 MeV/u$ on a $5mm$ carbon target with a magnetic field included in the configuration.

\subsection{GSI2021}
\subsubsection{Vertex position residuals}
The initial objective for analyzing the GSI simulations is to verify the accuracy of the vertex positions assigned by the reconstruction algorithm. MC simulations provide a unique advantage, as they grant access to both the true information (such as the exact fragmentation point of the primary particle) and the reconstructed positions determined by the algorithm. This dual insight enables a direct comparison between the true and reconstructed coordinates, facilitating a robust assessment of the algorithm's performance.

To this end, two types of plots are analyzed: correlation plots, which compare the true and reconstructed vertex coordinates, and residual plots that display the differences between these coordinates.

\begin{figure}[h!]
\begin{center}
\includegraphics[width=6.5 cm]{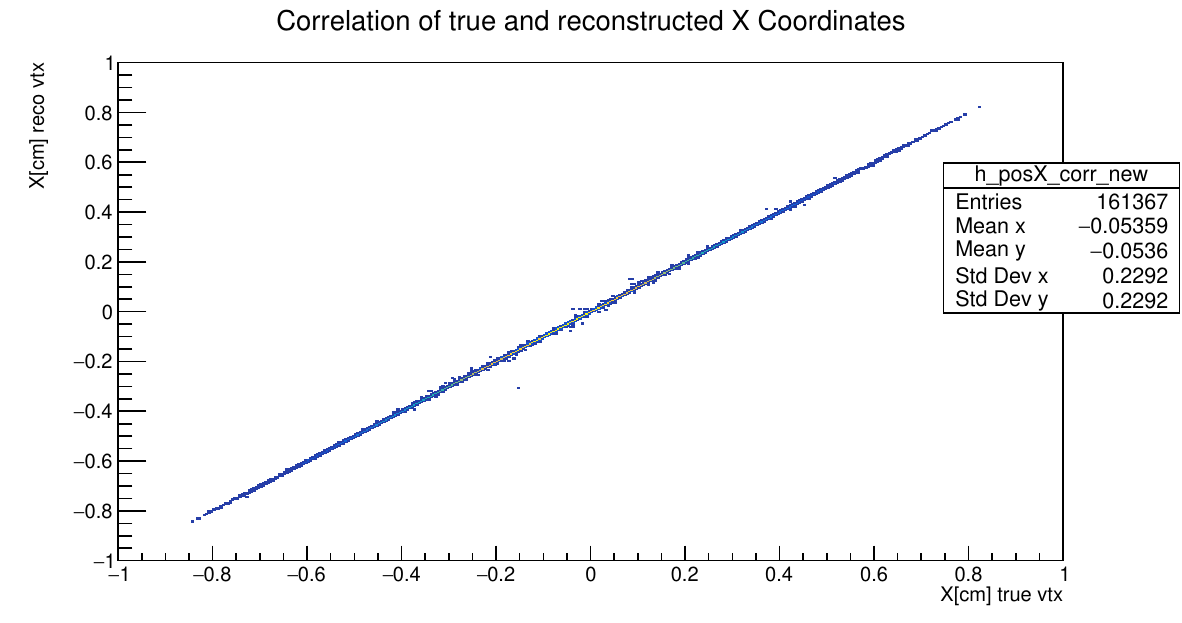} \quad 
\includegraphics[width=6.5 cm]{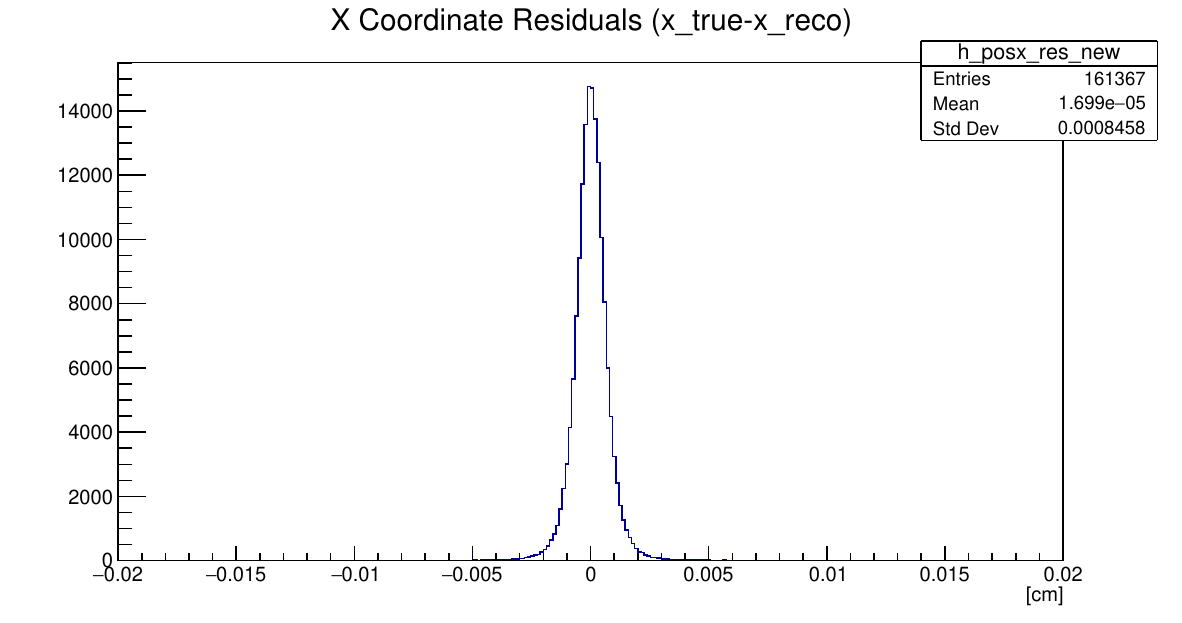} \quad\includegraphics[width=6.5 cm]{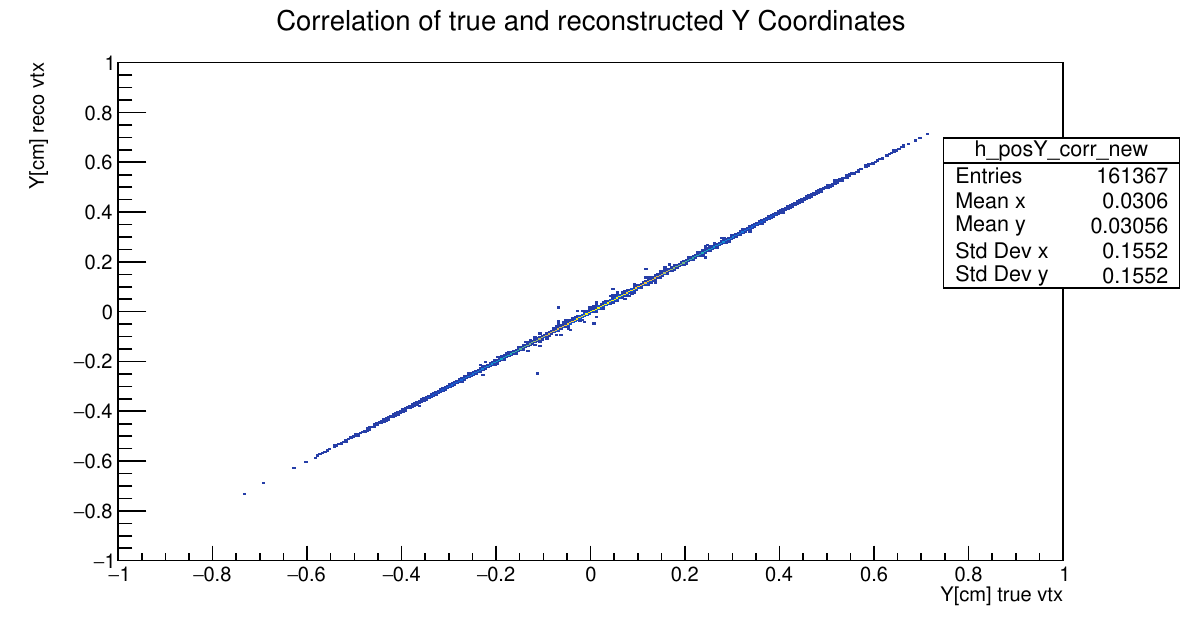} \quad 
\includegraphics[width=6.5 cm]{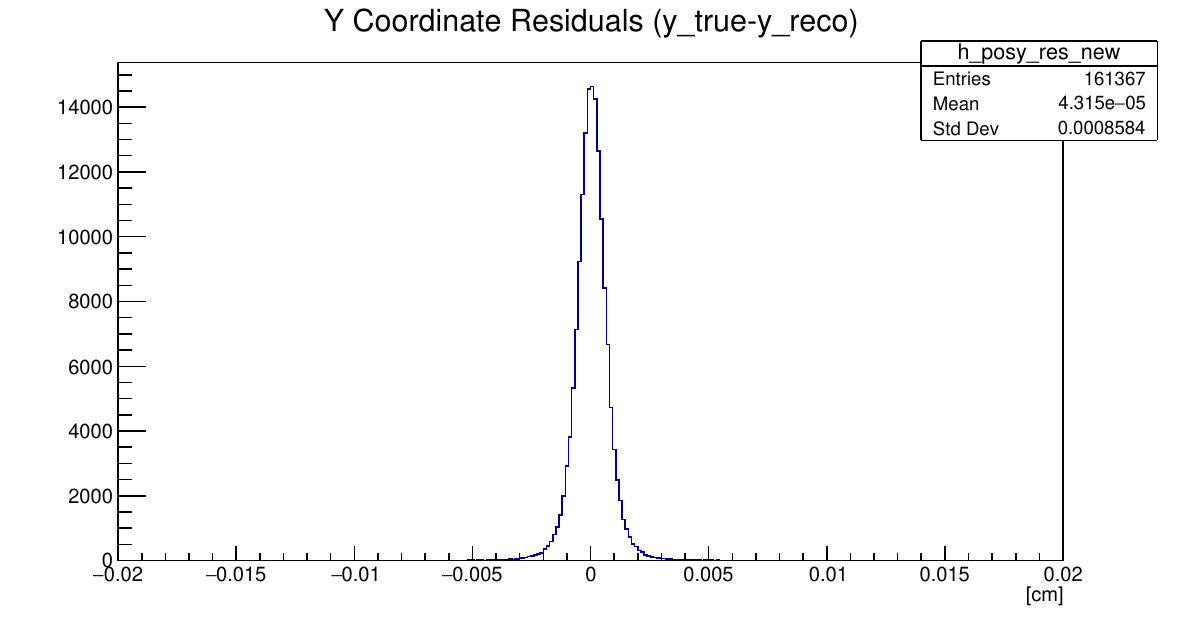} \quad\includegraphics[width=6.5 cm]{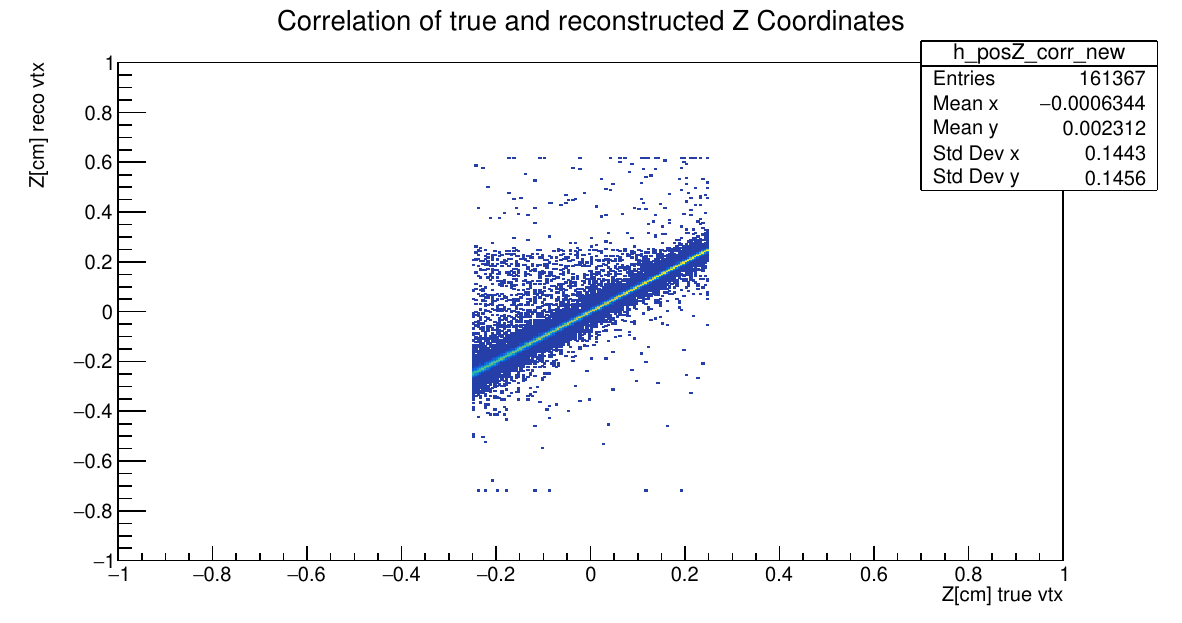} \quad 
 \includegraphics[width=6.5 cm]{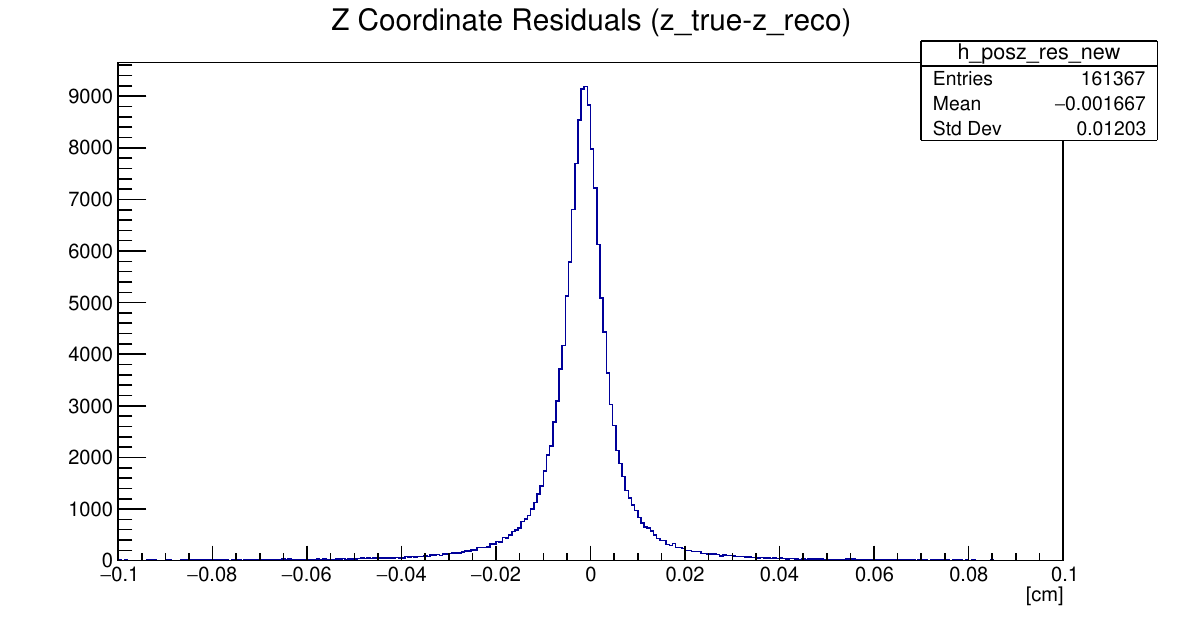} \quad
\caption{Correlation and residual plots for reconstructed versus true coordinates for the GSI2021 MC simulation. The top row shows the correlation and residuals  for the $x$ coordinates. The second row presents the same thing for the $y$ and last row for the $z$. Note that the residuals for the $z$ coordinates have a different scale compared to $x$ and $y$: the range for $x$ and $y$ is $[-0.02, 0.02]cm$, while for $z$ it is $[-0.1,0.1]cm$.}
\label{fig: MC_GSI2021}
\end{center}
\end{figure}

The correlation plots (Fig.: \ref{fig: MC_GSI2021}) show a clear alignment between true and reconstructed coordinates for all axes. For the x and y coordinates, a near-perfect correlation is observed, while the $z$-coordinate also demonstrates a strong correlation but with a slightly broader distribution.  Similarly, the residual analysis confirms the algorithm’s reliability in reconstructing accurate vertex positions, as the residual distributions are centered around zero. 
 
Table \ref{table: residual_MC_GSI21} provides the mean and $\sigma$ (standard deviation) values for the residuals in the $x$, $y$, and $z$ coordinates, obtained by fitting the central part of the distributions with a Gaussian function. 
The mean values for the $x$, $y$ and $z$ coordinates are compatible with zero within 1$\sigma$ of uncertainty. The $z$-coordinate exhibits a significantly broader distribution, with its $\sigma$ value increasing by an order of magnitude compared to those of the $x$ and $y$ coordinates, reflecting the greater difficulty in reconstructing the z-position. These findings are consistent with previous studies, which also highlighted the challenges of reconstructing the z-coordinate in fragmentation events \cite{ref: vtx performance}.

\begin{center}
\begin{tabular}{|c | c |c |} 
 \hline
- & mean $[\mu m]$ & $ \sigma $ $[\mu m]$\\
 \hline\hline
x & $(0.142 \pm 0.021) $ & $(5.406\pm 0.029)$ \\
y & $(0.423 \pm 0.021)$ &$(5.467 \pm 0.030)$\\
z & ($-13.27 \pm 0.14)$ & ($40.99 \pm 0.16)$ \\
 \hline
\end{tabular}
\captionof{table}{Mean and $\sigma$ for the residual histogram}\label{table: residual_MC_GSI21}
\end{center}

To select vertexes generated within the target region and reconstructed by the algorithm ($z$ in the range $[-0.25, 0.25]cm$ in the global reference frame, for the $5mm$ target), the selection range is extended by $5 \sigma_z$ to account for the reconstruction uncertainty in $z$. Based on this, the selection range is defined as $[-0.27, 0.27]cm$. For all future studies, both for data and simulations, any reference to vertexes generated within the target will consider the $z$-coordinate of vertexes within this extended range. This selection criterion will also be applied in the calculation of efficiency described in the following subsection.

\subsubsection{Vertex reconstruction efficiency}

Another powerful tool of the MC is the study of the efficiency of the vertex reconstruction algorithm employed in this analysis. By leveraging the true tracks from MC, a study is conducted to evaluate the algorithm’s performance.

The first measure of efficiency is calculated by considering events where at least two true tracks are generated within the target, have a positive charge less than that of the beam, and escape the target. Additionally, the mother particle of these tracks must be a primary particle, meaning the tracks represent fragments, and so the event contains a valid vertex. The total number of such events forms the denominator, denoted as ($true\_ValidVtx$). The numerator, ($reco\_ValidVtx$), is the number of valid vertexes reconstructed by the algorithm. For this analysis only events with a single track in the BM are considered. This efficiency measurement accounts for both the reconstruction efficiency of the algorithm and the geometric acceptance of the detector. The latter arises from the fact that some valid vertexes cannot be reconstructed due to the absence of at least two tracks within the geometric acceptance of the VTX. The efficiency is calculated as:

\begin{equation}\label{eq:eff}
    \epsilon_{tot}=\epsilon_{reco} \times \epsilon_{geo} =\frac{reco\_ValidVtx}{true\_ValidVtx}=(91.09 \pm 0.07)\%
\end{equation}

The reported efficiency indicates a good performance of the selection process, but a more detailed and precise study could be carried out to capture only the algorithm's efficiency. To address this, a new metric is introduced, focusing exclusively on reconstructable vertexes ($reconstructable\_ValidVtx$). A vertex is considered reconstructable if at least two tracks are within the acceptance of the detector's third layer. Specifically, a track within the third layer’s acceptance, even if outside the fourth layer, still has the minimum three clusters required for reconstruction. This ensures that any track within the third layer’s acceptance is reconstructable. This new metric, accounting only for the algorithm reconstruction efficiency, is defined as:


\begin{equation}\label{eq:eff:_acc}
    \epsilon_{reco}=\frac{reco\_ValidVtx}{reconstructable\_ValidVtx}=(97.83 \pm 0.04)\%
\end{equation}
This efficiency (\ref{eq:eff:_acc}), indicate a very high performance. This demonstrates that the algorithm is highly effective in reconstructing valid vertexes. Furthermore, as expected, this efficiency is higher than the previous one (\ref{eq:eff}), which was smaller because it also included the impact of geometrical efficiency, accounting for unreconstructed vertexes outside the geometric acceptance.

A further consideration involves the scenario where one of the layers in the VTX is non-functional. This situation is particularly relevant, as data from GSI and subsequent campaigns suggest a decline in the performance of one layer (the third layer for GSI2021 and CNAO2022, and the second layer for CNAO2023). Having a non-functional layer can impact the number of reconstructable vertexes, primarily due to acceptance issues. In particular, some vertexes may consist of tracks that are within the acceptance of the first three layers but not the last. If one of these earlier layers becomes non-functional, such vertexes cannot be reconstructed because they lack the minimum number of clusters required for track reconstruction. To assess the impact of having only three functional layers, it is necessary to study the fraction of reconstructable vertexes in such cases compared to the scenario where all layers are functional. When all layers are functional, a vertex is reconstructable if at lest two tracks are within the acceptance of the third layer ($reconstructable\_ValidVtx$). This ensure the minimum requirement of three clusters to reconstruct a track. However, when one layer is non functional, the criterion changes: a vertex is reconstructable only if at least two tracks are within the acceptance of the last functional layer (the fourth layer). This ensures that even with a missing layer, each track has the minimum three clusters required for reconstruction.  The total number of such vertexes is denoted as $reconstructable\_ ValidVtx_{4}$, and the fraction of reconstructable vertexes ($frV$), is given by:

\begin{equation}\label{eq: frv}
    frV=\frac{reconstructable\_ ValidVtx_{4}}{reconstructable\_ValidVtx}=(98.89 \pm 0.03)\%
\end{equation}

This $frV$, indicating that, in a situation where everything is perfectly aligned, like in MC, the vast majority of vertexes remain reconstructable even with only three functional layers. 

\subsubsection{Proportion of vertexes}
The final quantity we extract from the simulation is the percentage of fragmentation vertexes. Specifically, we select only events where there is a single track in the BM and ensure that this track is within the acceptance of the VTX. A track is considered within acceptance if, when projected onto the target, its x-coordinate lies between $[-0.8,0.8] cm$, and similarly for the  y-coordinate. Next, we count the number of events where there is a fragmentation vertex, produced in the target, that lies within the same region ($- 0.8\le x\le 0.8$ and $- 0.8\le y\le 0.8$) and is matched to the BM track. This spatial selection is used in order not to have border effects in which a vertex can be lost, reminding the VTX sensor transversal dimensions of (-1,1) cm in both X and Y. The percentage of valid vertexes is then calculated as the ratio of these two numbers:

\begin{equation}\label{eq: p_vv gsi2021}
    p_{vv}= \frac{\textit{events with a matched valid vtx}}{\textit{events with 1 BM track}}= (3.440 \pm 0.008)\%
\end{equation}

This quantity is particularly interesting because it is proportional to the total fragmentation cross section and it provides a basis for comparison with experimental data. 

\subsection{CNAO2022}
The analysis for CNAO2022 follows the same methodology as that for GSI2021. Applying the same metrics to the CNAO2022 simulation it is interesting to find out if there are any changes in the efficiencies considering that the geometry of the VTX has remained the same, but the beam used is now $200MeV/u$ carbon.

\begin{figure}[h!]
\begin{center}
\includegraphics[width=6.5 cm]{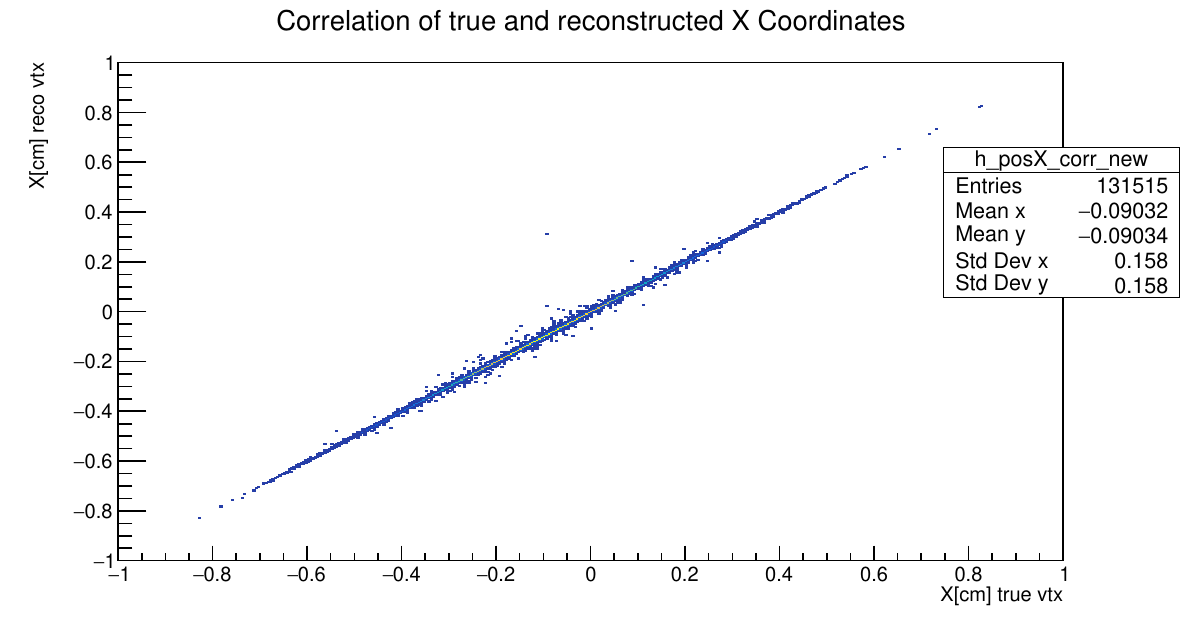} \quad 
\includegraphics[width=6.5 cm]{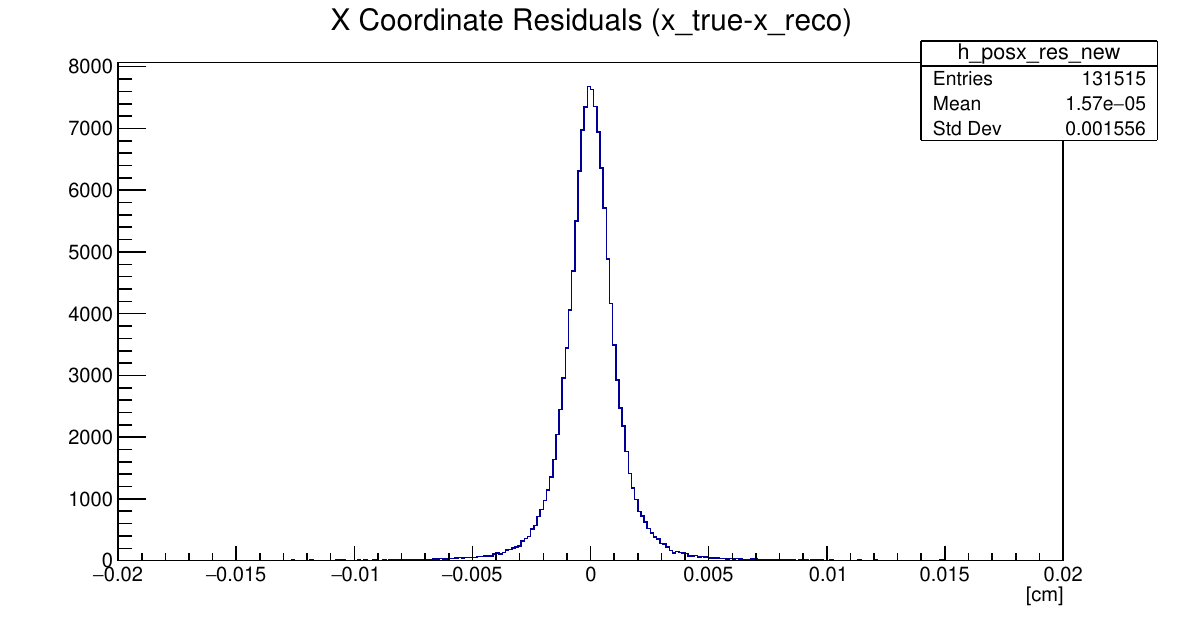} \quad\includegraphics[width=6.5 cm]{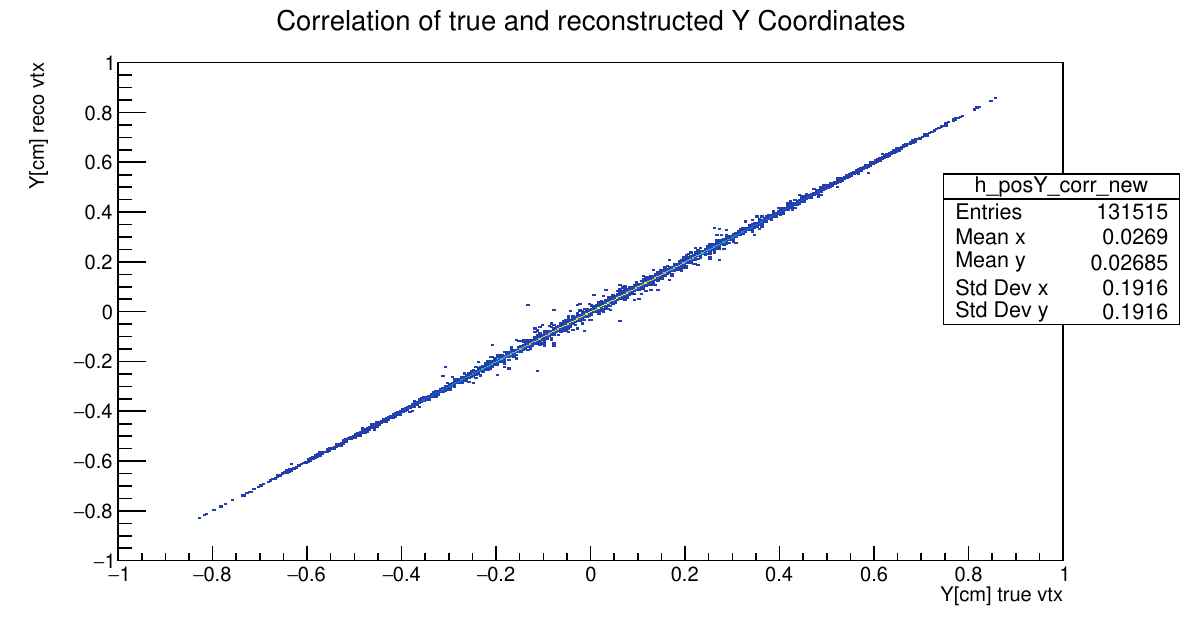} \quad 
\includegraphics[width=6.5 cm]{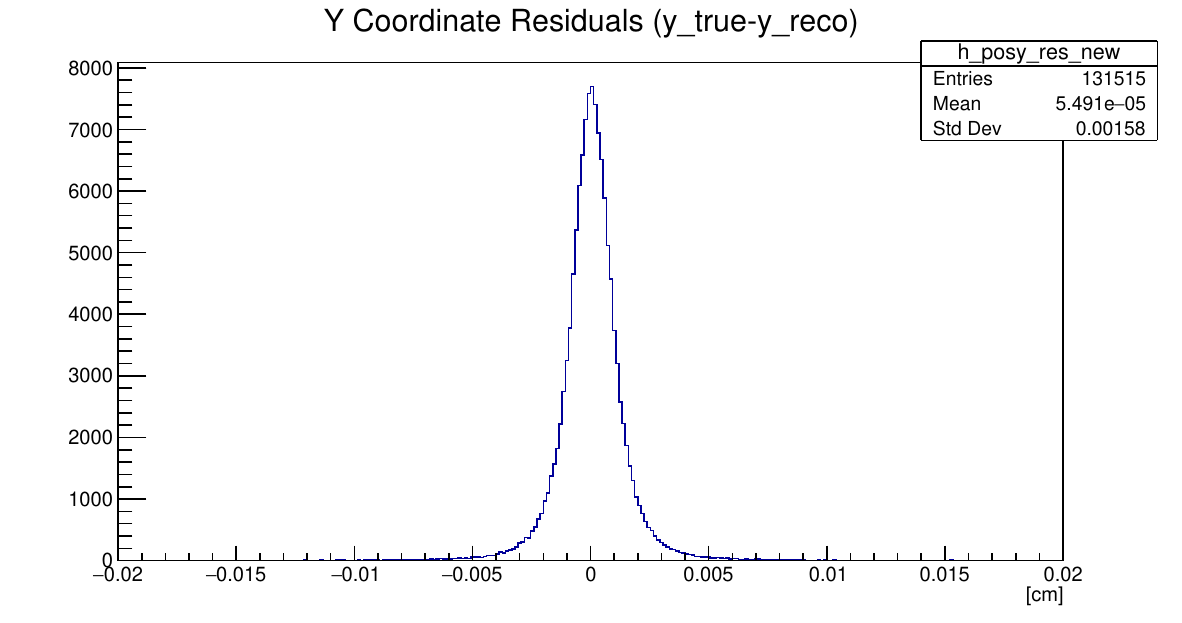} \quad\includegraphics[width=6.5 cm]{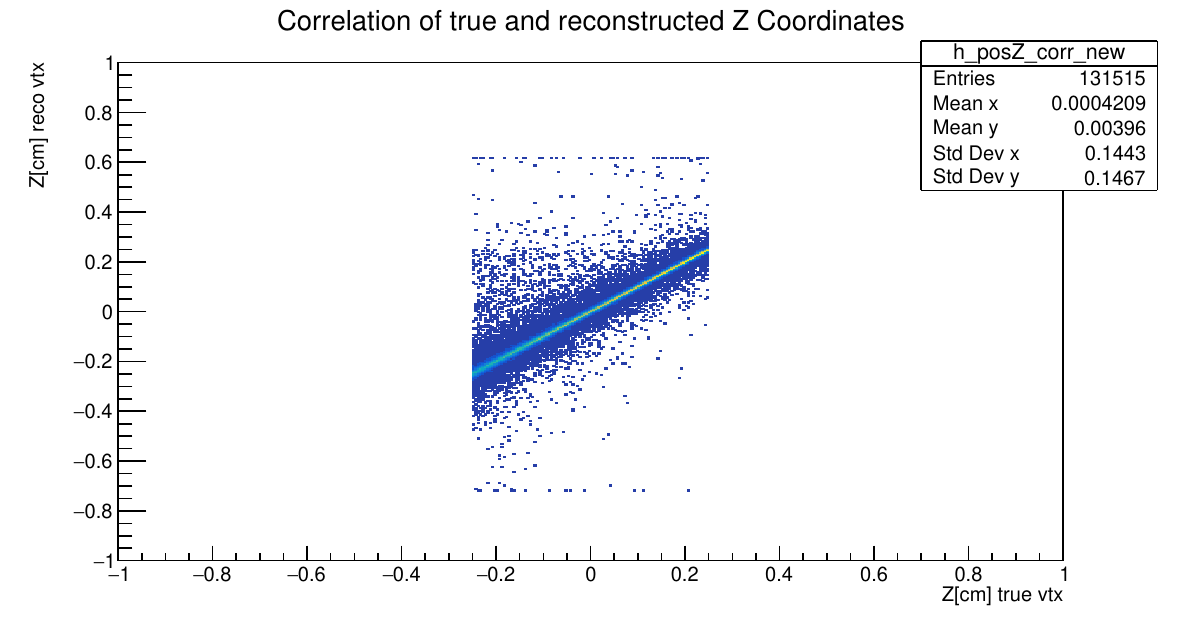} \quad 
 \includegraphics[width=6.5 cm]{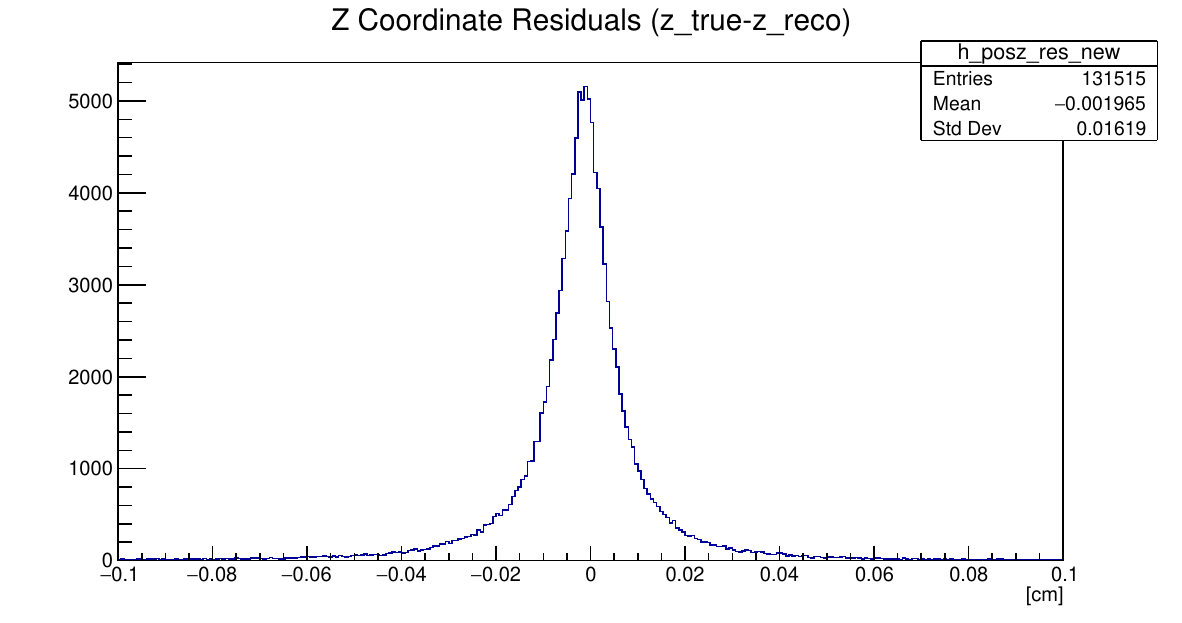} \quad
\caption{Correlation and residual plots for reconstructed versus true coordinates for the CNAO2022 MC simulation. The top row shows the correlation and residuals  for the $x$ coordinates. The second row presents the same thing for the $y$ and last row for the $z$. Note that the residuals for the $z$ coordinates have a different scale compared to $x$ and $y$: the range for $x$ and $y$ is $[-0.02, 0.02]cm$, while for $z$ it is $[-0.1,0.1]cm$}
\label{fig: MC_CNAO2022}
\end{center}
\end{figure}

\subsubsection{Vertex position residual}
The residual and correlation plots (Fig.:\ref{fig: MC_CNAO2022}) for the true and reconstructed coordinates confirm the results obtained in GSI2021 analysis. 

\begin{center}
\begin{tabular}{|c | c |c |} 
 \hline
- & mean $[\mu m]$ & $ \sigma $ $[\mu  m]$\\
 \hline\hline
x & $(9.66 \pm 0.31) \times 10^{-2}$ & $(8.628 \pm 0.033)$ \\
y & $(0.432 \pm 0.036)$ &$(8.478 \pm 0.041)$ \\
z & $(-14.01 \pm 0.25)$ & $(51.92 \pm 0.36)$ \\
 \hline
\end{tabular}
\captionof{table}{Mean and $\sigma$ for the residual histogram}\label{table: residual_MC_CNAO2022}
\end{center}
 
In table: \ref{table: residual_MC_CNAO2022} are reported the mean and $\sigma$ of the Gaussian fits applied to the residual distributions. Although the widths of the residual distributions for CNAO2022 exhibit a slight increase, they remain of the same order of magnitude of GSI2021 simulation (Table: \ref{table: residual_MC_GSI21}), meaning that the overall precision of the reconstruction process remains consistent and comparable across different beam types and experimental setups.

Also in this case, the broader residuals in the $z$-coordinate must still be taken into account. To ensure accurate selection of fragmentation vertexes occurring within the target, the region of interest of the target is extended by $5\sigma_z$. For this simulation, this extension results in a range of $[-0.28, 0.28]cm$, ensuring that the reconstruction uncertainty in z is properly accounted for.

\subsubsection{Vertex reconstruction efficiency}
Turning to the efficiencies ($\epsilon_{tot}$ ) is found to be:

\begin{equation}\label{eq:eff_2}
    \epsilon_{tot}=(85.16 \pm 0.09)\%
\end{equation}

This efficiency is lower than the one observed in GSI2021 (Eq.: \ref{eq:eff}). Since the geometry of the VTX remains unchanged, the decrease is attributed to differences in the physical processes associated with the different beam. CNAO2022 uses carbon ions at $200 MeV/u$, whereas GSI2021 used oxygen ions at $400 MeV/u$. This difference in beam species and energy impacts the fragmentation process, which in turn affects the reconstruction efficiency of valid vertexes.

When considering the efficiency $\epsilon_{reco}$, the value is:
\begin{equation}\label{eq:eff_acc_2}
    \epsilon_{reco}=(91.01 \pm 0.08)\%
\end{equation}
As expected $\epsilon_{reco}$ is higher than $\epsilon_{tot}$, as it accounts only for the reconstruction efficiency of the VTX algorithm, whereas $\epsilon_{tot}$ also includes geometric efficiencies. This efficiency also decreases compared to GSI2021 (Eq.:\ref{eq:eff:_acc}). Again, this reduction highlights the influence of the beam species on reconstruction performance. Despite this it remains reasonably high, demonstrating that the reconstruction algorithm still performs well under the conditions of CNAO2022.

The study then examines the impact of a missing detector layer for CNAO2022. Also in this case, the slightly deteriorated layer is the third layer. To analyze the impact on reconstructable vertexes when this layer is missing, we again consider the fraction of reconstructable vertexes . This quantity allows us to account for the geometric acceptance in both scenarios: when all layers are functional and when one layer is non-functional. The fraction of reconstructable vertexes ($frV$), when one layer is out, is:
\begin{equation}
    frV=(97.88 \pm0.04)\%
\end{equation}
This value is comparable to that observed for GSI2021 (Eq.: \ref{eq: frv}), confirming that the absence of a single layer has a minimal impact on the fraction of reconstructable vertexes. 

\subsubsection{Proportion of vertexes}
The final quantity analyzed for this campaign is the percentage of valid vertexes, with the calculated value reported below: 

\begin{equation}\label{eq:p_vv_cnao2022}
    p_{vv}=(2.897 \pm 0.008)\%
\end{equation}
As expected this value is different from the one of GSI2021, depending on the fragmentation cross sections implemented in MC and expected to be different for two different ions ($^{12}C$ and $^{16}O$) of two different energies (200 and 400 MeV/u respectively), impinging on the same TG (5 mm carbon). 

\subsection{CNAO2023}

The final MC campaign analyzed is CNAO2023. In this case, the beam is the same of CNAO2022 (carbon ions at $200MeV/u$); however, as discussed in the next section (\ref{sect: vtx geometry}), VTX geometry is different, and a magnetic field is now present.  Specifically, the geometry has changed in the following way: a $180^{\circ}$ rotation along the $y-axis$ has caused the first layer in the previous campaigns to become the last layer, and so on for the other layers. This rotation has also resulted in a slight displacement of the VTX center away from the target. How do these changes influence the results?

To better understand the impact of the magnetic field and geometry differences, two dedicated simulations were produced. Both simulations share the same characteristics, including beam and geometry, but one was conducted without the magnetic field ($\slashed{B}$). 

\begin{figure}[h!]
\begin{center}
\includegraphics[width=6.5 cm]{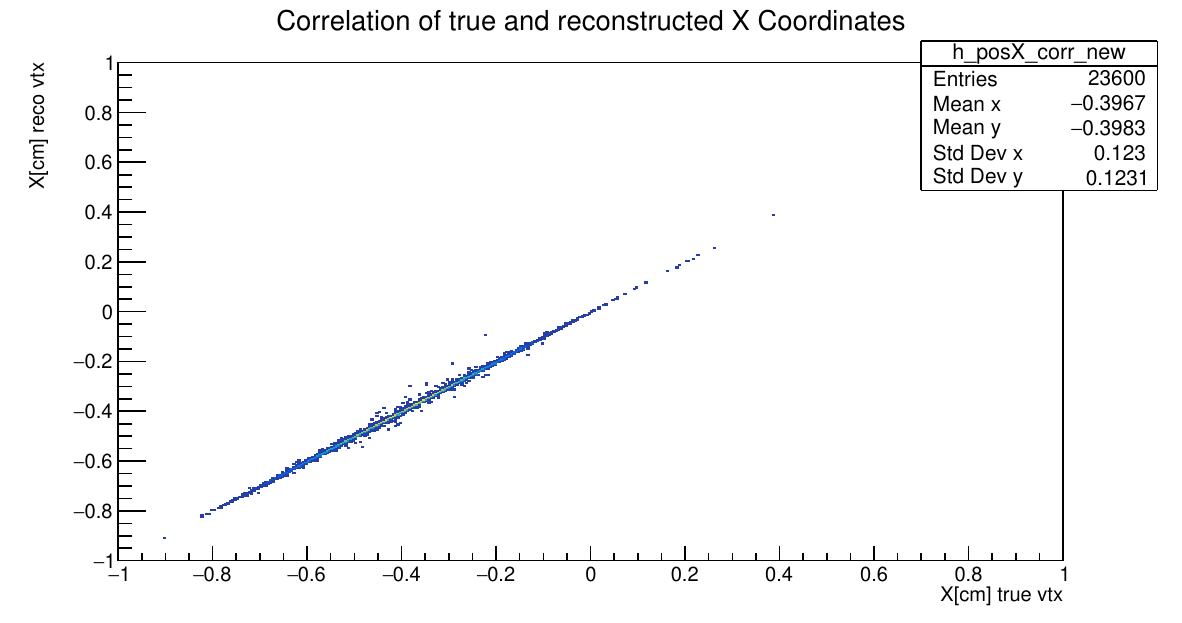} \quad 
\includegraphics[width=6.5 cm]{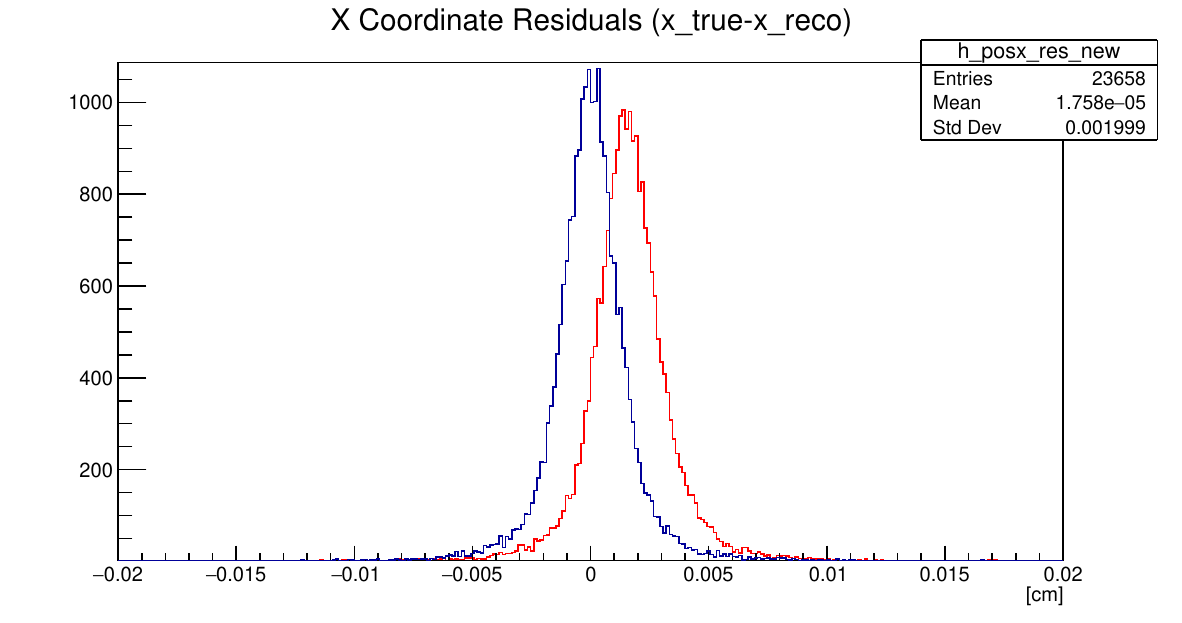} \quad\includegraphics[width=6.5 cm]{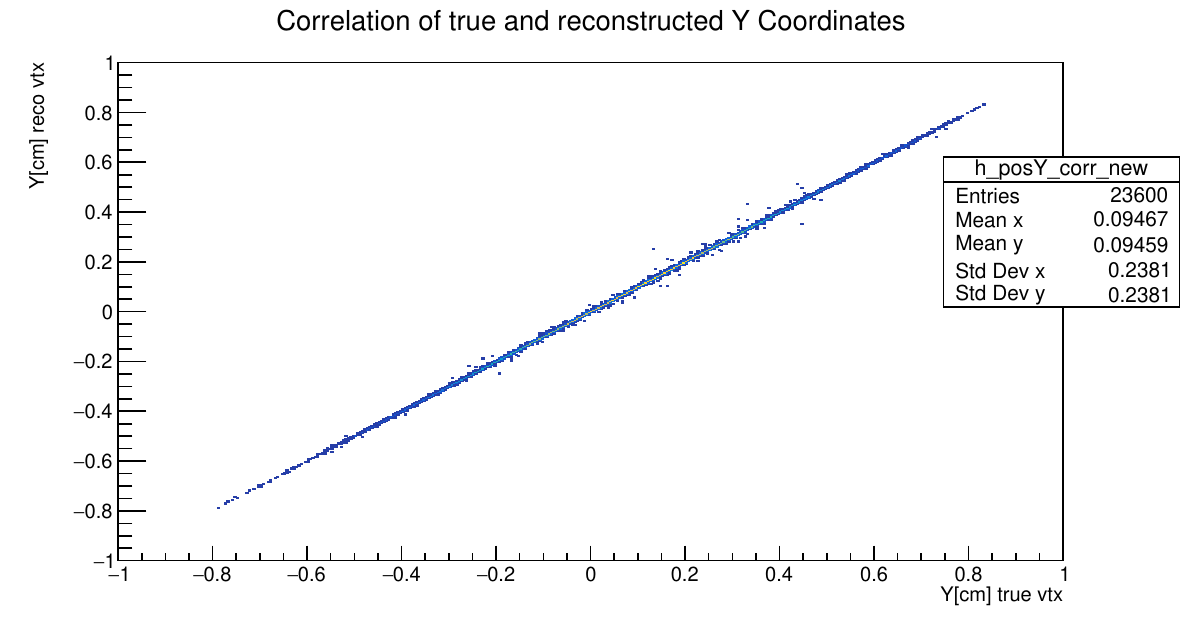} \quad 
\includegraphics[width=6.5 cm]{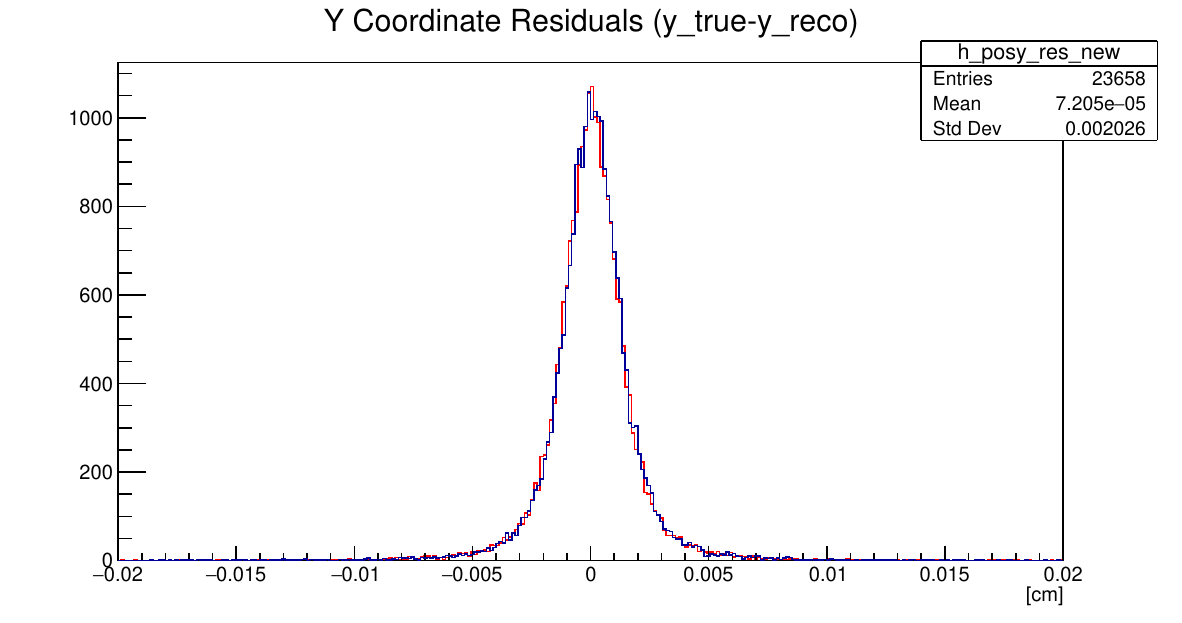} \quad\includegraphics[width=6.5 cm]{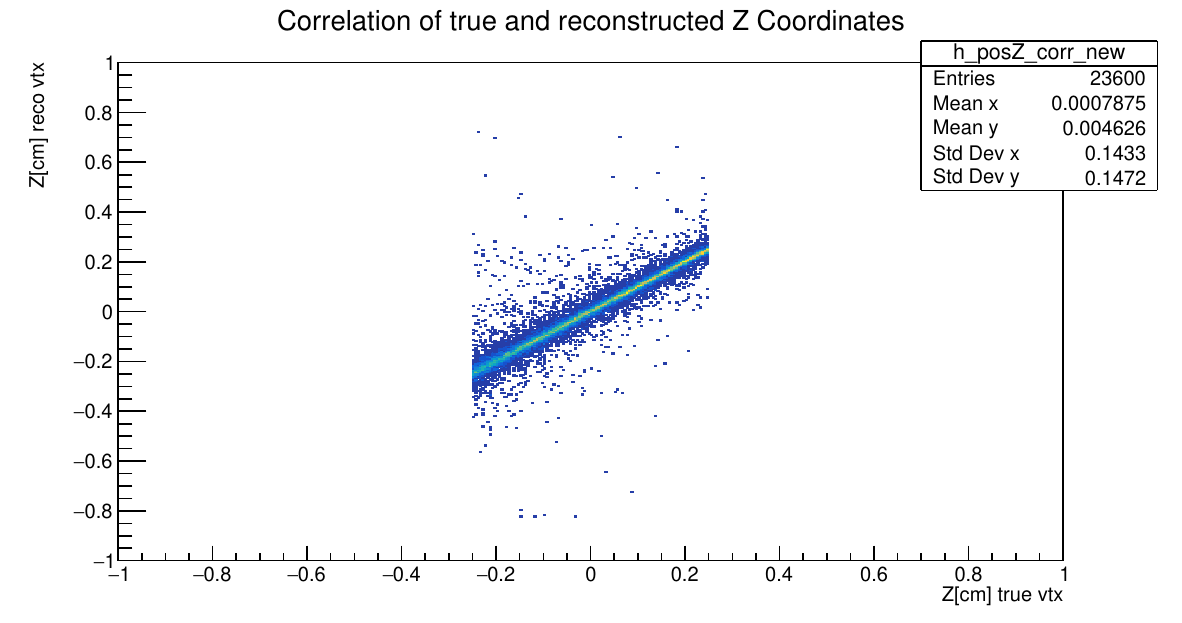} \quad 
 \includegraphics[width=6.5 cm]{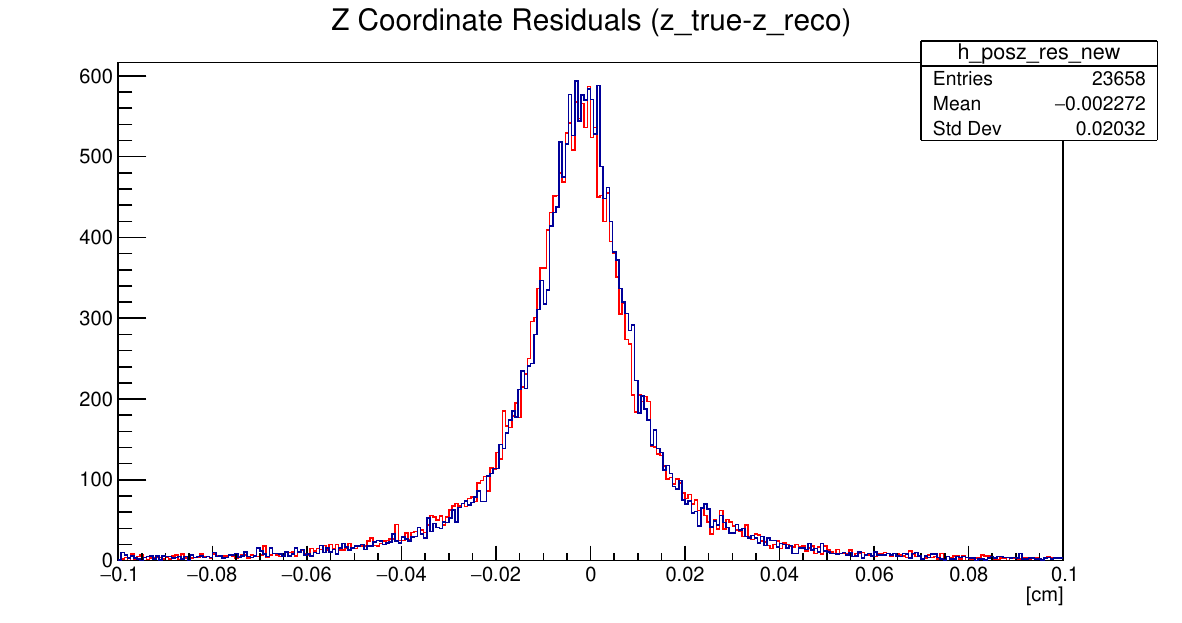} \quad
\caption{Correlation and residual plots for reconstructed versus true coordinates for the CNAO2023 MC simulation. The top row shows the correlation and residuals  for the $x$ coordinates. The second row presents the same thing for the $y$ and last row for the $z$. Note that the residuals for the $z$ coordinates have a different scale compared to $x$ and $y$. Correlation plots are shown only for the campaigns with the magnetic field, while residual plots are presented for both campaigns with the magnetic field (in blue) and without the magnetic field (in red)}
\label{fig: MC_CNAO2023}
\end{center}
\end{figure}

\subsubsection{Vertex position residual}
Starting with the correlation and residual plots (Fig. \ref{fig: MC_CNAO2023}), it is evident that the results are similar to the previous campaigns. 

\begin{figure}[h!]
\begin{center}
\includegraphics[width=8 cm]{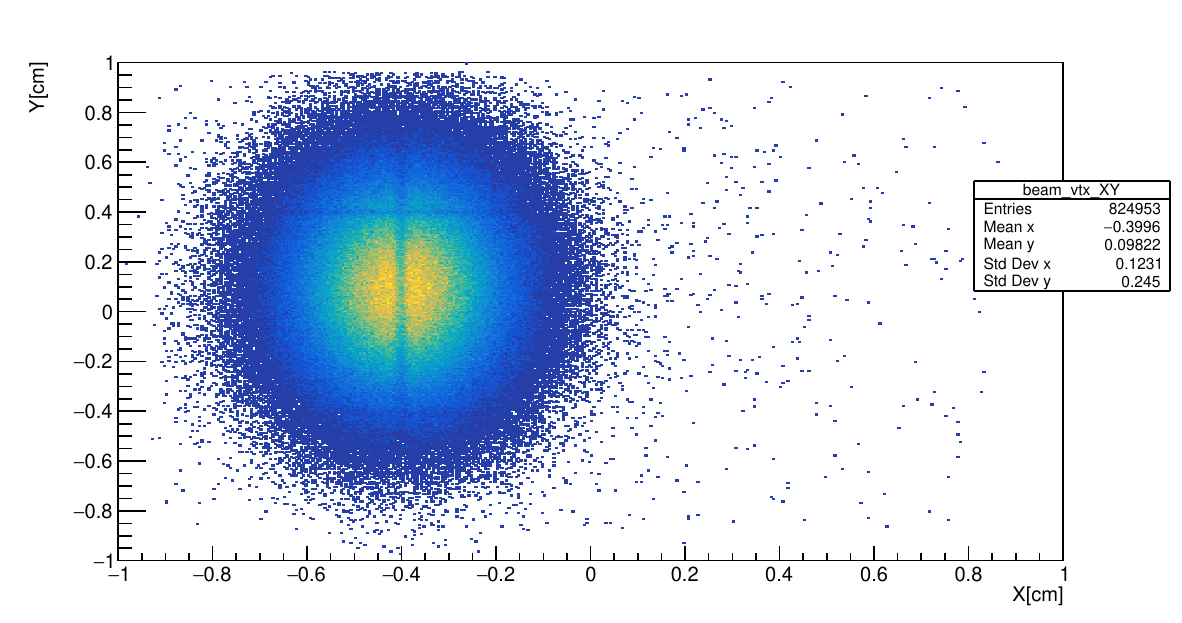}
\caption{Beam profile for the CNAO2023 campaign, reconstructed using primary tracks from the VTX. The visible lines superimposed on the beam are the shadows of the BM wires.}
\label{fig: beam CNAo23_ps}
\end{center}
\end{figure}

One notable observation is related to the correlation plot for the $x$ coordinates. Unlike CNAO2022, where the correlation spans the entire range of $[-1, 1]cm$, in CNAO2023, it is predominantly confined to the range of negative $x$ values, with only few entries in the positive $x$ region. This shift can be attributed to the beam being significantly more decentered along the negative $x$ direction compared to CNAO2022. As shown in the figure \ref{fig: beam CNAo23_ps}, the beam profile, obtained through the reconstructed primary tracks from the VTX, clearly illustrate a beam completely shift toward negative $x$
 values.

Another observation concerns the residuals in the $x$-coordinate. Examining the residuals with the magnetic field, they appear to be a little bit decentered compared to those without the magnetic field. This effect is observed only along the $x$-axis, confirming that is due to the influence of the magnetic field. For a more detailed analysis, the mean and $\sigma$ values of the Gaussian fit for these plots, reported in the table (Tab.:\ref{table: residual_MC_CNAO2023} and Tab.:\ref{table:residual_MC_CNAO2023_noB}), can be examined. Compared to CNAO2022 (Table: \ref{table: residual_MC_CNAO2022}), this residuals are slightly larger, both with and without the magnetic field. Since the two campaigns used the same beam, this small increase is attributed to the change in geometry. Specifically, the rotation of the vertex shifted the layers slightly farther from the target, leading to a modest broadening of the residual distributions.
Additionally, the fits confirm that the center of the Gaussian for the $x$-coordinate residuals worsens in the presence of the magnetic field. The mean of the residual distribution in x is not anymore compatible with zero within 1$\sigma$ uncertainty. This is due to the fact that the vertex reconstruction implemented in FOOT so far doesn't account for magnetic field effects.  This is confirmed by the fact that without the magnetic field, the Gaussian center aligns well with the results from previous campaigns. For all the other quantities, the simulations with and without the magnetic field are in agreement.

In the case of CNAO2023 (with magnetic field) the z-range of vertexes to be considered generated within the target, adding again $5 \sigma_z$ to the target dimensions (Tab.: \ref{table: residual_MC_CNAO2023}), results in a range of $[-0.29,0.29]$~cm.

\begin{center}
\begin{tabular}{|c | c |c |} 
 \hline
        - & mean $[\mu m]$ & $ \sigma $ $[\mu m]$\\
         \hline\hline
        x & $(15.52 \pm0.12) $ & $(11.68 \pm 0.14) $ \\
        y & $(0.476\pm 0.098)$ &$(11.33 \pm0.11)$ \\
        z & $(-23.3 \pm 1.1)$ & $(86.7 \pm 1.4)$ \\
 \hline
\end{tabular}
\captionof{table}{Mean and $\sigma$ for the residual histogram for the CNAO2023 campaigns with magnetic field }\label{table: residual_MC_CNAO2023}
\end{center}

\begin{center}
\begin{tabular}{|c | c |c |} 
 \hline
        - & mean $[\mu m]$ & $ \sigma $ $[\mu m]$\\
         \hline\hline
        x & $(-7.81 \pm0.11) \times10^{-2}$ & $(11.07\pm 0.13) $ \\
        y & $(0.703\pm 0.096)$ &$(11.24 \pm0.11) $ \\
        z & $(-19.19 \pm 0.88)$ & $(80.2 \pm 1.1)$ \\
 \hline
\end{tabular}
\captionof{table}{Mean and $\sigma$ for the residual histogram, for the CNAO2023 campaigns without magnetic field} \label{table:residual_MC_CNAO2023_noB}
\end{center}

\subsubsection{Vertex reconstruction efficiency}

The efficiency ($\epsilon_{tot}$) for the CNAO2023 campaign is:

\begin{equation}
    \epsilon_{tot}=(83.83\pm 0.22)\%  \quad  \quad \quad \epsilon_{tot}(\slashed{B})= (84.40 \pm 0.22)\%
\end{equation}

The efficiencies with and without the magnetic field are found to be compatible within the uncertainties, indicating that the presence of the magnetic field does not significantly impact the algorithm's ability to reconstruct vertexes. When compared to CNAO2022, the efficiencies are of the same order of magnitude, with the CNAO2023 efficiency being slightly lower but only by a marginal amount. Overall, the efficiencies can be considered compatible, and the small difference observed can be attributed to the updated VTX geometry in CNAO2023. 

\begin{equation}
    \epsilon_{reco}= (92.95 \pm 0.16)\%   \quad  \quad \quad  \epsilon_{reco}(\slashed{B})= (92.99 \pm 0.16)\%
\end{equation}

The $\epsilon_{reco}$ with and without the magnetic field is also found to be compatible, further confirming that the presence of the magnetic field does not interfere with the algorithm's ability to identify vertexes. Moreover, these values are of the same order of magnitude of CNAO2022. This further supports the conclusion that that previous discrepancy are linked to the geometry of the VTX, when accounting only for the reconstruction efficiency, there is no differences between the two campaigns.

For CNAO2023, studying the fraction of reconstructible vertexes when a layer is not operational becomes even more relevant. The problematic layer (in this case, the second) appears to have progressively deteriorated across the campaigns, performing the worst in CNAO2023.
\begin{equation}
    frV= (96.99 \pm 0.11)\%   \quad  \quad \quad frV(\slashed{B})= (96.81 \pm 0.11)\%
\end{equation}
The values of $frV$ confirm that, even in this campaign, the percentage of reconstructible vertexes with one layer out of operation remains high. 
\subsubsection{Proportion of vertexes}
To conclude the analysis of this campaign, the focus shifts to examining the percentages of valid vertexes.
\begin{equation} \label{eq:p_vv cnao2023}
    p_{vv}= (2.78\pm 0.02)\% \quad  \quad \quad p_{vv}(\slashed{B})=(2.79\pm 0.02)\%
\end{equation}

The percentages of valid vertexes with and without the magnetic field, are fully compatible with each other. Moreover, they are in excellent agreement with the result obtained for CNAO2022 \ref{eq:p_vv_cnao2022}, as expected given the identical beam and target configurations. The slightly lower values observed in this case may again be related to the rotation of the VTX, which shifted its center slightly backward, increasing the number of unreconstructed vertexes as they fall outside the geometric acceptance of the vertex detector.

\section{Vertex geometry} \label{sect: vtx geometry}
In this section, I will presents the analysis in chronological order, highlighting the steps taken, insights gained, and their contribution to the final results.

Having demonstrated the excellent performance of the vertex reconstruction algorithm on MC simulations, I proceeded to apply it to the experimental data.

The first observed quantity is the proportion of valid vertexes with respect to the primaries impinging on the target, denoted as $p_{vv}$ and described in the previous paragraph. This quantity is of interest because, as already mentioned, the total fragmentation cross section is proportional to it. 


 

The first dataset examined consisted of data collected at CNAO2023, specifically runs with a $^{12}C$ beam of $200MeV/u$, a $5 mm$ carbon target and a MB trigger.

The percentage of valid vertexes obtained is $p_{vv}=(0.066 \pm 0.006)\%$, a result consistent across all runs within the uncertainties. However, this percentage appeared notably low. Confronting it with the result from MC simulation \ref{eq:p_vv cnao2023}, it becomes evident that there is a significant discrepancy between the two. It is clear that the experimental data would not perfectly match the simulations, as the latter relied on cross-sections implemented within the model, which is precisely what the experiment aimed to measure. 
However, the total cross sections used in simulations should not deviate too far from reality, because they are based on models constrained to measured data (which are abundant for total cross sections, while are missing for the differential ones). Thus, the expectation is that the experimental results should not diverge significantly from the simulated ones. 

\begin{figure}[h!]
\begin{center}
\includegraphics[width=6.7 cm]{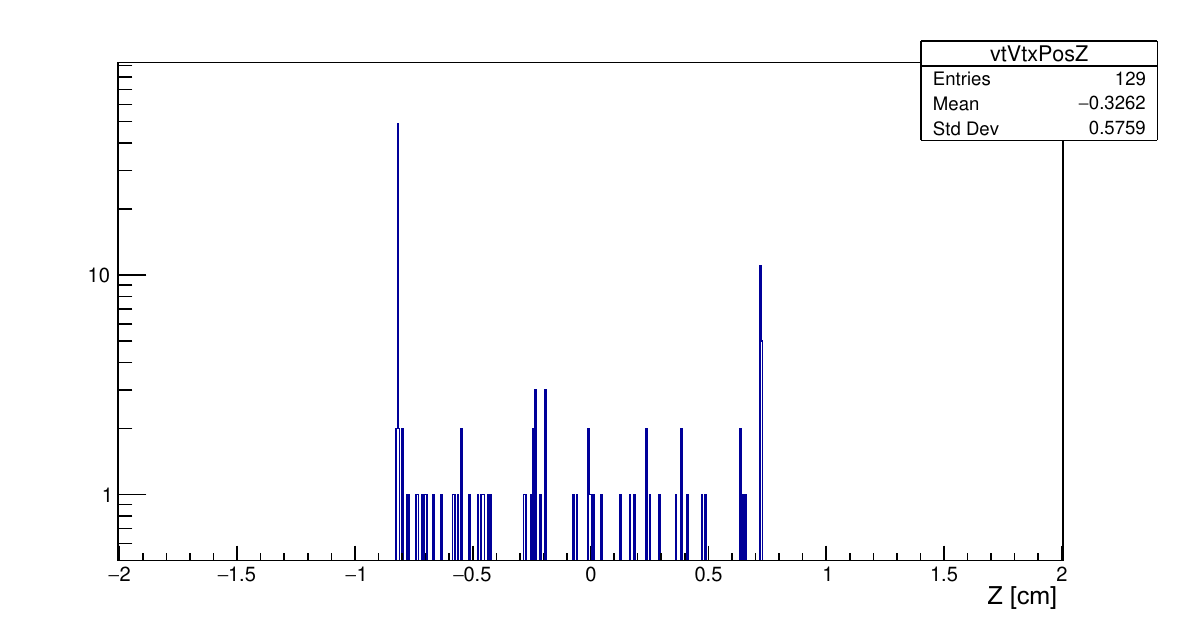} \quad \includegraphics[width=6.7 cm]{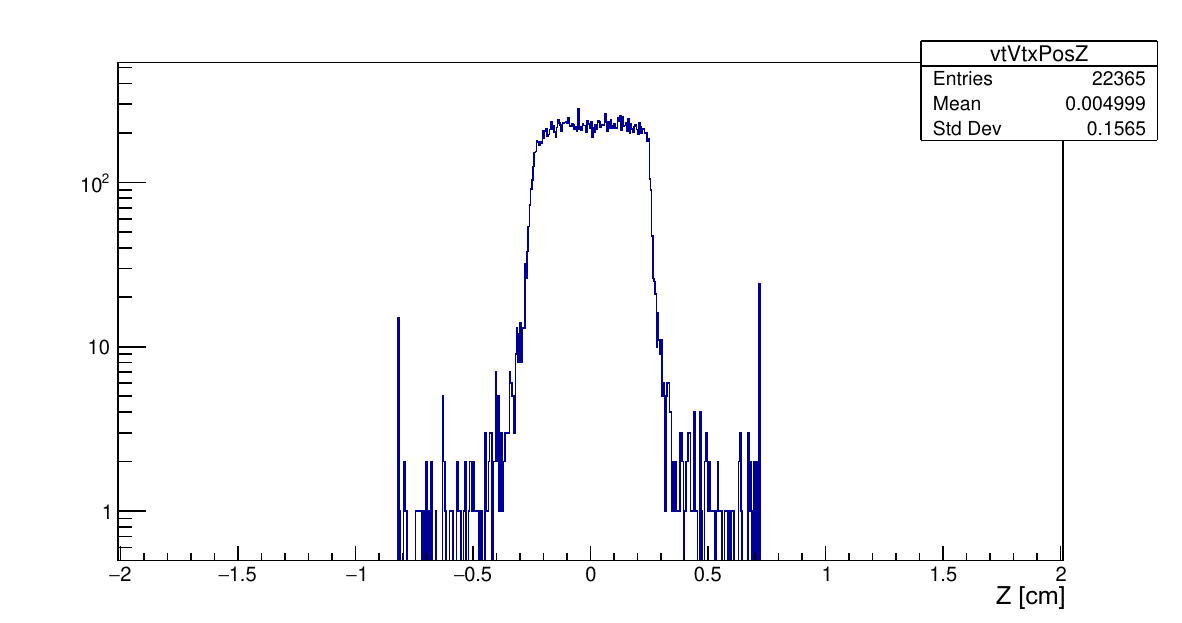} \quad \includegraphics[width=6.7 cm]{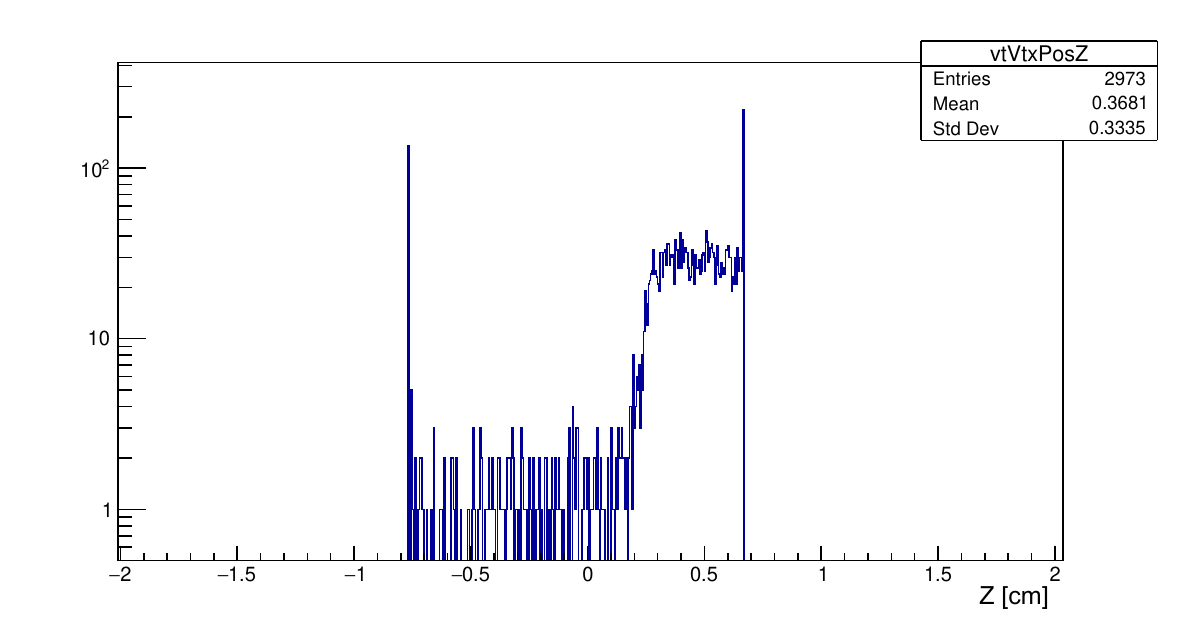} \quad \includegraphics[width=6.7 cm]{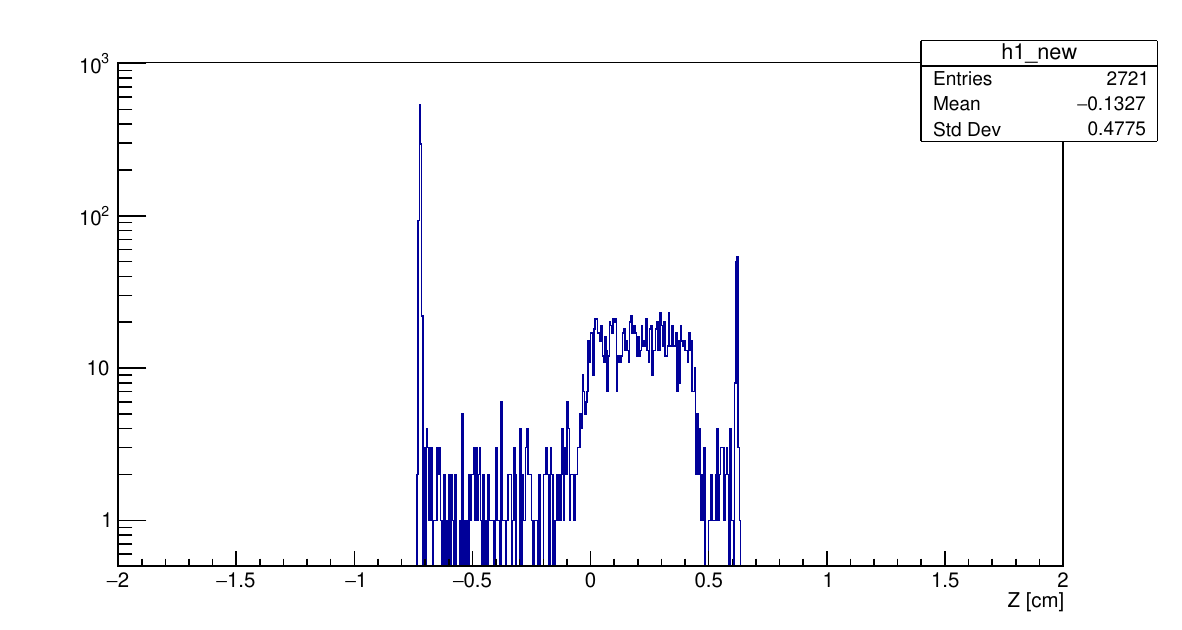} 
\caption{Histograms of z-coordinate of the fragmentation vertexes. In order are reported the histogram of campaign CNAO2023 (run 6136), MC simulation of CNAO2023 with passive materials (CNAO23PS$\_$MC), CNAO2022 (run 5468) and GSI2021 (run 4305).}
\label{fig: posz}
\end{center}
\end{figure}

The presence of an issue specific to this data was further supported by the histograms in Figure \ref{fig: posz}, which display the z-coordinate distribution of valid vertexes for different campaigns.
As expected for all the campaigns the vertexes are distributed within the region [$1.2 \cdot minZ$, $0.8 \cdot MaxZ$]. Examining the histograms from MC simulations (top right plot in Fig. \ref{fig: posz} ), as well as those from the CNAO2022 and GSI2021 campaigns ( bottom left and bottom right plots respectively in Fig. \ref{fig: posz}), a clear peak corresponding to the target is evident. This is a direct result of the higher fragmentation probability within the target compared to the surrounding air. However, for CNAO2023 (top left in Fig. \ref{fig: posz} ), there is no such evidence of the peak corresponding to the target.

The first hypothesis was a shift causing the peak to fall outside the expected range. As the target is not centered at zero in other campaigns, a more significant shift in CNAO2023 could place the peak outside the valid vertex region. Expanding this region, however, shows no evidence of the target.

\begin{figure}[h!]
\begin{center}
\includegraphics[width=6.7cm]{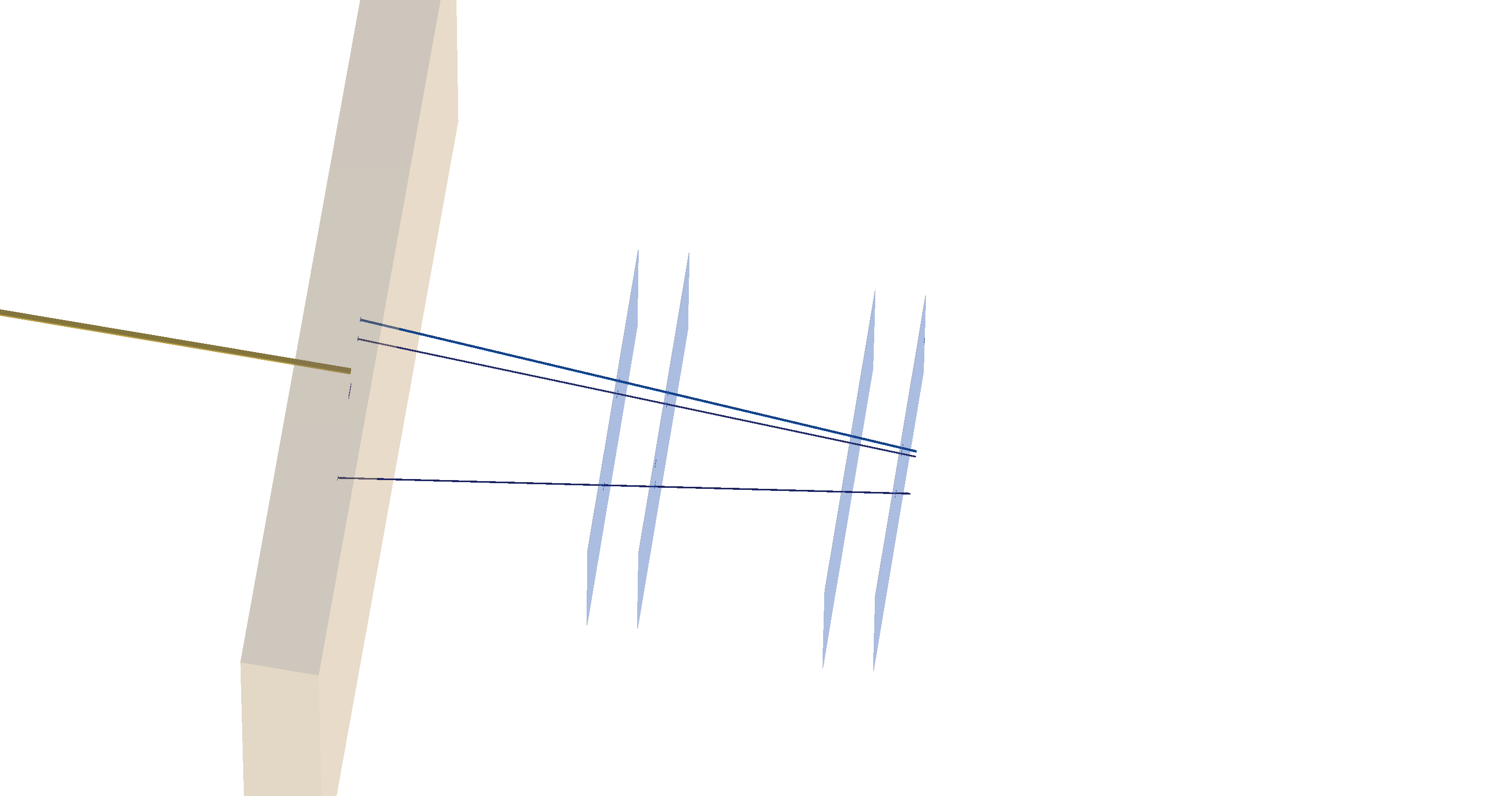} \quad \includegraphics[width=6.7cm]{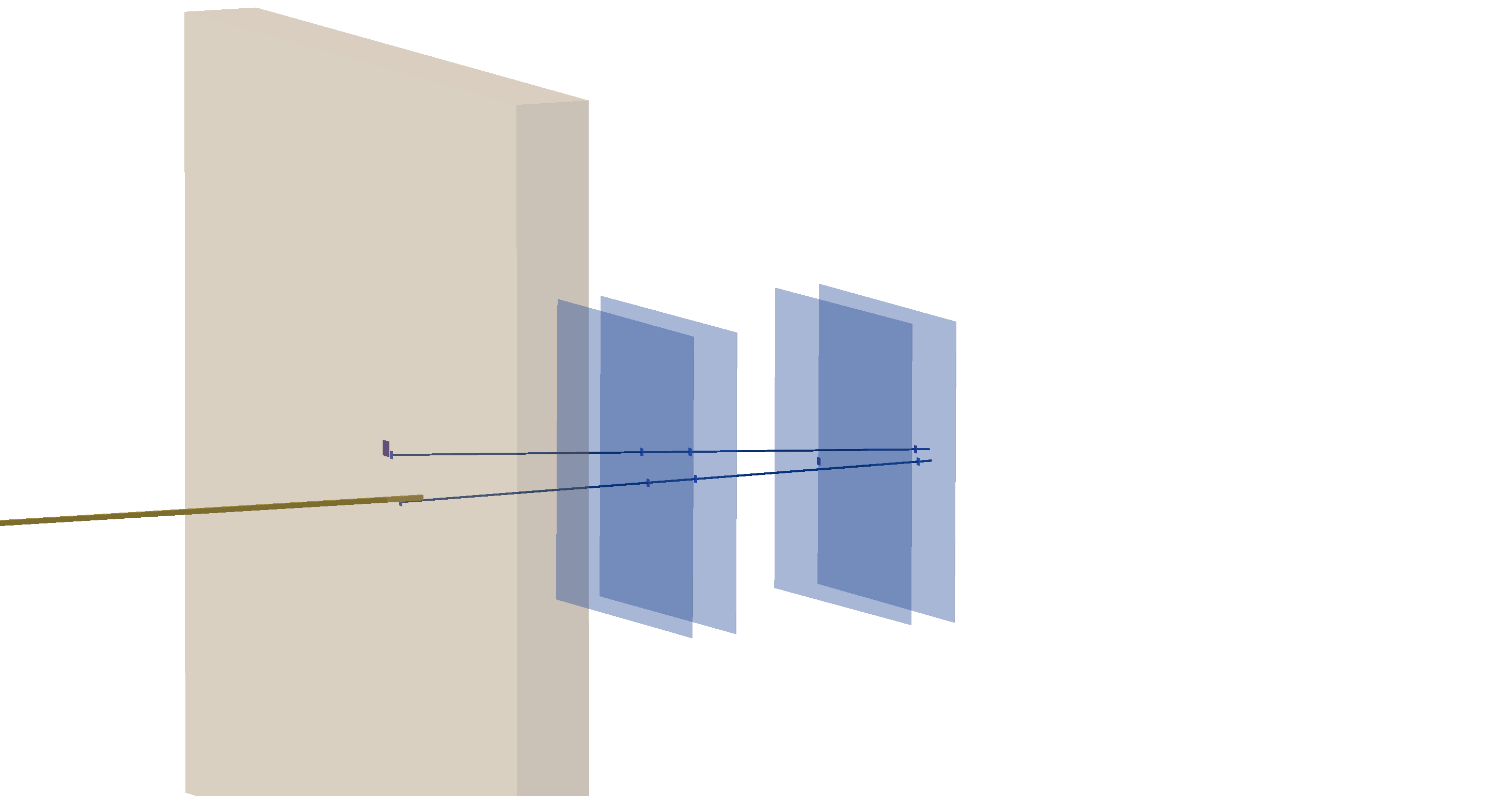} 
\caption{Event display. The 4 planes in blue are the layers of the VTX, the box in gray is the target, the track in yellow is the track reconstructed by the BM and those in blue are those reconstructed by the VTX. For both images is possible to see how the tracks converge on the opposite side to the target.}
\label{fig: opposit convergence}
\end{center}
\end{figure}

An analysis of the event display revealed an inconsistency: in several events, such as the two depicted in the Figure \ref{fig: opposit convergence}, it appears that the tracks converge in the opposite direction to the target. This observation suggests that the layers are not positioned in accordance with the data at the software level. Specifically, what is the first layer encountered by the beam in the experimental setup is positioned as the last layer in the software, the second layer is placed as the third, and so on. This bad positioning results from the $180 ^ \circ $ rotation of the VTX detector box around the Y-axis during the CNAO2023 data acquisition, which is not accounted for in the software configuration. Therefore, to achieve the correct configuration, it is necessary to change the order of the layers and also apply a left-right inversion.

\begin{figure}[h!]
\begin{center}
\includegraphics[width=9cm]{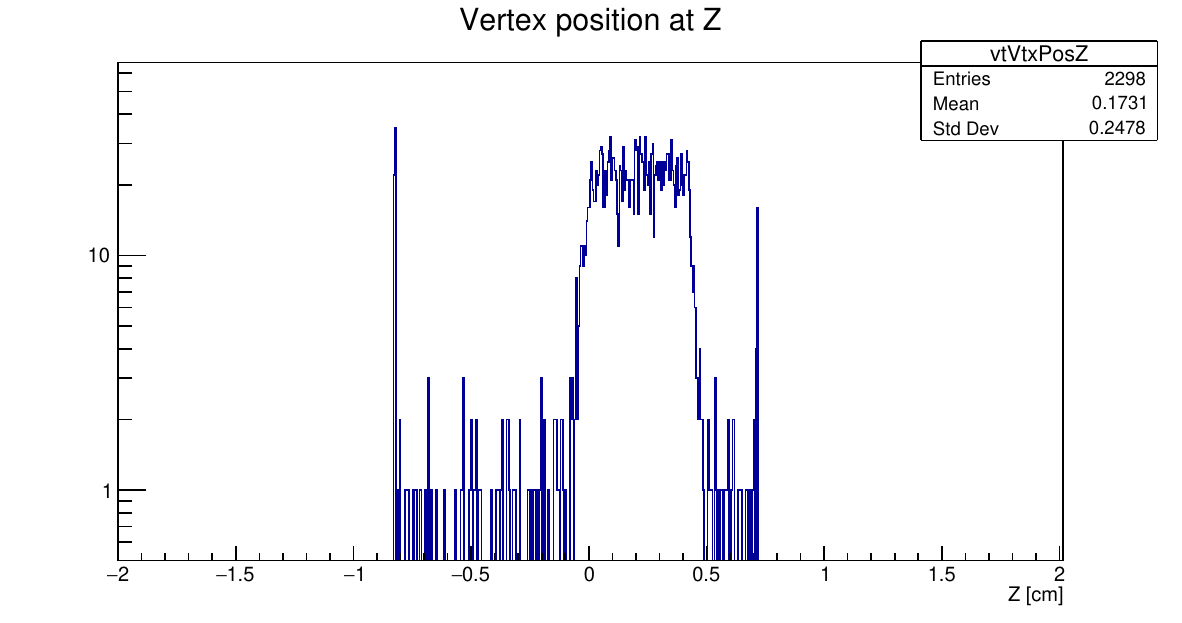}
\caption{Histograms of z-coordinate of the fragmentation vertexes, for campaign CNAO2023 (run 6136), after the rotation of the box of the VTX.}
\label{fig: posz CNAO2023 after rotation}
\end{center}
\end{figure}

After this adjustment, the 
z-coordinate distribution for valid vertexes revealed the expected peak corresponding to the target, confirming the resolution of the issue (Figure: \ref{fig: posz CNAO2023 after rotation}).

\subsection{Layer and Global Positioning of the Vertex Detector}
It is necessary to understand why the target is not centered at zero, as expected from the global reference frame coordinates used in the plot. 

In the experimental data, the target appears not centered at zero, and this shift is consistent across different runs with the same target for each campaign. In contrast, the MC simulations consistently show the target perfectly centered at zero for all campaigns. Top right figure in \ref{fig: posz} shows the histogram for the CNAO2023 simulation, with a peak centered at zero as expected. The same behavior is observed for the other MC campaigns.




Regarding the shift of the peak present on the date, having ruled out software errors, since this issue does not appear in the MC simulations, it is concluded that the shift originate from an error in the geometry implemented in the reconstruction code. The first hypothesis considered is that the error stemmed  from an incorrect measurement of the global z-position of the VTX in the software geometry files. Given the high precision of the VTX detector, it can be used to determine the correct z-position.  The Table \ref{table: shift 1} presents both the old and new position values, showing differences of few millimeters.
\begin{center}
\begin{tabular}{|c |c| c| } 
 \hline
Campaign & $Pos_Z$ old $[cm]$ & $Pos_Z$ new $[cm]$\\
 \hline\hline
CNAO2023 & $2.10$ & $1.90$\\
CNAO2022 & $2.35$ & $1.90$\\
GSI2021 & $2.60$ & $2.40$\\
 \hline
\end{tabular}
\captionof{table}{Z-coordinates for the VTX detector position in the global reference frame, showing original values and updated values after correcting for the observed shift.}\label{table: shift 1}
\end{center}



\begin{figure}[h!]
\begin{center}
\includegraphics[width=1.9cm]{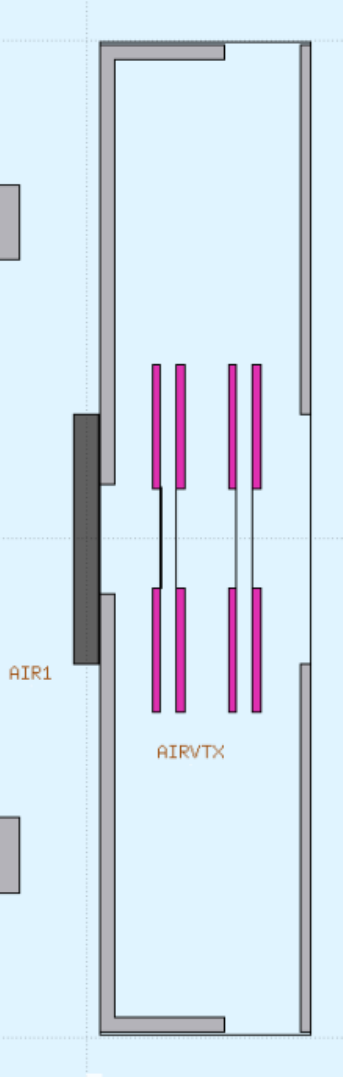} \quad \includegraphics[width=2.4cm]{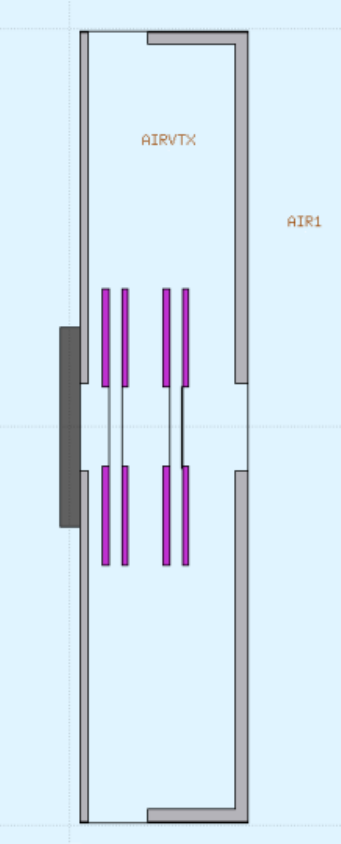}
\caption{Simulated VTX detector box configurations for CNAO2023 and CNAO2022, shown in order. The figure shows the target and the VTX box containing the four layers. Note that the layers are nearly centered in CNAO2023, while they are visibly offset to the left in CNAO2022. The only expected difference between the two configurations should be the box-rotation.}
\label{fig: box MC}
\end{center}
\end{figure}

By utilizing this new z-positions of the VTX, it is possible to achieve a target centered at zero in all campaigns. 
However, when attempting to align the MC geometry with the updated configuration, discrepancies arise. Specifically, in CNAO2023, the four layers are nearly centered within the box, whereas in CNAO2022, the configuration appears highly asymmetric (Fig.:\ref{fig: box MC}). Since the only difference between the two campaigns is the $180^{\circ}$ rotation of the VTX, no change in the relative distances between the layers should be observed. This unexpected asymmetry suggests potential inaccuracies in the obtained values.

At this point, it is crucial to confirm the dimensions and relative distances of the VTX box. A measurement reveals slight discrepancies in the positioning of the layers within the software (see Table \ref{table: layers pos}). Given the high precision of the VTX, even a misalignment of $\sim 1 mm$ can have a significant impact.

\begin{center}
\begin{tabular}{|c |c| c| } 
 \hline
layer & $Pos_Z$ old $[cm]$ & $Pos_Z$ new $[cm]$\\
 \hline\hline
1 & $-0.92$ & $-1.0625$\\
2 & $-0.60$ & $-0.5975$\\
3 & $0.60$ & $0.5975$\\
4 & $0.92$ & $1.0625$\\
 \hline
\end{tabular}
\captionof{table}{Relative positions of the layers with respect to the VTX center in the previous geometry configuration (old) and the updated geometry based on the new measure (new) }\label{table: layers pos}
\end{center}

Using the updated measurements, the VTX global position for the different campaigns are :

\begin{center}
\begin{tabular}{|c |c|} 
 \hline
Campaign & $Pos_Z$ $[cm]$ \\
 \hline\hline
CNAO2023 & $2.61$\\
CNAO2022 & $2.09$ \\
GSI2021 & $2.09$ \\
 \hline
\end{tabular}
\captionof{table}{Z-coordinates for the VTX detector position in the global reference frame.}\label{table: vtx_pos}
\end{center}

With this corrected vertex geometry, the target is now centered in all the analyzed campaigns except for CNAO2023. This small discrepancy of $\sim$2~mm in CNAO2023 has not been yet understood (as studied with the use of the MC it is not an effect of the magnetic field). 

With the correct geometry now implemented in both the reconstruction software and the MC simulations, the analyses are consistent with the experimental configuration. This improvement allows for more accurate studies of the system and ensures the reliability of the results presented in this thesis. Also the results presented in the previous section are obtained using these new simulations. 

\section{Parameter tuning for the different campaigns}
After analyzing the efficiencies and the percentages of vertexes in the MC simulations for the various campaigns, and after correcting the detector geometry, the focus now shifts to the analysis of the experimental data. For each campaign, a dedicated study is performed on the different parameters of the algorithm responsible for track and vertex reconstruction. The parameters on which the study is carried out are the selection of the algorithm (Full or Std), the SHD, and the potential impact of including or not the deteriorating layer. For this last parameter the objective is to determine whether adding the layer enhances the algorithm performance in terms of number of reconstructed vertexes or introduces complications, such as degrading track quality or making vertex identification more difficult.

This study must be carried out on the experimental data rather than the MC simulations. In data are present other effects not present in MC which affect the vertexing performance: the misalignment between the layers of the VTX and also between the VTX and the others detectors, and also the pile-up. It is therefore crucial to optimize the algorithm's parameters to ensure it can identify the highest possible number of vertexes while effectively managing pile-up and noise. At the moment the MC simulations use a perfectly aligned system, and the pile-up is not simulated, consequently the impact of changing some parameters of tracking algorithm, like the SHD, is almost zero. On the contrary in data an established procedure to align offline the detectors with respect to the global reference frame is performed for each campaign. Anyhow small mis-alignments can remains and affects the results. The analysis described in this paragraph is done downstream the alignment procedure.

However, this study will be reported only for the GSI2021 and CNAO2023 campaigns. For the CNAO2022 campaign, after correcting the layer positions as detailed in the previous paragraph, no alignment procedures was performed. Moreover, the TW calibration required for Z identification has not yet been finalized, preventing a consistent analysis.

The study will monitor the number of identified vertexes generated within the target region and successfully matched with the BM, with the aim of maximizing this number. The entire analysis will be conducted using a single physics run (beam + target + minimum bias trigger) for each campaign, carefully selecting the run with the highest available statistics. 

After completing the analysis and identifying the best configuration, the next step is to compare the percentage of valid vertexes with the corresponding results from the MC simulations to check for consistency.

\subsection{Algorithm optimization}
The first aspect we aim to address is the algorithm itself, specifically the two possibility are Full or Std. The objective is to determine whether the choice between these two options is inconsequential or if one offers superior performance in terms of vertex reconstruction.
\subsubsection{GSI2021}
The first campaign analyzed is GSI2021. For this study, the SHD is kept at $0.06$~cm and all 4 layers of the VTX are active. These settings serve as the baseline for assessing the algorithm's performance and understanding the impact of the algorithm.

The percentage of valid vertex generated within the target and successfully matched with the BM  for the two respective configurations are:

\begin{equation}
    p_{vv}(Full)=(1.63\pm 0.02)\% \quad\quad\quad p_{vv}(Std)=(2.34\pm 0.02)\%
\end{equation}

This result shows that the Std configuration identifies approximately $43.5\%$ more vertexes than the Full configuration. To ensure that the additional vertexes identified in the standard configuration are not false positives or low-quality artifacts, two plots are analyzed.

The first plot examines the distribution of the angles ($\Delta \theta$) between the tracks of valid vertexes. Particular attention is given to small angles ($\Delta \theta <0.02 rad$), because it is observe that under suboptimal conditions, the algorithm may erroneously reconstruct vertexes composed of pile-up from primary particles. In such cases, since both tracks originate from primary particles, the angle between them tends to be small. 

\begin{figure}[h!]
\begin{center}
\includegraphics[width=8 cm]{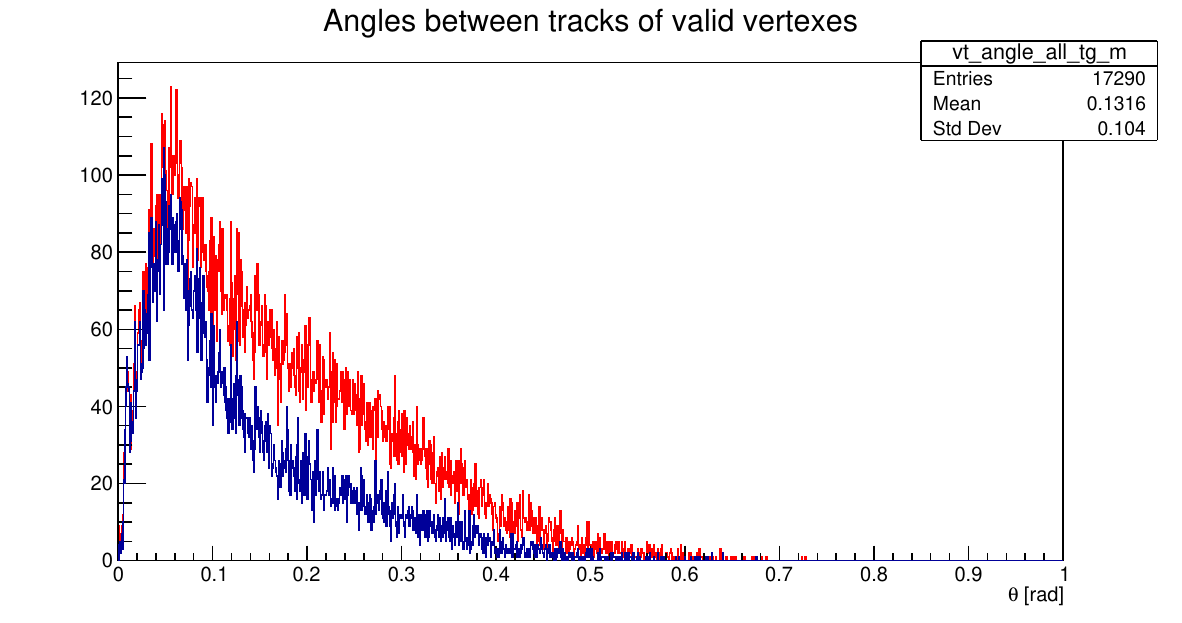} 
\caption{Distribution of angles between the tracks of valid vertexes generated in the target and matched with the BM. In red is reported the \textit{Std} configuration, in blue the \textit{Full} configuration.  }
\label{fig: angles GSI2021}
\end{center}
\end{figure}

The plot \ref{fig: angles GSI2021} reveals that at small angles, the two configurations exhibit a similar behavior, remaining largely compatible. However, for larger angles, the \textit{Std} configuration shows a higher number of entries. This observation suggests that the $Std$ algorithm is effectively recovering valid vertexes with tracks forming realistic angles, rather than misidentifying pile-up events.

The second plot analyzed focuses on the charge distribution of TW points for events where a valid vertex is generated within the target region and successfully matched with the BM. For these events, a histogram is filled with the charge associated with the TW points. By analyzing this distribution, it is possible to identify any anomalies or systematic differences that might indicate issues with the reconstruction process in the two configuration.

\begin{figure}[h!]
\begin{center}
\includegraphics[width=8 cm]{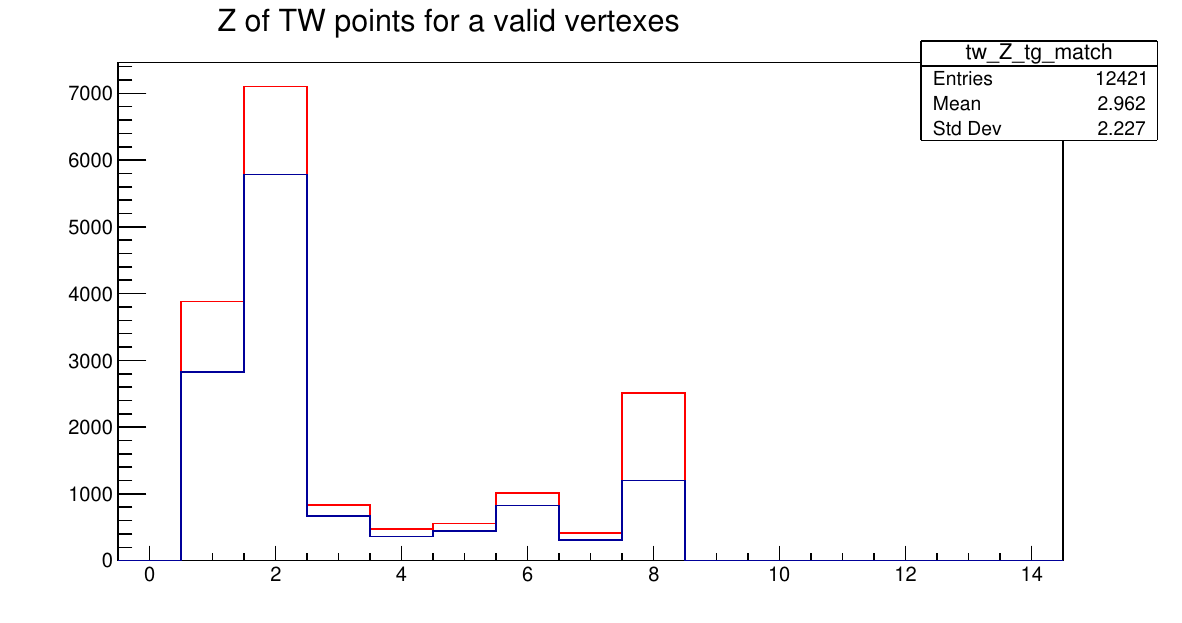} 
\caption{Charge distribution of TW points for events with a matched vertex generated in the target. In red is reported the \textit{Std} configuration, in blu the \textit{Full} configuration. }
\label{fig: charge GSI2021}
\end{center}
\end{figure}

The plot \ref{fig: charge GSI2021} demonstrates that in the \textit{Std} configuration, there is a noticeable increase in entries with TW point charges less than 8. This indicates that the $Std$ algorithm has effectively included events where fragments, reach the TW. However, there is also an increase in entries in the bin corresponding to a charge of 8, which suggests that events involving primary particles reaching the TW have also been included.

\begin{figure}[h!]
\begin{center}
\includegraphics[width=8.5 cm]{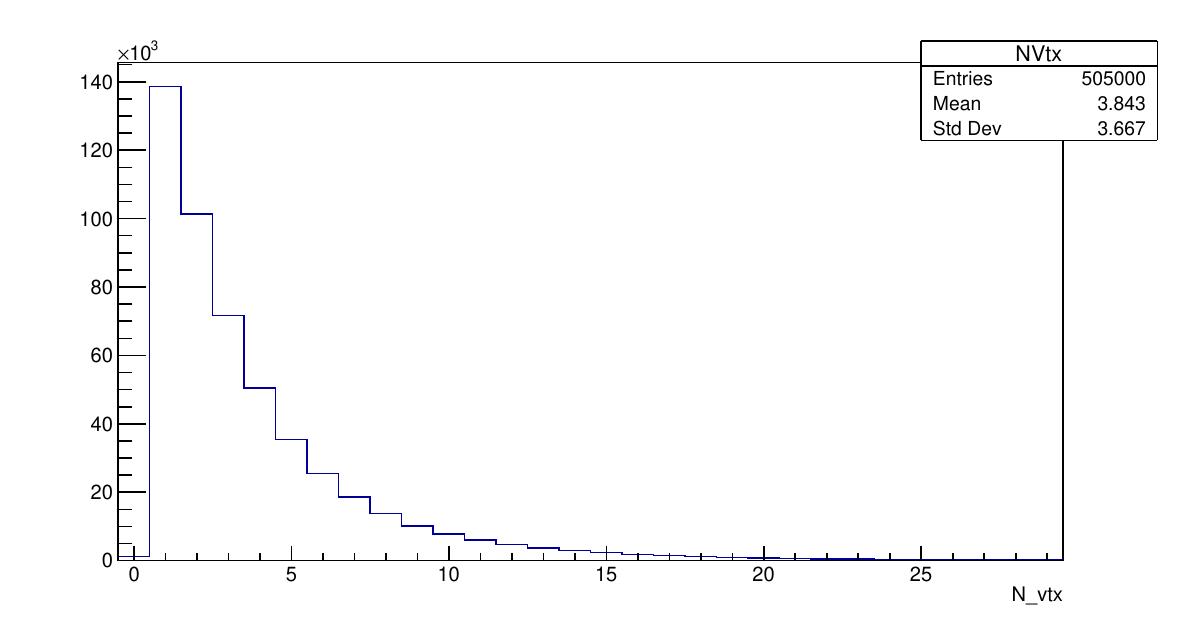}
\caption{Vertex multiplicity (valid and not-valid) per event. Only events with one BM track within the VTX acceptance are selected. }
\label{fig: pileup GSI2021}
\end{center}
\end{figure}

In general, entries in the charge-8 bin are not expected or should remain minimal. Only a small fraction is expected to arise from TW mis identification of the fragment charge Z (charge-7 points being reconstructed as charge-8). However the most important contribution to these vertexes comes from pile-up. For instance, in an event with pile-up, the primary particle reconstructed by the BM might reach the TW without fragmenting, while a second primary from the pile-up may not be detected by either the BM or the TW. If this second primary fragments and its vertex is mistakenly matched to the BM signal, the event could be incorrectly labeled as good. In reality, it would be a pile-up event with one primary particle contributing to the TW, contrary to expectations. This is expected in GSI where in $70\%$ of the events there are at least two vertex in the VTX detector, that means that in the $70\%$ of the cases there is pile-up to deal with. In Fig. \ref{fig: pileup GSI2021} is shown the multiplicity of vertexes per event ("valid+not valid") which show the impact of pile-up at GSI. To better understand these charge-8 entries, enabling global tracking could help analyze cases where a fragmentation vertex is reconstructed, but a primary particle is associated with the TW signal.

At this stage, the \textit{Std} configuration has allowed us to recover events where fragments correctly reach the TW, confirming that some of the additional events identified are indeed valid vertexes. Moving forward, the \textit{Std} configuration will be retained, and other parameters will be studied to determine if their optimization can help reduce the number of entries in the charge-8 bin, thereby improving the accuracy of the reconstruction process. In addition, in the cross-section chapter, we will discuss how to account for and clean the sample of events of this type, where a primary particle reaches the TW.

\subsubsection{CNAO2023}
The second campaign analyzed is CNAO2023. Similarly to GSI2021 the starting parameters are: SHD$=0.06 cm$ and all the 4 layers active.

Also in this case there is the confirmation that \textit{Std} algorithm give better result. Starting with the percentage of valid vertexes:

 \begin{equation}
    p_{vv}(Full)=(1.69\pm 0.03)\%\quad\quad\quad p_{vv}(Std)=(1.85\pm 0.03)\%
\end{equation}

\begin{figure}[h!]
\begin{center}
\includegraphics[width=6.7 cm]{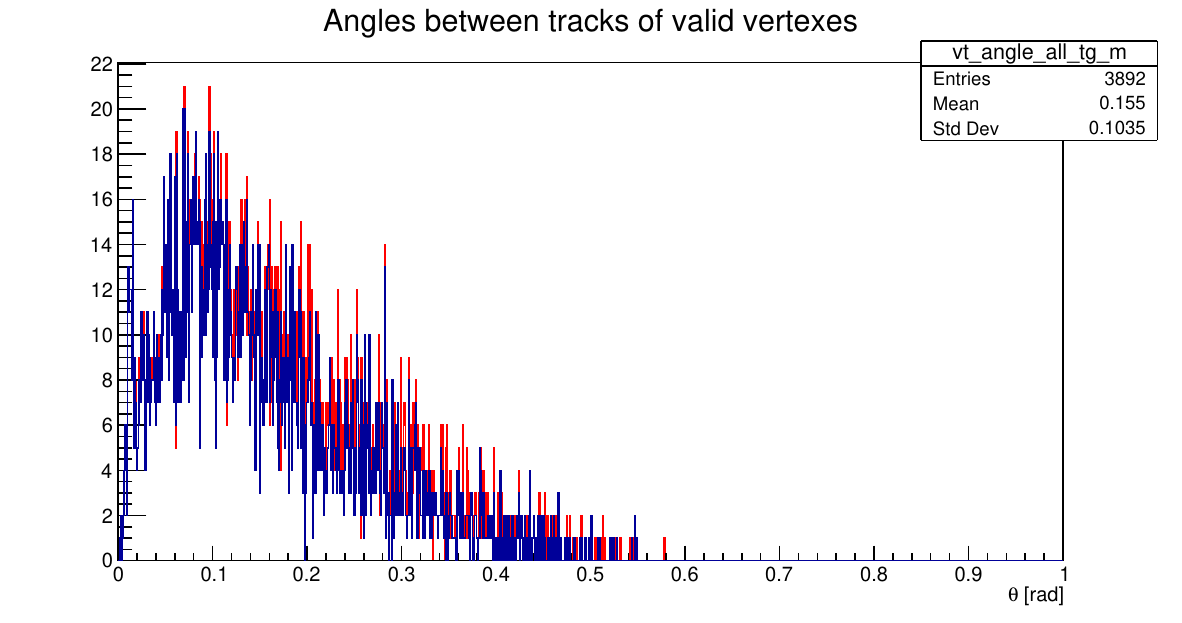} \quad \includegraphics[width=6.7cm]{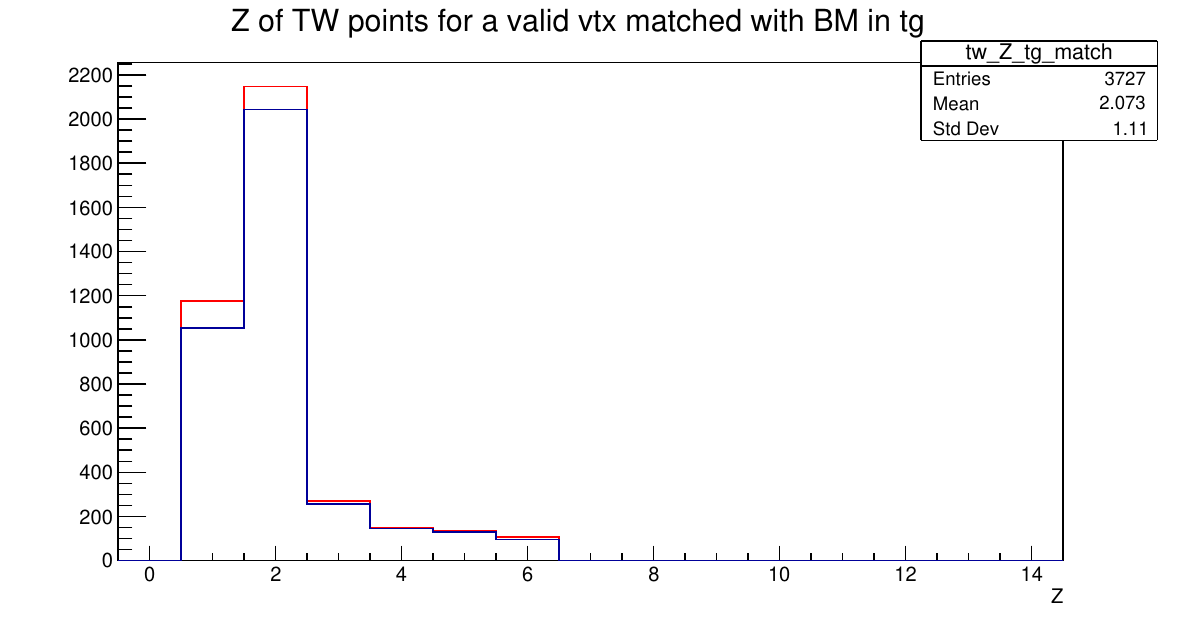} 
\caption{\textbf{Left:} Distribution of angles between the tracks of valid vertexes generated in the target and matched with the BM. \textbf{Right:}Charge distribution of TW points for events with a matched vertex generated in the target. 
In red is reported the \textit{Std} configuration, in blu the \textit{Full} configuration. }
\label{fig: angles and z of CNAO2023}
\end{center}
\end{figure}

For the CNAO2023 campaign, the \textit{Std} configuration is again more effective than the \textit{Full} configuration. This difference between this two is of approximately $9.3\%$, which is smaller compared to the differences observed in the previous campaign.

\begin{figure}[h!]
\begin{center}
\includegraphics[width=8.5 cm]{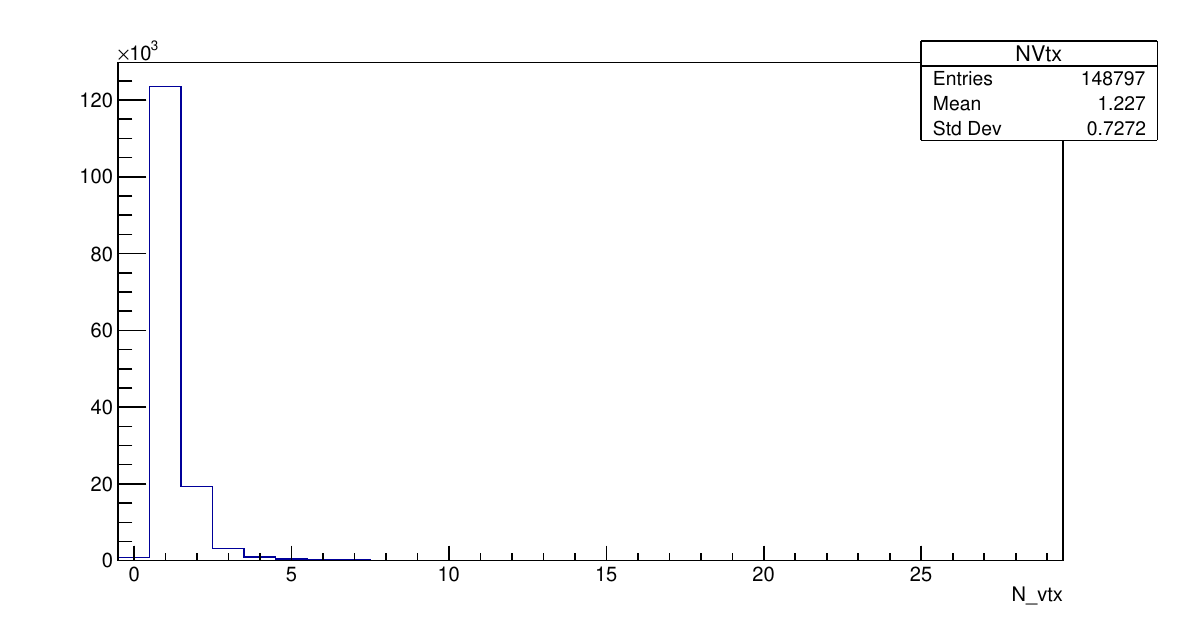}
\caption{Vertex multiplicity (valid and not-valid) per event. Only events with one BM track within the VTX acceptance are selected. }
\label{fig: pileup CNAO2023}
\end{center}
\end{figure}

Examining the angular distribution plot(\ref{fig: angles and z of CNAO2023}), the two configurations appear nearly compatible across the entire angular range, indicating that the additional vertexes identified by Std are distributed throughout the range. Looking at TW points charge distribution, the additional events recovered by \textit{Std} are actually associated with fragments reaching the TW, confirming the improved performance of this configuration. Like GSI2021, entries appear at charge $Z=6$ (primary particles) in this campaign. While pile-up remains lower than in GSI2021, but still not negligible as shown in Fig \ref{fig: pileup CNAO2023}. 

Overall, \textit{Std} continues to demonstrate superior performance, and efforts will focus on tuning additional parameters to minimize events where primary particles reach the TW.

\subsection{SHD optimization}
After selecting the algorithm to be used, the next step is the optimization of the SHD parameter. This parameter plays a crucial role in the tracking process. 
For both the campaigns it is used the Std algorithm and all the 4 layers of the VTX are active.

\subsubsection{GSI2021}

\begin{figure}[h!]
\begin{center}
\includegraphics[width=6.7 cm]{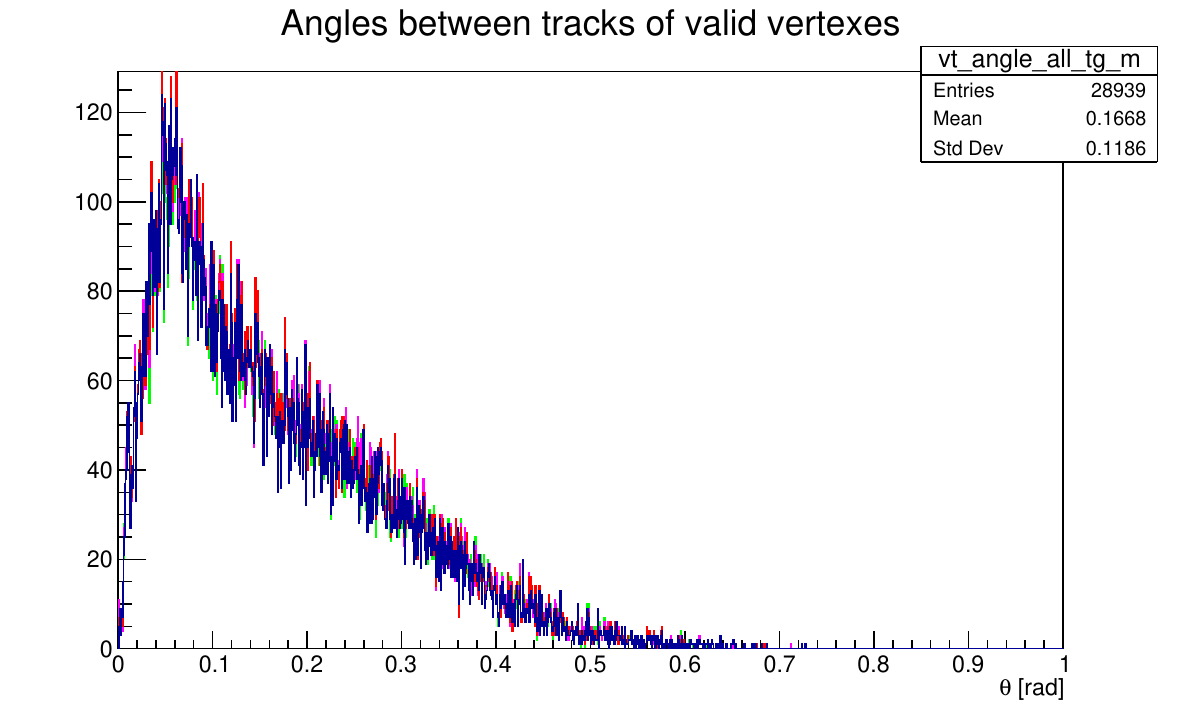} \quad \includegraphics[width=6.7cm]{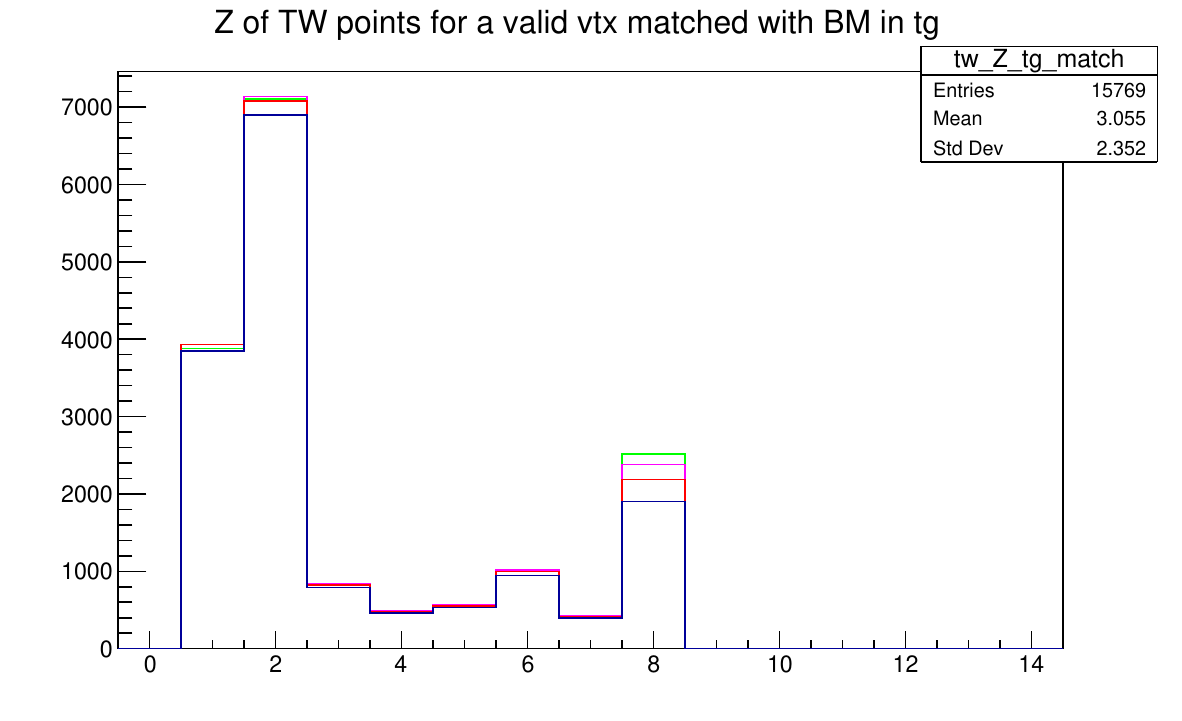} 
\caption{\textbf{Left:} Distribution of angles between the tracks of valid vertexes generated in the target and matched with the BM. \textbf{Right:}Charge distribution of TW points for events with a matched vertex generated in the target. 
In green is reported the configuration with $SHD=0.06$, in pink $SHD=0.05$, in red $SHD=0.04$ and in blue $SHD=0.03$. }
\label{fig: angles and z of GSI2021 SHD}
\end{center}
\end{figure}

Analyzing the results for different SHD configurations (Table: \ref{table: shd}), we observe that the optimal configuration appears to be the one with SHD set to 0.06, as it identifies the highest number of vertexes. But is not so simple. Examining the angular distribution (\ref{fig: angles and z of GSI2021 SHD}), all configurations exhibit consistent results. However, a closer look at the charge distribution for the TW points reveals that the 0.06 configuration shows a notable increase in entries within the charge-8 bin, whereas the increase in other charge bins is negligible. This suggests that the additional events identified by the algorithm with SHD set to 0.06 are predominantly "contaminated" events, where the TW is reached by a primary particle. The contamination comes from pile-up as already discussed and as will be shown in the chapter dedicated to the preliminary cross section measurement at GSI2021 [\ref{chapter 5}].

Given these results, we choose to focus on the cleaner configuration, corresponding to SHD set at 0.03. This setting minimizes contamination from dirty events with primaries reaching the TW, while still providing reliable number of vertexes.

\begin{center}
\begin{tabular}{|c |c|} 
 \hline
        SHD & $p_{vv}[\%]$ \\
         \hline\hline
         $0.07$ & $2.32\pm 0.02$\\
         $0.06$ & $2.34 \pm 0.02$\\
         $0.05$ & $2.33 \pm 0.02$\\
         $0.04$ & $2.28 \pm 0.02$\\
         $0.03$ & $2.16 \pm 0.02$\\
 \hline
\end{tabular}
\captionof{table}{Percentage of valid vertex within the target and matched with the BM for the different SHD values}\label{table: shd}
\end{center}

\subsubsection{CNAO2023}

\begin{figure}[h!]
\begin{center}
\includegraphics[width=6.7 cm]{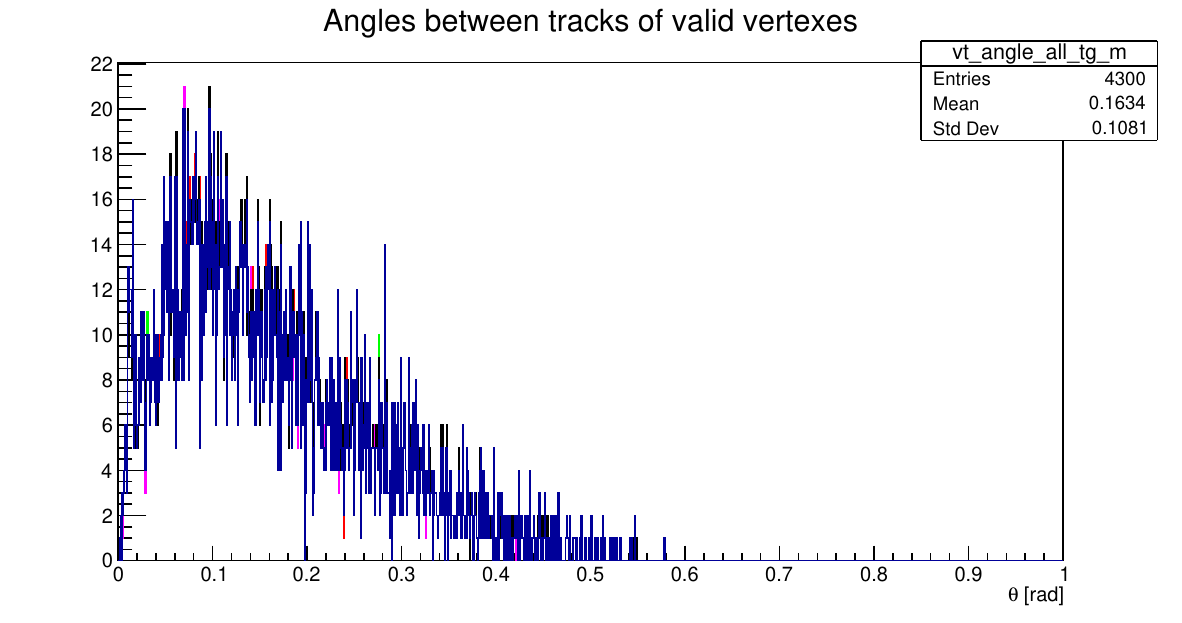} \quad \includegraphics[width=6.7cm]{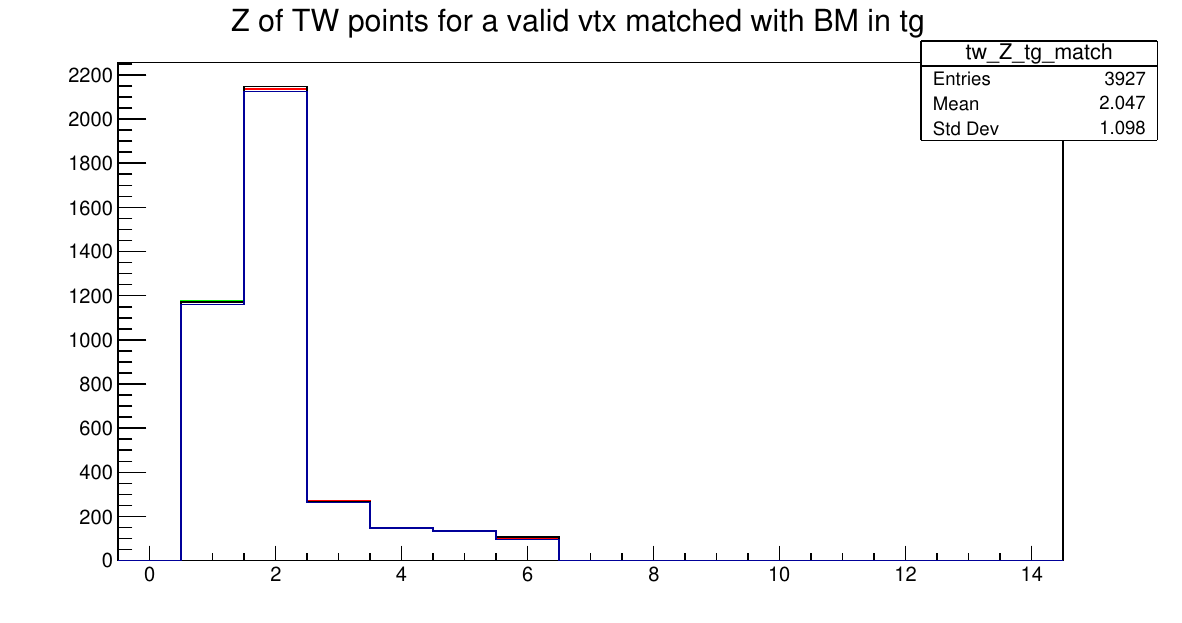} 
\caption{\textbf{Left:} Distribution of angles between the tracks of valid vertexes generated in the target and matched with the BM. \textbf{Right:}Charge distribution of TW points for events with a matched vertex generated in the target. 
In green is reported the configuration with $SHD=0.06$, in pink $SHD=0.05$, in red $SHD=0.04$,in blue $SHD=0.03$ and in black $SHD=0.07$. }
\label{fig: angles and z of CNAO2023 SHD}
\end{center}
\end{figure}

Analyzing the results for the different SHD configurations in the CNAO2023 campaign (Table: \ref{table: shd CNAO2023}), we find that SHD set at 0.06 identifies the highest number of valid vertexes. However, the difference compared to the SHD configuration at 0.03 is minimal, with 0.06 yielding only about $1.7\%$ more vertexes. Examining the angular distributions (\ref{fig: angles and z of CNAO2023 SHD}), all configurations remain consistent and show agreement regardless of the SHD setting. Turning to the charge distribution for the TW points a small amount of primary particles reach the TW for all SHD configurations. Among them, SHD set to 0.03 has the fewest entries at charge 6, indicating fewer primary particles reaching the TW. While the differences across configurations remain really small, we prefer to select the cleaner option, which minimizes the number of primaries on the TW. Thus, SHD set at 0.03 is chosen for this campaign as well.
\begin{center}
\begin{tabular}{|c |c|} 
 \hline
        SHD & $p_{vv} [\%]$ \\
         \hline\hline
         $0.07$ & $1.84 \pm 0.03$\\
         $0.06$ & $1.85 \pm 0.03$\\
         $0.05$ & $1.84 \pm 0.03$\\
         $0.04$ & $1.83 \pm 0.03$\\
         $0.03$ & $1.81 \pm 0.03$\\
 \hline
\end{tabular}
\captionof{table}{Percentage of valid vertex within the target and matched with the BM for the different SHD values}\label{table: shd CNAO2023}
\end{center}

\subsection{Usable layers}
\label{sec:usableLayer}

The final aspect to evaluate is whether to keep the deteriorated layer enabled or disabled, in order to understand its impact on the results. This layer appears to have experienced some degradation, and it is important to assess how its inclusion or exclusion affects the overall performance of the reconstruction algorithm.
\subsubsection{GSI2021}
For GSI2021 the optimal configuration has been found so far with the Std algorithm, and the SHD at $0.03 cm$.

\begin{figure}[h!]
\begin{center}
\includegraphics[width=8 cm]{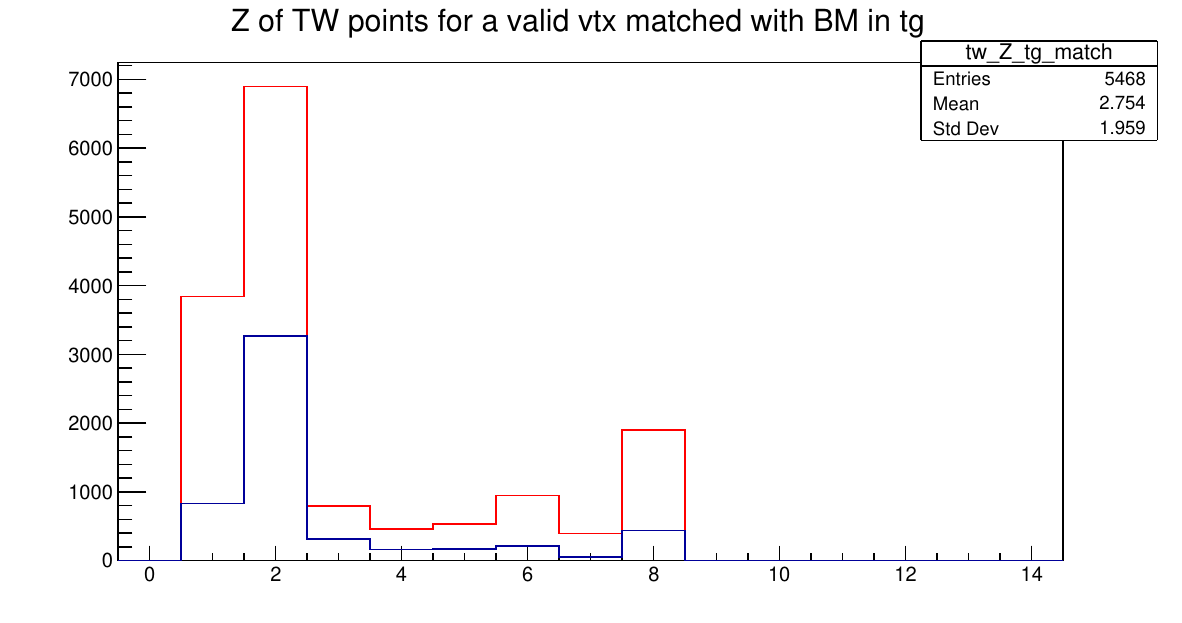} 
\caption{Charge distribution of TW points for events with a matched vertex generated
in the target. In red is reported the configuration with layer 3 activate, in blue the configuration with layer3 disactivate. }
\label{fig: tw points GSI2021 layer}
\end{center}
\end{figure}

Looking at the percentage of valid vertex generated within the target and successfully matched with the BM  for the two respective configurations we obtained:
\begin{equation}
    p_{vv}(layerON)=(2.16\pm 0.02)\% \quad\quad\quad p_{vv}(layerOFF)=(0.67 \pm 0.01)\%
\end{equation}
Analyzing the impact of layer 3 being active or inactive during the GSI campaign reveals a significant difference in the number of reconstructed vertexes. The configuration with layer 3 active identifies a noticeably higher number of vertexes compared to the configuration where it is inactive. Additionally, examining the charge distribution of the TW points (Fig.: \ref{fig: tw points GSI2021 layer}), the shape of the distribution remains unchanged, but the configuration with layer 3 inactive simply has fewer entries.

This indicates that keeping layer 3 active is clearly advantageous for vertex reconstruction. Interestingly, these findings do not align with prior MC studies, which suggested that the absence of a layer would have minimal impact on the number of reconstructable vertexes. The discrepancy is likely due to the significant pile-up observed during this campaign, which amplifies the importance of having all layers operational. Another contributing factor could be related to alignment issues. Unlike simulations, real data are subject to uncertainties stemming from the alignment of detector layers and their alignment with the rest of the experiment. These mis-alignments can introduce systematic biases that are not accounted for in the Monte Carlo simulations. A check not yet performed but useful for further and future analysis is to introduce some pile-up and mis-alignment in MC and look for an effect on the reconstruction of the vertexes. The data suggest that the effects of pile-up and also of some mis-alignment in the experimental environment make the absence of a layer far more significative than previously anticipated based on MC simulations.

\subsubsection{CNAO2023}
Passing at CNAO2023 the optimal configuration has been founded to be with the Std algorithm, and the SHD at $0.03 cm$.

The percentage of valid vertex generated within the target and successfully matched with the BM  for the two respective configurations are:
\begin{equation}
    p_{vv}(layerON)=(1.81\pm 0.03)\% \quad\quad\quad p_{vv}(layerOFF)=(1.68\pm 0.03) \%
\end{equation}

\begin{figure}[h!]
\begin{center}
\includegraphics[width=6.7 cm]{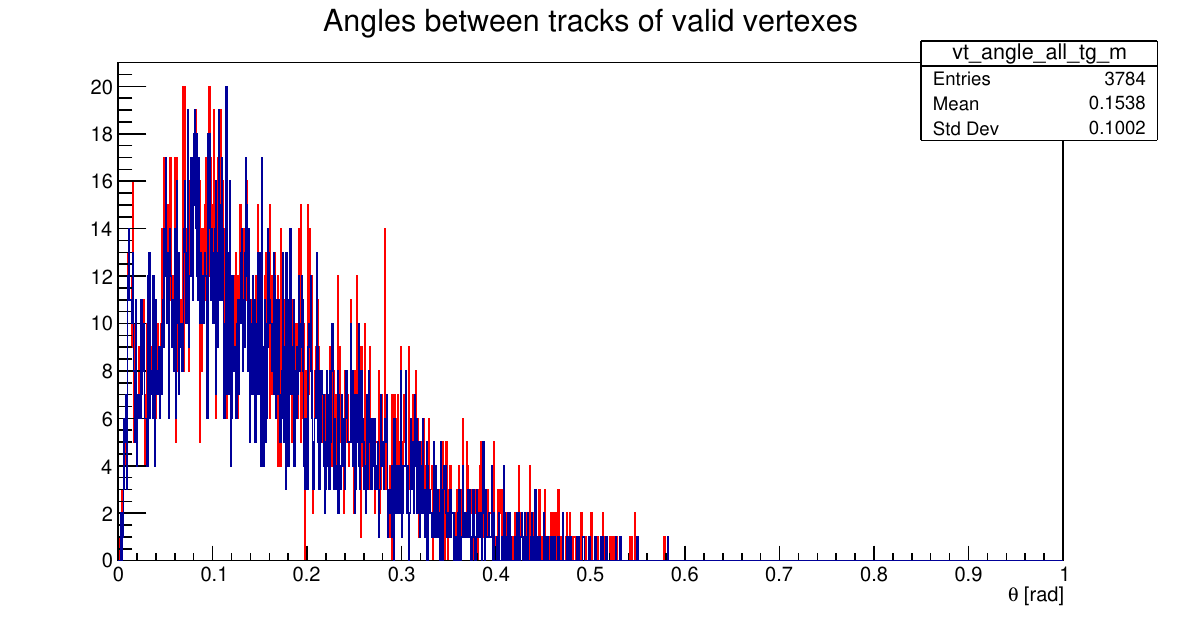} \quad \includegraphics[width=6.7cm]{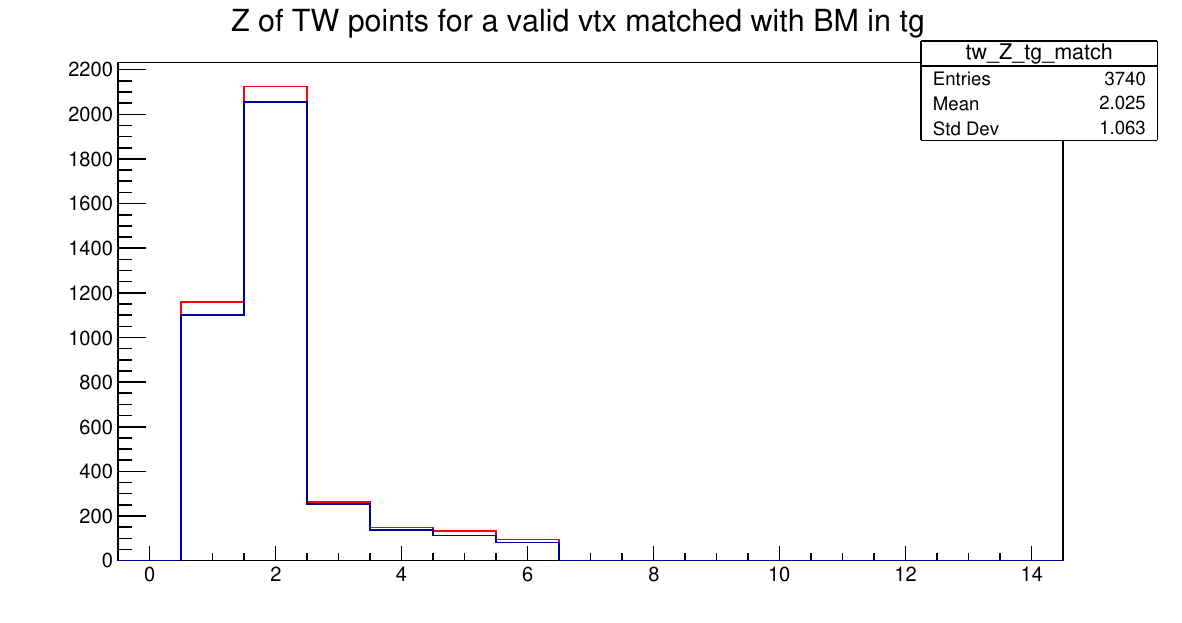} 
\caption{\textbf{Left:} Distribution of angles between the tracks of valid vertexes generated in the target and matched with the BM. \textbf{Right:}Charge distribution of TW points for events with a matched vertex generated in the target. 
In red the configuration with layer 2 active, and in blue the one with layer 2 non active. }
\label{fig: angles and z of CNAO2023 layer}
\end{center}
\end{figure}

The difference between the two configurations is negligible, and both exhibit the same angular distribution (Fig.: \ref{fig: angles and z of CNAO2023 layer}). Similarly, the analysis of the TW points reveals that the overall situation remains largely unchanged. Activating the layer allows for the recovery of some events where fragments with TW points of charge of 1 and 2 are observed. Consequently, the decision was made to operate with layer 2 activated. However, it is important to note that the difference between the activated and deactivated states is minimal, in contrast to previous campaign, where the absence of this layer had a more pronounced impact. This observation further confirms that the performance of the layer has been deteriorating over time, but also that with a lower level of pileup, the absence of one layer has a lesser impact.

\subsection{ Results}

After determining the optimal configuration for each campaign, given the consistency across different runs with the same target and trigger, the average value of the $p_{vv}$ parameter is reported. This average is calculated over all available runs for each campaign to ensure a comprehensive representation of the results.


Starting with GSI2021:
\begin{equation}
    p_{vv}=(2.18 \pm 0.01) \%
\end{equation}
Comparing this result with the MC (\ref{eq: p_vv gsi2021}) reveals a noticeable difference, with the value obtained from the data being significantly lower than that predicted by the simulation.

Passing to CNAO2023:


\begin{equation} \label{eq: p_vv mean cnao2023}
    p_{vv}=(1.76\pm 0.02)\% 
\end{equation}

Comparing this result with the value predicted by the MC simulation (\ref{eq:p_vv cnao2023}) reveals again a clear discrepancy. 

What could be the origin of this data-MC discrepancy? Although we do not expect the data and MC to yield identical values (since the MC is based on the cross section we aim to measure) the observed difference is nevertheless significant. This raises the question: is this discrepancy purely related to the cross section, or is there another contributing factor?

Examining the event display, the noise appears to be well under control. While pile-up is present only in the data, and not simulated in the MC, it does not seem sufficient to cause failures in the vertex reconstruction algorithm, but seems that the algorithm is able to reconstruct the valid vertex and also the pile-up primaries of the same event. Furthermore, MC studies have demonstrated excellent algorithmic performance. Even accounting for challenges introduced by pile-up or small mis-alignments, such effects are unlikely to cause a dramatic degradation in performance. For future studies, it could be considered to implement these features (pile-up and misalignment) in the MC to investigate their effects.

\begin{figure}[h!]
\begin{center}
\includegraphics[width=12 cm]{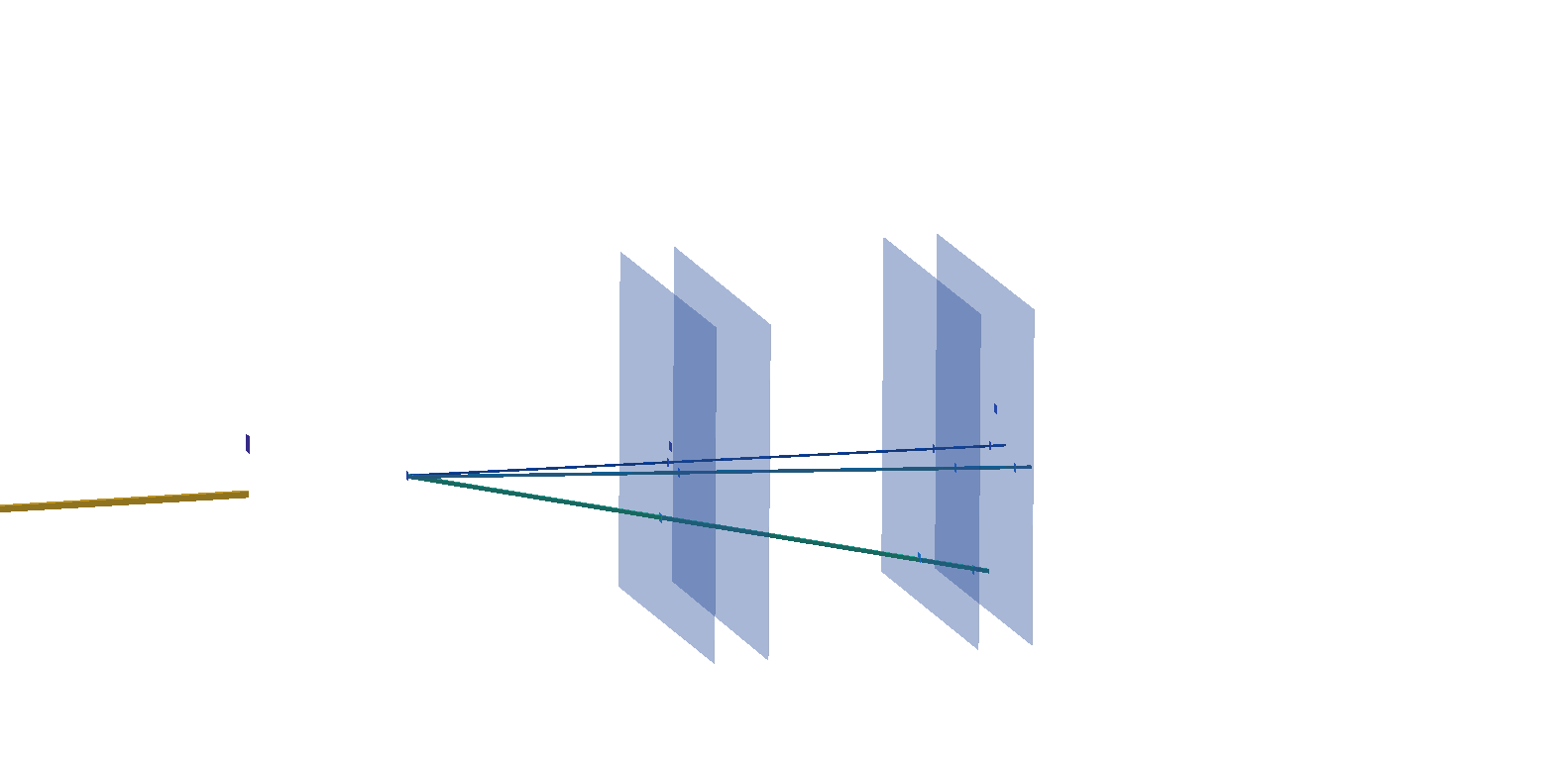} 
\caption{Event display of a CNAO2023 event showing a reconstructed vertex composed of three tracks, with two additional clusters that appear aligned with each other. The imaginary line connecting these clusters seems to points back toward the reconstructed vertex, suggesting a possible correlation.}
\label{fig: missing track}
\end{center}
\end{figure}

But a closer look at the event display has raised an additional concern about the particles that the algorithm can detect. In different events like the one shown in the figure \ref{fig: missing track}, there is a reconstructed vertex composed of three tracks, additionally, two extra clusters are visible. Upon closer inspection, these clusters appear to be aligned with each other and the line connecting these two clusters seems to point back toward the reconstructed vertex, suggesting a possible correlation with it. This raises the hypothesis that these clusters could correspond to a particle that is being lost in the reconstruction process. It seems possible that particles with lower ionization, such as protons, are not detected because they are unable to create clusters in at least three layers, which is a requirement for the track reconstruction. If such particles might be missed, it lead to the loss of certain vertexes in processes like $C \to B+p$ or $C \to Be+p+p$ and similar fragmentation process, where the missing proton would result in the loss of the vertex.

This raises a critical question: what is the detector’s efficiency for protons? Addressing this point is crucial to understanding whether tracking inefficiencies, particularly for low-ionization particles, might be contributing to the observed discrepancy between data and MC predictions.
M28 sensors can be highly efficient to mip particles, after a proper threshold setting, and protons at 200 MeV ionize more than a factor 2 with respect to a mip.

\chapter{Threshold optimization and efficiency analysis for the VTX} \label{chapter 4}

To address the open question raised in the previous chapter of this work \textit{How efficient is the VTX at detecting protons of the energies of interest for FOOT?} a dedicated study was conducted during the data-taking campaign in November 2024. Two different tasks were planned:
\begin{enumerate}
    \item 
    The first task was to measure the VTX efficiencies for protons in the energy range accessible at CNAO (E$_p <$~250~MeV) at the threshold values of the VTX used in the previous campaigns (GSI2021, CNAO2022, CNAO2023). The study of these efficiencies is crucial for measuring cross sections with the VTX detector in those campaigns. Actually in the reactions of interest for FOOT, like $^{16}$O at 400 MeV/u and $^{16}$O and $^{12}$C at 200 MeV/u impinging on thin targets, the produced protons can have energies up to 400-800 MeV (see Fig.~\ref{fig: fluka.foot}). So also a measurement in laboratory with cosmic ray is of fundamental importance in order to address the VTX efficiency to mip, having not access to facilities with p energy > 250 MeV in the next future.
    \item Subsequently, the focus shifted to optimize the VTX thresholds in order to take physics data specifically for the CNAO2024 campaign, with the VTX detector fully optimized. The goal in this case was to maximize detection efficiency while controlling noise levels, ensuring the VTX operates at its best.

\end{enumerate}

The plan of the CNAO2024 data taking, regarding the VTX detector, was completely triggered by this thesis work, which highlight the problems of this detector in previous data takings, as shown in previous chapter~\ref{chapter 3}.

\begin{figure}[h!]
\begin{center}
\includegraphics[width=6.5 cm]{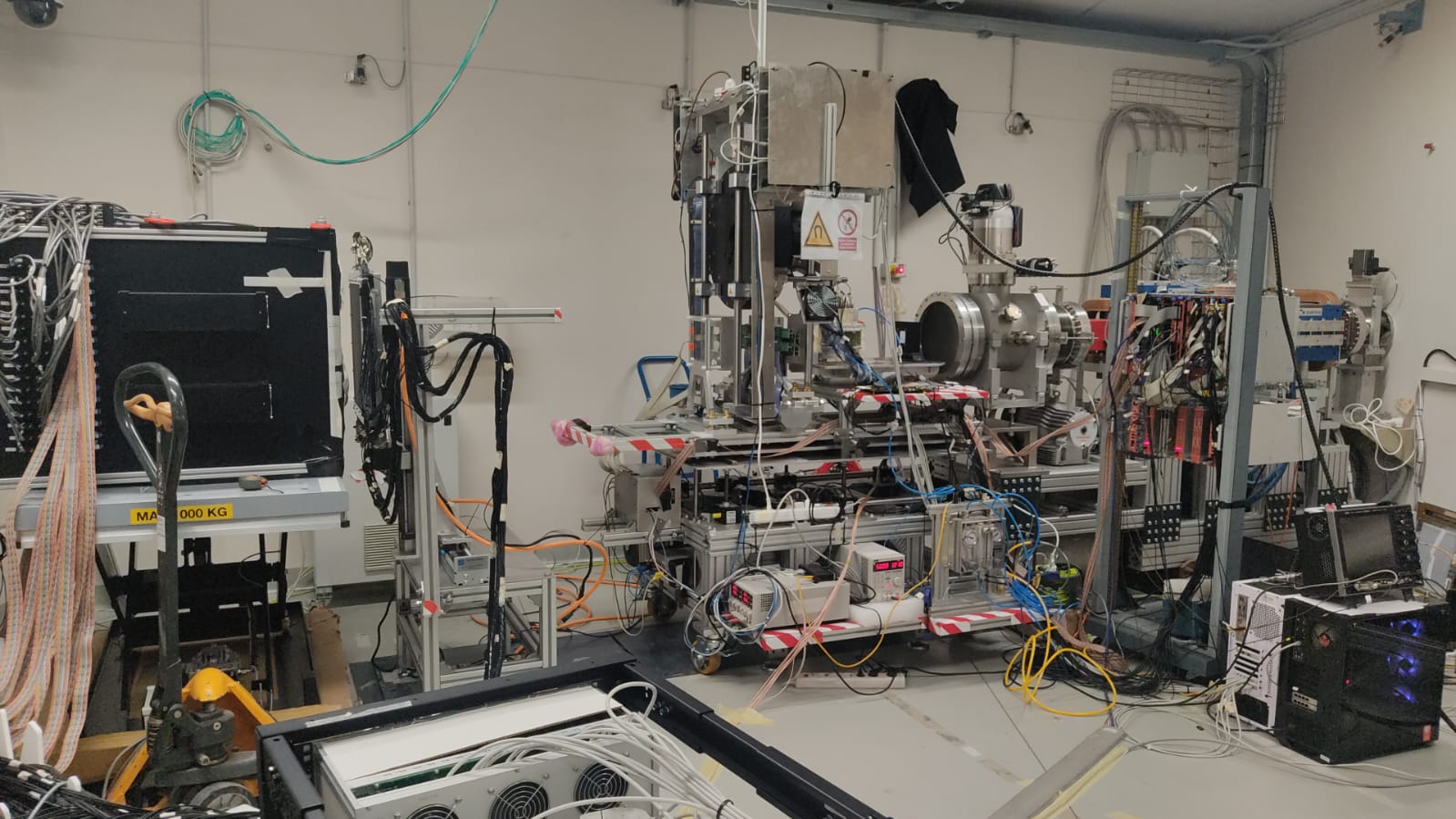} \quad \includegraphics[width=3.5cm]{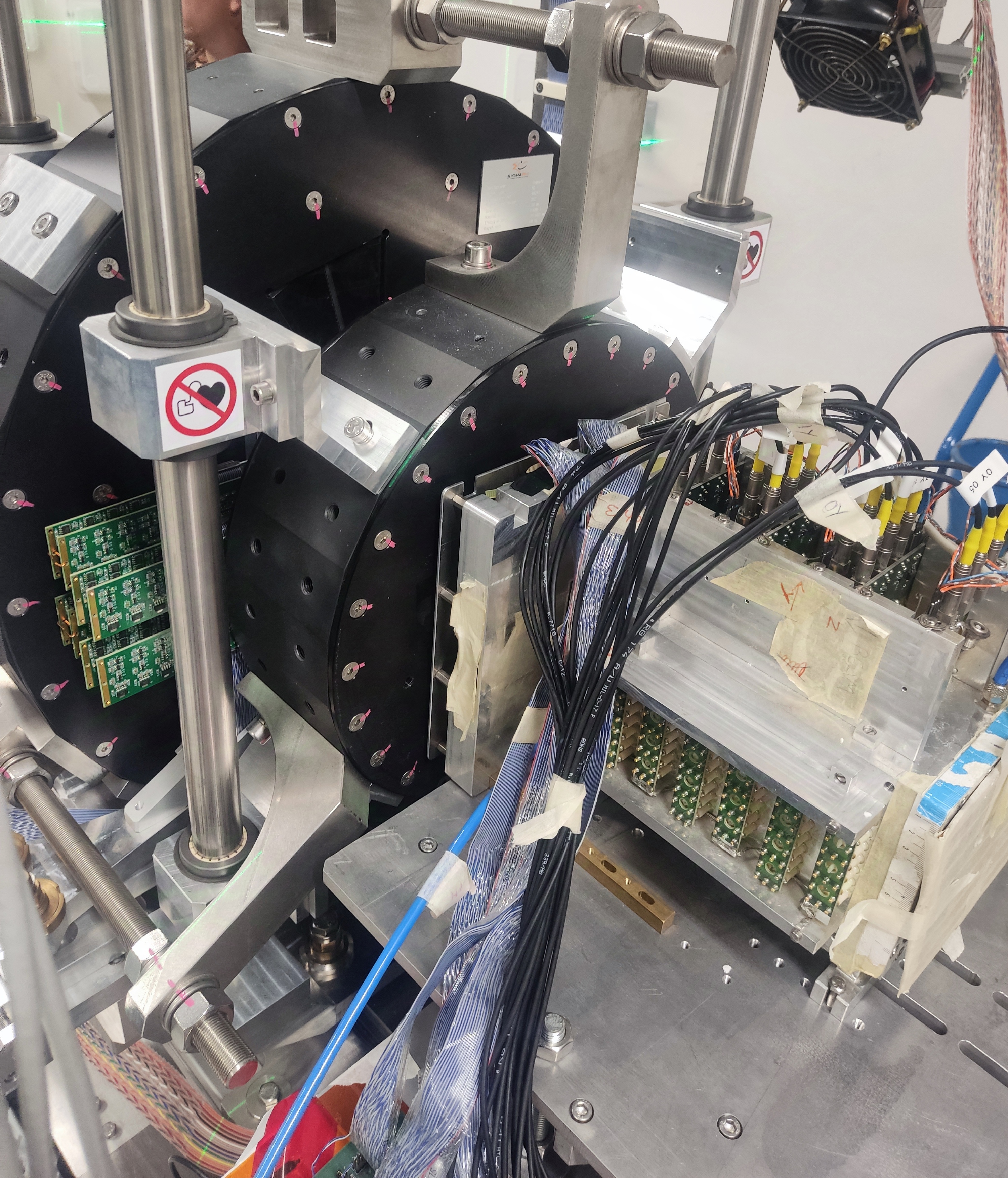} 
\caption{ \textbf{Left:}, a photograph of the full experimental setup at CNAO in November 2024 is shown. \textbf{Right:}, a closer view highlights only the BM, VTX, magnet, and IT..}
\label{fig: setup}
\end{center}
\end{figure}

The setup of the experiment, shown in Figure \ref{fig: setup}, highlights the main components, including a focus on the VTX detector.

\begin{figure}[h!]
\begin{center}
\includegraphics[width=8 cm]{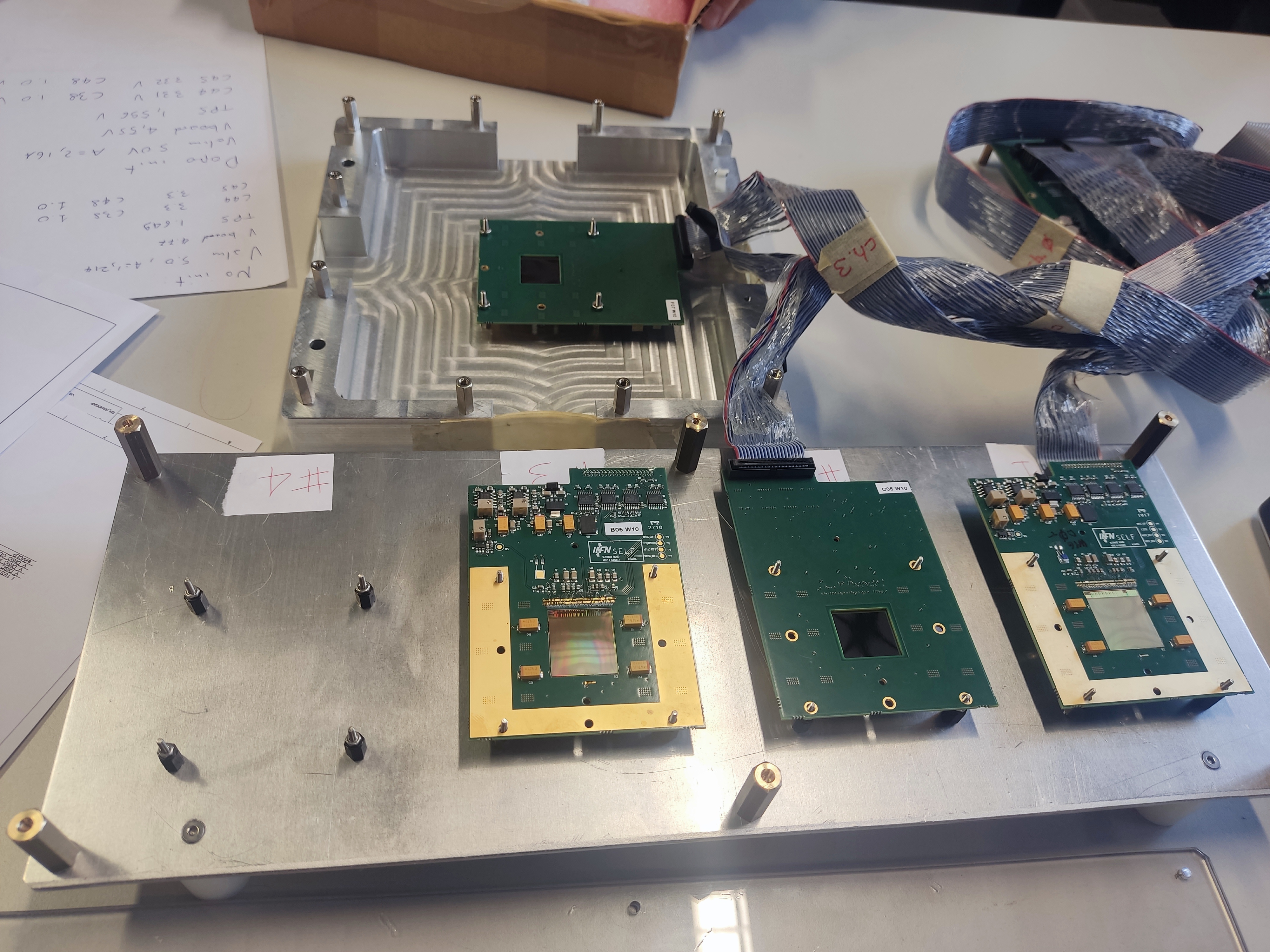}
\caption{The VTX detector during the replacement of the deteriorated layer. All four sensors are visible, showcasing the detector's structure.}
\label{fig: replecement}
\end{center}
\end{figure}

Given the deteriorating performance of the second layer in the last campaign, it was deemed critical to replace it. Since spare layers were available, allowing the substitution to proceed ( Fig.: \ref{fig: replecement} is taken during the substitution). Unfortunately, during the replacement process, another issue was identified with a different layer (layer 3 in the CNAO2023 geometry), necessitating its replacement as well. Consequently, the VTX now features two new layers. This means that while the efficiency studies performed for the new data-taking remain reliable and consistent, they may not perfectly match the results from previous campaigns, as half of the detector has been replaced. Nevertheless, meaningful conclusions about the system's overall efficiency can still be drawn.

\section{Analysis for previous campaigns}
The primary objective of this analysis is to study the efficiency of the VTX detector in detecting protons. In the GSI2021 and CNAO2022 campaigns, no data with protons were collected. During the CNAO2023 campaign, some proton data at energies of $200 MeV$ and $100 MeV$ were acquired. However, the BM was not optimized for detecting these particles, resulting in very limited statistics. Having a track in the BM is essential to ensure that a track reconstructed by the VTX corresponds to an actual particle arriving at the VTX. Due to the insufficient statistics from the 2023 campaign, it was decided to allocate a dedicated slot of time during the new data-taking campaign (CNAO2024) to address this study.

It was found that in all the previous campaigns (GSI20221, CNAO2022, CNAO2023) the VTX thresholds were set to their maximum value. The idea at the base of this choice was to maximize VTX performance for the fragments with charge Z$\geq$2 and at the same time minimize noise in the VTX detector. So also the  study  of the efficiency for protons in CNAO2024 campaign was done with the VTX detector thresholds set at maximum, in order to recover this information useful to measure cross sections in the previous campaigns. An energy scan was then performed using proton beams with energies ranging from $230 MeV$ to $150 MeV$.

The first observed quantity is the tracking efficiency as a function of the beam energy. Tracking efficiency is defined as :
\begin{equation}
    trck\_eff=\frac{\textit{\#evt with a matched VTX}}{\textit{\#evt with 1 BM track in the VTX acceptance}}
\end{equation}

\begin{figure}[h!]
\begin{center}
\includegraphics[width=8 cm]{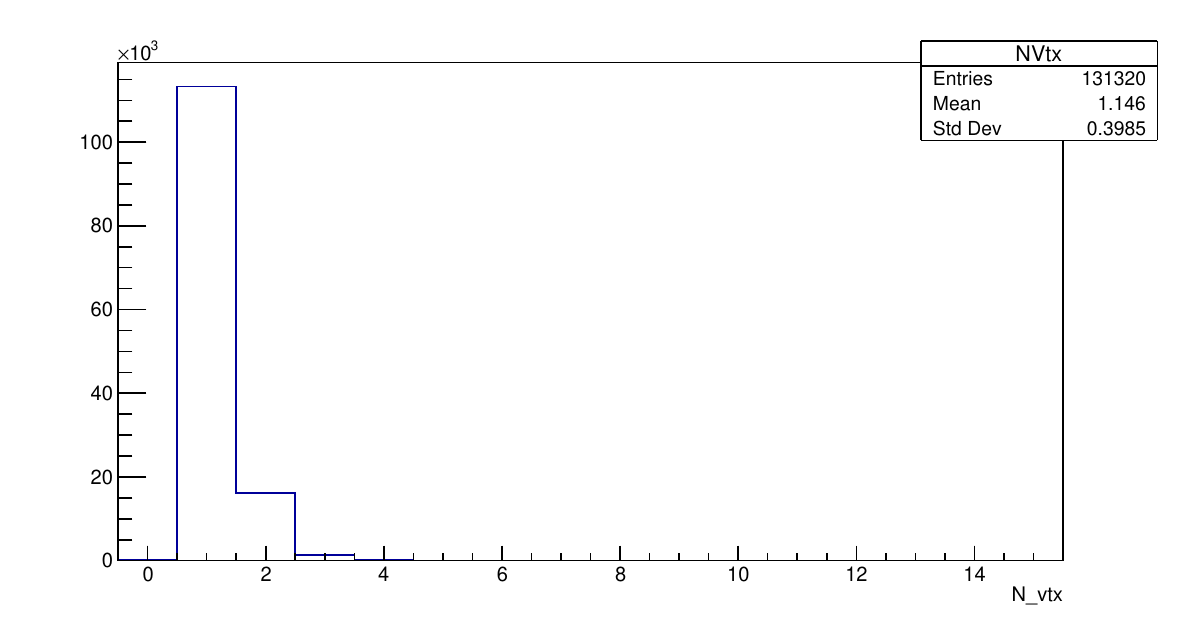}
\caption{Vertex multiplicity (valid and not-valid) per event. Only events with one BM track within the VTX acceptance are selected.}
\label{fig: pileup CNAO2024}
\end{center}
\end{figure}

\begin{figure}[h!]
\begin{center}
\includegraphics[width=8 cm]{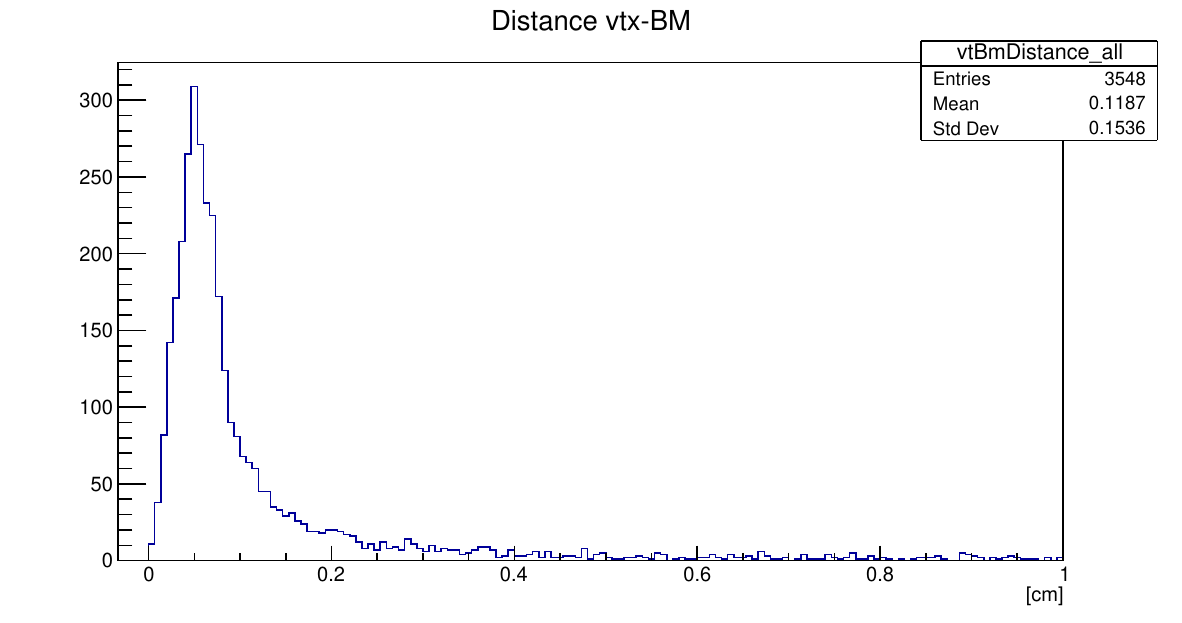}
\caption{Distance distribution in the $x-y$ plane between the projection of the BM track onto the target and the matched vertexes, for a run with a $230 MeV$ proton beam.}
\label{fig: distance CNAo2024}
\end{center}
\end{figure}

Specifically, an additional request has been added to the match, namely that the distance between the reconstructed vertex and the BM track is smaller than $2mm$. This is because, due to beam structure at CNAO and VTX dead time, pile-up of primaries in VTX is frequent. With the beam condition of CNAO2024 the number of events with at least two primary tracks in the VTX is between 15-20\% (Fig.: \ref{fig: pileup CNAO2024}). So without the request of matching the BM track and VTX vertex within a distance selection (2 mm was our choice) the pile-up could bias our final result on the efficiency measurement. The choice of $2mm$ is reasonable based on the analysis of the distance distribution in the $x-y$ plane between the projection of the BM track onto the target and the matched vertexes (Fig. \ref{fig: distance CNAo2024}). This distribution exhibits a clear peak at distance smaller that $2mm$, confirming that this cut effectively minimizes the impact of pile-up while ensuring accurate vertex to BM matching.  

\begin{figure}[h!]
\begin{center}
\includegraphics[width=8 cm]{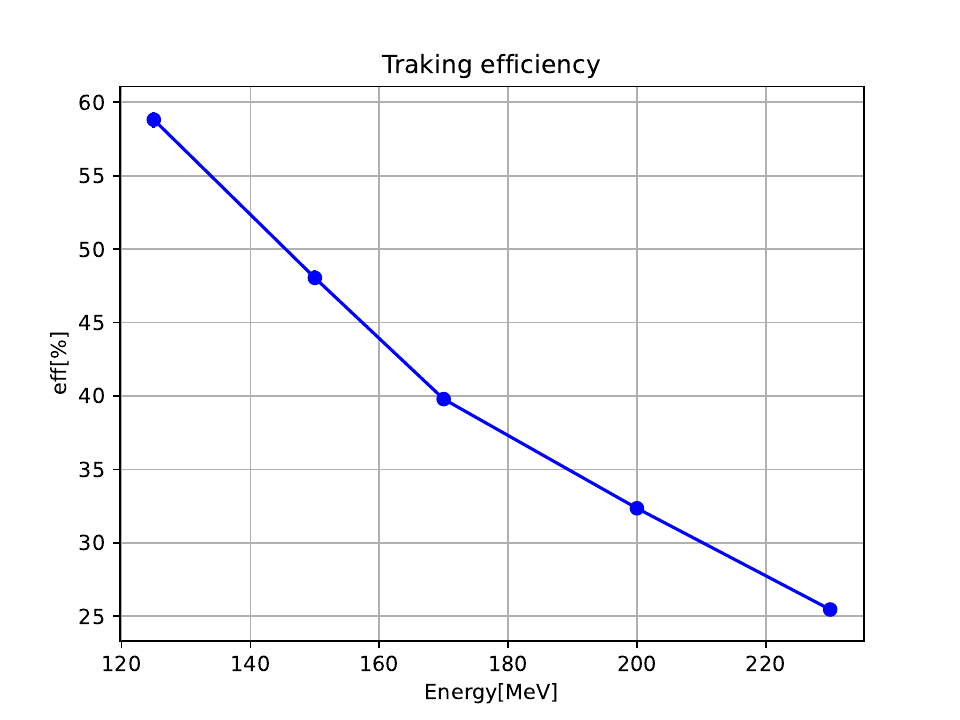}
\caption{Tracking efficiency as a function of the proton beam energy.}
\label{fig: tracking_eff_default1}
\end{center}
\end{figure}

The plot of tracking efficiency (\ref{fig: tracking_eff_default1}) shows the expected trend, with efficiency decreasing as the particle energy increases. This is due to the fact that higher-energy particles are less ionizing and, consequently, produce a lower tracking efficiency. In general, the efficiencies observed are not optimal for a detector of this type. For example, the efficiency is around $25\%$ for protons at $230 MeV$, which is relatively low. This is a consequence of the choice to set the VTX thresholds at their maximum value. 

\begin{figure}[h!]
\begin{center}
\includegraphics[width=6.5 cm]{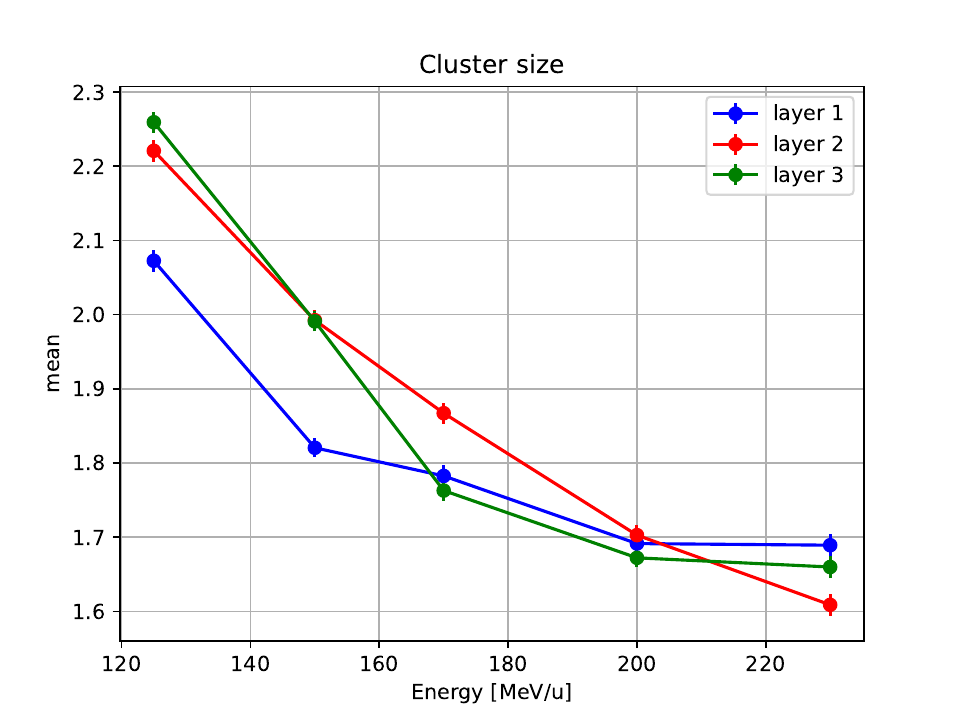} 
\caption{ Cluster size of the clusters forming the reconstructed tracks as a function of beam energy for the different sensors.}
\label{fig: cluster size default1}
\end{center}
\end{figure}

Another thing to observe is the cluster size of the clusters that contribute to the reconstructed tracks as a function of the energy for the different sensors (\ref{fig: cluster size default1}). The cluster size represents the number of sensor pixels grouped into clusters contributing to primary track formation. To isolate primary tracks, we analyzed only vertexes composed by a single track, constructed a histogram of the cluster sizes for the relevant layer, and used the mean of this histogram as the representative cluster size.

As expected, the cluster size decreases slightly with increasing beam energy, and this trend is consistent across all sensors. In general, however, the cluster size does not appear to be particularly optimal. For instance, a cluster size of 1.7 for protons at 230 MeV is relatively low, considering that electronic noise exhibits a cluster size of 1. So in these conditions (thresholds at maximum) the dynamical range of the sensor is narrow. As a result, it would not be possible to apply a cut, such as selecting clusters larger than 1, to reliably exclude noise.

It is important to note that the cluster size for sensor 4 is not reported. This omission is due to the fact that sensor 4 exhibited significant issues during data acquisition, with nearly one-quarter of the sensor completely non-functional and the remaining area operating with very low efficiency. Similarly, the previously reported tracking efficiency reflects this limitation, as the reconstructed tracks were predominantly formed using clusters from only the first three sensors.

It is important to emphasize that in the new campaign, half of the detector has been replaced (2 out of 4 layers). Despite this modification, the analysis of the cluster size indicates that the different layers exhibit a uniform behavior. This observation gives us confidence that, with thresholds set to their maximum values, all layers behave in a consistent manner and, therefore, are expected to perform similarly to the replaced ones. However, due to the changes in the detector configuration, applying the efficiencies obtained here to data from previous campaigns can still only provide an estimation.  In general, however, we can affirm that the efficiency for protons is low.

\subsubsection{What about helium?}

After studying the efficiency of protons, we extended our investigation to other particles, beginning with helium nuclei. Unfortunately, a dedicated helium beam was not available for direct measurements; therefore, we estimated the efficiency using the results obtained for protons. 

By referencing the NIST \cite{ref: NIST} tables for alpha particles, we calculated the energy loss in silicon for helium nuclei with a kinetic energy of $E_k=800MeV$. With an energy loss of $E_l=14.45MeV \frac{cm^2}{g}$, and considering the silicon density of $\rho=2.329 \frac{g}{cm^3}$  and a thickness of $d=0.0025cm$ of silicon (which corresponds to the epitaxial layer of the M28 sensors), we obtained an energy loss of :
\begin{equation}
    E_{loss}=E_l \times \rho \times d=84.135 keV
\end{equation}

For protons, the energy required to produce a comparable energy loss is approximately $30MeV$, corresponding to $E_l =14.69MeV \frac{cm^2}{g}$, and obtaining:
\begin{equation}
    E_{loss}=E_l \times \rho \times d=85.532 keV
\end{equation}

\begin{figure}[h!]
\begin{center}
\includegraphics[width=6.6 cm]{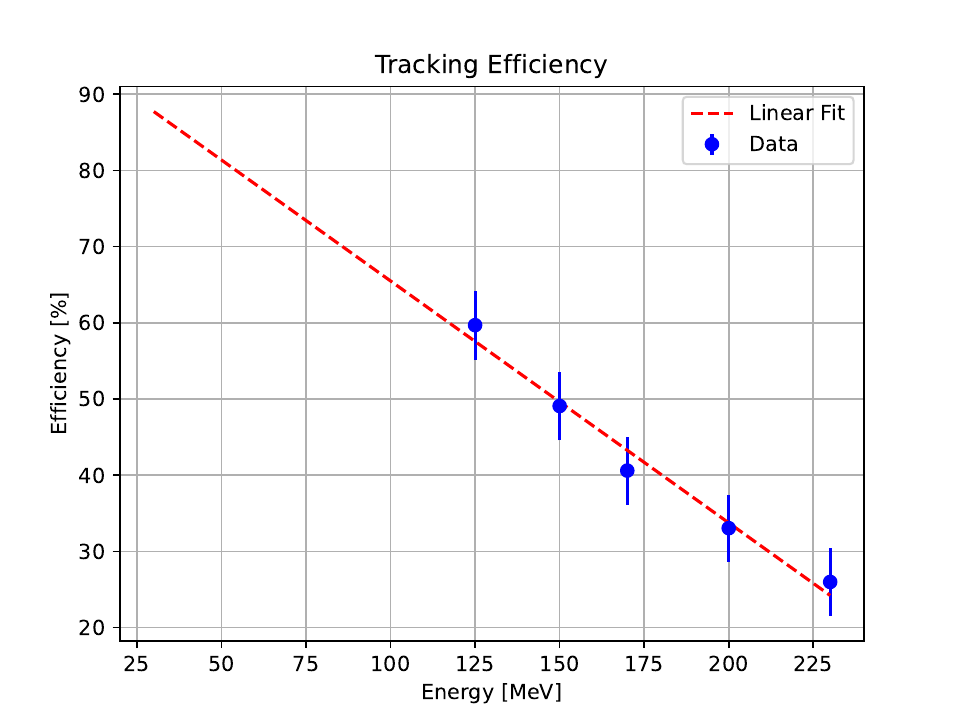}  
\caption{linear fit on the tracking efficiency to extrapolate the efficiency for protons at $30MeV$.}
\label{fig: fit lin1}
\end{center}
\end{figure}

The proton efficiency at $30MeV$ is not directly available, but assuming a linear trend in efficiency, through a linear fit we estimate it to be around $85\%$ (Fig.: \ref{fig: fit lin1}). Consequently, the efficiency for helium nuclei is expected to be significantly high. Clearly this is only an estimation, but for sure helium nuclei are more ionizing than protons at $125MeV$, which exhibit an efficiency of $ \sim 60\%$. Moreover, as the energy of helium decreases, the efficiency of the layers increases, further reducing the probability of losing helium nuclei.

Given the high estimated efficiency for helium, we are confident that the efficiency for subsequent, more ionizing particles (e.g., heavier ions) will also be very high. 

In comparison, the situation for protons is slightly less favorable. However, it is worth noting that fragmentation events with $Z<3$ are studied using emulsions. The primary concern arises from events in which is produced at least a proton, i.e. events like $C\to B+p$, where losing the proton results in the loss of the fragmentation vertex. To address this it could be studied the angular deviation (the kink) of boron with respect to the primary track as detected by the BM. Boron is expected to exhibit a larger angle compared to primary particles, helping in the identification of such events. Another approach worth considering is the use of post-tracking methods, trying to recover proton inefficiencies. For example, in events such as those presented in the previous chapter, where two aligned clusters appear to correspond to a single particle, it may be feasible, given the very low noise levels, to attempt constructing tracks using clusters that are not already part of existing tracks. If such a track intersects with the identified vertex, it could then be added to the vertex as well.  This is possible, however, only if the particle is detected by at least two layers, forming at least two clusters. Additionally, downstream detectors can verify the presence of the proton, enabling recovery of these events. These studies are for future investigations and have not been included in this thesis work.

In conclusion, with maximum thresholds the VTX detector is not fully optimized, but the measured efficiencies, together with some additional recovery strategies (post-tracking to recover proton efficiency and the study of the kink of the single tracks in the VTX detector to recover vertexes) suggested that cross section measurements in previous campaigns using VTX detector are still feasible.

\section{Optimization for the new campaign}

After analyzing the efficiencies related to the VTX configuration from previous campaigns, the focus shifted to optimize the thresholds for the new data-taking, aiming to strike a balance between minimizing noise and maximizing efficiency. 

This optimization began with a detailed pedestal study of the VTX system. Using dedicated configuration scripts, a threshold scan was performed to evaluate the response of each sensor under varying threshold values. Each sensor is divided into four quadrants, with each quadrant corresponding to a group of 240 columns (out of the 960 total columns per sensor). Thresholds can be individually adjusted for each of these four region. During the scan, the fraction of active columns within a quadrant was measured as a function of the applied threshold. At low thresholds, nearly all columns remain active, while at high thresholds, activity significantly decreases. Ideally, this relationship would appear as a step function; however, due to thermal noise and fluctuations, it exhibits a smoother, sigmoid-like behavior. Fitting this sigmoid function provides the mean transition point and its standard deviation ($\sigma$). The operational threshold for each region is set at the mean value plus a chosen multiple of $\sigma$. A dedicated study will be performed to determine the optimal multiple of $\sigma$ that minimizes noise while maintaining high efficiency. 

\begin{figure}[h!]
\begin{center}
\includegraphics[width=6.6 cm]{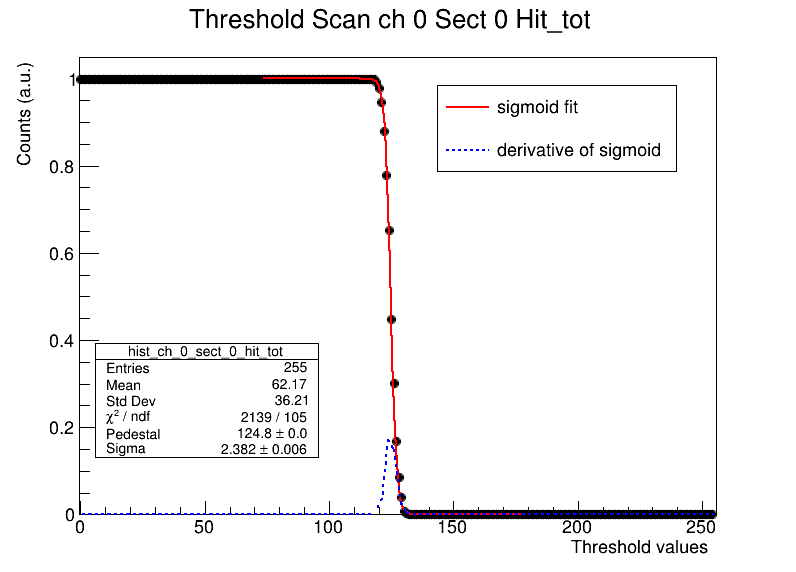} \quad
\includegraphics[width=6.6 cm]{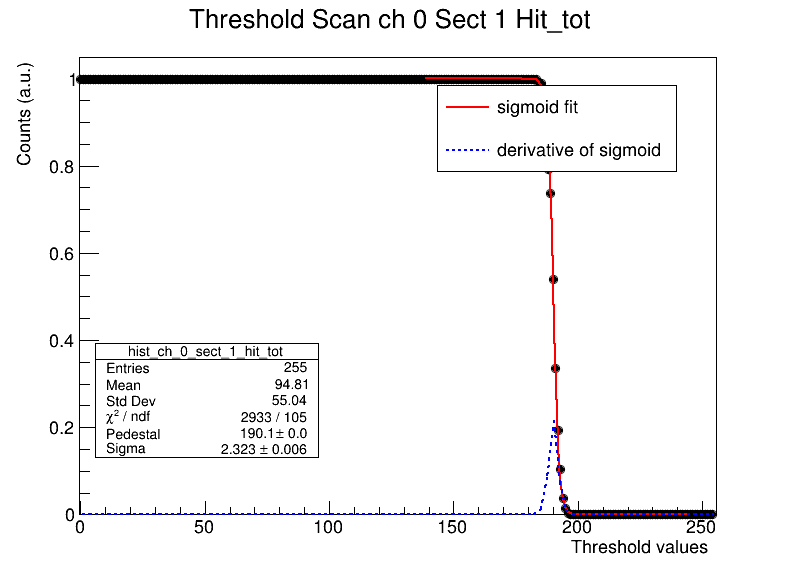}
\quad
\includegraphics[width=6.6 cm]{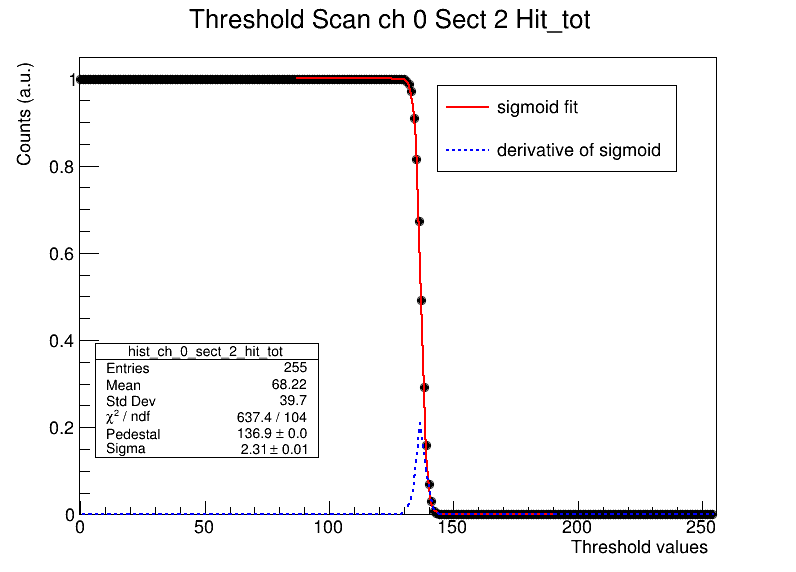}
\quad
\includegraphics[width=6.6 cm]{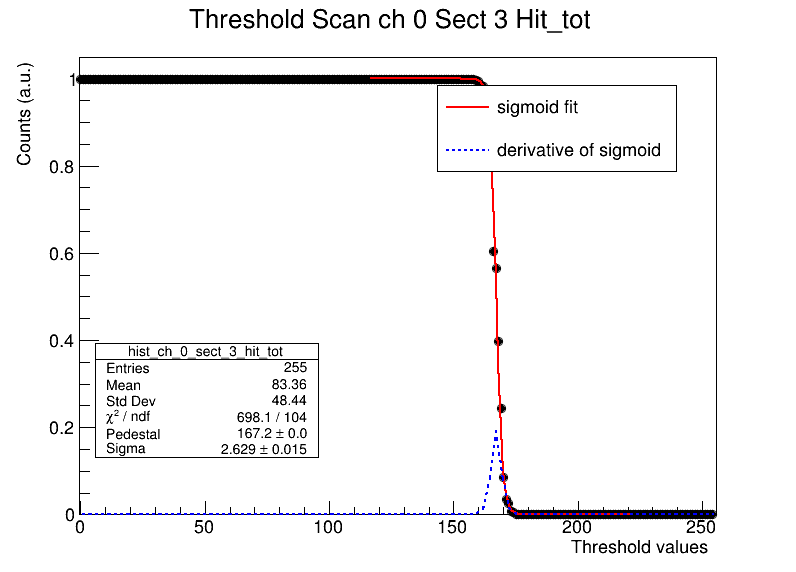} 
\caption{Threshold scans for the four regions of the first sensor (Channel 0) are shown. The scans illustrate the fraction of active columns as a function of the applied threshold for each region.}
\label{fig: pedestal scan1}
\end{center}
\end{figure}

In the figure \ref{fig: pedestal scan1}, the threshold scans for the four regions of the first layer are presented, the other sensors exhibit comparable behavior. To verify the stability of the chosen threshold values, pedestal measurements were taken both before and after data-taking. This was done to ensure that the conditions remained consistent during the experiment. Indeed, the pedestal values for the first three sensors remained stable, both in terms of the mean and $\sigma$. However, the last sensor, which turned out to be problematic, showed small variations only in its $\sigma$ values, indicating potential instability in its response. Furthermore, from the figure, it can be observed that the mean value varies from region to region, while the $\sigma$ remains consistent across the different regions. This suggests that while the transition points may differ between regions, the noise characteristics are stable throughout the sensor.

\subsection{Threshold scan}
Based on the results obtained from the pedestal threshold scans, some data acquisition,  with protons at $230MeV$, the least ionizing particles available, were performed with thresholds set at 4, 4.5, 5, 5.5, 6, and 7$\sigma$ . For these thresholds, the tracking efficiency, sensor efficiency, and cluster size were analyzed. The tracking efficiency and cluster size have been described earlier. Regarding the sensor efficiency, it is obtained by projecting the BM track onto each VTX layer for events with a single track in the BM. The closest cluster to the projected intersection was identified, and if the distance was less than $2mm$, the layer was considered efficient for that event.

\begin{figure}[h!]
\begin{center}
\includegraphics[width=6.5 cm]{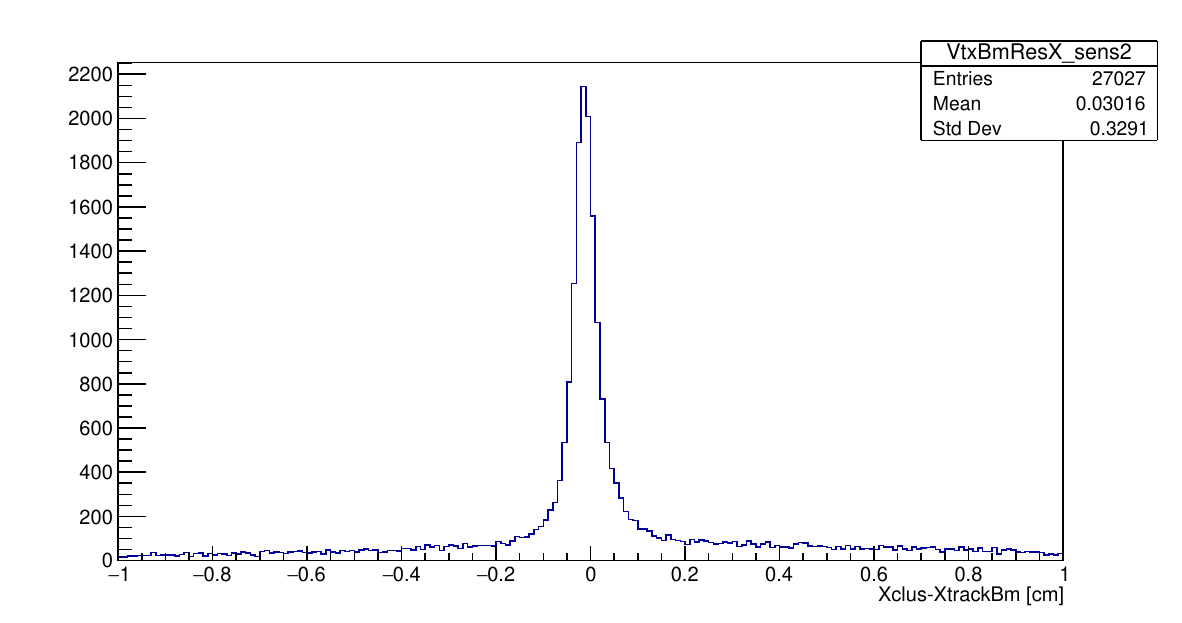} \quad \includegraphics[width=6.5 cm]{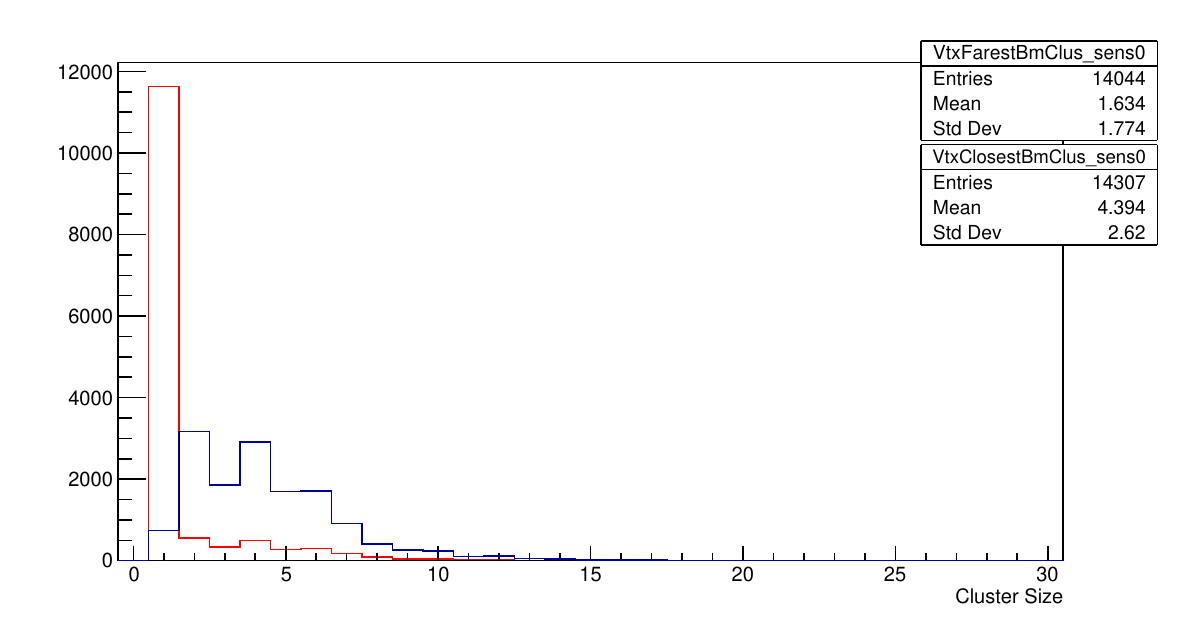}
\caption{\textbf{Left:} Residuals in x between the BM intersection on layer 1 and the clusters of this layer layer . \textbf{Right:} the cluster size distribution for the "farest" (red) and "closest" (blue) cluster. Both plots correspond to layer 1, though similar behavior is observed across all layers.}
\label{fig: res and clst size sensor eff1}
\end{center}
\end{figure}

\begin{figure}[h!]
\begin{center}
\includegraphics[width=6.5 cm]{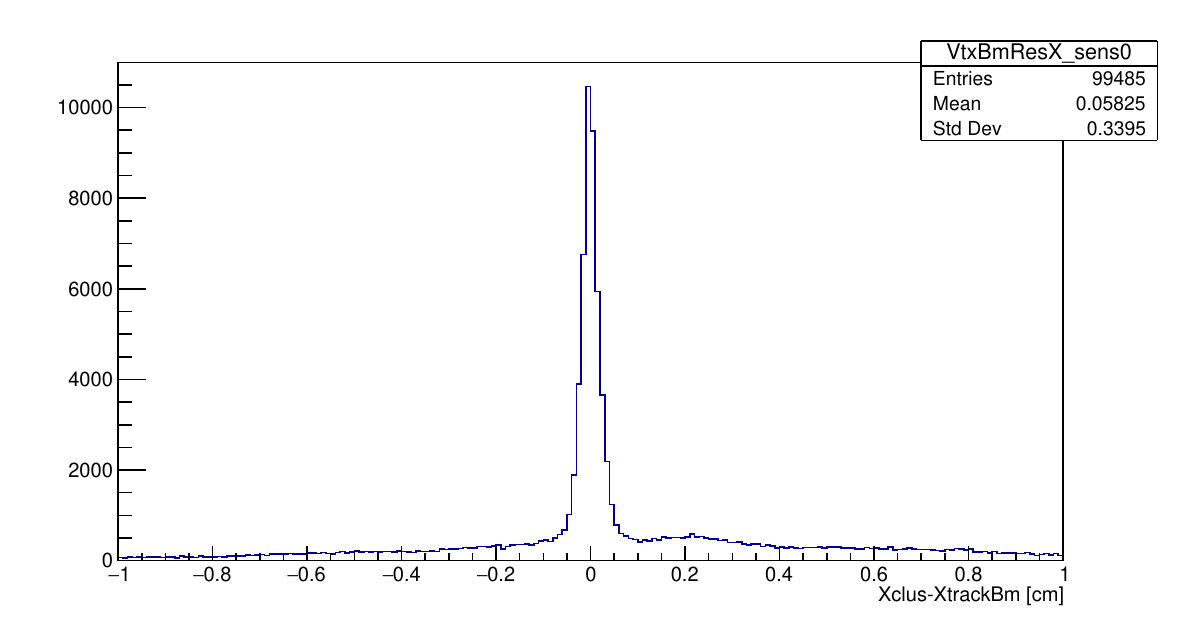} \quad \includegraphics[width=6.5 cm]{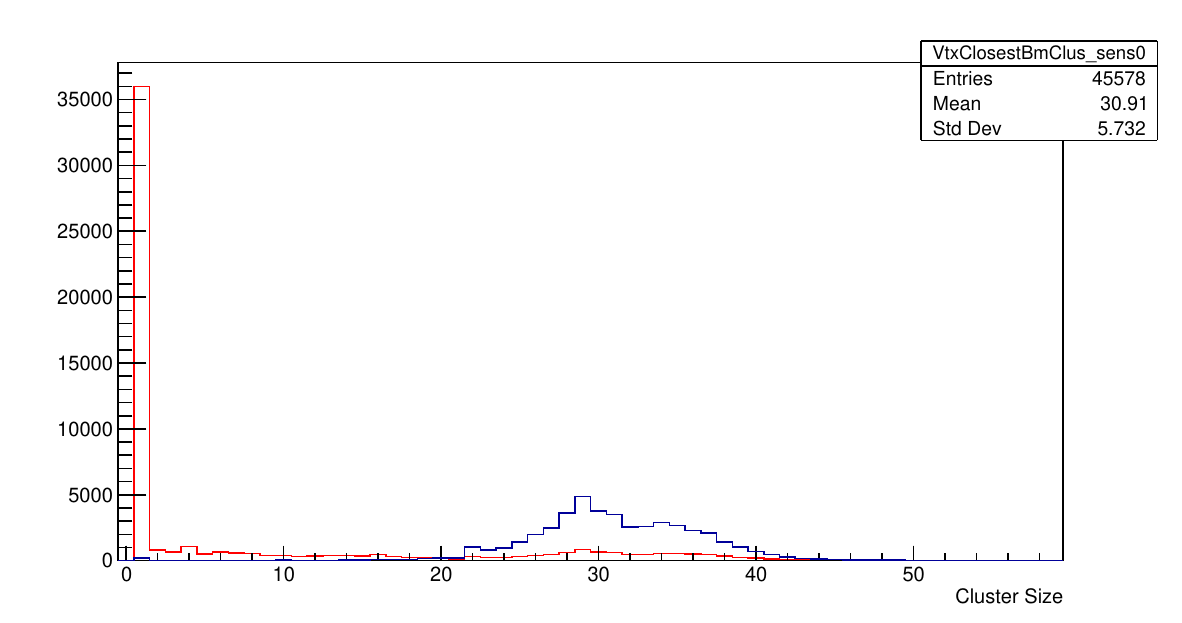}
\caption{\textbf{Left:} Residuals in x between the BM intersection on layer 1 and the clusters of this layer layer . \textbf{Right:} the cluster size distribution for the "farest" (red) and "closest" (blue) cluster. Both plots correspond to layer 1, though similar behavior is observed across all layers.}
\label{fig: res and clst size sensor carbon 200}
\end{center}
\end{figure}

To validate the implemented method for evaluating sensor efficiency, two plots are examined (fig.: \ref{fig: res and clst size sensor eff1}). The plots are reported only for layer 1, for the run at $6\sigma$ but a similar behavior is observed for all the layers and $\sigma$ values. The plot on the left of Fig.~\ref{fig: res and clst size sensor eff1} shows the residuals between the point of intersection of the BM with the considered layer and the clusters, which reveals a clear peak centered at zero. This confirms the presence of clusters near the intersection point of the BM track, indicating that the clusters are appropriately linked to the BM track. The plot on the right displays the cluster size distribution for the closest clusters, the ones that are identified as efficient for our metric, and for the clusters not considered efficient (the farest). For the farest clusters, the distribution shows a peak at a cluster size of 1, indicating that these are single-pixel clusters, which is consistent with noise. Additionally, there are a few entries at larger cluster sizes, corresponding to clusters that are likely due to pile-up events, where a second primary might contribute to a cluster further from the BM intersection point.  In contrast, the closest clusters do not show a peak at 1, but instead exhibit a larger cluster size, which is more consistent with proton interactions rather than noise. The same conclusions can be drawn for the runs with the carbon beam. Specifically, in figure \ref{fig: res and clst size sensor carbon 200}, are reported the same histograms for layer 1 with a beam of carbon at $200 MeV/u$ and  $6\sigma$. This pattern further supports the idea that the method is functioning correctly.

\begin{figure}[h!]
\begin{center}
\includegraphics[width=6.6 cm]{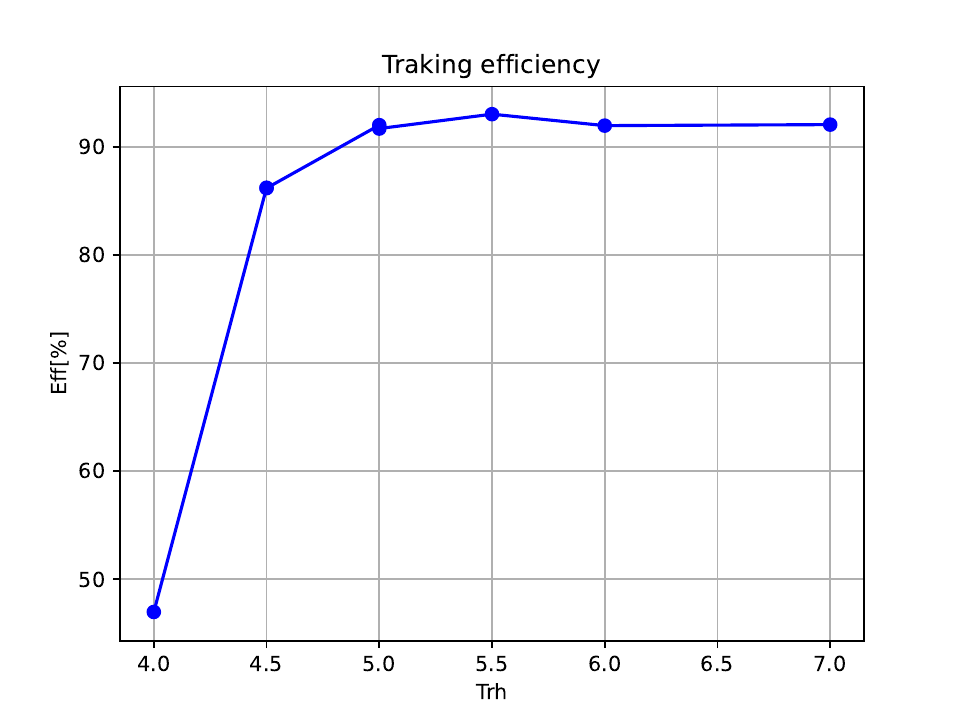}  \quad 
\includegraphics[width=6.6 cm]{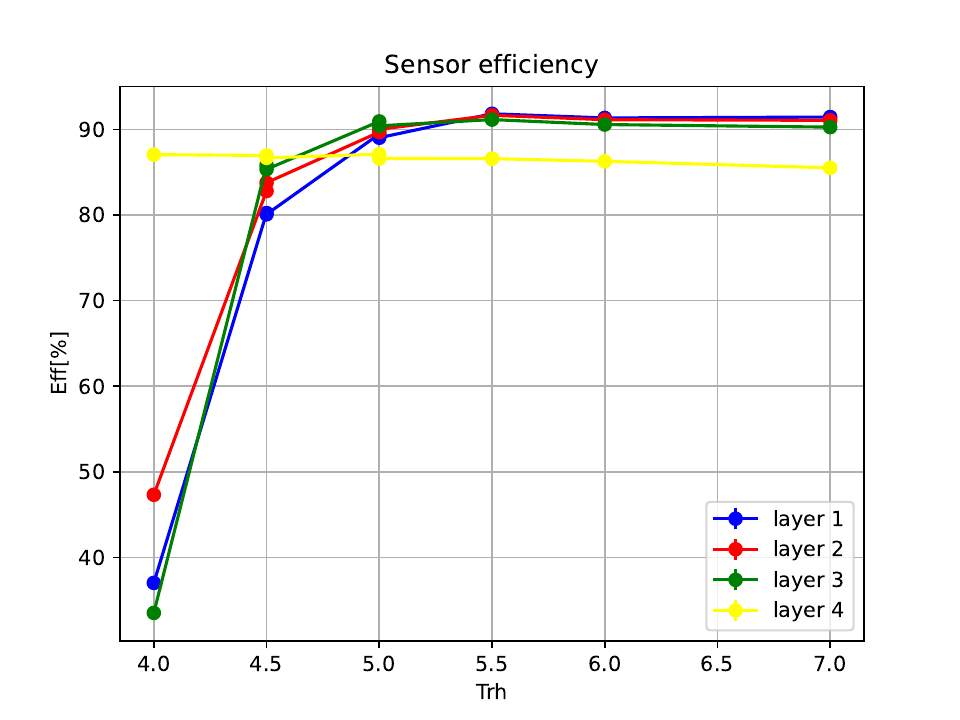} 
\caption{\textbf{Left:} Tracking efficiency, and \textbf{right:} sensors efficiencies as function of the threshold.}
\label{fig: efficiency threshold scan1}
\end{center}
\end{figure}

For the first three layers, the tracking and sensor efficiencies displayed similar behavior (Fig.: \ref{fig: efficiency threshold scan1}): the efficiency remained constant between $6\sigma$ and $7\sigma$ but dropped sharply at $4\sigma$ and was notably low even at $4.5\sigma$. This drastic decline appears to be linked to the noise.

\begin{figure}[h!]
\begin{center}
\includegraphics[width=6.6 cm]{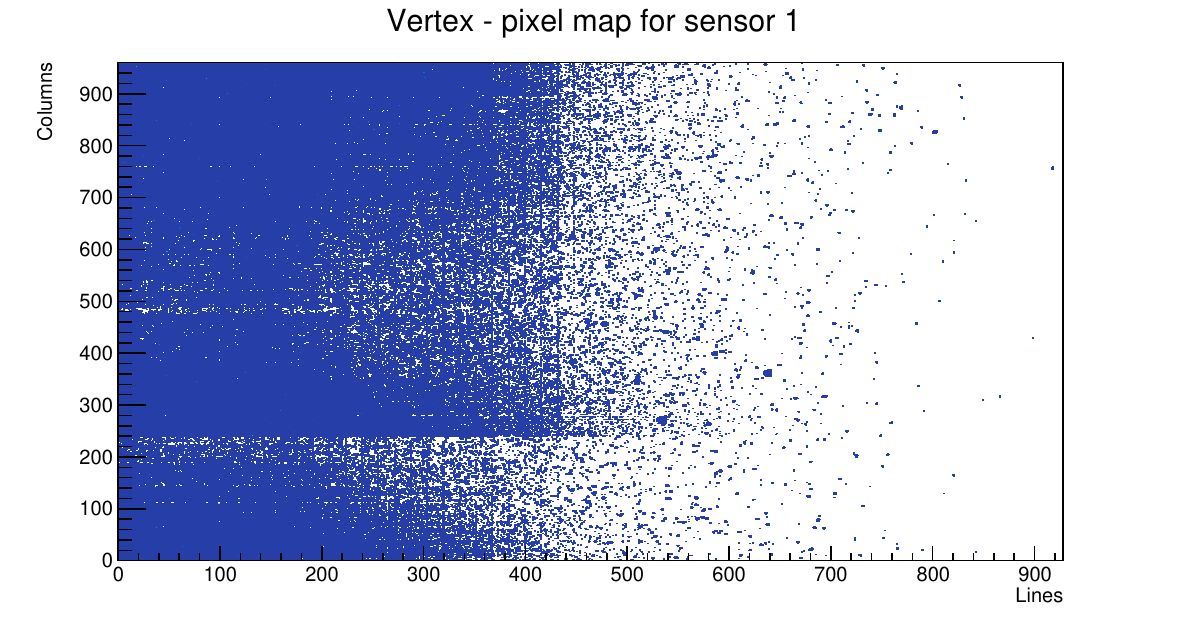}  
\caption{Pixel map of sensor 1 at $4\sigma$ threshold. The map shows that half of the sensor is fully activated and saturated, while the other half is completely inactive. A similar situation is observed for sensors 2 and 3 under the same threshold.}
\label{fig: pixel map 4 sigma1}
\end{center}
\end{figure}

Indeed, upon inspecting the pixel maps of the sensors, it became evident that the sensors were not functioning correctly at low thresholds. As shown in the accompanying figure \ref{fig: pixel map 4 sigma1}, half of the sensor appears fully activated, while the other half remains entirely inactive. Despite this noise, the tracking maps revealed that the detector was still able to identify the beam and reconstruct tracks correctly. However, operating with only half of the sensors functioning properly is clearly not viable.  Some anomalous features were also observed at $4.5 \sigma$ and $5\sigma$ thresholds, although less severe than those at 4. These observations led to the exclusion of these thresholds for further use.

\begin{figure}[h!]
\begin{center}
\includegraphics[width=5.8cm]{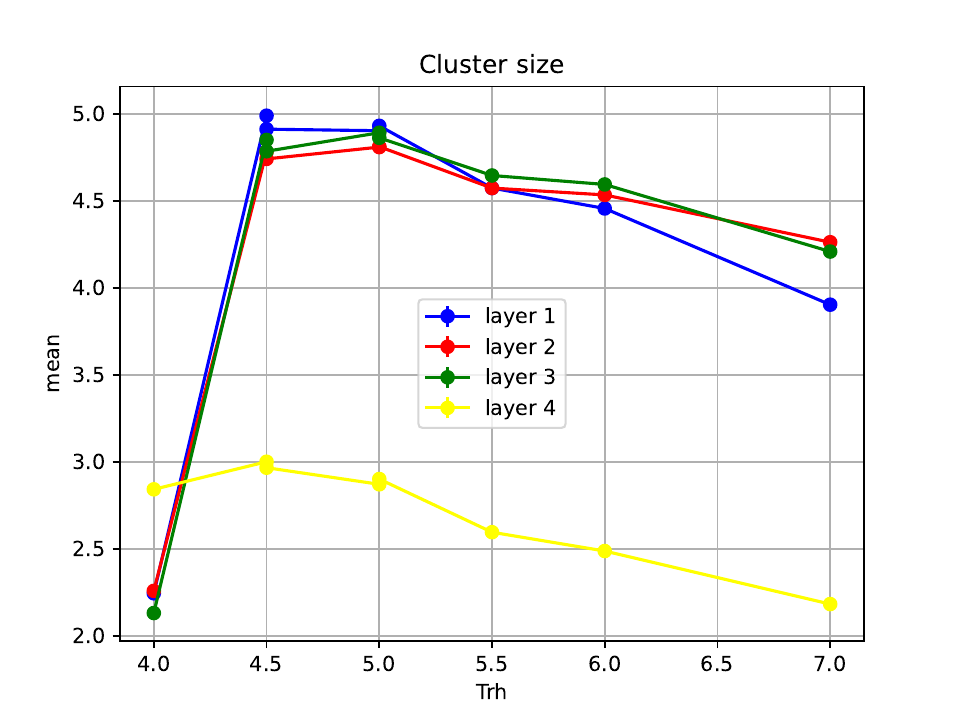}   \quad\includegraphics[width=7.2cm]{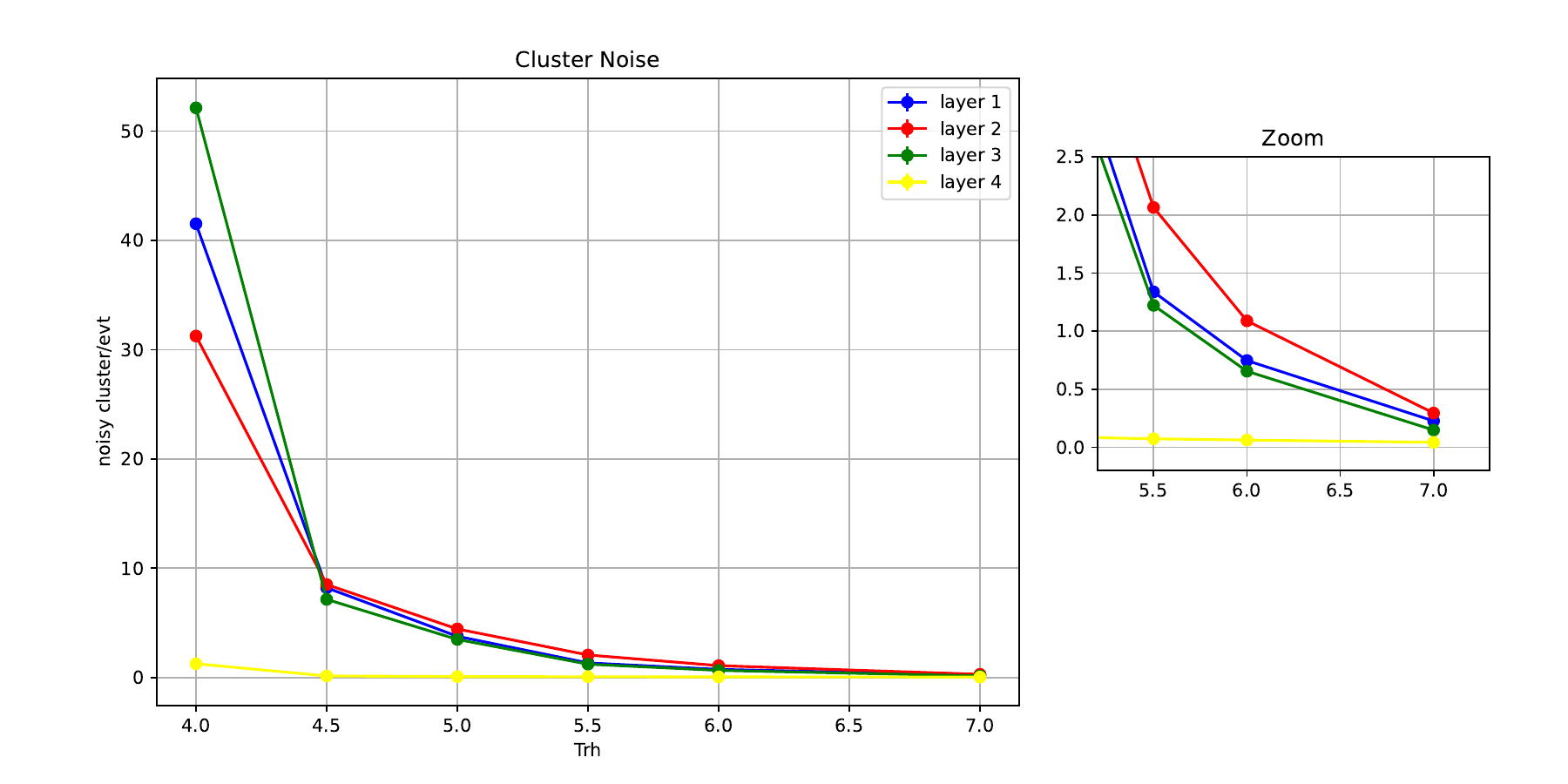}  
\caption{\textbf{Left:} Cluster size of the different sensors as function of the threshold. \textbf{Right:} Cluster noise of the different sensors as function of the threshold.}
\label{fig: cluster size threshold scan1}
\end{center}
\end{figure}

In terms of cluster size (Fig.: \ref{fig: cluster size threshold scan1}), looking at the first 3 sensors, aside from the problematic behavior at $4 \sigma$, it was observed that the cluster size slightly decreases with increasing thresholds, as expected. Furthermore, the cluster size values observed across different layers at each threshold are consistent with one another, indicating that the layers are functioning in a similar and reliable manner.

A specific observation can be made regarding sensor 4. Unlike the findings in the previous section, sensor 4 was not entirely inefficient during this test. On the contrary, it demonstrated a reasonably high efficiency, even though nearly a quarter of the sensor was completely inactive. Interestingly, its sensor efficiency remained constant between $4\sigma$ and $6\sigma$. Analyzing the cluster size of this sensor, it is evident that it is significantly smaller compared to the other sensors. This observation suggested that further lowering the threshold for this specific sensor might be beneficial, potentially bringing its cluster size closer to that of the other sensors and even slightly improving its efficiency. For this reason, a test was conducted with only this sensor operating at a $3\sigma$ threshold. However, the noise became unmanageable at this setting, rendering it unsuitable for use. The improved performance of sensor 4 during this phase can be attributed to the masking of certain highly noisy columns. It seems that these noisy columns were saturating the amount of information the detector could process, rendering it inefficient. By masking them, the detector’s performance are significantly enhanced.


The final aspect of the analysis focuses on noise. In the plot at right in Fig.: \ref{fig: cluster size threshold scan1}, is reported the number of noisy clusters per event. The calculation is performed as follows: only very clean events are considered, specifically those containing a single vertex composed of a single track (primary). For these events, we count the number of clusters not belonging to the track in each layer and divide this by the total number of events under consideration. As expected, the noise is significantly high for thresholds of 4, 4.5, and 5 $\sigma$. However, starting from 5.5 $\sigma$, the noise becomes more reasonable. Specifically, from 6$\sigma$ the noise go down to one noisy clusters per layer per event, decreasing slightly to higher threshold values. Anyhow the noisy level found is always greater than the one expected only from thermal noise computable from the tails of the Gaussian at a fixed sigma threshold. This is because part of the noise comes from electronics, and this is observable looking at some noisy columns of the sensors, that provide data also with no beam. In the M28 sensor single noisy pixels cannot be masked, but only columns, so the choice to mask some columns is a compromise between keeping a noisy column or a dead column if masked. As a consequence only the really noisy columns are masked. The choice done is good reminding that the remaining noisy columns (not masked) account for a noise of about 1~cluster/sensor/event, which correspond about to a fake hit rate of 10$^{-6}$ noisy pixels/sensor/Total pixels/event, so it's really low.

From this observation, we can conclude that the only thresholds worth considering are 5.5, 6, and $7\sigma$. At these thresholds, tracking efficiency and sensor performance remain almost identical. The noise decreases slightly as the threshold increases. Similarly, the cluster size shows a slight decrease as the threshold increases. In the end, a threshold of $6 \sigma$ was chosen as the working point. This represents a balanced compromise, offering a slightly larger cluster size while maintaining low noise levels. The differences between these three thresholds are not substantial, but $6\sigma$ appeared to be the optimal choice for the analysis.

\subsection{Energy scan}
After setting the thresholds to work at a $6\sigma$ level, an energy scan was performed using protons beam, with an energy range spanned from 70 MeV to 230 MeV. 

\begin{figure}[h!]
\begin{center}
\includegraphics[width=6.6 cm]{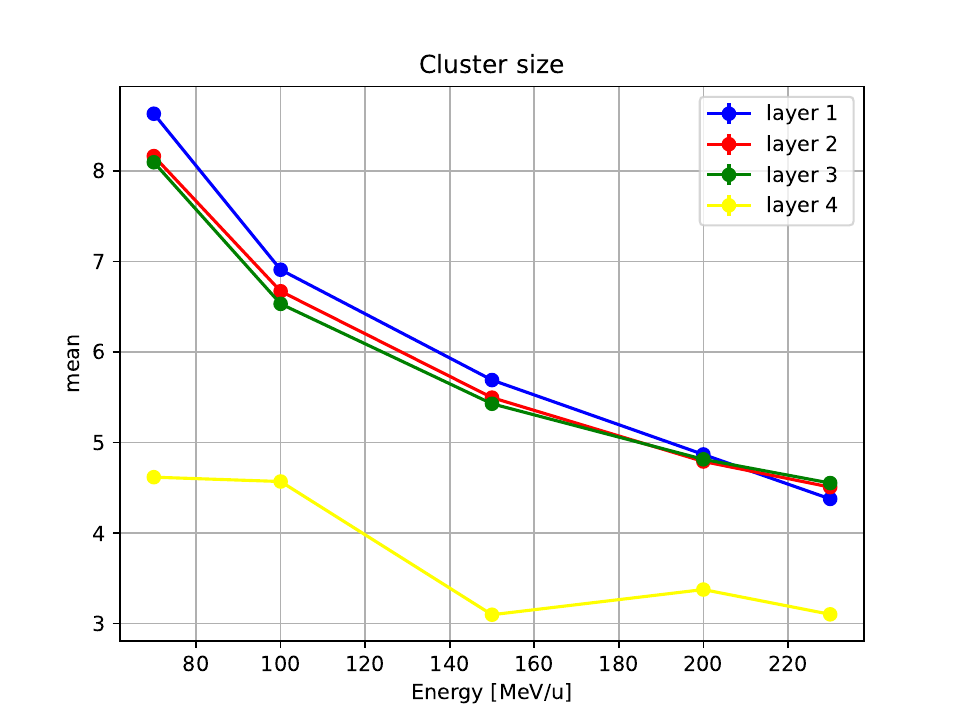}  
\caption{Cluster size as function of the energy of the proton beam.}
\label{fig: cluster size energy scan1}
\end{center}
\end{figure}

The results are presented in the cluster size plot of Fig.~\ref{fig: cluster size energy scan1}, where the expected behavior is observed: as the kinetic energy of the beam increases, the cluster size decreases. This trend is consistent with the reduced ionizing power of the particles at higher energies. Notably, the cluster sizes for the first three layers are consistent across the different energies, confirming that these layers are operating under the same regime. However, the fourth layer shows a deviation from this behavior, but we know this layer is problematic and not functioning optimally.

\begin{figure}[h!]
\begin{center}
\includegraphics[width=6.6 cm]{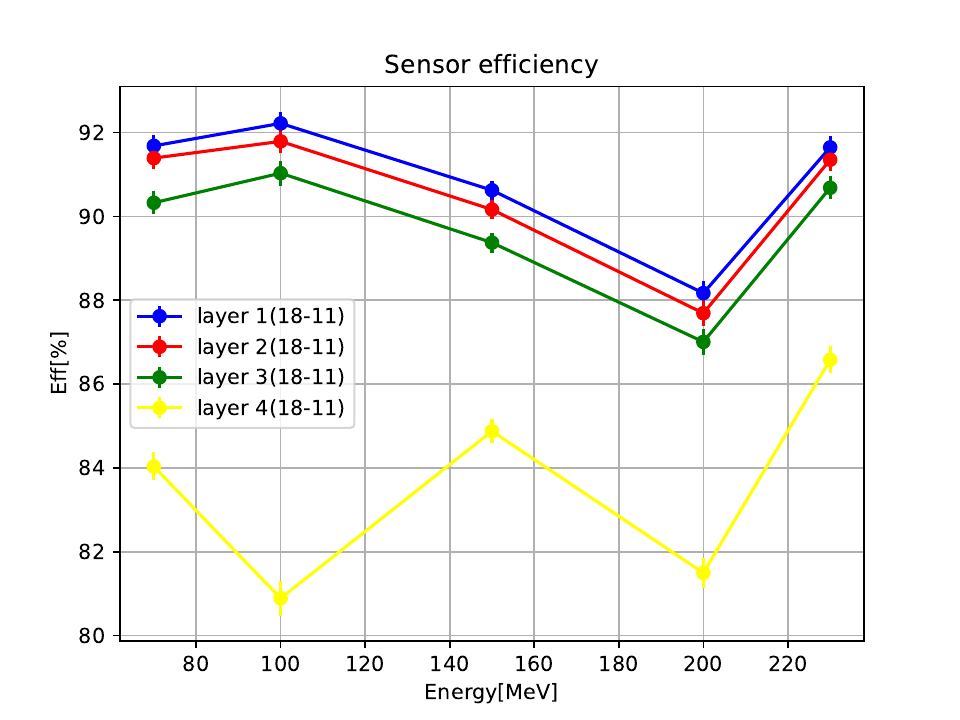}   \quad \includegraphics[width=6.6 cm]{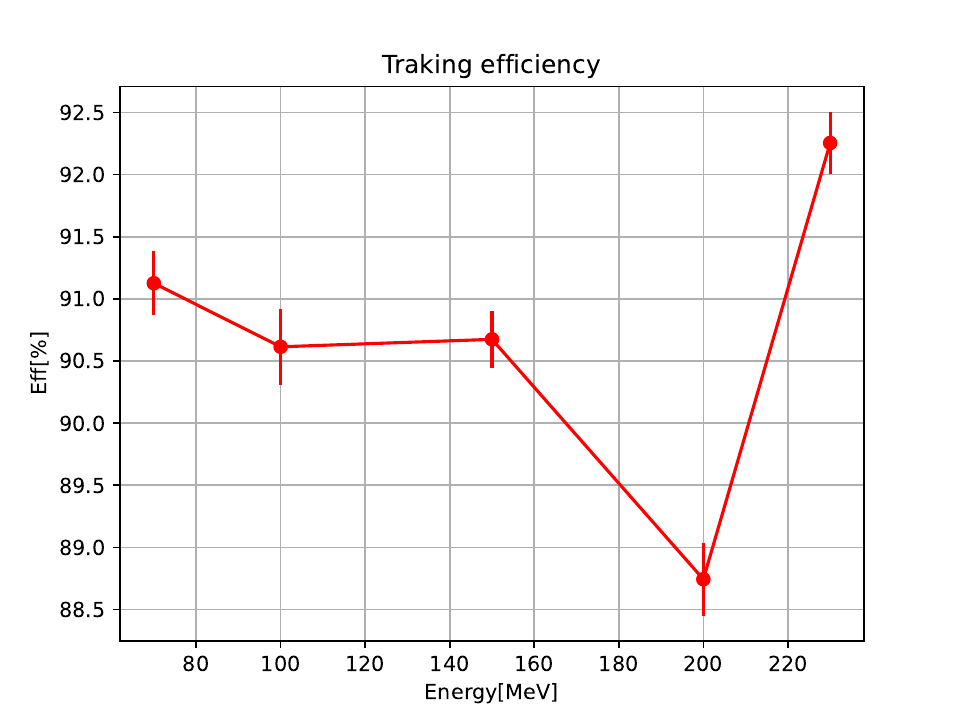}
\caption{\textbf{Left:} Sensors efficiency as function of the energy of the proton beam. \textbf{Right:} Tracking efficiency efficiency as function of the energy of the proton beam.}
\label{fig:sensor and tracking eff energy scan1}
\end{center}
\end{figure}

Examining the sensor efficiency plot of Fig.~\ref{fig:sensor and tracking eff energy scan1}, a similar pattern emerges. The first three sensors exhibit comparable efficiencies, confirming their consistent performance across the energy range.  The fourth layer displays lower efficiency. This could be attributed to the fact that a quarter of the sensor is damaged. Interestingly, the efficiency of layer 4 remains nearly constant across all energy levels, with fluctuations of approximately $\sim 4\%$, suggesting it is functioning in a regime where its performance is stable over the entire energy range, and the fluctuation are more linked to beam's dependencies. Looking at the first three layers, it is apparent that they show higher efficiency.  For this layers, the efficiency appears slightly energy-dependent at first glance, with a potential decrease at intermediate energies. However, an increase in efficiency is observed at $230MeV$.  Initially, this was attributed to extended time spent optimizing BM track reconstruction at this energy. Since sensor efficiency estimation relies on the BM tracks, improved tracks would lead to better results. Further analysis of the residuals between BM tracks and clusters revealed no significant changes with varying energies, indicating that the observed variations are due to systematic effects such as changes in the beam with energy, edge effects, and pile-up. These systematics on efficiency have been studied moving the geometrical selections used in the matching of the tracks between BM and VTX and it was estimated an systematic uncertainty of about 2\%. Overall, the efficiency of the first three layers is effectively constant across the energy range, with maximum fluctuations within $\sim 4\%$.

The tracking efficiency plot \ref{fig:sensor and tracking eff energy scan1} confirms these observations. It aligns closely with the sensor efficiencies of the first three layers, further validating the coherence of the results.

In general, with the thresholds chosen, the system operates with high proton detection efficiencies, achieving values around $92\%$.

\subsection{Noise study}

The final aspect to investigate concerns noise levels, particularly in datasets acquired with carbon ions, protons, and pedestal runs (the latter being acquired with an external trigger in the absence of a beam). The goal is to verify that the noise levels are comparable across these configurations. In beam runs, cluster noise is reported as previously described. However, in pedestal runs, where no tracks are present, the same method cannot be applied. Instead, noise is studied as the number of active pixels divided by the number of triggers.

\begin{figure}[h!]
\begin{center}
\includegraphics[width=6.6 cm]{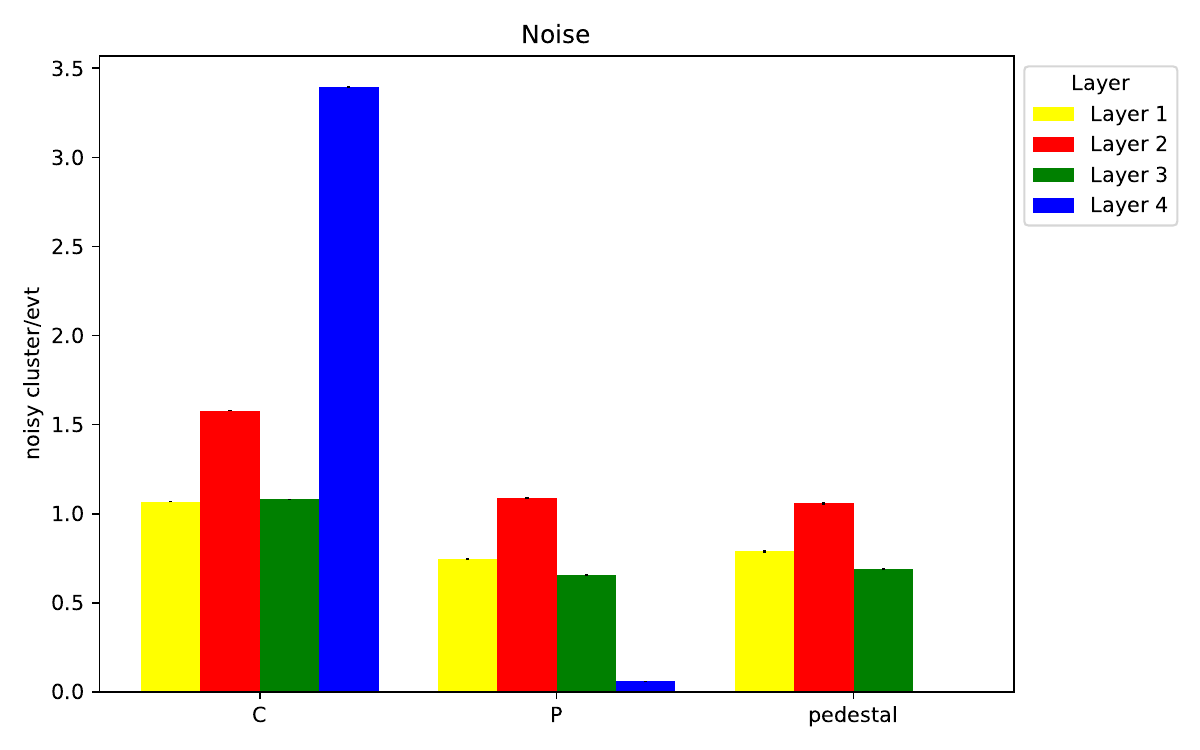}   
\caption{Noise for run with a carbon beam, a proton beam and pedestal.}
\label{fig: noise1}
\end{center}
\end{figure}

Initial observations (Fig.: \ref{fig: noise1}) reveal significantly higher noise levels for the carbon runs compared to others.

\begin{figure}[h!]
\begin{center}
\includegraphics[width=8.5cm]{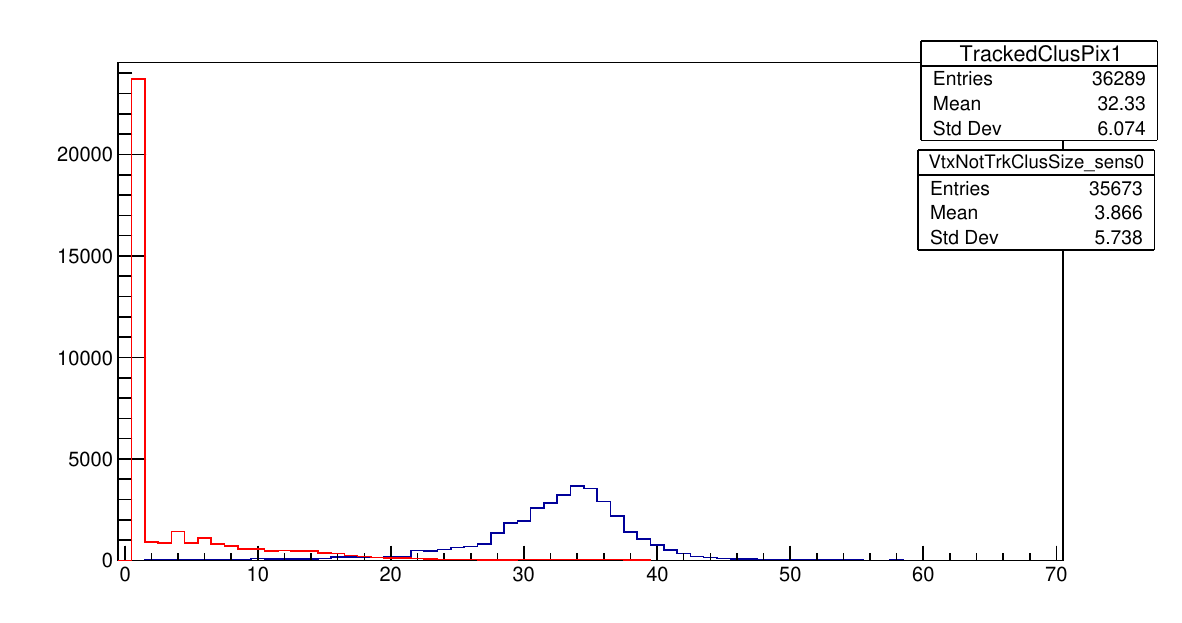}   
\caption{Cluster size distribution for the untracked cluster(noise)(red), and tracked
(blue) cluster.}
\label{fig: nois C200}
\end{center}
\end{figure}

To better understand this discrepancy, we analyzed the cluster size of untracked clusters (classified as noise). We observed, for run with the carbon beam, entries exceeding 5, reaching values as high as 18 (Fig.: \ref{fig: nois C200}). Clearly, clusters of this size cannot be attributed to electronic noise; rather, they correspond to real particles. In carbon runs, the presence of the graphite target ($5mm$) introduces the possibility of actual particles or fragments scattering off passive materials and re-entering a layer, as well as fragments traversing only two layers. These fragments, that don't cross at least three layers, are not reconstructed as tracks and are therefore incorrectly categorized as noise. Such cases are absent in pedestal and proton runs. Proton runs were conducted without a target, and the cluster size distribution for untracked clusters consistently shows a size of 1, indicating that only genuine noise is being detected.

\begin{figure}[h!]
\begin{center}
\includegraphics[width=6.6 cm]{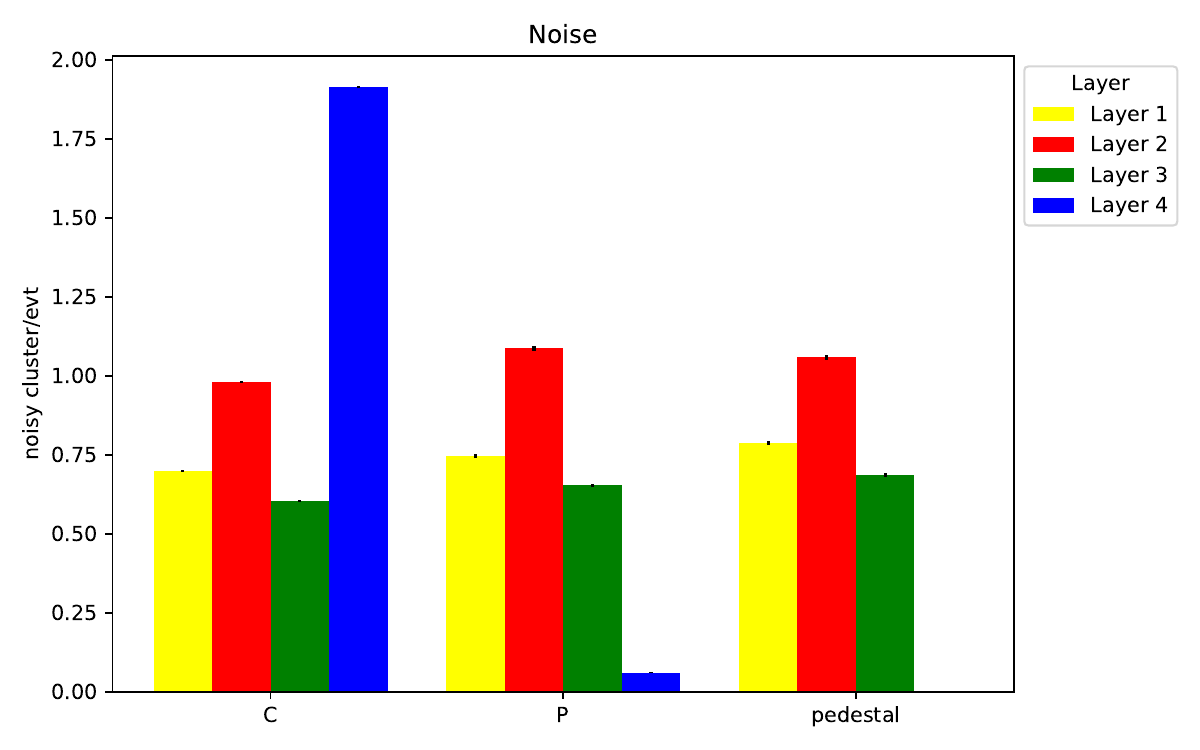}   
\caption{Noise for run with a carbon beam, a proton beam and pedestal. For the run with the carbon only not tracked cluster with a clsuter size of 1 are considered.}
\label{fig: noise 1clst1}
\end{center}
\end{figure}

When reanalyzing the data for carbon runs, filtering for untracked clusters with cluster size of 1 (to isolate noise and exclude real particles), (Fig.:\ref{fig: noise 1clst1}) the noise levels become comparable across all configurations. 

The only exception is observed in layer 4, where the noise level is zero in pedestal runs but increases for proton runs and further escalates in carbon runs. Unfortunately, this layer exhibits signs of instability. For example, pedestal runs show temporal variations in the sigma threshold for this specific layer. Consequently, it remains unclear whether the observed differences in noise levels are due to physical effects or an inherent instability of the layer. This instability casts doubt on the reliability of a fixed 6-sigma threshold for this sensor, suggesting that its behavior requires further investigation.

Overall, the noise levels appear consistent across configurations, which also confirms that the algorithm effectively controls noise without, for example, erroneously constructing tracks from noise. This is a strong indication that the chosen thresholds are optimal and that the performance, at least for of the first three layers, is consistent over time and across different configurations.

\subsection{Impact of threshold optimization on physics runs}
At this point, it is interesting to evaluate the impact of the threshold optimization. This optimization allows for a very high efficiency, even for protons. As a result, we might now be recovering vertexes of the type $C \to B+p$.

To investigate this, we analyze the percentage of valid vertexes. Since the values for the different runs are compatible within uncertainties, we take the weighted average, which results in :
\begin{equation}
    p_{vv}=(2.38 \pm 0.01)\%
\end{equation}

This value is higher than what was observed at CNAO2023 (\ref{eq: p_vv mean cnao2023}), despite the physics being the same, with the same beam and target. It is worth noting that at CNAO2024, the target was positioned $2mm$ farther from the vertex center for structural reasons, respect to CNAO2023. However, this shift is expected to primarily increase the number of tracks falling outside the acceptance of the VTX so in the limit it decreases the vertexes found, but this should have only a minor impact similar to what was observed between CNAO2022 and CNAO2023.

Therefore, the observed increase in valid vertexes identified appears to be related to the higher efficiency for protons, enabling the recovery of previously missed vertexes. This was exactly the initial idea and hope of this work that triggered the CNAO2024 data acquisition schedule, related to VTX performance studies and physics runs with an optimized VTX detector.

\chapter{Preliminary cross section} \label{chapter 5}
After studying the VTX detector, its performance, refining its geometry, confirming its capability to reconstruct vertexes and studying its efficiencies we now aim to obtain a preliminary measurement of the cross section, including the VTX detector.

The focus of this analysis is the GSI2021 campaign, which was conducted without a magnetic field. In this configuration, the tracks associated to a vertex go straight and can be projected onto the TW, searching for close TW points, which provide the charge Z of the fragments. 

\section{Experimental setup}
The experimental setup employed during the GSI2021 campaign consisted of the Start Counter, the BM, the VTX detector, and the TW. The analysis focused on runs involving $400 MeV/u$ oxygen beams incident on a $5 mm$ thick graphite target. Only data acquired with the minimum bias trigger configuration were considered in this study. About $1.2 \times 10^6$ events were acquired in this configuration, but the request of only one BM track within the acceptance of the VTX reduces this number by $ 15-20\%$ depending on the run.

The SC, in conjunction with the TW, provided a precise measurements of the time of flight for each detected event (about $70 ps$ for $^{16}$O ions), as detailed in  \cite{ref: range v} . The BM played a crucial role in characterizing the direction of primary particles incident on the target. This information enabled the selection of events where a primary particle was within the acceptance of the VTX detector. Additionally, only events featuring a single track in the BM were included in the analysis, effectively suppressing pile-up events in the BM and filtering out potential contributions from fragmentation out of target, upstream the BM.

The VTX is essential in identifying events associated with fragmentation processes occurring within the target. It also provided information about the directions of the resulting fragments.

\begin{figure}[h!]
\begin{center}
\includegraphics[width=8 cm]{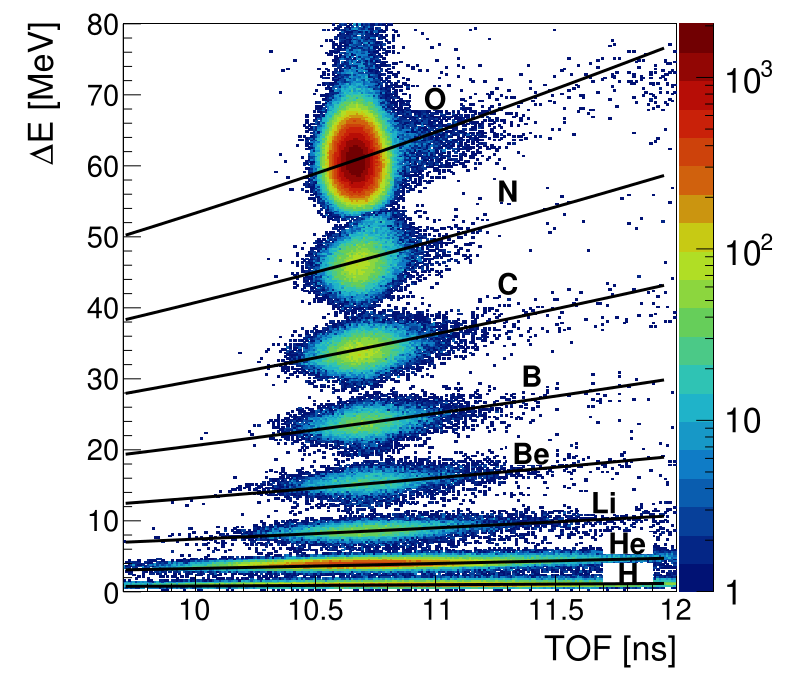}   
\caption{$\Delta E$ vs $ToF$ distribution for the data collected at GSI2021. The Bethe-Bloch curves are superimposed.}
\label{fig: Eloss vs ToF}
\end{center}
\end{figure}

The TW allows for the identification of fragment charges ($Z$) emitted within the angular acceptance of $\theta \le 5.7^\circ$. As detailed in \cite{ref:elemental}, the fragment charge Z was extracted by parametrizing the energy loss ($\Delta E$) as a function of ToF, based on a Bethe-Bloch curve (Fig. \ref{fig: Eloss vs ToF}). 

\section{ Event selection for VTX detector}
\label{sec:eventSelection}
The first step in this analysis is to carefully define the events to be studied. Specifically, the normalization is always performed using events with a single track in the BM within the acceptance of the VTX.

Next, we select events where the VTX reconstructs a fragmentation vertex generated within the target, ensuring that at least one track is within the acceptance of the TW.

Let's remember the definitions about the type of vertexes, given in Chapter~\ref{chapter 3}. A fragmentation vertex is called a "valid vertex" if two or more tracks, reconstructed by the VTX detector, converge in a point of the TG. A primary track (or a single track) crossing the VTX which doesn't interact in the TG provide a "not valid vertex".

One key consideration is whether to require that the reconstructed vertex is matched with the BM track. The match criterion is designed to select the vertex (whether it is a primary or a fragmentation vertex) closest to the BM track. This approach is primarily intended to remove pile-up vertexes from VTX detector using the spatial information of the relative position between the projection of the BM track on the TG and the vertexes positions. The BM dead time is more than 1000 times lower than the VTX one (675 $\mu$s), so the vertex closest to the BM projected track is the one "expected to be on time". Clearly the other vertexes are spatially distributed according to a Gaussian distribution with a FWHM of the beam XY profile. So this method doesn't prevent to associate a pile-up vertex spatially closer to the BM track than the vertex "on time".

As previously noted, the GSI2021 campaign is characterized by a high level of pile-up. Observations from earlier studies, particularly those analyzing charge distributions in Sec.~\ref{sec:usableLayer}, revealed that for some events, even when a valid vertex is matched with a BM track, only a TW point with charge Z=8 is identified, as shown in Fig.~\ref{fig: tw points GSI2021 layer}. This raises concerns that the high pile-up and so a large number of vertexes potentially located close to one another, with the finite resolutions of both the VTX and BM, might lead to errors in associating the match.
Two main kind of errors can be done.
A first case is the one of a pile-up vertex matched to the BM track, while a primary track (a not valid vertex), not matched with the BM track, is on time and the TW charge of the identified point is Z=8. 


A second case of error, more critical, arises when valid vertexes, whose fragments actually reach the TW, are erroneously discarded because not matched with the BM, because another primary track or fragmentation vertex not in time overlap between the vertex in time and the BM track. This scenario risks excluding genuine fragmentation events that are key to the study, and will be discussed in the section~\ref{sec:yields}.

\begin{figure}[h!]
\begin{center}
\includegraphics[width=8.5 cm]{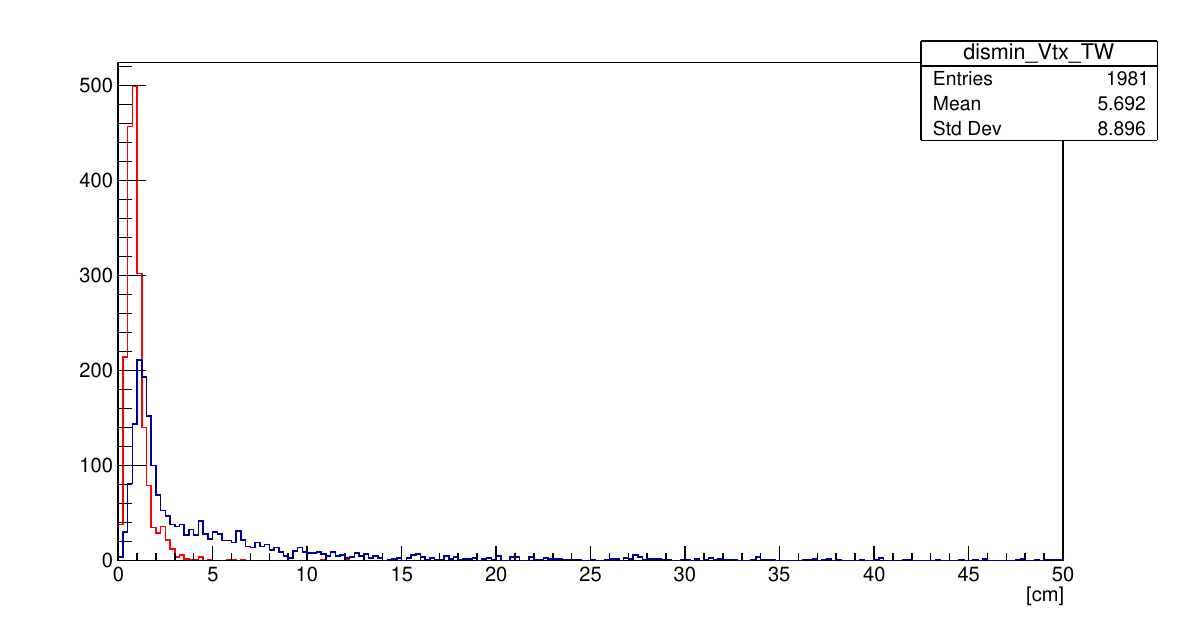}   
\caption{Histogram showing the distance between the TW point (charge 8) and the closest tracks of primary (unmatched) vertexes (in red) and the closest track of valid matched vertexes (in blue).}
\label{fig: match or not}
\end{center}
\end{figure}

To investigate on the first kind of error, which is the hypothesis of incorrect matching of a pile-up vertex while a primary with Z=8 reach the TW, the following approach was adopted. Events were selected where a valid vertex generated in the target and matched with the BM was identified. For the same events, a TW point with charge 8 was required. Two histograms were then filled:

\begin{itemize}
    \item The distance between the TW point and the closest of the projections on TW of the not valid vertexes (not matched with the BM) found in the VTX detector (Fig.~\ref{fig: match or not}, in red).
    \item The distance between the TW point and the closest of the vertex tracks projected on the TW detector, between the ones associated to the valid vertex matched with the BM (Fig.~\ref{fig: match or not}, in blue).
\end{itemize}

\begin{figure}[h!]
\begin{center}
\includegraphics[width=6.6 cm]{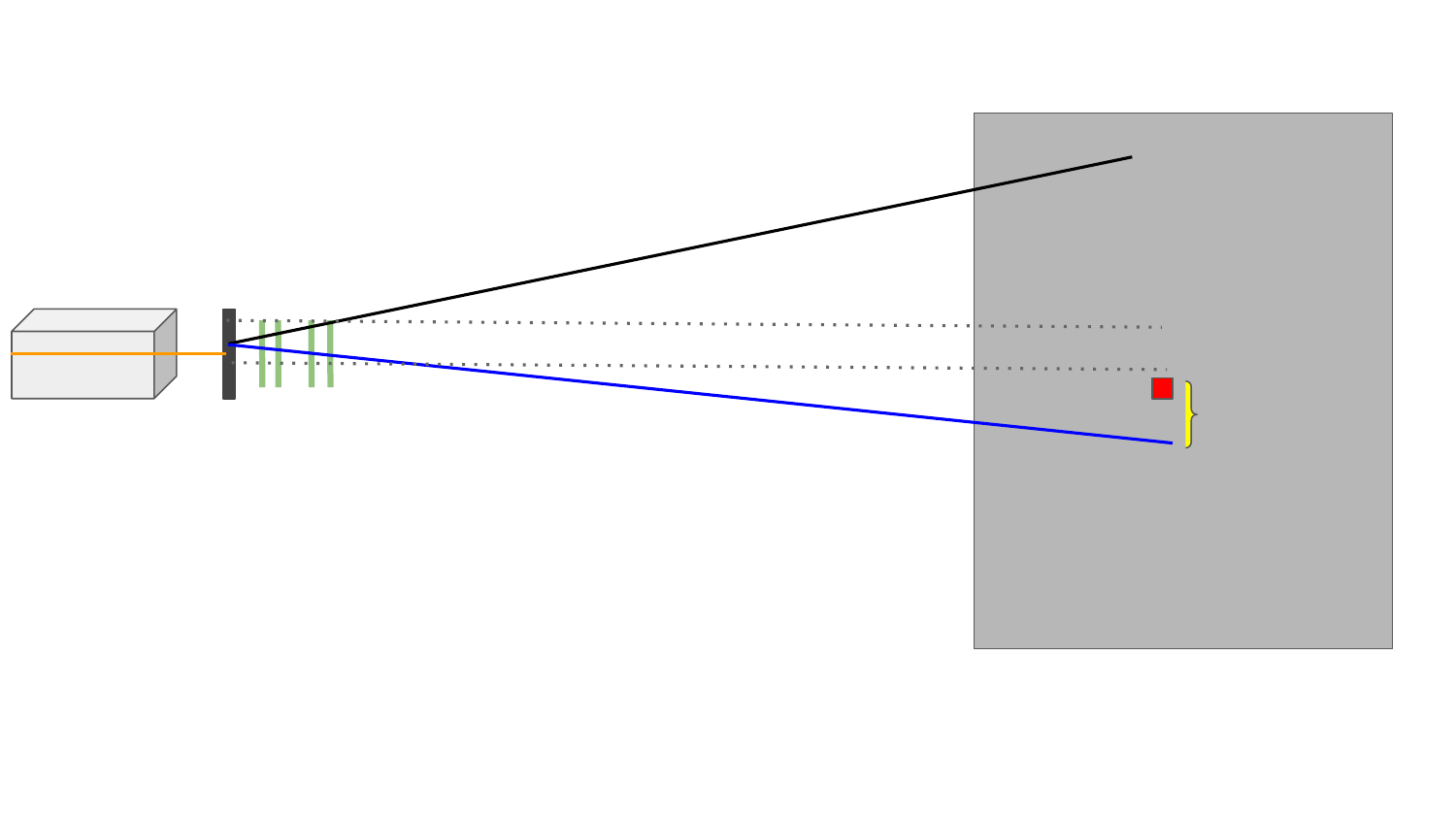}   \quad \includegraphics[width=6.6 cm]{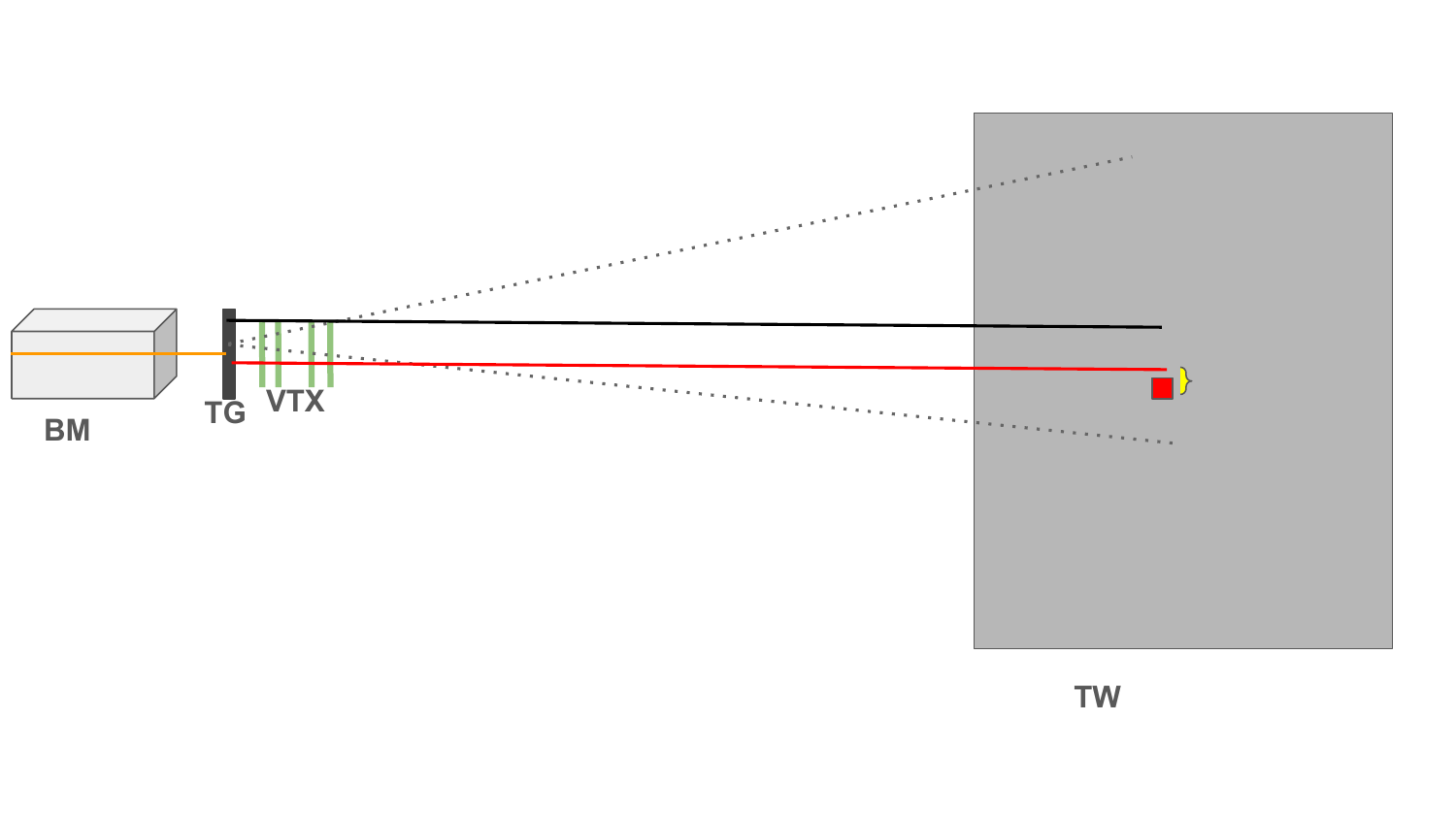}
\caption{Diagram of the analyzed events. The two images refer to the same event. On the left, the matched vertex is highlighted. The 2 tracks of the vertex are projected onto the TW, the one closest to the TW point with charge $Z=8$, in blue, is considered, and its distance form the TW point is measured. On the right, the not valid vertex (primary) and not matched with the BM are highlighted. The closest track respect to the TW point with charge $Z=8$ is identified in red, and its distance from the TW point is measured. }
\label{fig: schema dist min}
\end{center}
\end{figure}

In the figure \ref{fig: schema dist min}, a schematic of these two cases is provided to better illustrate the strategy used.

The results of these histograms reveal distinct patterns. The histogram of Fig.~\ref{fig: match or not}, in red, for the primary tracks (not valid vertexes, unmatched with BM) shows a peak at distances smaller than 4 cm, with entries only below this value, indicating a strong spatial correlation with the TW point at $Z = 8$. On the other hand, the histogram for matched vertexes also exhibits a smaller shifted peak at small distances but includes entries extending up to $50 cm$. This broader distribution for matched vertexes supports the hypothesis that 
the vertex selected, because matched with the BM, is a pile up fragmentation vertex. 
For each event, looking at the projection of the tracks of all the vertexes (valid and not valid) on TW, in which there is a TW point with Z=8 and a valid vertex, BM matched, the closest projected track comes always from a not valid vertex unmatched with BM.

For this reason in the following analysis, the match of the vertex with the BM track will not be requested, and only valid vertexes generated within the target will be considered. Among the selected valid vertex, the pile-up fragmentation vertexes which correspond to a TW point with Z=8, discussed above, will be discarded just looking at events with a valid vertex in TG and a TW point of Z=8. Between these discarded vertex could be present some event in which a Nitrogen fragment (Z=7) is misidentified in a oxygen (Z=8). The number of these events has been evaluated in the contest of the charge identification algorithm developed elsewhere and is negligible (only less than 0.001\% of the oxygens are actually wrongly identified).

With a lower probability with respect to the case discussed so far, it can happen that a primary particle fragments in the air between the vertex and the TW. In this case we still expect that the resulting fragments will arrive at positions different from those where the fragments of the reconstructed pile-up vertex are expected. Anyhow this kind of topology is included in the selection of tracks for cross section measurement shown in the section~\ref{sec:yields}. 


\subsection{Data Analysis}
The objective of this study is to determine the angular differential cross sections. The analysis focuses on the $He$, $Li$, $Be$, $B$, $C$, and $N$ nuclei, produced within an angular acceptance of $\theta\le 5.7^\circ$. The $H$ measurement is excluded because the TW detector is not optimized for the detection of these fragments. For each fragment characterized by a specific charge Z, the cross section is calculated using:
\begin{equation}\label{eq: diff cross sec}
    \frac{d\sigma}{d \Omega}(Z, \theta)= \frac{Y(Z,\theta)}{N_{prim}\cdot N_{TG} \cdot \epsilon(Z,\theta) \cdot \Delta\Omega}
\end{equation}
where $N_{prim}$ represents the number of primary particles incident on the target and within the acceptance of the VTX, when the number of tracks reconstructed by the BM is equal to 1. $\epsilon(Z, \theta)$ is the efficiency for a given charge $Z$ and angle $\theta$ , which is described in the following section. $\Delta \Omega$ is defined as the width of the solid angle bin; the minimum angular bin width used is set taking into account the TW granularity, which is $\sim0.6 ^{\circ}$. $Y(Z,\theta)$ corresponds to the number of reconstructed fragments with charge $Z$ at a given angle $\theta$. $N_{TG}$ denotes the number of interaction centers in the target per unit area and is expressed as:
\begin{equation}\label{eq: n_tg}
    N_{TG}=\frac{\rho \cdot d \cdot N_A}{A}
\end{equation}

with $\rho=1.83 g/cm^3$ the graphite target density, $d = 0.5 cm$ the target thickness, $N_A$ is the Avogadro number and $A = 12.01 g/mol$ is the graphite mass number.

\subsection{Yields}
\label{sec:yields}
The yields, $Y(Z,\theta)$, are calculated using fragmentation vertexes reconstructed by the VTX detector. It is important to note that no matching with the BM is required for these vertexes, in line with the considerations discussed earlier in Sec~\ref{sec:eventSelection}.

To determine the yields, the tracks originating from these reconstructed vertexes and falling within the acceptance of the TW are projected onto the TW plane. For each track projection, the closest TW point is identified, provided that the distance between the track projection and the TW point is less than or equal to $4 cm$. This threshold was chosen based on an analysis of the residuals between the track projections and the corresponding TW points, which showed a peak within $4cm$, consistent with the TW granularity of 2~cm. This distance criterion also helps to exclude events where fragments undergo secondary fragmentation in air before reaching the TW, as in such cases the TW points would be located farther from the projected track intersections. Clearly the impact of this choice and the background that can arise from this fragmentation in air can only be studied in MC not having other tracker detector between VTX and TW at the moment. As discussed in~\ref{sec:eventSelection}, events where the TW is reached by a primary particle, $Z=8$, despite the presence of a reconstructed fragmentation vertex, are not considered.

If a track is successfully associated with a TW point, the event is included in the yield calculation. The charge $Z$ of the fragment is assigned based on the $Z$ value of the associated TW point. The angular information, $\theta$, is determined by calculating the angle between the respective track from the VTX and the track reconstructed by the BM. These measurements allow for the classification of fragments into bins of angle $\theta$ and charge $Z$.

\subsection{Efficiency}
The efficiency for each charge and emission angle is calculated through MC simulations. The overall efficiency used in this analysis is composed of three components.

The first is the vertex efficiency, $\epsilon_{VTX}$, which is defined as the ratio of reconstructed vertexes to true vertexes: 
\begin{equation}
    \epsilon_{vtx}=\frac{reco_{vtx}}{true_{vtx}}= (90.95 \pm 0.07)\%
\end{equation}

This efficiency accounts for both the performance of the vertex reconstruction algorithm and the geometric acceptance of the VTX detector. Only true vertexes with at least one track within the acceptance of the TW are considered in this calculation. This requirement ensures that the selected events correspond to those contributing to the cross-section measurement of interest.

The reported value of $\epsilon_{vtx}$ is slightly lower than the one provided in the previous chapter, \ref{eq:eff}. While it represents the same quantity, the difference arises from the additional selection applied here, requiring at least one track to be within the TW acceptance. Nonetheless, this difference is negligible and they remain compatible between the uncertainties.

Next, the analysis incorporates the efficiency for protons and helium nuclei, which has been studied previously in Chap.~\ref{chapter 4}. Specifically, for all vertexes reconstructed by the algorithm (the numerator of the previous efficiency calculation), it is acknowledged that the detector is not completely efficient for protons and helium. 
For protons with energies less than or equal to $230$~MeV, the efficiency is modeled using the linear fit obtained in the previous section (\ref{fig: fit lin1}). For higher energies, the efficiency is assumed to remain constant at the value corresponding to $230$~MeV. This assumption is clearly very strong and affect the vertexes containing protons of high energy. It's clear that the efficiency need to be constrained better at higher energy values as discussed in Chap.~\ref{chapter 4}. Anyhow this is a preliminary study and it is the first "complete exercise" of computing cross sections in FOOT using a setup including the VTX detector and it is needed in order to highlight all the critical aspects of the analysis in order to understand how to improve in next steps.

A similar approach has been applied to helium nuclei. 
At CNAO there is no He beam. So the curve of efficiency obtained for protons of Fig.~(\ref{fig: fit lin1}) was used also for He converting the x-axes values of proton kinetic energy in values of protons energy loss in 25~$\mu$m of silicon (which corresponds to the epitaxial layer of the M28 sensors). Clearly due to the Z$^2$ of the Bethe-Bloch the impact of this efficiency on He is small with respect to protons, but still not negligible for He of high energy (400-800 MeV/u).


As already said for protons also for He (we're using the same efficiency curve) it is important to remember that this efficiency estimation is a very simplified model aimed at accounting for the detector response to these particles, whereas more data are needed to constrain the efficiency curve. For all other particle species, no efficiency corrections have been applied, as these are expected to exhibit very high efficiencies due to their higher ionization in silicon.

For vertexes containing protons or helium nuclei, inefficiencies for these particles can result in the loss of the entire vertex. The reconstructed vertexes, when accounting for the detector's efficiency for protons and helium nuclei, are referred to as $reco_{vtx+eff}$. Although this study assumes angular independence of the efficiency, it identifies a dependence on the fragment charge. Thus, the efficiency is defined for each fragment charge $Z$:

\begin{equation}
    \epsilon_{vtx+eff}(Z)=\frac{reco_{vtx+eff}(Z)}{reco_{vtx}(Z)}
\end{equation}

\begin{figure}[h!]
\begin{center}
\includegraphics[width=8 cm]{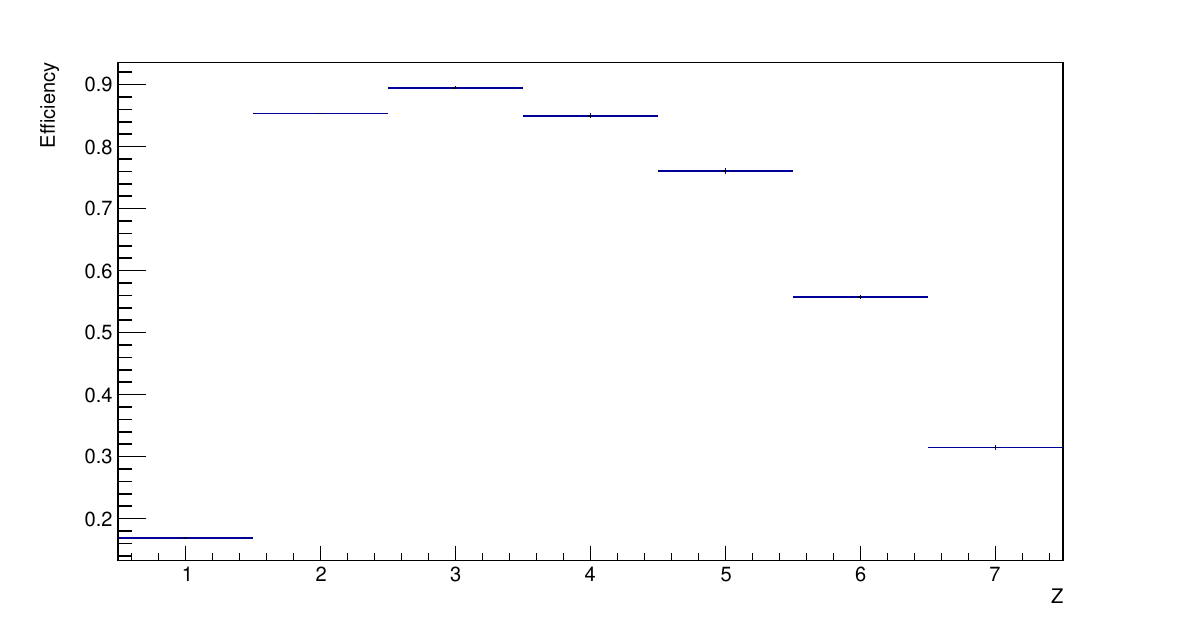}   
\caption{Efficiency of reconstructible vertexes as a function of fragment charge ($Z$) after accounting for the detector inefficiency for $p$ and $He$.}
\label{fig: eff_p}
\end{center}
\end{figure}

The efficiency $\epsilon_{vtx+eff}(Z)$ is shown in Figure \ref{fig: eff_p}. Notably, excluding helium and protons, the impact of the efficiency corrections becomes more pronounced for higher fragment charges. This trend can be understood by considering how the inefficiency of detecting specific particles, such as protons or helium nuclei, affects the reconstruction of vertexes and their associated fragments. For example, consider a case where an oxygen fragment produces a particle with charge 7 (nitrogen) and a proton. If the proton is not detected, the entire vertex is lost, including the nitrogen fragment. Similarly, for particles with charge 6 (carbon), these fragments are often produced in association with either a helium nucleus (He) or two protons. Since the probability of losing both protons or the helium nucleus is lower than missing a single proton, the resulting efficiency for carbon is higher respect to nitrogen, and so on. As mentioned in Chapter~\ref{chapter 3} no vertex identification with single track using the kink of the track has been implemented up to now. It will be something to be explore in the next future to improve the vertex reconstruction.

The final efficiency taken into consideration is that of the TW detector. In this case, the efficiency is calculated as a function of both charge ($Z$) and angle ($\theta$). For the vertexes reconstructed after applying the detector efficiencies for protons and helium nuclei, we consider how many of these fragments are successfully reconstructed by the TW with the correct charge $Z$ and angle $\theta$. This efficiency accounts for fragments that are not reconstructed by the TW, as well as fragments that fail to reach the TW due to multiple Coulomb scattering and fragmentation in air. However, the latter contribution is minimal. The significant effect arises from the TW reconstruction process itself, as detailed in \cite{ref:elemental} and this effect is expected to be more pronounced for light fragments, such as helium and lithium.
The TW efficiency, $\epsilon_{TW}$, is defined as:

\begin{equation}
    \epsilon_{TW}(Z, \theta)=\frac{N_{TW}(Z, \theta)}{N_{prod}(Z,\theta)}
\end{equation}
where  $N_{TW}(Z, \theta)$ rappresents the fragments reconstructed by the TW with correct $Z$ and $\theta$, and $N_{prod}(Z,\theta)$ are the number of true fragments produced in the target and within the acceptance of the TW, for which the VTX algorithm successfully reconstructed the corresponding vertex, after applying the efficiencies for protons and helium nuclei. 

\begin{figure}[h!]
\begin{center}
\includegraphics[width=8 cm]{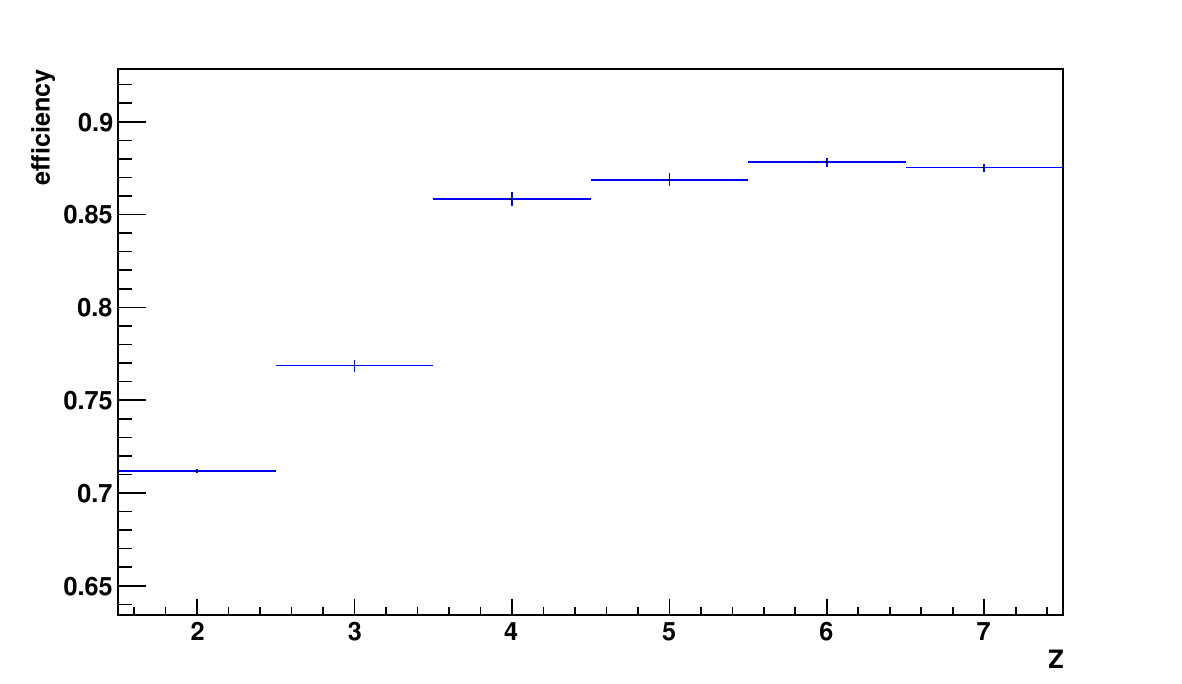}   
\caption{Efficiencies for each reconstructed fragment of charge Z emitted in the TW acceptance.}
\label{fig: eff_tw}
\end{center}
\end{figure}

In the figure \ref{fig: eff_tw}, is reported the TW efficiency for each charge $Z$, obtained by integrating over the angle $\theta$ the efficiencies $\epsilon_{TW}(Z, \theta)$.  
As anticipated the most significant contribution to this efficiency reduction comes from the TW reconstruction process itself which impact more on light fragments such He and Li. 

The efficiency that appears in the cross section calculation is given by the product of this efficiencies:
\begin{equation}
\label{eq:totEff}
    \epsilon(Z, \theta)= \epsilon_{vtx} \times \epsilon_{vtx+eff}(Z) \times \epsilon_{TW}(Z, \theta)
\end{equation}

\section{Closure test}
\label{sec:closureTest}
A MC closure test is performed to verify the consistency and reliability of the analysis strategy employed for the calculation of the differential cross-sections, discussed in previous sections.
In a MC closure test the true MC cross section is compared to the reconstructed cross section in MC in order to validate the analysis strategy used in the cross section reconstruction.

The true differential cross-section is determined by considering all fragments generated by a primary particle within the target and within the acceptance of the TW. The true yields, $Y_{true}(Z,\theta)$, represent the number of these fragments produced with a given charge $Z$ and emission angle $\theta$. The true differential cross-section is calculated using the expression:

\begin{equation} 
\frac{d\sigma}{d\Omega}(Z, \theta) = \frac{Y_{true}(Z, \theta)}{N_{prim} \cdot N_{TG} \cdot \Delta \Omega}
\end{equation}
where $N_{prim}$ denotes the number of incident primary particles  that are in the acceptance of the VTX detector, $\Delta \Omega$ is the width of the solid angle bin, and $N_{TG}$ represents the number of interaction centers in the target per unit area (see Eq.\ref{eq: n_tg}).

The true cross-sections are then compared with the reconstructed cross-sections, which are obtained using the reconstruction procedure described in the previous sections. Specifically, the vertexes reconstructed by the algorithm, after applying the detector efficiencies for protons and helium, are considered. Tracks associated with these reconstructed vertexes and within the TW acceptance are projected onto the TW. For each projected track, the nearest TW point is identified, provided that the distance between the track projection and the TW point does not exceed $4cm$. These TW points provide the charge information ($Z$) of the fragments, while the emission angle ($\theta$) is determined by calculating the angle between the track and the BM track. These reconstructed fragments constitute the yields, $Y_{reco}(Z, \theta)$. The yields are then corrected for the total efficiencies $\varepsilon(Z,\theta)$ of Eq.~\ref{eq:totEff} and the final differential cross-section is calculated using Eq.~\ref{eq: diff cross sec}.

\begin{figure}[h!]
\begin{center}
\includegraphics[width=6.5 cm]{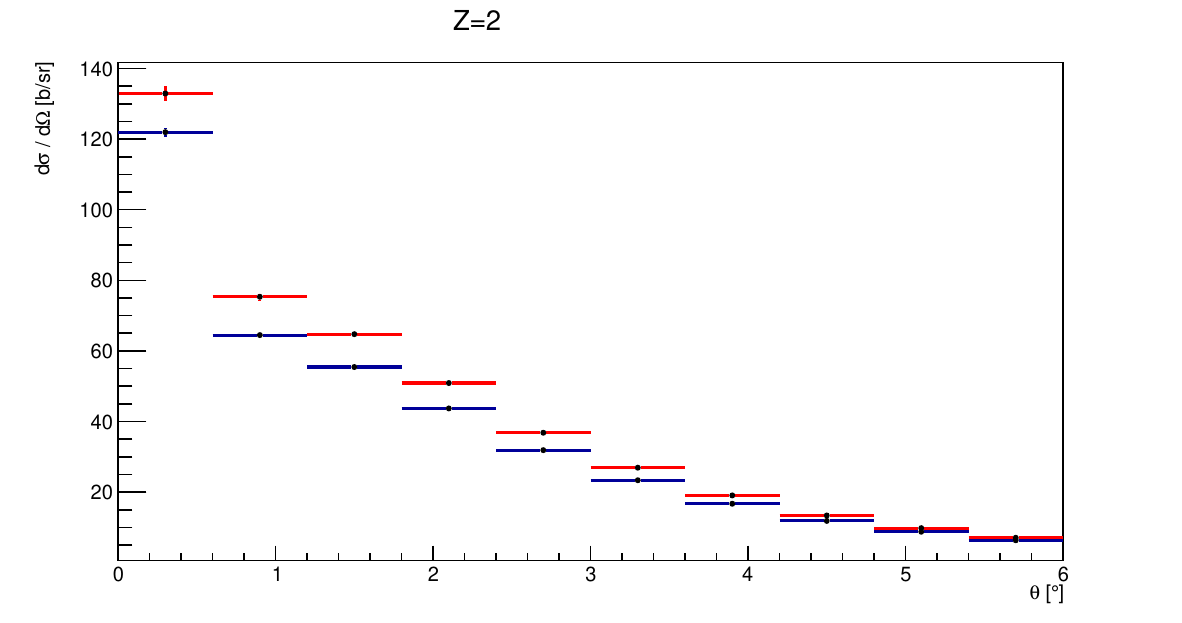} \quad  \includegraphics[width=6.5 cm]{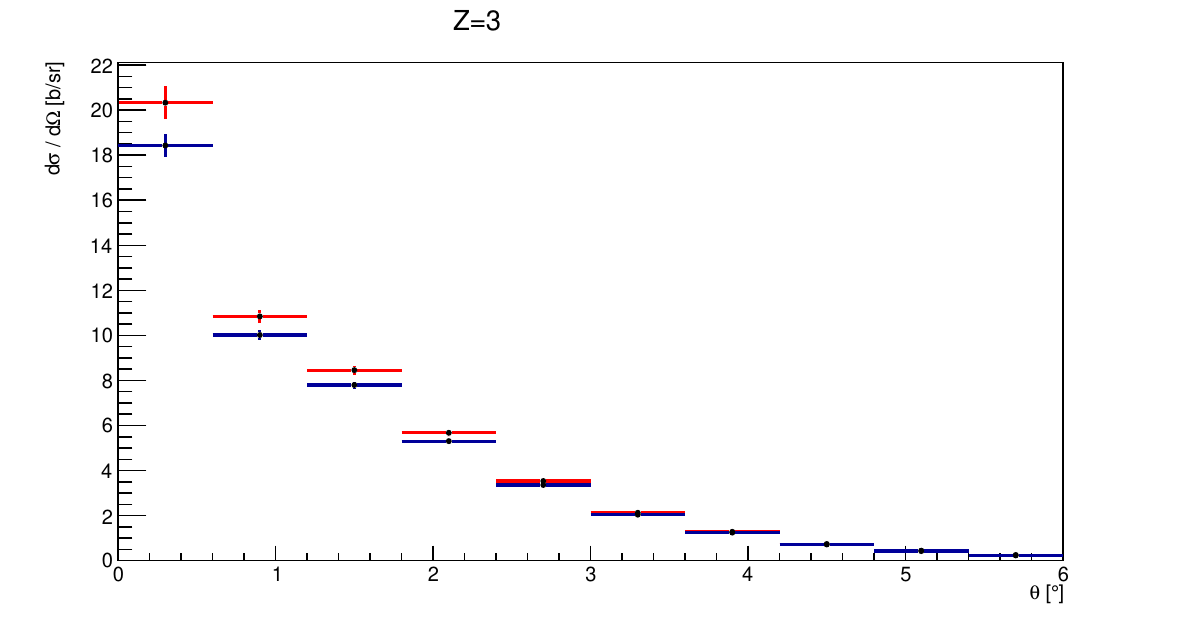} \quad \includegraphics[width=6.5 cm]{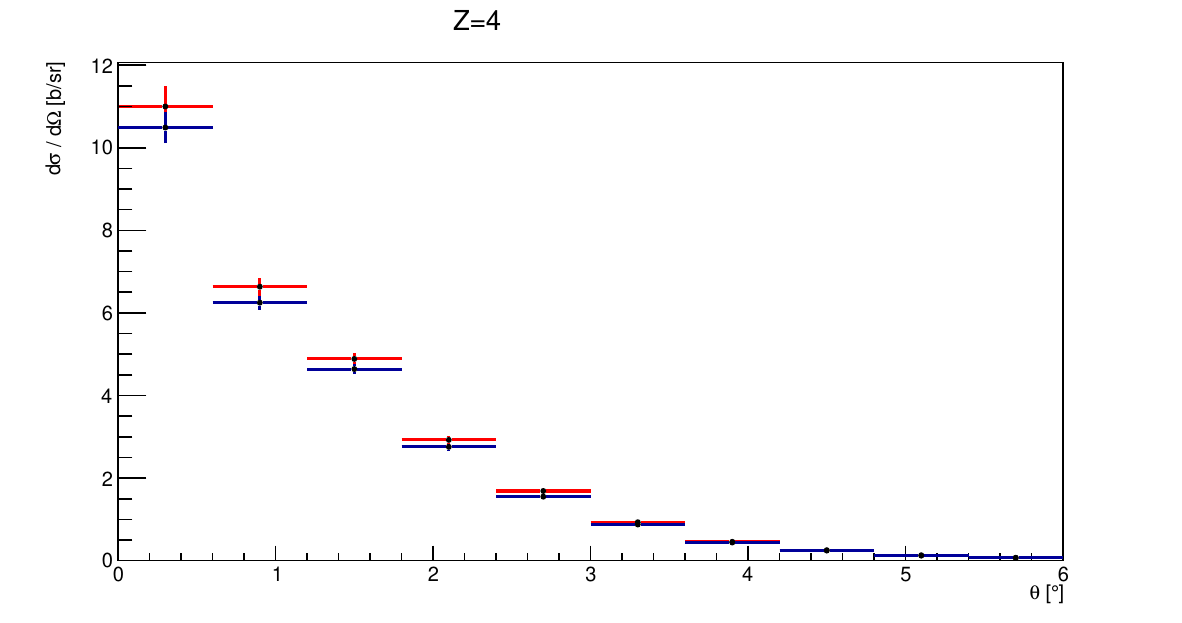} \quad  \includegraphics[width=6.5 cm]{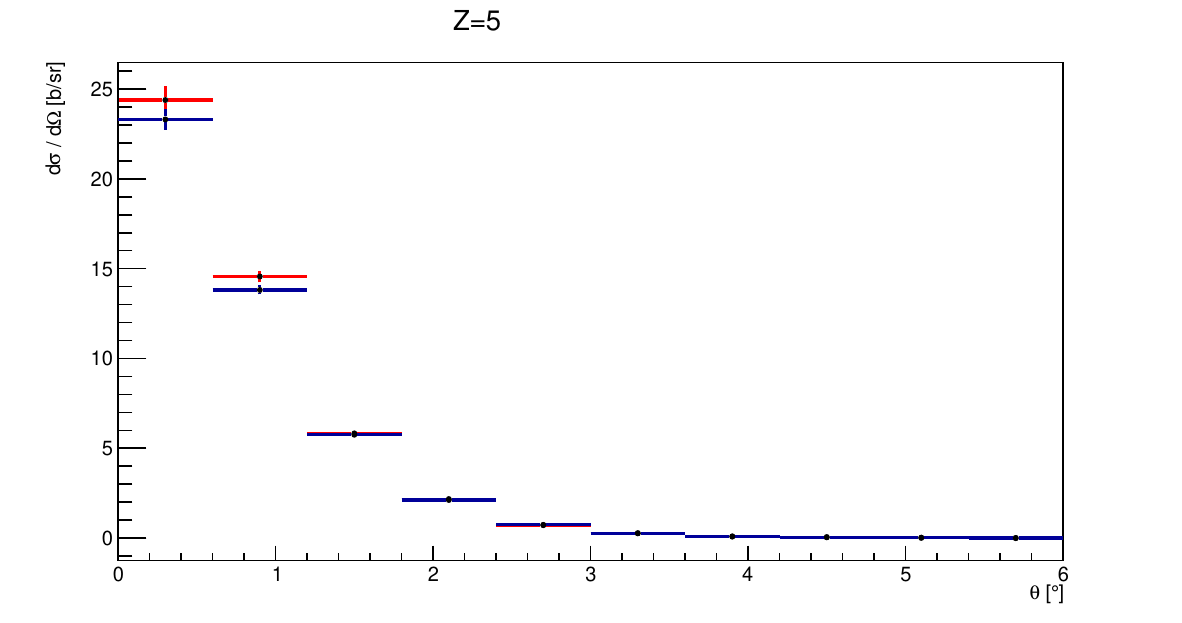} \quad
\includegraphics[width=6.5 cm]{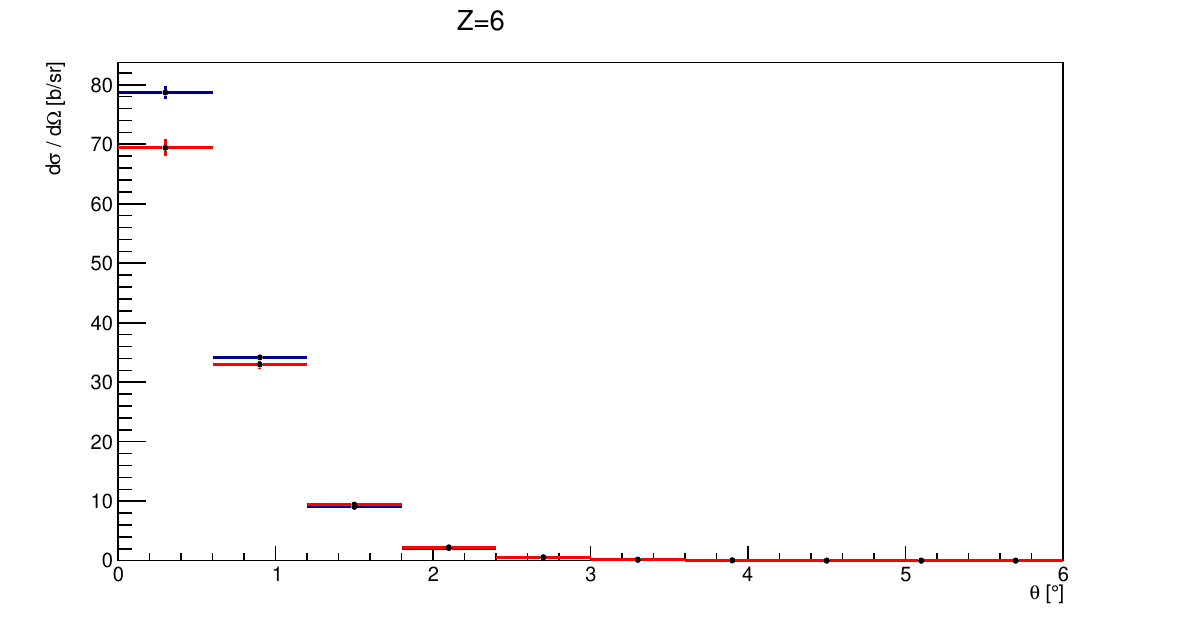} \quad  \includegraphics[width=6.5 cm]{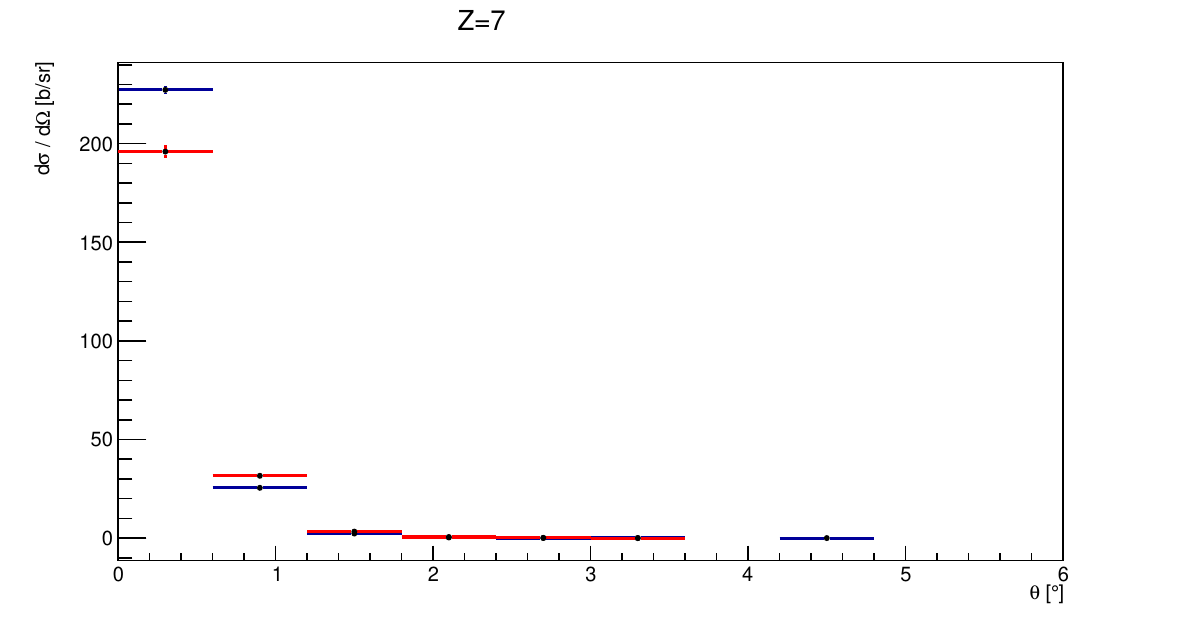} \quad
\caption{Angular differential cross sections for fragments $2 \le Z \le 7$. The reconstructed cross-sections are shown in red, while the true cross-sections are shown in blue. }
\label{fig: closure test}
\end{center}
\end{figure}

Figure \ref{fig: closure test} shows the differential cross-sections for all fragments with charges $Z\geq 2$ and $Z\le 7$, comparing the true cross-sections (blue) with the reconstructed ones (red). The results demonstrate a good agreement between the two sets of cross-sections, validating the reconstruction procedure and the applied efficiency corrections. 
It has been verified that the remaining discrepancy is mostly due to the proton and helium efficiencies tuned from data introduced in the MC reconstructed cross sections, but not in the true ones.
For this reason this discrepancy will be used as an estimation of the systematic uncertainties due to the protons and helium efficiency of Fig.~\ref{fig: eff_p}. In particular the relative error bin by bin of angle $\theta$ for each Z will be attributed as systematic uncertainties to the cross sections measured in data:

\begin{equation}
 \Delta_{sys}^{rel} = \Bigg|\frac{(d\sigma(Z,\theta)/d\Omega)_{rec}-(d\sigma(Z,\theta)/d\Omega)_{true}}{(d\sigma(Z,\theta)/d\Omega)_{true}}\Bigg|
\end{equation}


\section{Results}

\begin{figure}[h!]
\begin{center}
\includegraphics[width=6.5 cm]{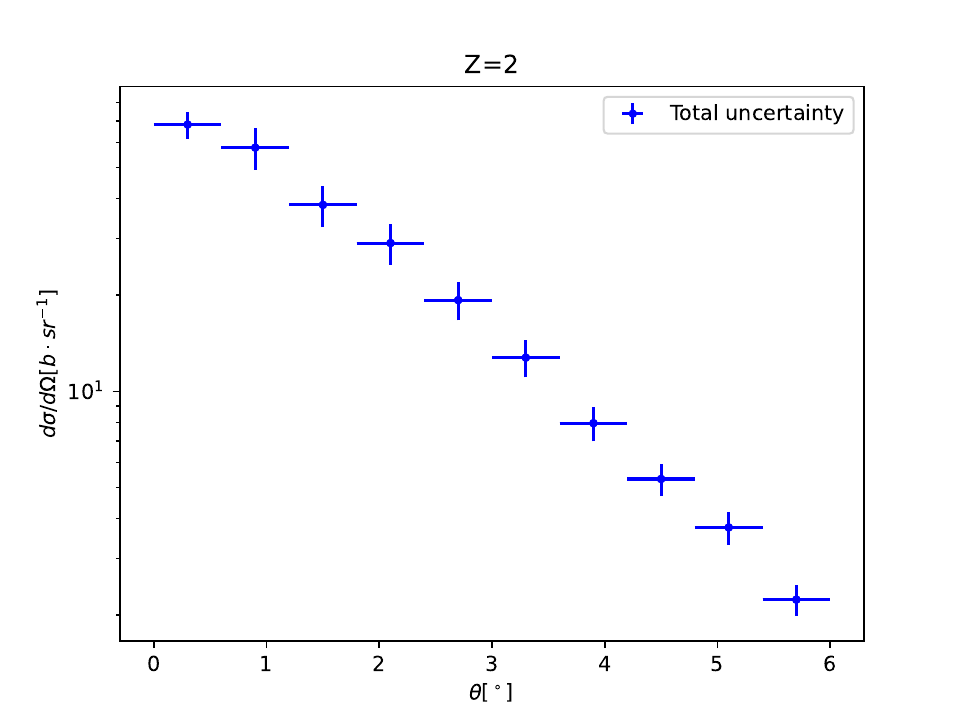} \quad  \includegraphics[width=6.5 cm]{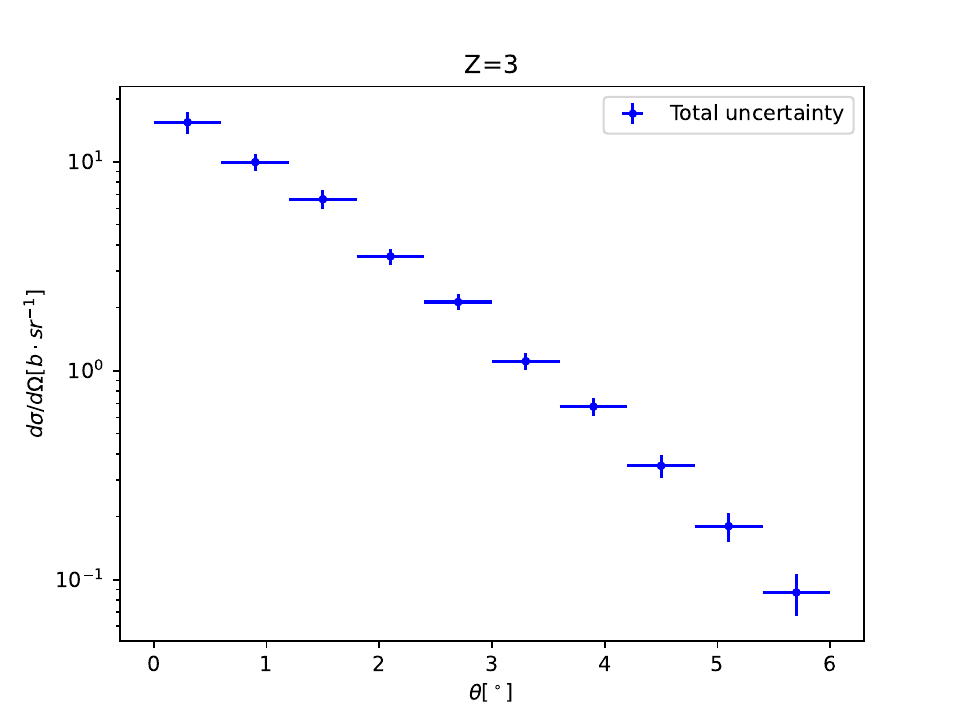} \quad \includegraphics[width=6.5 cm]{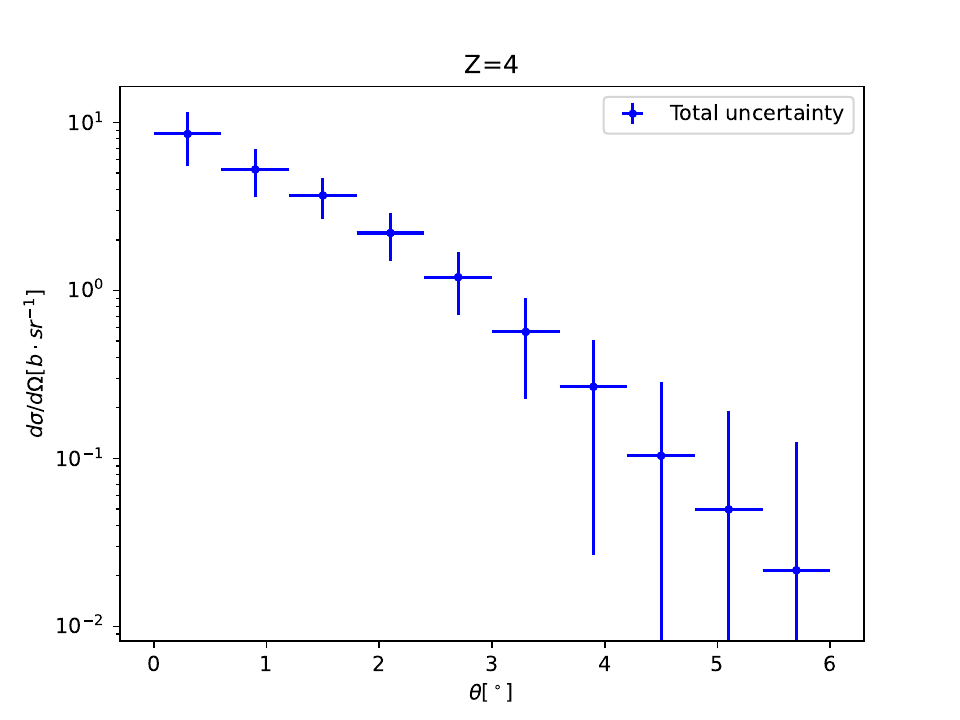} \quad  \includegraphics[width=6.5 cm]{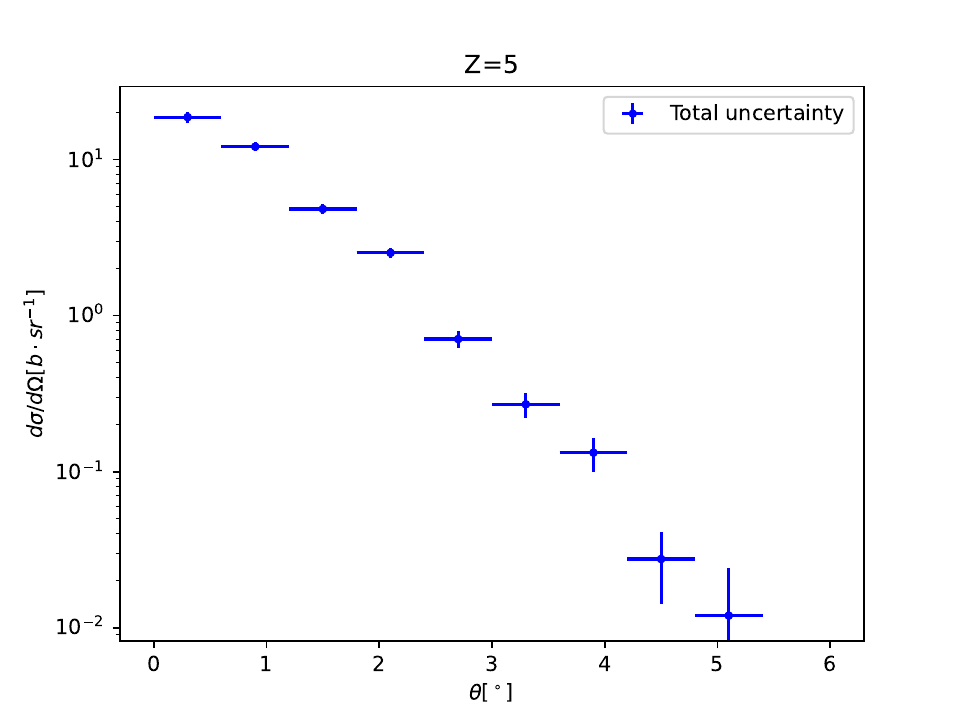} \quad
\includegraphics[width=6.5 cm]{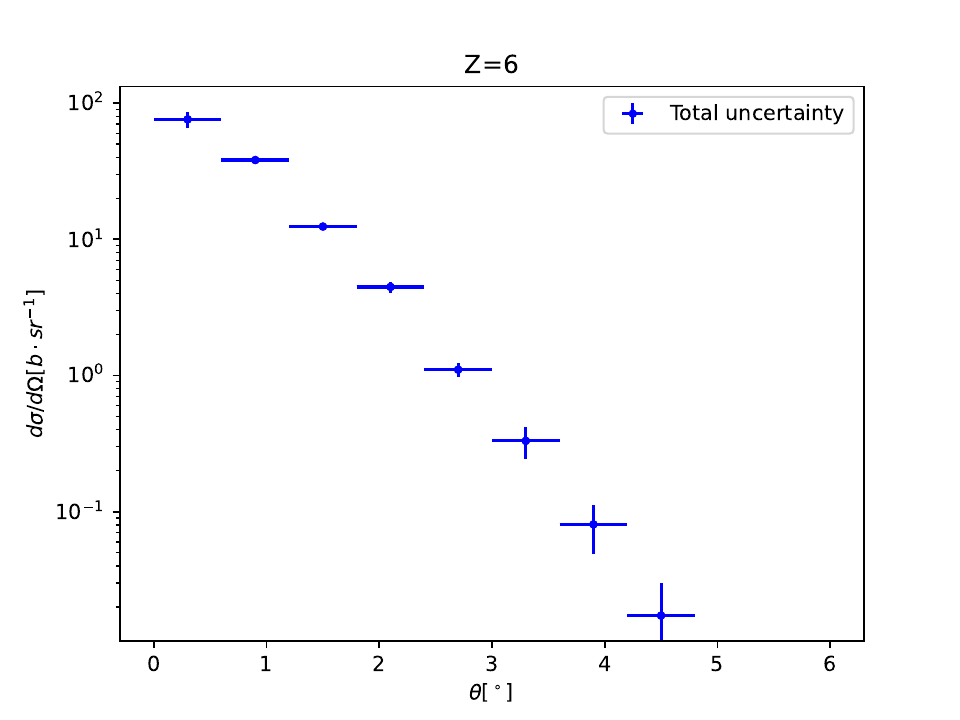} \quad  \includegraphics[width=6.5 cm]{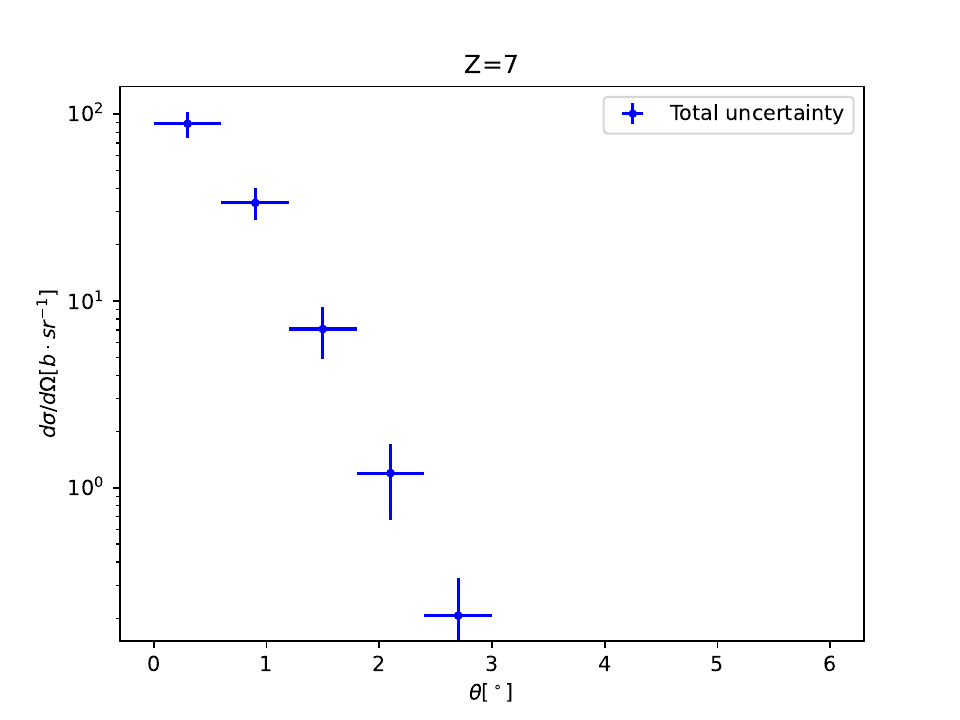} \quad
\caption{Angular differential cross sections for fragments $2 \le Z \le 7$ }
\label{fig: diiferential cros sec}
\end{center}
\end{figure}

The differential angular cross-sections are presented for all fragments with charges $Z\geq 2$ and $Z\le 7$ (see Figure \ref{fig: diiferential cros sec}). These results correspond to the application  to the experimental data of the analysis strategy described in the previous sections of this chapter for the cross sections calculation, providing a direct measurement of the angular distribution of the fragments produced in the target. The uncertainties shown in Fig~\ref{fig: diiferential cros sec} are the sum in quadrature between the statistical uncertainties and the systematic uncertainties extracted from the MC closure test, as described in the previous Sec.~\ref{sec:closureTest}.

The results obtained are preliminary for different reasons. Firstly, the detector efficiencies still need to be measured in the full energy range, as already discussed. 
The efficiency extrapolation on the whole energy range significantly impacts the cross section results obtained in the analysis. Their impact has been estimated from a MC closure test from which a systematic uncertainty has been extracted and assigned to the measured cross sections. In the future efficiency will be measured in a larger energy range, using the optimized thresholds found in the data analysis described in Chapter~\ref{chapter 4}. 

Moreover several factors have not yet been considered, such as the purity of the charge identification. In this analysis, the TW detector is the only one capable of identifying the charge of the fragments by utilizing the energy release $\Delta E$ in a TW bar and the ToF between the SC and TW detectors. However, the purity, which is linked to the performance of the charge identification algorithm in the TW detector, depends on the energy loss and ToF resolutions. To account for misidentification of charge, a purity correction can be applied, which can be calculated in MC simulations for each charge and angle.

Additionally, a study of systematic uncertainties related to both the detectors and the analysis method could be performed. Specifically, uncertainties associated with the charge reconstruction algorithm in the TW detector have not been analyzed. Moreover, the impact of cuts, such as the track-TW point distance being less than $4 cm$, and the robustness of the procedure for projecting VTX tracks onto the TW have not been fully investigated. These aspects will be explored in future work to enhance the quality and accuracy of the results.

\chapter*{Conclusions}
\addcontentsline{toc}{chapter}
{\numberline{}Conclusions} 
This thesis presents the work carried out within the FOOT experiment. The primary aim of the FOOT experiment is to measure double differential cross-sections with respect to kinetic energy and emission angle for nuclear fragmentation reactions relevant to Particle Therapy and Radiation Protection in Space. The results obtained from FOOT are intended to improve the accuracy of Treatment Planning Systems in Particle Therapy and contribute to the development of novel shielding materials for future long-term space missions. The experiment consists of two complementary setups: an Emulsion Spectrometer with a large angular acceptance, mainly focused on detecting light fragments ($Z \le 3$), and an Electronic Setup with a narrower acceptance, optimized for detecting heavier nuclei ($3 \le Z \le 8$).

The work described in this thesis focus on the VTX detector, a pixel silicon tracker essential for the high-precision reconstruction of fragmentation vertexes. The performance of the algorithm of this detector was evaluated through Monte Carlo simulations, which indicated its optimal behavior. The data collected by the detector in different FOOT data taking campaigns were analyzed. Studying the impact of the VTX sensors alignment and of the tuning of different parameters for tracking and vertexing algorithms, on the proportion of the reconstructed vertexes with respect to the number of primary particles, different problems and errors in detector geometry have been found and solved. From the study of the vertexes reconstructed, their topology, the fragment composition, the missing clusters and the efficiency measured for protons, a not fully efficient response of the VTX to the passage of light fragment has been understood.
Thank to this study a substantial time slot has been assigned to the measurement of the VTX efficiencies in the last data taking campaign at CNAO in november 2024. A large part of the work of this thesis has been done to plan the needed measurements and develop the analysis software to measure the efficiencies with different methods during the data taking itself, along with an optimization of the thresholds to ensure high efficiency while keeping noise under control. The study done also highlighted the worsening in performance of one of the 4 VTX layers and triggered the substitution of one of these layers in preparation for the CNAO 2024 data campaign.


This thesis also marks the first attempt to measure a cross-section using the VTX detector of FOOT, and the results obtained were highly promising. The preliminary differential fragmentation cross sections for the production of He, Li, Be, B, C and N in the interaction process of an oxygen beam of 400 MeV/u with a 5~mm graphite target were measured. The elemental fragmentation cross sections were measured as a function of the production angle measured by the VTX detector. The fragment charge Z was measured using the SC and the TW detectors.


Future efforts will be focused on mitigating the effects of the low efficiency in proton detection for the data collected prior to the optimization carried out in November 2024. To mitigate this issue, an algorithmic post-tracking method might be developed to recover protons that generated at least two clusters on the layers. Additionally, the integration of downstream detectors will be explored to ensure no protons, and most importantly, no valid vertexes are lost. Furthermore, a study of the particle angles emerging from the target could be conducted, measuring the kink of single tracks in the VTX detector. In cases where a vertex like $C\to B+p$ loses the proton, but the angle of particle $B$ can still be measured, a deviation in its angle from that of the primary particles could allow for the identification of such vertexes, even in the absence of the proton.

Finally only systematics coming from efficiency estimation have been considered. Other source of systematics, like systematics coming from TW calibration and Z identification of the fragments still need to be introduced in a final result.

\backmatter
\phantomsection

 \newpage
 \chapter*{Ringraziamenti}

Scrivere questa tesi è stato un viaggio pieno di grafici, calcoli, codici, addirittura viaggi e qualche momento di panico, ma non sarei mai arrivata alla fine senza il supporto di tante persone straordinarie.

Parto da te, \textbf{Marco}. Grazie. Grazie per avermi insegnato tantissimo e per avermi fatto sentire parte integrante di FOOT. Se mi sono divertita così tanto lavorando a questo progetto, se ho ritrovato la fiducia nelle mie capacità e soprattutto la passione per questo mondo, è merito tuo. Grazie anche per le mille chiacchierate sul presente, ma soprattutto sul futuro. 
Vorrei poi ringraziare tutto il team di \textbf{SBAI} per avermi accolto durante questo percorso. Grazie per avermi dato l’opportunità di entrare in contatto con il mondo della ricerca e con i tanti progetti che portate avanti. E, infine, grazie per il caffè, dose giornaliera di sopravvivenza!

Al di fuori del laboratorio, però, ci sono state tante altre persono che devo ringrazire.

Un grazie davvero speciale va alla mia famiglia, questo traguardo è anche vostro.

A \textbf{mamma} e \textbf{papà}, grazie per esserci sempre stati, per avermi incoraggiata a seguire i miei sogni e per i mille sacrifici che avete fatto per permettermi di essere qui. Grazie, mamma, per le tue parole che porto sempre con me: "Cadi, rialzati, salta, vola in alto, ma sempre con la testa alta e con il sorriso." È stato così. Qualche volta sono caduta, qualche volta le cose non sono andate come speravo, ma con il tuo aiuto sono sempre riuscita a rialzarmi. E grazie anche per ricordarmi di tutto, ormai sono grande, fingo di essere responsabile, eppure senza di te a volte dimenticherei persino la testa! Grazie, papà, perché nei giorni no, quelli in cui superare un esame sembrava impossibile, c’era sempre una tua chiamata. Dopo avermi preso un po’ in giro e fatto ridere, la tua proposta era sempre la stessa: "Non ti preoccupare, torna qui e impara a fare a maglia." Beh, papà, non penso che imparerò mai, ma continuerò ad aver bisogno delle tue chiamate, soprattutto nei momenti peggiori... e quando si rompe qualcosa!

A \textbf{Luca} e \textbf{Leonardo}. Non vivere più con voi non è stato facile. Le lacrime di Leo quando mi avete lasciata qui a Roma la prima volta e l’abbraccio di Luca quando sono tornata per la prima volta a casa sono ricordi che conserverò per sempre. Grazie per essermi sempre accanto, pronti a difendermi. Sono felice di avervi qui oggi e spero di avervi per sempre al mio fianco, anche quando cresceremo e prenderemo le nostre strade. Vi voglio bene e vi auguro il meglio. Sappiate che io ci sarò sempre per voi, per accompagnarvi nei vostri traguardi e sostenervi nelle vostre difficoltà, così come voi ci siete stati per me. La nostra famiglia è la mia forza, e voi due siete parte di tutto ciò che mi rende felice.

A \textbf{Francesco}, semplicemente grazie. Grazie per tutte le volte che volevo raccontarti quello che studiavo e la tua risposta era sempre: "Ma tanto non ne capisco nulla." Eppure, nonostante tutto, restavi lì ad ascoltarmi. Grazie per esserci stato nelle giornate no, quelle in cui ero certa di fallire, ma tu eri sicuro che ce l’avrei fatta, che l’esame per cui mi lamentavo da giorni, se non settimane, alla fine sarebbe andato bene... beh, avevi ragione. Grazie per farmi ridere e per riuscire sempre a farmi tornare il buon umore. Grazie per essere un porto sicuro, per essere la persona da chiamare nelle emergenze.  Forse al telefono non mi rispondi subito, ma se ho bisogno di qualcosa, non ti tiri mai indietro. Non è stato facile stare lontani, e non sono mancati litigi e momenti no, ma di certo non è mai mancato il tuo supporto. Sono certa che ci aspettano tante altre avventure, e sicuramente tanti altri, anzi, tantissimi altri litigi, ma alla fine ci auguro di ritrovarci sempre lì a ridere, scherzare e, soprattutto ad esserci l’uno per l’altro.

Grazie di cuore a tutta la mia famiglia, a zii, cugini, nonni e a chi, purtroppo, non c'è più ma che sento sempre vicino. Grazie per avermi insegnato cos’è la famiglia: un insieme di persone diverse, con caratteri e storie differenti, ma che, nonostante tutto, si accettano per quel che sono. La famiglia è essere presenti, esserci per festeggiare anche le cose più piccole, perché ciò che conta davvero è stare insieme. Ma, soprattutto, la famiglia è esserci nei momenti più difficili. Grazie per esserci stati per me, in ogni momento, con il vostro amore e supporto.

A casa Fiore, per avermi accolta e regalato mille avventure. Grazie per le due settimane senza acqua: ora ho davvero capito perché la chiamano "oro blu". Grazie per avermi insegnato cosa significa vivere senza lavatrice, lasciandoci senza per un mese intero. Grazie perché ora so come cambiare una lampadina, ho capito cosa sono POD e PDR nelle bollette, e so cosa succede quando si brucia la resistenza dello scaldabagno, spoiler: ti svegli senza corrente e rimani una settimana senza acqua calda. Ci sarebbe molto altro da raccontare, ma mi fermo qui, perché il grazie più grande va alle persone che hanno reso casa Fiore una vera casa. A \textbf{Teresa}, la mia prima coinquilina, alle nostre prime uscite a Roma e ai mille allagamenti di casa. Ad \textbf{Antonietta}, la coinquilina che non mi ha mai detto di no a un caffè o a qualsiasi cibo le proponessi. Nonostante focacce mai cresciute e diete a base di patate e formaggi, siamo sopravvissute. A \textbf{Silvia}, la coinquilina dei dolci, sempre pronta a preparare un dolcetto per festeggiare o semplicemente per farti una coccola. A \textbf{Giulia}, la coinquilina dei piccoli momenti che scaldano il cuore. Con te è sempre bastato poco per stare bene: una passeggiata, un film o una giornata in spiaggia con una piadina. A \textbf{Martina}, la coinquilina delle mille emozioni. Hai portato a casa non solo emozioni, ma anche il senso di libertà di esprimersi e fare ciò che si desidera, senza mai sentirsi giudicati. E infine, a \textbf{Marika}, la nuova arrivata, hai portato una boccata d'aria fresca in casa. Esci, rientri, studi, ti alleni, sempre con un buon umore contagioso. Un grazie speciale a tutte voi, per avermi sopportata e supportata tra lezioni, sessioni infinite e tesi. Grazie perché è sempre bello tornare a casa e trovare persone con cui scherzare e sopratutto chiacchierare.

Alle \textbf{mie amiche}, grazie. Grazie per esserci sempre state. Non è stato facile affrontare la distanza e non avervi più nella mia quotidianità. Non è stato facile smettere di vedervi e parlare con voi ogni giorno, eppure voi continuate ad esserci. È strano quanto poco riusciamo a sentirci o vederci ormai, ognuna nella propria città, con i propri impegni e orari. Eppure, con voi basta poco: un caffè, uno sguardo, e mi capite alla perfezione. Basta un messaggio per trovare le risposte che cercavo. Vi ringrazio per essere sempre al mio fianco, per il vostro supporto e per aver sempre creduto in me. Spero che resteremo così per sempre, continuando a crescere insieme e a raggiungere i nostri traguardi, proprio come abbiamo sempre fatto. Abbiamo condiviso tutto: 18 anni, il diploma in musical, quello delle superiori, ed eccoci qui, ognuna sulla propria strada, ma sempre unite. Vi auguro il meglio, e sono immensamente grata di avervi nella mia vita.

Infine, ma non per importanza, grazie alle persone che hanno condiviso con me questo percorso: Andrea, Camilla, Danilo, Davide, Elisa, Irene, Leonardo, Lorenzo, Marco, Niccolò (Nicco), Ottavia, Saverio e Susanna. Dopo tanti laboratori, esercizi risolti insieme e argomenti confusi a lezione che abbiamo cercato di decifrare, eccoci qui: ce l'abbiamo fatta! Grazie perché avete reso l'università più divertente, migliorando anche i pranzi in mensa, ma soprattutto quelli sul pratone. Vi auguro un grandissimo successo in tutto ciò che deciderete di fare, ma soprattutto aspetto aggiornamenti sulle incredibili scoperte che sono certa farete. Grazie per aver condiviso con me la vostra passione e dedizione al mondo della fisica. Vi sono davvero grata per tutto.

\textbf{A me stessa.} Più che un ringraziamento, voglio augurarmi di non perdere mai la passione per questo mondo, la curiosità di capire le cose e la dedizione nel portare avanti gli studi, le analisi o qualsiasi sfida fino ad arrivare a una risposta, una soluzione o almeno a comprendere il problema. Nei momenti no voglio ricordarmi che i traguardi raggiunti finora dimostrano che posso farcela, e che a volte basta solo avere un po’ più di fiducia in me stessa.  E soprattutto, mi auguro di non dimenticare mai di celebrare ogni piccolo successo, di sorridere per ogni conquista e di godermi il viaggio, perché è lì che si nasconde la vera bellezza.
\end{document}